\documentclass[preprint,12pt]{elsarticle}

\usepackage{amssymb}
\usepackage{amsmath}

\usepackage[english]{babel} 

\usepackage{array}

\usepackage[table]{xcolor}  

\usepackage{multirow} 

\usepackage{url} 

\journal{International Journal of Human-Computer Studies}

\begin{document}

\begin{frontmatter}



\title{What Should Explanations Contain?\\A Human-Centered Explanation Content Model for Local, Post-Hoc Explanations}
\author{Helmut Degen\corref{cor1}\fnref{label2}}
\ead{helmut.degen@siemens.com}
\affiliation{organization={Siemens Research \& Predevelopment\\
		755 College Road East},
			city={Princeton},
			postcode={08540},
			state={NJ},
			country={USA}}





\begin{abstract}
Which categories of explanation content are relevant for users of industrial AI systems, and how can those categories be organized for local, post-hoc explanations? To address these questions, a hybrid inductive-deductive qualitative content analysis was applied to 325 meaning units drawn from six user studies in building technology, manufacturing, AI software development, and hospital cybersecurity. The inductive phase produced an initial twelve-code structure. A theory-informed coverage assessment and expert review then added two further codes, Rule base and What-if backward, that were not instantiated in the corpus but correspond to system architectures documented in the XAI literature. The resulting fourteen-code model is organized into four groups: rule-based, causal, epistemic (actual), and epistemic (similar), with twelve codes grounded in the corpus and two as theoretical extensions. An eleven-member expert panel supported the content adequacy of all codes (I-CVI $\geq$ 0.82; scale-level agreement of 0.93 for relevance, 0.92 for boundary clarity, and 0.94 for understandability). A stratified subsample of 82 units (25\% of the corpus), coded independently by two researchers using the finalized codebook, yielded Krippendorff's $\alpha = 0.920$ and Cohen's $\kappa = 0.920$. The paper therefore establishes content adequacy and coding reproducibility for a content-level explanation model intended to support elicitation, specification, and later evaluation of explanation content in industrial AI systems. Behavioral validation of downstream effects remains future work.
\end{abstract}
		
\begin{graphicalabstract}
\end{graphicalabstract}
		
\begin{highlights}

	\item A hybrid inductive-deductive qualitative content analysis of 325 meaning units from six industrial studies produced a fourteen-code explanation-content model organized into four groups: rule-based, causal, epistemic (actual), and epistemic (similar).

	\item Twelve codes are empirically grounded; two codes (Rule base, What-if backward) are theoretically motivated extensions for system architectures not represented in the corpus.

	\item An eleven-member expert panel supported the adequacy of all codes (item-level content validity index, I-CVI $\geq$ 0.82), with excellent scale-level averaged agreement (S-CVI/Ave) for relevance (0.93), boundary clarity (0.92), and understandability (0.94).

	\item Reliability testing on a 25\% stratified subsample yielded Krippendorff's $\alpha$ = 0.920 and Cohen's $\kappa$ = 0.920, substantially exceeding the 0.800 threshold.

	\item The content-level granularity provides a structured basis for future theory-driven and behavioral XAI research on how different types of explanation content relate to user outcomes.
	
\end{highlights}
		
\end{frontmatter}
	




\section{Introduction}
\label{sec:Introduction}

Artificial intelligence (AI) systems\footnote{In this paper, AI systems refer to systems built using approaches including rule-based systems, machine learning, deep learning, and large language models.} are now used in many industrial and professional settings to support decisions with operational, financial, safety, and regulatory consequences. In this paper, industrial AI systems refer to AI systems used in organizational work settings to support or automate decisions, analyses, or actions with such consequences. In manufacturing and infrastructure management, AI supports anomaly detection, predictive maintenance, and process optimization at scales and speeds that exceed manual monitoring \cite{Cachada2019-XAI-AIIndustrialMaintenance-IEEE}. In cybersecurity, it supports threat assessment and prioritization \cite{Apruzzese2023-XAI-AIRealWorldCybersecurity-IEEE}. In clinical and financial decision support, it is used to assist diagnosis, triage, and risk evaluation \cite{McKinney2020-XAI-AIBreastCancerScreening-SpringerNature,Jain2023-XAI-StockMarketPredictions-MDPI}. These settings share a common property: human practitioners remain accountable for outcomes and therefore need more than a raw prediction or recommendation. They need information that helps them understand, assess, and, where necessary, contest an individual AI output \cite{Miller2019-XAI-ExplanationAI-Elsevier,Ploug2020-XAI-ContestableAIDiagnostics-Elsevier,Alfrink2023-XAI-ContestableAIByDesign-Springer,Yurrita2023-XAI-ContestabilityFairness-ACM}.
		
Even highly capable AI systems can produce outputs that are wrong or misleading in particular cases despite strong aggregate performance. In real work settings, such case-level failures can have operational, financial, and safety consequences. Explanations are widely considered an important means of supporting trustworthy AI use and appropriate human oversight \cite{Guidotti2018-XAI-SurveyBlackBox-ACM,Miller2019-XAI-ExplanationAI-Elsevier,VanderWaa2021-XAI-RuleExampleComparison-Elsevier}. However, not every explanation is equally useful, and explanation usefulness depends on the user, the task, and the deployment context \cite{Miller2019-XAI-ExplanationAI-Elsevier,VanderWaa2021-XAI-RuleExampleComparison-Elsevier}. This motivates a more specific question than whether explanations help in general: what should explanations actually contain?

Much of the explainable AI literature on case-level decision support centers on explanation types, including causal, contrastive, counterfactual, epistemic, example-based, and rule-based explanations, and on their effects on understanding, trust, and decision performance \cite{Guidotti2018-XAI-SurveyBlackBox-ACM,Miller2019-XAI-ExplanationAI-Elsevier,VanderWaa2021-XAI-RuleExampleComparison-Elsevier,Speith2022-XAI-TaxonomiesXAIMethods-ACM}. Explanation types characterize broad reasoning styles but do not specify which concrete information elements should appear in a system interface. This finer-grained layer matters for design and implementation. A designer needs to know not only that a system may provide a causal or epistemic explanation, but which elements users require: contextual information, input values, causal factors, mechanism descriptions, uncertainty qualifiers, similar cases, rule references, or projected future states.

This paper addresses the missing layer by treating explanation content as the primary design unit and proposing a human-centered explanation-content model for local, post-hoc explanations. Explanations here denote justifications that enable end users\footnote{The terms \emph{end user} and \emph{user} are used interchangeably throughout this paper to denote all people who use a deployed AI system to accomplish their work. End users include both non-technical domain practitioners (such as facility managers, hospital cybersecurity leads, clinicians, or financial analysts) and technical professionals (such as data scientists, software engineers, or system testers) when they use a deployed AI system in their work, as distinct from when they design, train, or maintain such systems.} in deciding whether to accept, modify, or reject the outcome of an AI system for a specific case, in order to achieve their specific goals. Explanation content, as used here, denotes the discrete information elements that together compose an explanation and that can be specified, reviewed, implemented, and evaluated independently. Local explanations concern individual AI system outputs for specific cases, not general model behavior. Post-hoc explanations are generated after a system output has been produced, as opposed to being built into the model architecture \cite{Guidotti2018-XAI-SurveyBlackBox-ACM,Miller2019-XAI-ExplanationAI-Elsevier}. This scope fits many industrial settings, where users receive a recommendation, alert, ranking, or diagnosis for a particular case and must decide whether to accept, modify, or reject it.

The model is called human-centered for two reasons. First, it is intended to support the design and evaluation of user interfaces with explanation content in relation to user tasks and reasoning demands. Second, the source material to identify and categorize explanation content is elicited from representatives of end users in six industrial studies. This type of source material is uncommon in XAI research. XAI research frequently derives explanations from explanation types, theoretical taxonomies, researcher-defined categories, or model-accessible explanation content. A further methodological gap concerns the population from which explanation content is elicited or evaluated: most XAI user studies are human-grounded rather than application-grounded \cite{DoshiVelez2017-XAI-RigorousInterpretability-ICML,Vilone2021-XAI-NotionsExplainabilityEvaluation-Elsevier,Nauta2023-XAI-AnecdotalEvidenceQuantitative-ACM}, relying on crowdsourced or convenience samples performing simplified tasks rather than representatives of target users performing realistic tasks in their deployment context. A secondary analysis of the 73 publications reviewed by Kim et al. \cite{Kim2024-HumanCenteredEvaluationXAI-Frontiers}, reported in~\ref{app:Analysis}, found end-user elicitation of explanation content fully in only three studies and partially in three further studies. Only a single study \cite{Anjara2023-ExplainableCDSSThinkAloud-PLOSONE} combined an explicitly defined context of use, defined user role, defined user goals and tasks, and end-user elicitation. The present study uses explanation content elicited from target users across six studies as the empirical basis for the model.

The paper addresses two research questions:
\begin{itemize}

	\item RQ1: What categories of explanation content are needed for local, post-hoc explanations of AI systems in industrial settings?

	\item RQ2: How can explanation content be organized in a structure compatible with established forms of human reasoning and aligned with user task demands?

\end{itemize}

To answer these questions, a hybrid inductive-deductive qualitative content analysis was applied to 325 meaning units from six industrial user studies across building technology, manufacturing, AI software development, and hospital cybersecurity. The inductive phase produced a twelve-code structure grounded in the corpus. A theory-informed coverage assessment and expert review then added two further codes for system architectures documented in the XAI literature but not represented in the corpus, resulting in a fourteen-code model. The model was evaluated by an eleven-member expert panel and by a two-coder reliability study on a stratified subsample of the corpus.

The paper contributes a reusable explanation-content model for local, post-hoc explanations in industrial AI settings. The model is derived from heterogeneous explanation artifacts instead of from theory alone. Initial evidence of content adequacy and coding reproducibility is reported for the resulting model. The model provides a content-level unit of analysis between broad explanation types and concrete user interface instances. This intermediate unit is useful for HCI and XAI research because it allows theory-driven and behavior-driven studies to manipulate, compare, and measure explanation content systematically. It remains to be shown in practice that user interfaces derived from the model improve outcomes. Throughout, three terms are used consistently: explanation-content model (the conceptual contribution), codebook (its operational specification for coding), and blueprint (its use in design and evaluation practice).

The remainder of the paper is structured as follows. Section~\ref{sec:RelatedWork} reviews related work and positions the paper within existing XAI research. Section~\ref{sec:Methodology} describes the methodological approach, including corpus construction, coding, expert review, and reliability testing. Section~\ref{sec:Results} reports the resulting explanation-content model and the associated content-adequacy and reliability findings. Section~\ref{sec:Discussion} interprets these findings and discusses their implications for design and research. Finally, Sections~\ref{sec:Limitations}, \ref{sec:FutureWork}, and \ref{sec:Conclusion} address the study limitations, outline future work, and summarize the contribution. The appendix provides the study's supporting materials and traceability record, including a secondary analysis of human-centered XAI publications, an overview of the six user studies, detailed stepwise results, domain-of-origin and explanation-type definitions, draft and revised codebooks, change logs, and preregistration-related material.


\section{Related work}
\label{sec:RelatedWork}

The review is organized around three criteria that follow from the aim of producing a reusable explanation-content model: whether prior work involved end users in the creation of explanations, whether it identified reusable explanation-content categories, and whether it organized such categories into a reusable structure. Five bodies of work are covered in turn: XAI evaluation frameworks, explanation-need elicitation, participatory and co-design approaches to XAI, methodological work on content analysis, and adjacent content-oriented representations.


\subsection{XAI evaluation frameworks}
\label{sec:RelatedWork:XAIEvaluationFrameworks}

A substantial body of XAI work addresses how to evaluate explanations and what properties good explanations should have. Doshi-Velez and Kim distinguish application-grounded, human-grounded, and functionally grounded evaluation, clarifying that explanation quality cannot be judged at a single level \cite{DoshiVelez2017-XAI-RigorousInterpretability-ICML}. Hoffman et al. proposed metrics centered on explanation goodness, satisfaction, and trust calibration \cite{Hoffman2018-XAI-MetricsExplainableAI-arXiv}. Mohseni et al. surveyed human-AI interaction evaluation and organized relevant properties across user, model, and system dimensions \cite{Mohseni2021-XAI-MultidisciplinaryHAI-ACM}. Speith categorized explanation types which can be used for evaluation \cite{Speith2022-XAI-TaxonomiesXAIMethods-ACM}. These contributions are foundational for evaluation design but largely agnostic about what explanation content should be present; they clarify how explanations may be evaluatednd leave open what should be evaluated.

Empirical work reaches a similar limit. Studies in clinical, public-sector, and financial settings show that explanation effectiveness depends on the fit between explanation form and decision context, and that some forms fail to improve understanding or appropriate reliance \cite{Naiseh2023-ExplanationClassesTrustCalibration-IJHCS,Maltbie2021-XAIToolsPublicSector-FSE,Bertrand2023-FeatureBasedExplanationsFinance-FAccT}. This work establishes that explanation content must be elicited and specified in relation to human use, but it typically evaluates whole explanation types instead of deriving a reusable set of content elements.


\subsection{Explanation-need elicitation}
\label{sec:RelatedWork:ExplanationNeedElicitation}

A second body of work examines how explanation needs can be elicited from practitioners. Liao et al. developed a question bank capturing recurring explanation needs from practitioners, many of which reflect the concerns of technical stakeholders \cite{Liao2020-XAI-QuestioningAI-CHI-ACM}, and subsequently proposed a question-driven design process linking user questions to explanation strategies across development stages \cite{Liao2021-XAI-QuestionDrivenDesign-arXiv}. Chromik and Butz offer a human-XAI interaction taxonomy for describing explanation interactions and request types \cite{Chromik2021-XAI-HumanXAIInteractionTaxonomy-IUI}. These works move the elicitation and design of explanation content closer to user needs but remain focused on questions, requests, or explanation forms, not on the information elements that address the explanation needs of end users and that should appear in a user interface.

Empirical studies reinforce this point. Cai et al. showed that clinicians' explanation needs are task-specific and shift across stages of a decision process \cite{Cai2019-XAI-HumanCenteredToolsClinic-ACM}. Anjara et al. identified themes in how oncologists assess explainable clinical decision support systems \cite{Anjara2023-ExplainableCDSSThinkAloud-PLOSONE}. Explanation needs are contextual and domain practitioners articulate them concretely, but the outputs of such studies are theme sets or domain-specific insights, not a validated reusable content structure.

A secondary analysis of the 73 human-centered XAI publications reviewed by Kim et al. \cite{Kim2024-HumanCenteredEvaluationXAI-Frontiers} coded each publication on four criteria: whether the context of use was defined, whether the user role was defined, whether user goals and tasks were defined, and whether explanation content was elicited directly from end users. The four criteria were seldom jointly satisfied. Explanation content was elicited from end users fully in only three studies \cite{Nazaretsky2022-TeachersAI-LAK,Anjara2023-ExplainableCDSSThinkAloud-PLOSONE,Khodabandehloo2021-HealthXAI-FGCS} and partially in three further studies \cite{Selten2023-JustLikeIThought-PAR,Conati2021-PersonalizedXAI-ITS-AI,Maltbie2021-XAIToolsPublicSector-FSE}. Only a single study \cite{Anjara2023-ExplainableCDSSThinkAloud-PLOSONE} combined all four conditions, and the majority of publications satisfied none or only one. The full analysis is reported in~\ref{app:Analysis}. None of the studies proposes an explanation content model.


\subsection{Participatory and co-design approaches to XAI}
\label{sec:RelatedWork:ParticipatoryApproaches}

Participatory and co-design research places end users at the center of elicitation and design. Jin et al. proposed the EUCA framework, using co-design to help non-expert users articulate what they want to know about AI outcomes \cite{Jin2021-XAI-EUCA-arXiv}. Panigutti et al. reported a participatory process for clinical decision support that aligned explanation presentation with clinicians' workflow constraints and decision responsibilities \cite{Panigutti2023-XAI-CoDesignClinicalXAI-ACM}. Nazaretsky et al. showed that co-designed explainable learning analytics can improve domain-specific outcomes \cite{Nazaretsky2022-TeachersAI-LAK}. The primary outputs of this work are forms, schemes, or domain-specific design artifacts. They don't propose a explanation content model that can be applied across application domains.


\subsection{Content analysis in explanation research}
\label{sec:RelatedWork:ContentAnalysis}

Qualitative content analysis is widely used in HCI and related fields to derive structured coding systems from heterogeneous materials \cite{Mayring2014-XAI-QualContentAnalysis-Klagenfurt,Schreier2012-XAI-QualContentPractice-Sage,Saldana2013-XAI-CodingManual-Sage}. Hybrid inductive-deductive variants are useful when the phenomenon is not sufficiently understood for a fully deductive approach, but where prior theory can serve as a sensitizing lens and later coverage check \cite{Mayring2014-XAI-QualContentAnalysis-Klagenfurt}. Despite the availability of these methods, empirical elicitation of explanation content from users remains relatively uncommon in XAI research. The most influential contribution is the XAI Question Bank by Liao et al. \cite{Liao2020-XAI-QuestioningAI-CHI-ACM}, who interviewed UX practitioners to identify user needs as prototypical questions organized into ten categories. Sipos et al. \cite{Sipos2023-ExplanationNeeds-MuC} applied and extended this bank through think-aloud sessions with subject matter experts, surfacing gaps between developer-facing taxonomies and actual end-user needs. Syed et al. \cite{Syed2023-NotionXAI-arxiv} used focus group interviews to identify user-preferred explanation content. Chromik et al. \cite{Chromik2021-HumanXAI-IJHCS} elicited explanation requirements from clinical users to derive design patterns for explanation interfaces. Across these studies, qualitative methods have identified explanation content needs that technically derived taxonomies tend to overlook. To the best of the author's knowledge, there is no research that applied a qualitative content analysis to derive an explanation content model. 


\subsection{Adjacent content-oriented representations}
\label{sec:RelatedWork:Adjacent}

A further adjacent strand concerns content-oriented representations developed outside mainstream XAI evaluation and elicitation research. In the semantic web and knowledge-representation community, explanation ontologies formalize explanations, user needs, system attributes, and explanation types. The Explanation Ontology, for instance, offers a general-purpose representation intended to help system designers connect user-centered explanation requirements to system capabilities and literature-derived explanation types \cite{Chari2024-XAI-ExplanationOntology-SemanticWeb,Chari2020-XAI-ExplanationOntology-Springer}. In educational AI and tutoring research, structured taxonomies capture tutoring moves, feedback strategies, and instructional actions, representing how tutoring systems intervene during learning interactions \cite{Zhou2026-XAI-TutorMoveTaxonomy-arXiv,Khosravi2022-XAI-XAIED-ComputersEducationAI}. 
Both strands demonstrate that structured representations of explanation-related content and action are feasible and useful, but neither is grounded in explanation content elicited from end users, nor intended to specify what information should appear in a user-facing explanation interface.


\subsection{Summary of the gap}
\label{sec:RelatedWork:GapSummary}

Prior work provides important building blocks but leaves two gaps relative to the aims of this paper. First, despite semantic ontologies and pedagogical taxonomies from adjacent fields, no reusable empirically derived explanation-content model exists for local, post-hoc explanations in industrial AI settings that specifies which explanation elements should appear in user interfaces. Second, no such model is grounded in end-user-elicited artifacts and evaluated through explicit codebook development, expert review, and reproducibility testing.


\section{Methodology}
\label{sec:Methodology}


\subsection{Study aim and methodological rationale}
\label{sec:Methodology:StudyAim}

The study aimed to produce a reusable explanation-content model for identifying, specifying, and reviewing the explanation content needed for industrial AI systems. The target outcome was a structured representation of explanation-content elements suitable as a design and evaluation reference across systems. This distinguishes the goal from describing how participants make sense of explanations or developing a substantive theory of explanatory behavior. The source material comprised explanation content elicited in six user studies with representatives of industrial target user groups \cite{Degen2021-inproceedings,Degen2022-inproceedings,Degen2023-inproceedings,Degen2024-inproceedings,Degen2025-XAI-ExplainDataScientists-Springer-InProceedings}\footnote{The content validation study for Study~6 (importance ratings for user groups) is ongoing and will be reported separately; it does not affect the meaning-unit extraction or coding reported here.}. This material contained concrete explanation-content expectations and artifacts that could be analyzed to produce a reusable content structure. Because the material was elicited with end-user representatives, the resulting abstraction reflects end-user expectations and reasoning needs, not researcher-defined categories or model-accessible content alone.

A hybrid inductive-deductive qualitative content analysis was selected because it supports the transformation of heterogeneous artifacts into a stable set of categories and additionally permits a structured theoretical coverage check \cite{Mayring2014-XAI-QualContentAnalysis-Klagenfurt,Schreier2012-XAI-QualContentPractice-Sage,Saldana2013-XAI-CodingManual-Sage}. The inductive phase derived categories from the corpus without a pre-existing coding frame. The deductive phase used established explanation types, specifically rule-based, causal, contrastive, counterfactual, epistemic, and example-based explanations, as a coverage lens for assessing whether the emerging structure omitted theoretically important classes of explanation content \cite{Byrne2019-XAI-CounterfactualsXAI-IJCAI,Guidotti2018-XAI-SurveyBlackBox-ACM,Miller2019-XAI-ExplanationAI-Elsevier}.

The inductive phase was researcher-led. The source material consisted primarily of user interfaces and diagrammatic mental models. Qualitative content analysis provides explicit support for segmenting such material into meaning units and abstracting those units into a category system, and it aligns with the later need for explicit code definitions, examples, inclusion criteria, and reliability testing. The study therefore prioritized transparent codebook construction, external expert review, and independent reproducibility testing over parallel multi-coder category generation.


\subsection{Why content analysis was preferred to alternative approaches}
\label{sec:Methodology:ContentAnalysis}

Two alternatives were considered. Grounded theory was rejected because the study did not aim to generate a substantive theory of social process through iterative theoretical sampling and constant comparison \cite{Charmaz2014-XAI-ConstructingGT-Sage}. The corpus was fixed, multimodal, and assembled from prior studies, not generated through sampling decisions guided by emergent theory. Much of the material consisted of structured artifacts, including user interfaces and diagrammatic mental models, as opposed to extended textual conversations or narrative interview accounts. The analytic task was therefore segmentation, abstraction, and stabilization of explanation-content categories, not theory generation.

Thematic analysis was rejected because the corpus was not narrative in character and because the intended outcome was a structured codebook capable of supporting later specification and review \cite{Braun2006-XAI-UsingTA-TQR,Braun2019-XAI-ReflexiveTA-CounselPsych}. Qualitative content analysis was better aligned with that goal because it emphasizes explicit segmentation, category definition, codebook construction, and reproducibility testing \cite{Mayring2014-XAI-QualContentAnalysis-Klagenfurt,Schreier2012-XAI-QualContentPractice-Sage}.

The methodological choice orients the resulting categories toward design and evaluation practice, not toward understanding users' subjective experience of AI explanations. This produces explicit, reusable content units suited to empirical, comparative, and implementation-focused work.


\subsection{Corpus and analytical scope}
\label{sec:Methodology:Corpus}

The corpus comprises six previously conducted user studies from four industrial domains: building technology, manufacturing, AI software development, and hospital cybersecurity vulnerability management (see~\ref{app:Userstudies}). Explanation content was elicited with representatives of target user groups in each study \cite{Degen2021-inproceedings,Degen2022-inproceedings,Degen2023-inproceedings,Degen2024-inproceedings,Degen2025-XAI-ExplainDataScientists-Springer-InProceedings}. The six studies represent the full set of prior project studies that yielded externalized explanation-content material suitable for secondary analysis, and together they contain 325 meaning units.

The studies vary along dimensions relevant to the intended scope of the model. User roles include facility managers, energy engineers, model managers, system testers, data scientists, and hospital IT and cybersecurity experts. Lifecycle phases covered include development, operation, and maintenance. Elicitation formats differ: user-interface representations were used in Studies 1, 2, 3, and 5, and diagrammatic mental models in Studies 3, 4, 5, and 6. This variation was retained deliberately. The aim was broad coverage of explanation situations likely to arise in industrial AI settings, not statistical representativeness \cite{Mayring2014-XAI-QualContentAnalysis-Klagenfurt,Schreier2012-XAI-QualContentPractice-Sage}. Table~\ref{tab:UserStudies:ExplanationContentVariations} in~\ref{app:Userstudies} provides an overview about the six user studies.

Study 6 was added after the coverage assessment identified rule-related explanation content as theoretically important but absent from Studies 1 to 5. Three aspects of this decision warrant discussion. First, the inclusion was transparent and documented: the assessment produced a concrete gap finding, and a study was identified to address it. Study 6 differs methodologically from Studies 1 to 5. Explanation content was elicited from three domain experts (two cybersecurity experts and one hospital IT expert), and the 14 meaning units coded for this paper were derived from that content. A UX mockup based on the same content is being evaluated through a usability test and a content-relevance survey, both reported separately and not used to identify or refine the coded explanation content. The inclusion does not create circularity in the empirical claim. The 14 meaning units from Study 6 were assigned to Applied rule (PoRC 1.2) using the same segmentation and coding rules applied throughout; zero units were assigned to Rule base (PoRC 1.1), which remains a theoretically motivated extension without corpus assignments. Study 6 therefore strengthens empirical support for the rule-based group without introducing it artificially; the Rule base code is flagged as theoretically motivated in all relevant tables and figures.

The target scope is local, post-hoc explanation content. Global explanations, purely model-internal interpretability diagnostics, and broader governance documentation fall outside scope unless they appear as user-facing explanation content for a specific case. This reflects the practical decision situations behind the corpus studies, in which users received a particular output, recommendation, alert, or ranking and needed content to assess and act on it \cite{Guidotti2018-XAI-SurveyBlackBox-ACM,Miller2019-XAI-ExplanationAI-Elsevier}.


\subsection{Meaning-unit definition and coding logic}
\label{sec:Methodology:MeaningUnitDefinition}

All source materials were segmented into meaning units before coding. A meaning unit is the smallest independently interpretable element that expresses explanatory meaning yet preserves sufficient context for consistent coding \cite{Mayring2014-XAI-QualContentAnalysis-Klagenfurt,Schreier2012-XAI-QualContentPractice-Sage}. In user interfaces and mockups this typically meant one group of elements, label, indicator, chart annotation, or rule statement. In diagrammatic mental models it typically meant one node or one explicitly labeled relation.

Each unit received one primary explanation-content code along with two parallel annotations. The domain-of-origin annotation indicated whether the content belonged to the AI domain, the surrounding system domain, the application domain, or mixed domains, and was included to help interpret what information sources an implemented system would require to generate it. The explanation-type annotation linked each unit to established explanation classes, including rule-based, causal, counterfactual, epistemic, contrastive, and example-based explanations \cite{Byrne2019-XAI-CounterfactualsXAI-IJCAI,Guidotti2018-XAI-SurveyBlackBox-ACM,Miller2019-XAI-ExplanationAI-Elsevier}. Because content codes and explanation types operate at different levels of abstraction, mappings were not assumed to be one-to-one.

A small number of units required special treatment at one boundary. Some units described both an actual input value and a hypothetical variation of it, making them defensibly codable as either Input or What-if forward. The equivalence was defined before reliability coding began and later treated as agreement in the reliability computation.


\subsection{Analytic procedure}
\label{sec:Methodology:AnaylticProcedure}

\begin{figure}[htbp]
	\centering
	\includegraphics[width=1.0\linewidth]{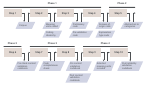}
	\caption{Analytic procedure}
	\label{fig:Method_Flow}
\end{figure}

The analysis proceeded in four phases (see Figure~\ref{fig:Method_Flow}) \cite{Mayring2014-XAI-QualContentAnalysis-Klagenfurt,Schreier2012-XAI-QualContentPractice-Sage,Saldana2013-XAI-CodingManual-Sage}. Phase 1 (Steps 1 through 4) segmented and open-coded the corpus. Meaning units were reviewed repeatedly and assigned short descriptive labels that captured their functional meaning, treated as preliminary codes \cite{Saldana2013-XAI-CodingManual-Sage}. Semantically equivalent labels were consolidated into a normalized set of initial content codes. This reduced superficial wording variation without losing distinctions relevant for later inclusion and exclusion decisions.

Phase 2 (Steps 5 and 6) abstracted the normalized initial codes into higher-level categories. Abstraction was based on functional similarity in terms of what each code contributes to the elicitation, specification, and evaluation of explanation content, not on similarity of wording \cite{Mayring2014-XAI-QualContentAnalysis-Klagenfurt,Schreier2012-XAI-QualContentPractice-Sage}. This phase produced the four-group structure: rule-based content, causal content, epistemic content for the actual case, and epistemic content for similar cases. A draft codebook was developed alongside the structure, specifying labels, definitions, inclusion and exclusion criteria, typical evidence indicators, and examples (see~\ref{app:DraftCodebook}) \cite{Schreier2012-XAI-QualContentPractice-Sage}.

Phase 3 (Steps 7 and 8) assessed the draft structure against established theory. Established explanation types served as sensitizing concepts during interpretation and as a lens for assessing completeness \cite{Blumer1954-XAI-WrongSocialTheory-ASR,Mayring2014-XAI-QualContentAnalysis-Klagenfurt}. The assessment showed that rule-related content, though theoretically important, was absent from the original studies; Study 6 was added for this reason. The assessment also clarified that the initial What-if code spanned both causal and counterfactual reasoning, which later motivated the distinction between What-if forward and What-if backward \cite{Byrne2019-XAI-CounterfactualsXAI-IJCAI}. The structured output of the coverage assessment, a mapping of the final codes to established explanation types, is reported in Section~\ref{sec:Results:TheoryInformedCoverageAssessment}. A revised draft (the \textit{Pre content validation codebook}) then entered expert review.

Phase 4 (Steps 9 and 10) comprised expert review followed by reproducibility testing. The expert panel evaluated all codes for relevance, boundary clarity, and understandability; their qualitative comments were consolidated into revision themes and led to refinements of labels, definitions, examples, and decision rules \cite{Lynn1986-XAI-DetermQuantContentValidity-LWW}. The most consequential changes were the splits of one rule-related code into Rule base and Applied rule and of the initial What-if code into What-if forward and What-if backward.

The resulting post-expert-review codebook (\textit{Post content validation codebook}, see~\ref{app:PostExpertReviewCodebook}) was frozen and used in a reliability study with two independent coders applying it to a stratified subsample. Disagreements were analyzed, and a limited set of additional decision rules were incorporated into the post-reliability codebook, which is the final codebook reported in this paper (see~\ref{app:PostReliabilityCodebook}). The reliability procedure for the frozen post-expert-review codebook was preregistered before independent coding began; details are reported in Section~\ref{sec:Methodology:ReliabilityEvaluation} and in~\ref{app:Preregistration}.


\subsection{Expert review}
\label{sec:Methodology:ExpertReview}

An eleven-member expert panel reviewed the codebook. The panel was recruited to cover AI or XAI expertise alongside application-domain expertise relevant to the corpus, reflecting the settings in which the model is intended to support elicitation, specification, and review of explanation content. Experts assessed each code on three four-point scales for relevance, boundary clarity, and understandability; ratings of 3 or 4 counted as acceptable for item-level agreement \cite{Lynn1986-XAI-DetermQuantContentValidity-LWW,Polit2006-XAI-ContentValidityIndex-Wiley}.
Item-level content-adequacy indices were computed for each code and criterion as the proportion of experts assigning an acceptable rating, with scale-level averages computed across codes. With eleven experts, the predefined adequacy threshold required at least nine acceptable ratings, corresponding to I-CVI $\geq$ 0.82 \cite{Polit2006-XAI-ContentValidityIndex-Wiley}. Open-text comments explaining low ratings, noting ambiguity, suggesting improved examples, or identifying potentially missing distinctions were also elicited and consolidated into concrete codebook revisions.

The expert panel served as a structured codebook refinement step, not only a confirmatory check \cite{Schreier2012-XAI-QualContentPractice-Sage,Lynn1986-XAI-DetermQuantContentValidity-LWW}. Experts reviewed whether each code captured content meaningful for explanation work, whether category boundaries were sufficiently clear for consistent use, and whether definitions and examples were understandable to the intended audience. A codebook can be theoretically plausible yet fail as a practical instrument when its distinctions are too subtle, too opaque, or too dependent on insider interpretation; the expert panel surfaced precisely these weaknesses.

The qualitative comments revealed where a plausible conceptual structure still required clearer coding rules. The resulting revisions strengthened the codebook at the points where later design teams or researchers would be most likely to encounter ambiguity, improving not only the adequacy of the content categories but the usability of the codebook as an operational specification \cite{Schreier2012-XAI-QualContentPractice-Sage}.


\subsection{Reliability evaluation}
\label{sec:Methodology:ReliabilityEvaluation}

Reliability was assessed after expert-review-driven revisions had been incorporated and the revised codebook had been frozen. The reliability procedure was preregistered on the Open Science Framework on 30 March 2026, before any independent coding activity; independent coding was conducted on 3 April 2026. The preregistration covered the frozen post-expert-review codebook, the subsample composition, the equivalence rule for Input versus What-if forward, the acceptance threshold of Krippendorff's $\alpha$ $\geq$ 0.800, and the stopping rule for codebook revision and re-testing if the threshold was not met. A summary is provided in~\ref{app:Preregistration}; the full registration is publicly available at \url{https://osf.io/up2sh}.

A stratified purposive subsample of 82 of the 325 meaning units, approximately 25\% of the corpus, was selected. The subsample was deliberately demanding. It included 32 borderline cases at three pre-identified code boundaries (Applied rule vs Causal factor, Causal factor vs Causal mechanism, and Set vs Individual element), 32 dual-valid cases at the Input vs What-if forward boundary, and 18 non-borderline routine cases. Coding-difficulty cases constituted 78\% of the subsample. Reliability obtained under these conditions functions as a conservative lower bound relative to expected full-corpus agreement, because the subsample concentrates the disagreements the codebook was most likely to produce \cite{OConnor2020-XAI-IntercoderReliabilityQualResearch-SAGE}.

Two coders independently coded the subsample using the frozen post-expert-review codebook. Reliability was quantified primarily with Krippendorff's alpha and supplemented by Cohen's kappa and percent agreement \cite{Krippendorff2004-XAI-ReliabilityMisconceptions-HCR,Cohen1960-XAI-AgreementNominal-EdPsychMeas}. Krippendorff's alpha served as the primary measure because it is suitable for nominal coding and robust to skewed code distributions \cite{Krippendorff2004-XAI-ReliabilityMisconceptions-HCR}. The statistics therefore provide evidence of reproducibility under conditions that deliberately included substantial coding difficulty. Full codebooks, change logs, survey instruments, extended procedural materials, and the preregistration summary are provided in the appendix.


\subsection{Structural decisions during abstraction}
\label{sec:Methodology:StructuralDecisions}

Two criteria justify the four-group structure: the kind of data required to generate each group's content, and the form of reasoning each group supports. Each group corresponds to a qualitatively different data source required to generate its content: rule-based content depends on explicit rule representations, causal content on feature-attribution or causal-inference outputs for the current case, epistemic-actual content on current-case probabilistic or ranking outputs, and epistemic-similar content on a historical case base. This distinction is supported by the domain-of-origin annotations described in Section~\ref{sec:Methodology:MeaningUnitDefinition} and reported in~\ref{app:Results}. Each group also aligns with a distinct reasoning form recognized in the cognitive and AI literatures: rule-based content supports deductive reasoning \cite{Russell2020}, causal content supports causal reasoning \cite{Pearl2009-XAI-Causality-Cambridge}, epistemic-actual content supports inductive and contrastive reasoning \cite{Lipton1990,Mill1843}, and epistemic-similar content supports analogical reasoning \cite{Gentner1983,Aamodt1994}. Together, data source and reasoning form provide a principled basis for the four-group structure that is independent of surface similarity in the corpus material.

Further structural decisions during abstraction shaped the final model. The analysis retained four groups instead of collapsing all epistemic content into one. Epistemic content for the actual case and epistemic content for similar cases differ in reference object, reasoning task, and technical infrastructure. Actual-case content draws on information about the current recommendation, ranking or probabilities; similar-case content requires retrieval and presentation of historical or analogous cases. The causal group was kept as a single explanatory chain instead of being divided into separate antecedent and consequence groups. A finer split was considered but rejected: context, input, factor, mechanism, outcome, and future state are most useful as connected parts of a single progression, and separating them would have made the model harder to use as a practical specification template \cite{Schreier2012-XAI-QualContentPractice-Sage}.

Rule-related content was treated as distinct from causal content. Some theoretical framings would treat explicit rules as a kind of causal mechanism, but in the corpus and in expert feedback, rule-related content answered what governs the decision or what rule was applied, not only what factor or relationship influenced the outcome. The distinction also matters technically, because queryable rule content depends on access to explicit rule representations, not on feature importance or current-case reasoning outputs.

The resulting structure was neither imposed deductively from explanation theory nor arrived at through induction alone. It emerged from iterative abstraction in dialogue with the empirical material and was then sharpened through theoretical coverage assessment and external review \cite{Mayring2014-XAI-QualContentAnalysis-Klagenfurt,Schreier2012-XAI-QualContentPractice-Sage}.


\subsection{Researcher position and quality assurance}
\label{sec:Methodology:ResearcherPosition}

The inductive phase of category development was conducted by a single researcher and required interpretive judgment. The resulting structure reflects one researcher's judgments during segmentation, abstraction, and boundary formation. A second researcher conducting an independent inductive analysis might have produced different higher-level categories or group boundaries.

The quality-assurance steps in the study should be read accordingly. The expert panel evaluated the adequacy, clarity, and boundary definition of an already developed category structure and informed its revision; it did not independently derive categories from the raw material. The reliability study tested whether independent coders could apply the finalized codebook consistently once the structure had been defined and refined; it did not test whether independent researchers would generate the same category system inductively.

The methodological contribution is a documented and externally refined explanation-content model whose codebook proved reproducibly applicable under independent coding, not a claim of independently replicated category discovery. 


\section{Results}
\label{sec:Results}


\subsection{Overview of the final model}
\label{sec:Results:FinalModel}

The analysis produced a fourteen-code explanation-content model organized into four groups: rule-based, causal, epistemic for the actual case, and epistemic for similar cases (Table~\ref{tab:Result:CodebookSummary}). Twelve codes are empirically grounded in the corpus; two codes, Rule base and What-if backward, were added as theoretical extensions because they correspond to system architectures documented in the XAI literature, although they were not instantiated in the available studies. The model therefore has mixed empirical and theoretical grounding. Support is uneven across the four groups: the causal and epistemic-actual groups are strongly grounded in the corpus, whereas the rule-based and epistemic-similar groups rest on narrower empirical material and warrant more cautious interpretation.

\begin{table}[h]
	\scriptsize
	\caption{Post-reliability explanation content codes (PoRC) - Summary}
	\label{tab:Result:CodebookSummary}
	\begin{tabular}
		{>{\raggedright\arraybackslash}m{2.5cm}
		>{\raggedright\arraybackslash}m{2.5cm}
		>{\raggedright\arraybackslash}m{7.5cm}}
		
		\hline
	
		\textbf{Code} & \textbf{Label} & \textbf{Brief definition} \\
	
		\hline
		\multicolumn{3}{l}{Rule-based explanation group} \\
		\hline
	
		PoRC 1.1* & Rule base & Repository or set of available rules for case assessment \\
		PoRC 1.2 & Applied rule & Rule evaluated and applied to current case \\
	
		\hline
	
		\multicolumn{3}{l}{Causal explanation group} \\
		\hline
	
		PoRC 2.1 & Context & Surrounding situation that helps interpret the case \\
		PoRC 2.2 & Input & Data provided to AI system to generate outcome \\
		PoRC 2.3 & Causal factor & Condition believed to have influenced the outcome \\
		PoRC 2.4 & Outcome & Result produced for current case \\
		PoRC 2.5 & Future state & Predicted state after outcome has been applied \\
		PoRC 2.6 & Causal mechanism & Description of causal relationship within causal chain \\
		PoRC 2.7 & What-if forward & Hypothetical input change and expected effect on outcome \\
		PoRC 2.8* & What-if backward & Desired outcome and required input change to achieve it \\
	
		\hline
		
		\multicolumn{3}{l}{Epistemic explanation (actual) group} \\

		\hline
				
		PoRC 3.1 & Set of ranked elements (actual) & Multiple candidate elements in ranked order \\
		PoRC 3.2 & Individual element (actual) & One candidate element with properties and evidence \\
	
		\hline
	
		\multicolumn{3}{l}{Epistemic explanation (similar) group} \\

		\hline
	
		PoRC 4.1 & Set of similar elements (similar) & Multiple similar past cases with success-score information \\
		PoRC 4.2 & Individual similar element (similar) & One similar past case with similarity property and score \\
	
		\hline
	
	\end{tabular}
	\textit{Note:} Codes marked with an asterisk (*) are theoretically proposed extensions without empirical corpus assignments (see Limitation~4 in Section~\ref{sec:Limitations}). Full definitions, inclusion and exclusion criteria, decision rules, and examples are provided in the supplementary materials (Appendix~\ref{app:PostReliabilityCodebook}).
\end{table}
		
The rule-based group contains Rule base and Applied rule. Rule base refers to a repository of rules available for case assessment. Applied rule refers to the specific rule evaluated and used in the current case. Expert review sharpened this distinction: a broader rule repository and a case-specific application answer different user questions and imply different system capabilities.

The causal group contains Context, Input, Causal factor, Outcome, Future state, Causal mechanism, What-if forward, and What-if backward. Context, Input, Causal factor, Future state, Causal mechanism, What-if forward, and What-if backward serve as explanations (explanans) for the outcome (explanandum). These codes describe the current case context, identify what entered the system, describe influential factors and relationships, report the current result, project what may happen after action, and support hypothetical reasoning. The model treats them as a coherent chain of explanatory content instead of unrelated fragments. The term causal is used to describe content that helps users understand what presumably led to an outcome, not in the strict sense of causal inference.

The causal group requires a terminological qualification. The term is used in the functional sense, in which explanation content helps users reason about what produced an outcome and what might follow from action, not in the sense of established causal inference. The AI systems represented in the corpus typically infer relationships from statistical associations that may be correlational instead of causally identified. The explanation content generated by such systems inherits this limitation. Codes such as Causal factor and Causal mechanism should therefore be read as diagnostic or attributional content that functions causally in users' reasoning, not as causal claims established through intervention or randomization. Interfaces that present such content should avoid language implying stronger causal warrant than the underlying model provides, and evaluations of downstream effects should treat claims of causal understanding with corresponding caution.

The epistemic group for the actual case contains two codes. Set of ranked elements refers to a ranked collection of candidate elements or recommendations. Individual element refers to the detailed representation of one candidate with ranking properties, evidence, or attributes. The epistemic group for similar cases mirrors this structure with two analogous codes for historical or analogous cases. The distinction between current-case and similar-case epistemic support is retained because each relies on a different reference object and typically a different technical infrastructure.

Full definitions, inclusion and exclusion criteria, decision rules, and examples are provided in~\ref{app:PostReliabilityCodebook}.


\subsection{Corpus-level patterns}
\label{sec:Results:Patterns}

Explanation content was multi-domain. Coded units drew from the AI domain, the surrounding system domain, and the application domain, with some units combining information from more than one. User interfaces with useful explanation content therefore depend on data integration, not on access to an AI model alone.

Explanation content was multi-type. No single explanation type adequately described the explanation space of the studied systems. Content across the six studies mapped to causal, epistemic, counterfactual, contrastive, example-based, and rule-based types, though the distribution was uneven. Every user study instantiated at least three explanation types, with Studies 1 through 4 each instantiating three, Study 6 four, and Study 5 five; the full distribution is reported in Table~\ref{tab:Result:ExplanationTypes} in~\ref{app:Results}.

The applicability of some content categories depended strongly on system architecture. What-if content appeared only where inputs were user-modifiable, rule-related content required explicit decision logic, and similar-case content required a case base or a comparable retrieval mechanism. The model is a reference structure, not a checklist that every interface must implement in full.


\subsection{Corpus composition and variation}
\label{sec:Results:Composition}

The corpus varies substantially in user role, elicitation format, and system architecture. One of the six studies contributed 211 of the 325 meaning units, because it involved a detailed user-interface design in which conceptual content was decomposed into many individually addressable elements. The remaining studies yielded lower counts, with explanation expectations represented through mental models (Studies 3, 4, 5, 6) and less detailed user interfaces (Studies 1, 2, 3). This asymmetry matters analytically: the model is a structural specification of explanation-content categories, not a frequency-based characterization of industrial explanation practice.

The variation nonetheless served the central purpose of the study. It exposed the analysis to explanation content from development, operation, and maintenance phases across multiple industrial domains, and to different levels of granularity. Some meaning units captured broad contextual or causal concepts; others captured narrowly specified user interface groups. The model therefore had to accommodate both high-level conceptual content and fine-grained implementation-oriented content.

Elicitation format and code frequency are related. Detailed user interfaces produce larger numbers of meaning units for epistemic details, ranking properties, and input variations, because each may be represented by a separate user interface group. Mental models compress the same conceptual space into fewer units. The model is therefore best interpreted in terms of its structural distinctions, not raw code-frequency distributions.

A sensitivity analysis compared code frequencies in Study 5 (n=211) against Studies 1–4 and 6 combined (n=114) to assess whether the structural distinctions in the model are driven by the dominant study; the full analysis is reported in~\ref{app:Results}. In the comparisons below, the first percentage in each pair refers to Study 5 and the second to the remaining five studies combined. Frequencies for Context (6.2\% vs.\ 5.3\%), Individual element (30.3\% vs.\ 30.7\%), and Set of ranked elements (3.8\% vs.\ 8.8\%) are comparable across sub-corpora, indicating that these distinctions are not specific to the UI-design context of Study 5. Frequencies for What-if forward (25.6\% vs.\ 15.8\%) and Input (26.1\% vs.\ 23.7\%) are higher in Study 5 but present across the remainder, consistent with the architecture-driven scoping criterion that What-if content arises when inputs are user-modifiable (as in Studies 4 and 5). Frequencies for epistemic-similar codes concentrate in Study 5 (1.9\% and 0.9\% vs.\ 0.0\% and 0.0\%) because only Study 5 involved a system with an accessible historical case base. The structural distinctions in the model, as opposed to their frequencies, are therefore supported across sub-corpora for causal and epistemic-actual content; cross-sub-corpus grounding is weaker for rule-based, epistemic-similar, and backward counterfactual content, consistent with the limitations noted in Section~\ref{sec:Limitations}.


\subsection{Prevalence and structural asymmetries}
\label{sec:Results:Asymmetries}

The distribution of codes across the corpus was uneven. The most frequent codes were Input, What-if forward, and Individual element. This pattern partly reflects genuine differences in explanation content frequency and partly results from segmentation granularity: one large user-interface study contributed a large proportion of the meaning units, and interface-level segmentation produces finer-grained counts than mental-model or interview material. Code frequencies reflect the corpus and segmentation logic, not the prevalence of explanation content in industrial AI systems generally.

Within this caveat, the frequency patterns remain informative. Causal and epistemic-actual content dominated the corpus; rule-based and example-based content were sparse. This sparsity reflects the architecture of the systems in the corpus, not the unimportance of the content types. Many systems exposed current-case reasoning and ranking information, but few exposed queryable rule repositories or retrieved analogous prior cases as a primary explanation mechanism.

The main empirical claim of the paper therefore concerns the structural distinctness, adequacy, and reproducibility of the categories, not the prevalence of specific content types across industrial AI systems. Empirical support is stronger for causal and epistemic-actual content than for rule-based, epistemic-similar, and backward counterfactual content, which remain provisional parts of the model.


\subsection{Domain-of-origin implications}
\label{sec:Results:DesignOfOrigin}

The domain-of-origin annotation shows that explanation generation in practice depends on multiple information sources that are often institutionally and technically separate. AI-domain content depends on model-side outputs such as rankings, probabilities, or confidence information. System-domain content depends on operational data, logs, configurations, or status information. Application-domain content depends on contextual knowledge about assets, environments, goals, constraints, and work practices.

This matters because many explanation failures are not attributable to the model alone. A system may produce a technically adequate ranked recommendation yet lack the application context needed for safe and efficient interpretation. Alternatively, it may have rich contextual data but expose too little epistemic content for users to judge uncertainty. The model therefore frames the elicitation, specification, and generation of explanation content as a multi-source integration problem, not only as a post-processing problem applied to model outputs.


\subsection{Theory-informed coverage assessment}
\label{sec:Results:TheoryInformedCoverageAssessment}

The theory-informed coverage assessment produced a mapping of the fourteen final codes to the six established explanation types that served as sensitizing concepts (Table~\ref{tab:MainText:CodeToTypeMapping}). The mapping is not one-to-one: multiple codes can instantiate a single explanation type, and the What-if codes span two types, since forward and backward reasoning over inputs correspond to causal and counterfactual reasoning respectively. The fourteen codes together cover all six explanation types, providing the completeness evidence on which the coverage assessment was based. Because the codes operate at a finer granularity than explanation types, the model functions as an intermediate layer between high-level reasoning styles and concrete interface content.

\begin{table}[!ht]
	\scriptsize
	\caption{The mapping is not one-to-one: multiple codes can instantiate a single explanation type. PoRC~2.7 spans both causal and counterfactual reasoning because counterfactual reasoning is a subclass of causal reasoning in which the antecedent condition is hypothetical instead of observed.}
	\label{tab:MainText:CodeToTypeMapping}
	\begin{tabular}
		{>{\raggedright\arraybackslash}m{4.5cm}|
		>{\raggedright\arraybackslash}m{8.0cm}}
	
		\hline
	
		\textbf{Explanation type} & \textbf{Post-reliability code(s)} \\
	
		\hline
	
		Rule-based explanations &
		PoRC~1.1 Rule base, PoRC~1.2 Applied rule \\
	
		\hline
	
		Causal explanations &
		PoRC~2.1 Context, PoRC~2.2 Input, PoRC~2.3 Causal factor, PoRC~2.4 Outcome, PoRC~2.5 Future state, PoRC~2.6 Causal mechanism, PoRC~2.7 What-if forward \\
	
		\hline
	
		Counterfactual explanations &
		PoRC~2.7 What-if forward, PoRC~2.8 What-if backward \\
	
		\hline
	
		Contrastive explanations &
		PoRC~3.1 Set of ranked elements (actual) \\
	
		\hline
	
		Epistemic explanations &
		PoRC~3.2 Individual element (actual) \\
	
		\hline
	
		Example-based explanations &
		PoRC~4.1 Set of similar elements (similar), PoRC~4.2 Individual similar element (similar) \\
	
		\hline
	
	\end{tabular}
\end{table}

Beyond the mapping, the coverage assessment played two roles in the development of the model: it acted as a completeness check against established explanation types, and it had corrective influence during corpus construction. The assessment identified rule-related explanation content as theoretically important but absent from the initial set of studies. Study 6 was included to ensure that at least one rule-related content type was represented, and a rule-related theoretical extension was added on the basis of its recognized status in the XAI literature. Anchors provides local rule-like conditions for predictions \cite{Ribeiro2018-XAI-AnchorsHighPrecision-AAAI}, and Local Rule-Based Explanations (LORE) generates local rule-based explanations for black-box decisions \cite{Guidotti2018-XAI-LORE-arXiv}. LORE's output structure is particularly relevant to the distinction between Rule base and Applied rule: the factual rule extracted for the target instance corresponds to Applied rule (PoRC 1.2), while the surrounding local tree from which that rule is drawn corresponds to Rule base (PoRC 1.1). The distinction between the two codes is therefore consistent with the structure of an established post-hoc rule-based explanation method.

The assessment also clarified that the initial What-if code spanned both causal and counterfactual reasoning. In the corpus, What-if content took a forward form, connecting a hypothetical input change to an expected outcome change. The backward form, in which a desired outcome is specified first and the system or user reasons backward to identify required input changes, did not appear in the available studies. Its addition was nevertheless theoretically justified: counterfactual explanations have been defined as statements of what would need to change to obtain a desired outcome \cite{Wachter2018-XAI-CounterfactualExplanations-HarvardJOLT}. Rule base and What-if backward were therefore added as theoretical extensions for explanation content meaningful to documented system architectures not represented in the present corpus.

Contrastive and example-based explanation content were present in the corpus but weakly represented. Their place in the broader explanation space rests on limited corpus grounding but stronger theoretical support; the model includes them, though their empirical basis is weaker than that of the causal and epistemic-actual groups. The same caution applies to rule-based and epistemic-similar content. Their inclusion is justified by limited corpus evidence and stronger architectural and theoretical plausibility, but the data do not yet establish robust recurrence across multiple independent domains or corpora. Their current status is preliminary, not fully generalized.


\subsection{Expert evaluation and codebook revision}
\label{sec:Results:ExpertEvaluation}

The expert panel supported the content adequacy of the codebook. All codes met the predefined acceptability threshold for relevance, boundary clarity, and understandability, with item-level I-CVI values at or above 0.82. Scale-level agreement was excellent: S-CVI/Ave was 0.93 for relevance, 0.92 for boundary clarity, and 0.94 for understandability. The code definitions were broadly judged meaningful, distinguishable, and comprehensible for the intended use context.

The qualitative comments identified recurring boundary problems and produced material improvements. The most consequential revisions were the split of the former rule-related code into Rule base and Applied rule, the split of the former What-if code into What-if forward and What-if backward, and clearer decision rules for set-versus-individual distinctions in the epistemic groups. 


\subsection{Reliability evaluation}
\label{sec:Results:ReliabilityEvaluation}

Reliability testing on the frozen post-expert-review codebook yielded strong evidence of reproducibility. On the stratified subsample of 82 units, percent agreement was 93.9\%, Krippendorff's $\alpha$ was 0.920, and Cohen's $\kappa$ was 0.920, values that substantially exceed commonly used reliability thresholds.

The reliability study establishes that the codebook can be applied consistently by independent coders even when the subsample deliberately includes ambiguous material, and it identifies where residual disagreement remained. Disagreements concentrated at a small number of boundaries, mainly the set-level versus element-level epistemic distinction and a small number of mechanism-related cases. They did not require structural changes to the code system, but they motivated a limited set of additional post-reliability decision rules.


\subsection{From the draft codebook to the post-reliability codebook}
\label{sec:Results:FromDraftToPostReliabilityCoding}

The final fourteen-code structure was reached through a sequence of revisions, not through a single static coding pass. The initial structure comprised twelve empirically grounded codes. Expert review produced two conceptually important changes: the split of the former rule-related code into Rule base and Applied rule, and the split of the former What-if code into What-if forward and What-if backward. Several decision rules were also clarified at boundaries that had generated ambiguity, particularly the distinction between set-level and individual-element epistemic content and the distinction between causal factor and causal mechanism. The post-reliability codebook is therefore a stabilized instrument, not a descriptive summary of the first coding round.

\begin{figure}[htbp]
	\small
	\includegraphics[width=1.0\linewidth]{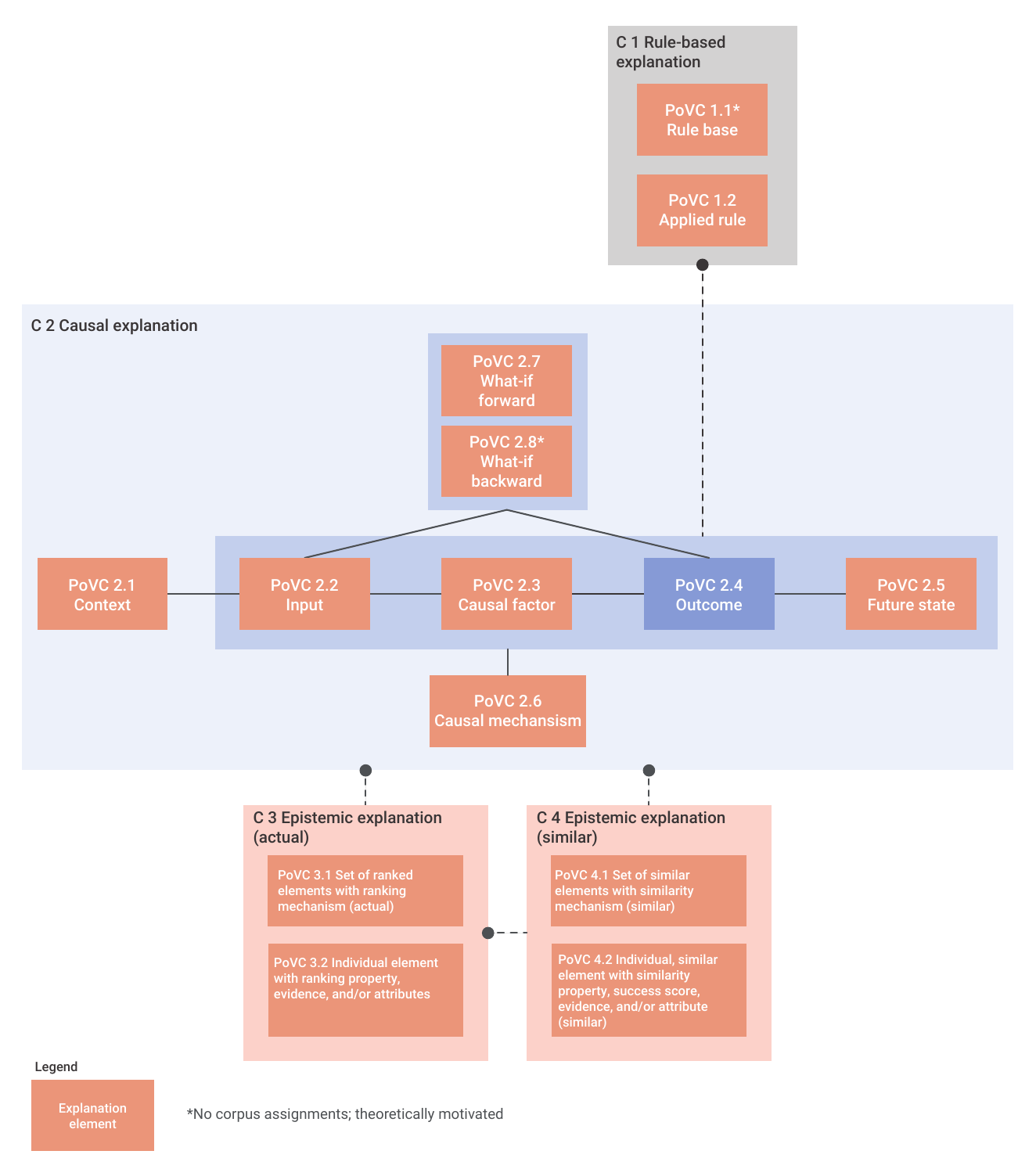}
	\caption{Explanation content model for local, post-hoc explanations (post-reliability, reflecting revisions introduced during theory-informed coverage assessment and expert review). PoRC~1.1 Rule base and PoRC~2.8 What-if backward are not represented in the corpus and were added as theoretically motivated extensions.}
	\label{fig:XAIModel}
\end{figure}

The final model with the explanation content categories and codes is depicted in Figure~\ref{fig:XAIModel}.

\subsection{Boundary sensitivity and residual ambiguity}
\label{sec:Results:BoundarySensitivity}

The reliability results were strong, but the disagreement analysis is informative about which content distinctions are cognitively and operationally demanding. The most sensitive boundaries involved set-level versus individual-level epistemic content and a small number of mechanism-related assignments. Design and evaluation practice should therefore pay attention to the representational level of explanation elements, since interfaces often combine set-level and element-level information visually and the distinction can be overlooked even when it is theoretically clear.

The pre-defined equivalence rule at the Input versus What-if forward boundary illustrates a broader property of explanation content: some elements serve more than one reasoning function at once. A user-provided input value can be both the current factual anchor and the starting point of a hypothetical forward variation. The model preserves the categorical distinction, acknowledging that some boundary cases are genuinely dual-valid.


\subsection{Key findings}
\label{sec:Results:KeyFindings}

The first substantive finding is that explanation content in industrial AI systems cannot be reduced to one explanation type or one information source. Users often need a combination of contextual information from the application domain, technical information from the system domain, and uncertainty or ranking information from the AI domain within the same decision episode. A content model that treats explanations as monolithic outputs risks overlooking this heterogeneity.

The second finding is that the four-group structure reflects functionally distinct explanation-design requirements. Rule-based content answers what governs the decision. Causal content answers what factors and relationships produced the current result and what may happen next. The group is termed causal because it captures content used functionally to explain why an outcome occurred and what may happen next, even when the underlying AI system does not establish causality in a strict scientific sense. In many AI systems, the relationships represented within this group are inferred from statistical associations and may therefore be correlational instead of truly causal. Epistemic content for the actual case answers how the system ranks or qualifies the current options. Epistemic content for similar cases answers what can be learned from analogous prior situations. 

The third finding is that the model should be understood as a set of explanation-content candidates, not as a universally applicable package. Which of these candidates are relevant depends on the target user group, the users' goals and tasks, and the context of use. The analysis did not support the view that all codes are equally relevant for all systems. Instead, the presence or absence of specific content types was predictably linked to architecture and deployment context. This makes the model practically useful because designers, ideally with the participation of end users, can decide which parts apply and which do not. The selection should be based on eliciting users' goals, tasks, context of use, and needed task-specific explanation content, not on assuming a one-size-fits-all explanation package.

The fourth finding is that some theoretically important content types are weakly represented in the available corpus and therefore remain only preliminarily supported as cross-domain categories. Rule-based content, similar-case content, contrastive content, and backward counterfactual content were less prevalent than causal and epistemic-actual content. Their inclusion in the model is justified, but the present study does not yet establish robust cross-domain generalizability for these parts of the structure.


\subsection{Proposed use of the blueprint}
\label{sec:Results:Blueprint}

The final blueprint suggests three plausible uses in future design and evaluation work. First, it is intended to support the elicitation, specification, design, and review of user-role-specific explanation content. A design team can examine the code groups systematically and determine which forms of content are relevant for a given user role, context of use, and user task. Second, it can support explanation-content audit by enabling existing systems to be reviewed code by code to identify whether relevant content is present, absent, or only partially instantiated. Third, it can support cross-system comparison, for example across AI systems or across design alternatives for the same type of AI system. To support learning of the coding scheme, Appendix~\ref{app:AnnotatedExplanationContentModel} presents the explanation-content model together with coding hints.

The current study does not behaviorally validate these uses with practitioners. However, the validity and reliability evidence reported here is important because such uses require a codebook that is both interpretable and reproducible. However, the blueprint should not yet be treated as equally mature across all categories. Its practical use is currently best supported for causal and epistemic-actual content, whereas rule-based, epistemic-similar, and backward counterfactual content should be treated as more provisional until broader cross-domain evidence is available.


\subsection{Illustrative applicability across system architectures}
\label{sec:Results:Applicability}

The context-dependent applicability of the blueprint can be illustrated without reproducing a full domain case. Consider three generic system architectures. In a rule-based or policy-driven system, the rule-related group becomes particularly relevant because users may need either the specific applied rule or access to a broader rule base that explains why the current case was assessed in a particular way. In a recommendation system with ranked alternatives for the present case, the epistemic-actual group becomes particularly relevant because users need both the ranked set and at least some detailed content about individual elements. In a case-based or retrieval-supported system, the epistemic-similar group becomes particularly relevant because analogous prior cases and their success properties can support comparison and calibration.

By contrast, not all architectures make all code groups equally relevant. Systems with machine-generated inputs and no practical user control over those inputs often gain little from exposing what-if reasoning over raw inputs, especially when the input space is continuous or high-dimensional. Systems that do not store or expose historical analogous cases cannot meaningfully instantiate similar-case content. Systems that rely on opaque statistical models without explicit rule repositories may still instantiate applied constraints or thresholds, but they may not instantiate a browsable rule-base code. These examples underscore the intended reading of the blueprint. It defines a structured space of possible explanation content whose relevance depends on system architecture and context of use, not a universal package that every system should implement in full. An illustrative example based on User Study~6 is provided in Section~\ref{sec:Example}.


\section{Discussion}
\label{sec:Discussion}

The results answer both research questions. RQ1 is answered by the fourteen codes that specify the explanation-content categories identified in the corpus and refined through expert review. RQ2 is answered by the structure of the model, which organizes these categories into four groups aligned with distinct user questions and information needs.


\subsection{From explanation types to explanation content}
\label{sec:Discussion:ExplanationContent}

The study implies a shift from explanation types to explanation content as the primary design unit. Explanation types remain useful as a vocabulary for reasoning style, but they are too coarse to specify what should appear on a user interface. The model contributes that missing layer by defining bounded information elements and their relationships, at a granularity that is directly actionable for design specification, implementation planning, and evaluation.

Research on explanation types, trust calibration, and explanation effectiveness remains essential to the field \cite{Hoffman2018-XAI-MetricsExplainableAI-arXiv,Miller2019-XAI-ExplanationAI-Elsevier,Naiseh2023-ExplanationClassesTrustCalibration-IJHCS}. The present contribution extends that literature by decomposing explanation types into content units that are tractable for specification, comparison, implementation, and evaluation. The question a designer can now ask is not only which explanation type to provide, but which content elements users need for a particular task and whether those elements are present, absent, or only partially instantiated.

A further methodological implication follows from the shift to content as the primary design unit. In the corpus, the explanation content elicited for a given user, task, and system context was frequently associated with more than one explanation type, a pattern referred to in the following as a \textit{hybrid explanation type}. Every user study in the corpus instantiated at least three explanation types (Studies 1, 2, 3, and 4 instantiating three, Study 6 four, and Study 5 five; see Table~\ref{tab:Result:ExplanationTypes} in~\ref{app:Results}). As the secondary analysis in Section~\ref{sec:RelatedWork:ExplanationNeedElicitation} showed, explanation content is rarely elicited from end users in current XAI research \cite{Kim2024-HumanCenteredEvaluationXAI-Frontiers}. Much of the behavioral XAI literature nevertheless investigates explanation effects at the level of \textit{single explanation types}, treating a given type (for example, counterfactual, rule-based, or feature-based) as a fixed experimental condition instead of deriving it from prior elicitation \cite{Naiseh2023-ExplanationClassesTrustCalibration-IJHCS,VanderWaa2021-XAI-RuleExampleComparison-Elsevier,Bertrand2023-FeatureBasedExplanationsFinance-FAccT,Speith2022-XAI-TaxonomiesXAIMethods-ACM}. This absence of elicitation is consistent with the broader pattern documented in the Kim et al. review.

This type-first framing raises a methodological concern. Ecological validity \cite{Brunswik1956-XAI-RepresentativeDesign-UCPress} requires that experimental conditions reflect the explanation content users expect or actually encounter. The present findings suggest that explanation content elicitation should be added as a prior step in XAI research. Content is first elicited from representatives of the target users for the target task and context of use. The elicited content is then categorized as a single explanation type or a hybrid explanation type, not selected in advance. If the elicited content is a single explanation type, a single-type study or system design is appropriate and the single type-based XAI literature applies directly. If the elicited content is a hybrid explanation type, a hybrid design or comparison of content combinations is the appropriate research object. Single and hybrid explanation types are therefore complementary; content elicitation determines which applies. Explanation types are a finding, not a starting point.


\subsection{Relation to prior work}
\label{sec:Discussion:PriorWork}

The findings align with prior work showing that explanation needs are task-dependent and context-sensitive \cite{Cai2019-XAI-HumanCenteredToolsClinic-ACM,Liao2020-XAI-QuestioningAI-CHI-ACM,Panigutti2023-XAI-CoDesignClinicalXAI-ACM}. The present study represents those needs as a structured set of explanation-content categories, not as questions, themes, or design suggestions. This provides the finer-grained object that implementation and evaluation work typically require. The evidence supports the model's adequacy as a content blueprint; it does not establish behavioral superiority over alternative design approaches.

The model complements co-design and participatory XAI research. Participatory approaches elicit what stakeholders want to know and shape explanation presentation in domain-sensitive ways \cite{Jin2021-XAI-EUCA-arXiv,Nazaretsky2022-TeachersAI-LAK,Panigutti2023-XAI-CoDesignClinicalXAI-ACM}. The present paper supplies a content-level structure that such approaches can use during elicitation, synthesis, or review, functioning as an intermediate layer between open-ended stakeholder input and user-interface-level specification.

Evaluation frameworks such as those of Doshi-Velez and Kim, Hoffman et al., and Mohseni et al. clarify what kinds of evidence are needed to justify claims about explanation quality and usefulness \cite{DoshiVelez2017-XAI-RigorousInterpretability-ICML,Hoffman2018-XAI-MetricsExplainableAI-arXiv,Mohseni2021-XAI-MultidisciplinaryHAI-ACM}. The present contribution adds a content-oriented complement by specifying the units of explanation content whose adequacy and effects can later be studied.


\subsection{Scoping explanation content by user needs and system architecture}
\label{sec:Discussion:ScopingExplanationContent}

The blueprint is not a checklist that every AI system must satisfy fully; it can be used from two complementary perspectives. From a user-needs perspective, the model provides a set of explanation-content candidates whose relevance depends on the target user group, the users' goals and tasks, and the context of use. From a system-architecture perspective, it provides a map of the explanation content a system can realistically produce given its data sources, interfaces, reasoning mechanisms, and implementation constraints. The process ideally begins with user needs and then examines architectural feasibility.

The two perspectives do not always align. Some content categories are highly relevant to users but feasible only under particular architectural conditions. What-if content is most useful when users can manipulate a bounded input space, similar-case content when analogous historical cases can be retrieved and presented, and rule-base content when the system exposes rule repositories or rule logic in user-meaningful form.

The blueprint is valuable because it makes these perspectives comparable. Teams can identify both the overlap between needed and feasible explanation content and the gap between them, supporting explanation-content design and explanation-system planning together.


\subsection{Explanations as socio-technical artifacts}
\label{sec:Discussion:SocioTechnicalArtifacts}

A substantial share of the explanation content in the corpus originated in the application domain and the surrounding system domain, not only in model rationale. Explanations frequently required content connecting AI outputs to operational conditions, domain constraints, available evidence, and projected consequences. This is consistent with work showing that explanation value depends on user role, workflow, and decision stage \cite{Cai2019-XAI-HumanCenteredToolsClinic-ACM,Panigutti2023-XAI-CoDesignClinicalXAI-ACM,Chromik2021-XAI-HumanXAIInteractionTaxonomy-IUI}. It also aligns with socio-technical traditions in HCI and CSCW that treat system use as situated in work practice, organizational context, and human-machine interaction, not as a property of interface mechanics alone \cite{Suchman1987-XAI-PlansSituatedActions-CUP,Bannon1991-XAI-HumanFactorsHumanActors-LawrenceErlbaum,Dourish2001-XAI-WhereTheActionIs-MITPress}. Within XAI specifically, Ehsan and Riedl have argued that explainable AI must be approached as a reflective socio-technical endeavor, not a purely algorithmic one \cite{Ehsan2020-XAI-ReflectiveSociotechnicalApproach-Springer}. Ehsan et al. have further proposed mapping the socio-technical gap between an XAI system's technical affordances and the social needs of its use context as a route to more effective explanation design \cite{Ehsan2023-XAI-SociotechnicalGap-ACM}. Explanations that support real decisions therefore often need to include context, operational constraints, evidence, projections, and uncertainty qualifiers alongside outcome justifications.

This has an engineering implication. Explanation generation is an integration problem requiring access to AI outputs, related system data, and application-domain information, not only a model-side problem. The present findings complement calls to evaluate explanations at the level of human-AI interaction and real-world use, not only at the level of model interpretability \cite{DoshiVelez2017-XAI-RigorousInterpretability-ICML,Mohseni2021-XAI-MultidisciplinaryHAI-ACM}. The domain-of-origin annotation was included to make this visible: it identifies which information sources are required for each content type and therefore where explanation quality may fail because the needed context is unavailable or fragmented.


\subsection{Implications for HCI research and design}
\label{sec:Discussion:ImplicationHCI}

The content-level granularity of the model creates a unit of analysis well suited to HCI theory and evaluation. Because the codes denote bounded information elements, they provide a unit of analysis compatible with established HCI theories of information foraging \cite{Pirolli1999-XAI-InformationForaging-PsychReview}, situation awareness \cite{Endsley1995-XAI-SituationAwareness-HF}, workload \cite{Hart1988-XAI-NASATLX-ElsevierNorthHolland}, verification behavior under automation \cite{Parasuraman2010-XAI-ComplacencyBiasAutomation-HumanFactors}, and appropriate reliance \cite{Lee2004-XAI-TrustAutomation-HumanFactors}. Set-level epistemic codes may support rapid comparison (information foraging; \cite{Pirolli1999-XAI-InformationForaging-PsychReview}); individual-element codes may support closer inspection (information foraging, situation awareness; \cite{Pirolli1999-XAI-InformationForaging-PsychReview,Endsley1995-XAI-SituationAwareness-HF}); causal-chain content may support comprehension and projection (situation awareness; \cite{Endsley1995-XAI-SituationAwareness-HF}); and similar-case content may support analogical reasoning and confidence calibration (appropriate reliance; \cite{Lee2004-XAI-TrustAutomation-HumanFactors}). When ranked options are present, set-level epistemic content is likely to be most useful as an entry point, because it helps users identify which elements warrant closer inspection. Once a promising or risky element has been identified, individual-element content can support more detailed assessment through evidence, attributes, or ranking properties, reducing unnecessary inspection effort.

The model suggests that explanation research need not remain tied to the larger units implied by explanation types alone. Explanation types are useful for describing broad reasoning forms, but they often combine several information elements within a single analytical category. Defining explanation-content elements as smaller units enables more precise analysis of how particular elements, alone or in combination, relate to outcomes such as efficiency, decision accuracy, confidence, trust calibration, verification behavior, or content-selection strategies. The present paper does not test these propositions behaviorally but provides a structured basis for doing so.

Four implications follow for design practice. Explanation content for user interfaces should be structured around the user decision, not around broad explanation types treated as undifferentiated units. Uncertainty should be communicated through explicit content elements tied to evidence, ranking properties, or analogous cases, not through a single generic confidence score. Application context should be treated as part of the explanation itself, because missing context can reduce explanation usefulness even when the model is technically strong. Finally, elicitation and specification of explanation content should begin with user needs. Teams, ideally with end-user participation, should:

\begin{enumerate}
	\item determine which content is relevant for the target user group, user goals, user tasks, and context of use;
	\item map the elicited content to the explanation-content model and, guided by the model, identify additional needed content;
	\item validate the project-specific model with representatives of the target user group for correctness and sufficiency;
	\item compare these needs against what the current system architecture can support or is intended to support, before deciding which content to include, how to structure it in the information architecture, and how to present it to end users.
\end{enumerate}

The immediate value of the model is the definition and structure of explanation-content types. It provides a vocabulary for asking whether a system exposes only an outcome or also the surrounding content users may need to interpret and act on that outcome: context, inputs, causal factors, causal mechanisms, future states, uncertainty qualifiers, and similar cases where relevant. This vocabulary supports concept development, interface critique, cross-system comparison, and the design of future behavioral evaluations.

The explanation-content model is not yet a field-validated practitioner tool. The underlying codebook is, however, content-adequate according to expert judgment and reproducible under independent coding, providing necessary preconditions for later design, behavioral evaluation, and practitioner use.


\subsection{Boundaries of the present claims}
\label{sec:Discussion:ClaimBoundaries}

The claims of the paper are bounded in three respects. The final explanation-content model is a plausible interpretation of the available corpus, and the associated codebook can be applied reproducibly in the industrial AI contexts represented in that corpus. Cross-domain generalizability is stronger for causal and epistemic-actual content than for rule-based, epistemic-similar, and backward counterfactual content, whose empirical grounding is currently narrower. The evidence supports content adequacy and reproducible application of the finalized codebook; it does not support independent replication of the original inductive category derivation, nor does it show that systems using the blueprint will necessarily produce better user outcomes. The two non-instantiated codes are treated as theoretically motivated extensions; their inclusion is justified on architectural and theoretical grounds, but they remain priorities for future empirical validation.

The model is therefore a starting point for further design and evaluation work, not a settled account of explanation quality.


\section{Limitations}
\label{sec:Limitations}

The inductive development of the category structure was conducted by a single researcher, and the resulting structure reflects that researcher's interpretive lens during segmentation, abstraction, and boundary formation. Expert review and independent reliability coding challenged and stabilized the category system after its construction, but neither step establishes that a second researcher working inductively from the raw material would have derived the same categories or group boundaries.

The corpus is imbalanced. One user-interface study contributed a large share of the meaning units because interface-level segmentation produces finer granularity than mental-model or interview materials. This affects both frequency distributions and the evidential basis for specific categories: causal and epistemic-actual content are represented across multiple studies; rule-based, epistemic-similar, and backward counterfactual content are supported by narrower material. Support for the core causal and epistemic-actual structure is therefore stronger than for the cross-domain generalizability of the less represented categories.

The reported reliability statistics apply to coding on pre-fixed meaning units, not to independent segmentation. Segmentation was rule-based and relatively determinate for much of the material, but a fuller reproducibility assessment would also test segmentation agreement.

Two codes in the final model, Rule base and What-if backward, are theoretical extensions without empirical corpus assignments. Their inclusion rests on architectural plausibility and the XAI literature, and they require empirical validation in future studies.

The paper establishes content adequacy and coding reproducibility but not behavioral utility. The study does not test whether using the blueprint produces better interfaces, faster verification, more appropriate reliance, or higher decision quality in practice.

The three proposed use modes for the model, that is, explanation requirements specification, explanation content audit, and cross-system gap analysis, are design-oriented propositions. They are derived from the codebook's structure and reliability properties, not empirically validated instruments. Independent coders can apply the codebook consistently; this does not establish that practitioners working without researcher mediation can apply the model successfully in design, audit, or comparison tasks. In the illustrative SHIELD example in~\ref{sec:Example}, explanation content was elicited from three user representatives: a hospital IT subject-matter expert and two cybersecurity subject-matter experts. The mapping of that content to the model's codes and groups was performed by the researcher. The example therefore shows how the model can organize elicited content and surface gaps; it does not show that practitioners can apply the model as a standalone design instrument. Behavioral validation of the use modes with practitioners applying the model without researcher mediation is a priority in Future Work.


\section{Future work}
\label{sec:FutureWork}

The corpus should be extended to additional domains and architectures, particularly settings where rule-based, epistemic-similar, and backward counterfactual content are expected to recur across multiple independent studies. This is necessary for determining whether the currently weakly represented categories generalize beyond the specific systems and artifacts of the present corpus.

The three proposed use modes for the model, that is, explanation requirements specification, explanation content audit, and cross-system gap analysis, should be evaluated as practical resources. Future studies should test whether teams using the model produce more complete or better-targeted explanation-content specifications than teams without such a structure, whether reviewers applying the model to existing systems reach consistent audit conclusions, and whether cross-system comparisons using the model surface meaningful design differences.

Behavioral validation is needed. Controlled and field studies should examine how different combinations of explanation-content elements affect understanding, verification behavior, workload, reliance, trust calibration, and decision quality.

Semi-automated support for coding and traceability could make the blueprint easier to apply at scale, particularly in large interface reviews and multi-system comparisons.

The blueprint should be examined in domains beyond the current industrial settings, including clinical decision support and other high-stakes environments, to test which parts transfer directly and which require adaptation.


\section{Conclusion}
\label{sec:Conclusion}

This paper has presented a human-centered explanation-content model for local, post-hoc explanations in industrial AI systems. The model was derived from a multimodal corpus of six user studies, refined through expert review, and evaluated through independent reliability testing. It organizes fourteen codes into four groups distinguishing rule-based, causal, epistemic-actual, and epistemic-similar explanation content, with twelve codes empirically grounded in the corpus and two theoretically motivated extensions that remain to be validated empirically. The empirical contribution lies in the structure, adequacy, and reproducibility of the model, not in frequency-based claims about explanation practice.

The model is not a prescription that every AI system should present every content type, nor does the paper claim behavioral benefits from interfaces built from the model. It offers a structured and reproducible basis for future explanation specification, comparison, and evaluation, and it supplies a missing content-level unit of analysis between high-level explanation types and concrete interface implementations. This intermediate unit allows theory-driven and behavioral XAI research to manipulate and assess explanation content element by element, not as a single undifferentiated treatment.


\section*{Acknowledgments}
\label{sec:Acknowledgements}

I thank my colleagues who participated in the expert validation studies. I am grateful to the named contributors, Christof C. Budnik, Christoph Kuhmuench, Enrico Lovat, Ziran Min, Adrienne Raglin, Robert G. Reynolds, Gaurav Kumar Srivastava, Matt Westveer, and Theresa Zimmermann, for their time, careful review, and constructive feedback. I also thank the additional participants who contributed anonymously and whose insights were essential for refining the codebook. I thank Christoph Kuhmuench for serving as the independent coder for the reliability validation. Finally, I would like to thank Christoph Brand for his support of the Explainable AI research for the last six years.


\section*{Declaration of competing interest}
The author declares that he has no known competing financial interests or personal relationships that could have appeared to influence the work reported in this paper.

\section*{Funding}
This research was funded, in part, by the Advanced Research Projects Agency for Health (ARPA-H).  The views and conclusions contained in this document are those of the author and should not be interpreted as representing the official policies, either expressed or implied, of the U.S. Government.

\section*{Declaration of Generative AI and AI-Assisted Technologies in the Writing Process}
During the preparation of this work the author used Claude (Anthropic, model: claude-sonnet-4-6) to assist with language improvement. The author reviewed and edited all AI-assisted content after use and takes full responsibility for the integrity and accuracy of the content of the published article. 

\section*{Data availability}
Data will be made available on request.


\bibliographystyle{elsarticle-harv} 
\bibliography{Bibliography}

@article{Jain2023-XAI-StockMarketPredictions-MDPI,
	author       = {Jain, Rahul and Vanzara, Rakesh},
	title        = {{Emerging Trends in AI-Based Stock Market Prediction: A Comprehensive and Systematic Review}},
	journal      = {{Engineering Proceedings}},
	volume       = {56},
	number       = {1},
	pages        = {254},
	year         = {2023},
	month        = nov,
	doi          = {10.3390/ASEC2023-15965},
	howpublished =  {\url{https://www.mdpi.com/2673-4591/56/1/254}},
	publisher    = {MDPI},
	address      = {Basel, Switzerland},
	issn         = {2673-4591},
}

@article{McKinney2020-XAI-AIBreastCancerScreening-SpringerNature,
	author    = {McKinney, Scott Mayer and Sieniek, Marcin and Godbole, Varun and Godwin, Jonathan and Antropova, Natalia and Ashrafian, Hutan and Back, Trevor and Chesus, Mary and Corrado, Greg S. and Darzi, Ara and Etemadi, Mozziyar and Garcia-Vicente, Florencia and Gilbert, Fiona J. and Halling-Brown, Mark and Hassabis, Demis and Jansen, Sunny and Karthikesalingam, Alan and Kelly, Christopher J. and King, Dominic and Ledsam, Joseph R. and Melnick, David and Mostofi, Hormuz and Peng, Lily and Reicher, Joshua Jay and Romera-Paredes, Bernardino and Sidebottom, Richard and Suleyman, Mustafa and Tse, Daniel and Young, Kenneth C. and De Fauw, Jeffrey and Shetty, Shravya},
	title     = {{International Evaluation of an AI System for Breast Cancer Screening}},
	journal   = {Nature},
	year      = {2020},
	month     = {Jan},
	volume    = {577},
	pages     = {89--94},
	doi       = {10.1038/s41586-019-1799-6},
}

@inproceedings{Speith2022-XAI-TaxonomiesXAIMethods-ACM,
	author       = {Speith, Timo},
	title        = {{A Review of Taxonomies of Explainable Artificial Intelligence (XAI) Methods}},
	booktitle    = {{Proceedings of the 2022 ACM Conference on Fairness, Accountability, and Transparency}},
	year         = {2022},
	month        = {Jun},
	pages        = {2239--2250},
	publisher    = {ACM},
	address      = {New York, NY, USA},
	doi          = {10.1145/3531146.3534639},
}

@article{Ploug2020-XAI-ContestableAIDiagnostics-Elsevier,
	author       = {Ploug, Thomas and Holm, S{\o}ren},
	title        = {{The Four Dimensions of Contestable AI Diagnostics: A Patient-Centric Approach to Explainable AI}},
	journal      = {{Artificial Intelligence in Medicine}},
	year         = {2020},
	month        = {Sep},
	volume       = {107},
	pages        = {101901},
	publisher    = {Elsevier},
	address      = {Amsterdam, Netherlands},
	doi          = {10.1016/j.artmed.2020.101901},
}

@article{Alfrink2023-XAI-ContestableAIByDesign-Springer,
	author       = {Alfrink, Kars and Keller, Ianus and Kortuem, Gerd and Doorn, Neelke},
	title        = {{Contestable AI by Design: Towards a Framework}},
	journal      = {{Minds and Machines}},
	year         = {2023},
	month        = {Dec},
	volume       = {33},
	number       = {4},
	pages        = {613--639},
	publisher    = {Springer},
	address      = {Dordrecht, Netherlands},
	doi          = {10.1007/s11023-022-09611-z},
}

@inproceedings{Yurrita2023-XAI-ContestabilityFairness-ACM,
	author       = {Yurrita, Mireia and Draws, Tim and Balayn, Agathe and Murray-Rust, Dave and Tintarev, Nava and Bozzon, Alessandro},
	title        = {{Disentangling Fairness Perceptions in Algorithmic Decision-Making: The Effects of Explanations, Human Oversight, and Contestability}},
	booktitle    = {{Proceedings of the 2023 CHI Conference on Human Factors in Computing Systems}},
	year         = {2023},
	month        = {Apr},
	pages        = {1--21},
	publisher    = {ACM},
	address      = {New York, NY, USA},
	doi          = {10.1145/3544548.3581161},
}

@article{Vilone2021-XAI-NotionsExplainabilityEvaluation-Elsevier,
	author    = {Vilone, Giulia and Longo, Luca},
	title     = {Notions of Explainability and Evaluation Approaches for 
	Explainable Artificial Intelligence},
	journal   = {Information Fusion},
	volume    = {76},
	pages     = {89--106},
	year      = {2021},
	doi       = {10.1016/j.inffus.2021.05.009}
}

@article{Nauta2023-XAI-AnecdotalEvidenceQuantitative-ACM,
	author    = {Nauta, Meike and Trienes, Jan and Pathak, Shreyasi and Nguyen, 
	Elisa and Peters, Michelle and Schmitt, Yasmin and 
	Schl{\"o}tterer, J{\"o}rg and van Keulen, Maurice and 
	Seifert, Christin},
	title     = {From Anecdotal Evidence to Quantitative Evaluation Methods: 
	A Systematic Review on Evaluating Explainable {AI}},
	journal   = {ACM Computing Surveys},
	volume    = {55},
	number    = {13s},
	pages     = {1--42},
	year      = {2023},
	doi       = {10.1145/3583558}
}

@inproceedings{Apruzzese2023-XAI-AIRealWorldCybersecurity-IEEE,
	author    = {Apruzzese, Giovanni and Laskov, Pavel and Schneider, Johannes},
	title     = {{SoK: Pragmatic Assessment of Machine Learning for Network Intrusion Detection}},
	booktitle = {Proceedings of the IEEE European Symposium on Security and Privacy},
	year      = {2023},
	month     = {Jul},
	pages     = {1--19},
	publisher = {IEEE},
	address   = {Delft, Netherlands},
	doi       = {10.1109/EuroSP57164.2023.00016},
}

@article{Byrne2019-XAI-CounterfactualsXAI-IJCAI,
	author    = {Byrne, Ruth M. J.},
	title     = {{Counterfactuals in Explainable Artificial Intelligence (XAI): Evidence from Human Reasoning}},
	journal   = {Proceedings of the International Joint Conference on Artificial Intelligence},
	year      = {2019},
	month     = {Aug},
	pages     = {6276--6282},
	doi       = {10.24963/ijcai.2019/876},
}

@inproceedings{Cachada2019-XAI-AIIndustrialMaintenance-IEEE,
	author    = {Cachada, Ana and Barbosa, José and Leitão, Paulo and Alves Teixeira, José and Teixeira, José Paulo and Moreira, A. Hélder and Moreira, Marta},
	title     = {{Maintenance 4.0: Intelligent and Predictive Maintenance System Architecture}},
	booktitle = {Proceedings of the IEEE International Conference on Industrial Informatics},
	year      = {2019},
	month     = {Jul},
	pages     = {173--178},
	publisher = {IEEE},
	address   = {Helsinki, Finland},
	doi       = {10.1109/INDIN41052.2019.8972134},
}

@inproceedings{Cai2019-XAI-HumanCenteredToolsClinic-ACM,
	author    = {Cai, Carrie J. and Winter, Samantha and Steiner, David and Wilcox, Lauren and Terry, Michael},
	title     = {{"Hello AI": Uncovering the Onboarding Needs of Medical Practitioners for Human-AI Collaborative Decision-Making}},
	booktitle = {Proceedings of the ACM Conference on Human Factors in Computing Systems},
	year      = {2019},
	month     = {May},
	pages     = {1--24},
	publisher = {ACM},
	address   = {Glasgow, Scotland},
	doi       = {10.1145/3290605.3300333},
}

@article{Chromik2021-HumanXAI-IJHCS,
	author    = {Chromik, Michael and Eiband, Malin and Buchner, Felix
	and Krüger, Antonio and Butz, Andreas},
	title     = {Human-Centered {XAI}: Developing Design Patterns for
	Explanations of Clinical Decision Support Systems},
	journal   = {International Journal of Human-Computer Studies},
	volume    = {154},
	pages     = {102684},
	year      = {2021},
	publisher = {Elsevier},
	doi       = {10.1016/j.ijhcs.2021.102684}
}

@inproceedings{Chromik2021-XAI-HumanXAIInteractionTaxonomy-IUI,
	author    = {Chromik, Michael and Butz, Andres},
	title     = {{Human-XAI Interaction: A Review and Design Principles for Explanation User Interfaces}},
	booktitle = {Proceedings of the ACM International Conference on Intelligent User Interfaces},
	year      = {2021},
	month     = {Apr},
	pages     = {1--13},
	publisher = {ACM},
	address   = {College Station, TX, USA},
	doi       = {10.1145/3397481.3450686},
}

@inproceedings{DoshiVelez2017-XAI-RigorousInterpretability-ICML,
	author    = {Doshi-Velez, Finale and Kim, Been},
	title     = {{Towards a Rigorous Science of Interpretable Machine Learning}},
	booktitle = {Proceedings of the ICML Workshop on Human Interpretability in Machine Learning},
	pages	  = {},
	year      = {2017},
	month     = {Aug},
	address   = {Sydney, Australia},
}

@article{Guidotti2018-XAI-SurveyBlackBox-ACM,
	author    = {Guidotti, Riccardo and Monreale, Anna and Ruggieri, Salvatore and Turini, Franco and Giannotti, Fosca and Pedreschi, Dino},
	title     = {{A Survey of Methods for Explaining Black Box Models}},
	journal   = {ACM Computing Surveys},
	year      = {2018},
	month     = {Aug},
	volume    = {51},
	number    = {5},
	pages     = {1--42},
	doi       = {10.1145/3236009},
}

@article{Guidotti2018-XAI-LORE-arXiv,
	author       = {Guidotti, Riccardo and Monreale, Anna and Ruggieri, Salvatore and Pedreschi, Dino and Turini, Franco and Giannotti, Fosca},
	title        = {{Local Rule-Based Explanations of Black Box Decision Systems}},
	journal      = {{arXiv}},
	year         = {2018},
	month        = {May},
	volume       = {abs/1805.10820},
	publisher    = {arXiv},
	address      = {Ithaca, NY, USA},
	doi          = {10.48550/arXiv.1805.10820},
}

@techreport{Hoffman2018-XAI-MetricsExplainableAI-arXiv,
	author      = {Hoffman, Robert R. and Mueller, Shane T. and Klein, Gary and Litman, Jordan},
	title       = {{Metrics for Explainable AI: Challenges and Prospects}},
	institution = {arXiv},
	year        = {2018},
	month       = {Dec},
	number      = {arXiv:1812.04608},
}

@techreport{Jin2021-XAI-EUCA-arXiv,
	author      = {Jin, Weina and Fan, Jiannan and Gromala, Diane and Pasquier, Philippe and Hamarneh, Ghassan},
	title       = {{EUCA: The End-User-Centered Explainable AI Framework}},
	institution = {arXiv},
	year        = {2021},
	month       = {Feb},
	number      = {arXiv:2102.02437},
}

@inproceedings{Liao2020-XAI-QuestioningAI-CHI-ACM,
	author    = {Liao, Q. Vera and Gruen, Daniel and Miller, Sarah},
	title     = {{Questioning the AI: Informing Design Practices for Explainable AI User Experiences}},
	booktitle = {Proceedings of the ACM Conference on Human Factors in Computing Systems},
	year      = {2020},
	month     = {Apr},
	pages     = {1--15},
	publisher = {ACM},
	address   = {Honolulu, HI, USA},
	doi       = {10.1145/3313831.3376590},
}

@techreport{Liao2021-XAI-QuestionDrivenDesign-arXiv,
	author      = {Liao, Q. Vera and Varshney, Kush R.},
	title       = {{Human-Centered Explainable AI (XAI): From Algorithms to User Experiences}},
	institution = {arXiv},
	year        = {2021},
	month       = {Oct},
	number      = {arXiv:2110.10790},
}

@book{Mayring2014-XAI-QualContentAnalysis-Klagenfurt,
	author    = {Mayring, Philipp},
	title     = {{Qualitative Content Analysis: Theoretical Foundation, Basic Procedures and Software Solution}},
	publisher = {SSOAR},
	year      = {2014},
	month     = {May},
	address   = {Klagenfurt, Austria},
}

@article{Miller2019-XAI-ExplanationAI-Elsevier,
	author    = {Miller, Tim},
	title     = {{Explanation in Artificial Intelligence: Insights from the Social Sciences}},
	journal   = {Artificial Intelligence},
	year      = {2019},
	month     = {Feb},
	volume    = {267},
	pages     = {1--38},
	doi       = {10.1016/j.artint.2018.07.007},
}

@article{Mohseni2021-XAI-MultidisciplinaryHAI-ACM,
	author    = {Mohseni, Sina and Zarei, Niloofar and Ragan, Eric D.},
	title     = {{A Multidisciplinary Survey and Framework for Design and Evaluation of Explainable AI Systems}},
	journal   = {ACM Transactions on Interactive Intelligent Systems},
	year      = {2021},
	month     = {Sep},
	volume    = {11},
	number    = {3--4},
	pages     = {1--45},
	doi       = {10.1145/3387166},
}

@inproceedings{Panigutti2023-XAI-CoDesignClinicalXAI-ACM,
	author    = {Panigutti, Cecilia and Beretta, Andrea and Fadda, Daniele and Giannotti, Fosca and Pedreschi, Dino and Perotti, Alan and Rinzivillo, Salvatore},
	title     = {{Co-design of Human-Centered, Explainable AI for Clinical Decision Support}},
	booktitle = {ACM Transactions on Interactive Intelligent Systems},
	year      = {2023},
	month     = {Mar},
	volume    = {13},
	number    = {4},
	pages     = {1--35},
	publisher = {ACM},
	doi       = {10.1145/3587271},
}

@book{Schreier2012-XAI-QualContentPractice-Sage,
	author    = {Schreier, Margrit},
	title     = {{Qualitative Content Analysis in Practice}},
	publisher = {SAGE Publications},
	year      = {2012},
	month     = {Feb},
	address   = {London, UK},
}

@inproceedings{Sipos2023-ExplanationNeeds-MuC,
	author    = {Sipos, Lars and Sch{\"a}fer, Ulrike and Glinka, Katrin
	and M{\"u}ller-Birn, Claudia},
	title     = {Identifying Explanation Needs of End-users: Applying and
	Extending the {XAI} Question Bank},
	booktitle = {Proceedings of Mensch und Computer 2023},
	series    = {MuC '23},
	year      = {2023},
	publisher = {ACM},
	address   = {New York, NY, USA},
	doi       = {10.1145/3603555.3608551}
}

@article{Syed2023-NotionXAI-arxiv,
	author    = {Syed, Tahir Abbas and others},
	title     = {Notion of Explainable Artificial Intelligence: An Empirical
	Investigation from a Users Perspective},
	journal   = {arXiv preprint arXiv:2311.02102},
	year      = {2023},
	url       = {https://arxiv.org/abs/2311.02102}
}

@article{VanderWaa2021-XAI-RuleExampleComparison-Elsevier,
	author    = {van der Waa, Jasper and Nieuwburg, Elisabeth G. I. and Cremers, Anita and Neerincx, Mark},
	title     = {{Evaluating XAI: A Comparison of Rule-Based and Example-Based Explanations}},
	journal   = {Artificial Intelligence},
	year      = {2021},
	month     = {Apr},
	volume    = {291},
	pages     = {103404},
	doi       = {10.1016/j.artint.2020.103404},
}

@article{Wachter2018-XAI-CounterfactualExplanations-HarvardJOLT,
	author       = {Wachter, Sandra and Mittelstadt, Brent and Russell, Chris},
	title        = {{Counterfactual Explanations Without Opening the Black Box: Automated Decisions and the GDPR}},
	journal      = {{Harvard Journal of Law \& Technology}},
	year         = {2018},
	month        = {Spring},
	volume       = {31},
	number       = {2},
	pages        = {841--887},
	publisher    = {Harvard Journal of Law \& Technology},
	address      = {Cambridge, MA, USA},
}

@incollection{Chari2020-XAI-ExplanationOntology-Springer,
	author       = {Chari, Shruthi and Seneviratne, Oshani and Gruen, Daniel M. and Foreman, Morgan A. and Das, Amar K. and McGuinness, Deborah L.},
	title        = {{A Model of Explanations for User-Centered AI}},
	booktitle    = {{The Semantic Web -- ISWC 2020}},
	editor       = {Allemang, Dean and Dumontier, Michel and Kellogg, Greg and Maleshkova, Maria and Pesquita, Catia and Stadler, Claus and Cruz, Isabel F.},
	year         = {2020},
	month        = {Nov},
	pages        = {244--259},
	publisher    = {Springer},
	address      = {Cham, Switzerland},
	doi          = {10.1007/978-3-030-62466-8\_15},
}

@article{Chari2024-XAI-ExplanationOntology-SemanticWeb,
	author       = {Chari, Shruthi and Seneviratne, Oshani and Ghalwash, Mohamed and Shirai, Sola and Gruen, Daniel M. and Meyer, Pablo and Chakraborty, Prithwish and McGuinness, Deborah L.},
	title        = {{Explanation Ontology: A General-Purpose, Semantic Representation for Supporting User-Centered Explanations}},
	journal      = {{Semantic Web}},
	year         = {2024},
	month        = {Jan},
	publisher    = {IOS Press},
	address      = {Amsterdam, Netherlands},
	doi          = {10.3233/SW-233282},
}

@article{Khosravi2022-XAI-XAIED-ComputersEducationAI,
	author       = {Khosravi, Hassan and Buckingham Shum, Simon and Chen, Guanliang and Conati, Cristina and Tsai, Yi-Shan and Kay, Judy and Knight, Simon and Martinez-Maldonado, Roberto and Sadiq, Shazia and Ga{\v{s}}evi{\'c}, Dragan},
	title        = {{Explainable Artificial Intelligence in Education}},
	journal      = {{Computers and Education: Artificial Intelligence}},
	year         = {2022},
	month        = {Jan},
	volume       = {3},
	pages        = {100074},
	publisher    = {Elsevier},
	address      = {Amsterdam, Netherlands},
	doi          = {10.1016/j.caeai.2022.100074},
}

@article{Zhou2026-XAI-TutorMoveTaxonomy-arXiv,
	author       = {Zhou, Zhuqian and Vanacore, Kirk and Thompson, Tamisha and St John, Jennifer and Kizilcec, Ren{\'e}},
	title        = {{Tutor Move Taxonomy: A Theory-Aligned Framework for Analyzing Instructional Moves in Tutoring}},
	journal      = {{arXiv}},
	year         = {2026},
	month        = {Mar},
	volume       = {abs/2603.05778},
	publisher    = {arXiv},
	address      = {Ithaca, NY, USA},
	doi          = {10.48550/arXiv.2603.05778},
}

@article{Blumer1954-XAI-WrongSocialTheory-ASR,
	author    = {Blumer, Herbert},
	title     = {{What Is Wrong with Social Theory?}},
	journal   = {American Sociological Review},
	year      = {1954},
	month     = {Feb},
	volume    = {19},
	number    = {1},
	pages     = {3--10},
	doi       = {10.2307/2088165},
	publisher = {American Sociological Association},
	address   = {Washington, DC, USA},
}

@inproceedings{Ribeiro2018-XAI-AnchorsHighPrecision-AAAI,
	author    = {Marco Tulio Ribeiro and Sameer Singh and Carlos Guestrin},
	title     = {Anchors: {H}igh-{P}recision Model-{A}gnostic Explanations},
	booktitle = {Proceedings of the Thirty-Second AAAI Conference on Artificial Intelligence},
	volume    = {32},
	number    = {1},
	pages     = {1527--1535},
	year      = {2018},
	publisher = {AAAI Press},
	doi       = {10.1609/aaai.v32i1.11491}
}

@inproceedings{Degen2021-inproceedings,
	author    = 	{Degen, Helmut and Budnik, Christof and Chitre, Kunal and Lintereur, Andrew},
	title     = 	{{How to Explain it to Facility Managers? A Qualitative, Industrial User Research Study for Explainability}},
	booktitle = 	{{HCI International 2021 - Late Breaking Papers: Multimodality, eXtended Reality, and Artificial Intelligence. 23rd HCI International Conference, HCII 2021, Virtual Event, July 24 -- 29, 2021, Proceedings.}},
	year      = 	{2021},
	address   = 	{Cham, Switzerland},
	editor    = 	{Stephanidis, Constantine and Kurosu, Masaaki and Chen, Jessie Y. C. and Fragomeni, Gino and Streitz, Norbert and Konomi, Shin'ichi and Degen, Helmut and Ntoa, Stavroula},
	pages     = 	{401--422},
	pagenum   = 	{22},
	publisher = 	{Springer Nature Switzerland AG},
	isbn      =       {978-3-030-90962-8},
	DOI = 		      {10.1007/978-3-030-90963-5\_31},
	howpublished =    {\url{https://doi.org/10.1007/978-3-030-90963-5\_31}},
}

@InProceedings{Degen2022-inproceedings,
	author    =     {Degen, Helmut and Budnik, Christof and Conte, Greg and Lintereur, Andrew and Weber, Seth},
	title     =     {{How to Explain it to Energy Engineers? A qualitative user study about trustworthiness, understandability, and actionability}},
	booktitle =     {{HCI} International 2022 - Late Breaking Papers: Multimodality, eXtended Reality, and Artificial Intelligence. 24th HCI International Conference, HCII 2022, Virtual Event, June 24 -- July 1, 2022, Proceedings.},
	year      =     {2022},
	address   =     {Cham, Switzerland},
	editor    =     {Stephanidis, Constantine and Kurosu, Masaaki and Chen, Jessie Y. C. and Fragomeni, Gino and Streitz, Norbert and Konomi, Shin'ichi and Degen, Helmut and Ntoa, Stavroula},
	pages     =     {1--23},
	publisher =     {Springer Nature Switzerland AG},
	howpublished =  {\url{https://doi.org/10.1007/978-3-031-21707-4\_20}},
	DOI		  =     {10.1007/978-3-031-21707-4\_20},
	isbn      =     {},
	pagenum   =     {23},
}

@inproceedings{Degen2023-inproceedings,
	author =        {Degen, Helmut and Budnik, Christof and Gross, Ralf and Rothering, Marcel},
	title = 	    {{How To Explain It To a Model Manager? A Qualitative User Study About Understandability, Trustworthiness, Actionability, and Action Efficacy}},
	year = 		    {2023},
	isbn = 		    {978-3-031-35890-6},
	publisher = 	{Springer Nature Switzerland AG},
	address = 	 	{Cham, Switzerland},
	howpublished =  {\url{https://doi.org/10.1007/978-3-031-35891-3\_14}},
	doi = 		    {10.1007/978-3-031-35891-3\_14},
	booktitle = 	{Artificial Intelligence in HCI: 4th International Conference, AI-HCI 2023, Held as Part of the 25th HCI International Conference, HCII 2023, Copenhagen, Denmark, July 23 -- 28, 2023, Proceedings, Part I},
	pages = 	   	{209--242},
	numpages = 	    {34},
	location = 	    {Copenhagen, Denmark},
}

@inproceedings{Degen2024-inproceedings,
	author = 		{Degen, Helmut and Budnik, Christof},
	title = 		{{How to Explain it to System Testers? A Qualitative User Study About Understandability, Validatability, Predictability, and Trustworthiness}},
	year = 			{2024},
	isbn = 			{978-3-031-60605-2},
	publisher = 	{Springer-Verlag},
	address = 		{Berlin, Heidelberg, Germany},
	howpublished =  {\url{https://doi.org/10.1007/978-3-031-60606-9\_10}},
	doi = 			{10.1007/978-3-031-60606-9\_10},
	booktitle = 	{{Artificial Intelligence in HCI: 5th International Conference, AI-HCI 2024, Held as Part of the 26th HCI International Conference, HCII 2024, Washington, DC, USA, June 29 -- July 4, 2024, Proceedings, Part I}},
	pages = 		{153--178},
	numpages = 		{26},
	location = 		{Washington DC, USA},
}

@inproceedings{Degen2025-XAI-ExplainDataScientists-Springer-InProceedings,
	author       = {Degen, Helmut and Min, Ziran and Nagaraja, Parinitha},
	editor       = {Degen, Helmut and Ntoa, Stavroula},
	title        = {{How to Explain It to Data Scientists?}},
	booktitle    = {{Artificial Intelligence in HCI: 6th International Conference, AI-HCI 2025, Held as Part of the 27th HCI International Conference, HCII 2025, Gothenburg, Sweden, June 22 -- 27, 2025, Proceedings, Part I}},
	year         = {2025},
	pages        = {41--66},
	publisher    = {Springer Nature Switzerland AG},
	address      = {Cham, Switzerland},
	isbn         = {978-3-031-93412-4},
	doi          = {10.1007/978-3-031-93412-4\_3},
	howpublished = {\url{https://doi.org/10.1007/978-3-031-93412-4\_3}},
	location = 		{Gothenburg, Sweden},
}

@misc{Degen2025-misc,
	author = 		{Degen, Helmut and Min, Ziran and Nagaraja, Parinitha},
	title = 		{{How to explain it to data scientists? A mixed-methods user study about explainable AI, using mental models for explanations}},
	year = 			{2025},
	eprint=         {2502.16083},
	archivePrefix=  {arXiv},
	primaryClass=   {cs.HC},
	howpublished =  {\url{https://arxiv.org/abs/2502.16083}},
}

@article{Lynn1986-XAI-DetermQuantContentValidity-LWW,
	author = {Lynn, Mary R.},
	title = {{Determination and quantification of content validity}},
	journal = {Nursing Research},
	year = {1986},
	month = {Nov},
	volume = {35},
	number = {6},
	pages = {382--386},
	doi = {10.1097/00006199-198611000-00017},
	publisher = {Lippincott Williams \& Wilkins},
	address = {Philadelphia, PA, USA},
}

@article{OConnor2020-XAI-IntercoderReliabilityQualResearch-SAGE,
	author       = {O'Connor, Cliodhna and Joffe, Helene},
	title        = {{Intercoder Reliability in Qualitative Research: Debates and Practical Guidelines}},
	journal      = {{International Journal of Qualitative Methods}},
	year         = {2020},
	month        = {Jan},
	volume       = {19},
	pages        = {1--13},
	publisher    = {SAGE Publications},
	address      = {Thousand Oaks, CA, USA},
	doi          = {10.1177/1609406919899220},
}

@article{Polit2006-XAI-ContentValidityIndex-Wiley,
	author = {Polit, Denise F. and Beck, Cheryl Tatano},
	title = {{The content validity index: are you sure you know what's being reported? Critique and recommendations}},
	journal = {Research in Nursing \& Health},
	year = {2006},
	month = {Oct},
	volume = {29},
	number = {5},
	pages = {489--497},
	doi = {10.1002/nur.20147},
	publisher = {Wiley Periodicals, Inc.},
	address = {Hoboken, NJ, USA},
}

@article{Cohen1960-XAI-AgreementNominal-EdPsychMeas,
	author       = {Cohen, Jacob},
	title        = {{A Coefficient of Agreement for Nominal Scales}},
	journal      = {{Educational and Psychological Measurement}},
	year         = {1960},
	month        = {Apr},
	volume       = {20},
	number       = {1},
	pages        = {37--46},
	publisher    = {SAGE Publications},
	address      = {Thousand Oaks, CA, USA},
	doi          = {10.1177/001316446002000104},
}

@article{Krippendorff2004-XAI-ReliabilityMisconceptions-HCR,
	author = {Krippendorff, Klaus},
	title = {{Reliability in Content Analysis: Some Common Misconceptions and Recommendations}},
	journal = {Human Communication Research},
	year = {2004},
	month = {Jul},
	volume = {30},
	number = {3},
	pages = {411--433},
	doi = {10.1111/j.1468-2958.2004.tb00738.x},
	publisher = {Oxford University Press},
	address = {Oxford, UK},
}

@book{Saldana2013-XAI-CodingManual-Sage,
	author    = {Salda{\~n}a, Johnny},
	title     = {{The Coding Manual for Qualitative Researchers}},
	year      = {2013},
	month     = {Mar},
	edition   = {2},
	publisher = {SAGE Publications},
	address   = {Thousand Oaks},
	isbn      = {978-1-4462-7434-8},
}

@book{Charmaz2014-XAI-ConstructingGT-Sage,
	author    = {Charmaz, Kathy},
	title     = {{Constructing Grounded Theory}},
	year      = {2014},
	month     = {Jan},
	edition   = {2},
	publisher = {SAGE Publications},
	address   = {London, UK},
}

@article{Braun2006-XAI-UsingTA-TQR,
	author    = {Braun, Virginia and Clarke, Victoria},
	title     = {{Using Thematic Analysis in Psychology}},
	journal   = {Qualitative Research in Psychology},
	year      = {2006},
	month     = {Jan},
	volume    = {3},
	number    = {2},
	pages     = {77--101},
	publisher = {Taylor \& Francis},
	address   = {Abingdon, UK},
	doi       = {10.1191/1478088706qp063oa},
}

@article{Braun2019-XAI-ReflexiveTA-CounselPsych,
	author    = {Braun, Virginia and Clarke, Victoria},
	title     = {{Thematic Analysis}},
	journal   = {Counselling and Psychotherapy Research},
	year      = {2019},
	month     = {Apr},
	volume    = {19},
	number    = {2},
	pages     = {77--83},
	publisher = {Wiley},
	address   = {Chichester, UK},
	doi       = {10.1002/capr.12362},
}

@misc{ARPA-H_UPGRADE_2024,
	author       = {{Advanced Research Projects Agency for Health (ARPA-H)}},
	title        = {{UPGRADE: Universal Patching and Remediation for Autonomous Defense}},
	year         = {2024},
	howpublished = {\url{https://arpa-h.gov/explore-funding/programs/upgrade}},
	note         = {Accessed: 2026-04-05}
}

@book{Russell2020,
	author    = {Stuart Russell and Peter Norvig},
	title     = {Artificial Intelligence: {A} Modern Approach},
	edition   = {4th},
	publisher = {Pearson},
	year      = {2020}
}

@book{Pearl2018,
	author    = {Judea Pearl and Dana Mackenzie},
	title     = {The Book of Why: {T}he New Science of Cause and Effect},
	publisher = {Basic Books},
	year      = {2018}
}

@book{Mill1843,
	author    = {John Stuart Mill},
	title     = {A System of Logic},
	publisher = {John W. Parker},
	year      = {1843}
}

@incollection{Lipton1990,
	author    = {Peter Lipton},
	title     = {Contrastive explanation},
	booktitle = {Explanation and Its Limits},
	editor    = {Dudley Knowles},
	publisher = {Cambridge University Press},
	year      = {1990},
	pages     = {247--266}
}

@article{Gentner1983,
	author  = {Dedre Gentner},
	title   = {Structure-mapping: {A} theoretical framework for analogy},
	journal = {Cognitive Science},
	volume  = {7},
	number  = {2},
	pages   = {155--170},
	year    = {1983},
	doi     = {10.1207/s15516709cog0702\_3}
}

@article{Aamodt1994,
	author  = {Agnar Aamodt and Enric Plaza},
	title   = {Case-based reasoning: {F}oundational issues, methodological variations, and system approaches},
	journal = {AI Communications},
	volume  = {7},
	number  = {1},
	pages   = {39--59},
	year    = {1994}
}

@book{Brunswik1956-XAI-RepresentativeDesign-UCPress,
	author    = {Brunswik, Egon},
	title     = {Perception and the Representative Design of Psychological Experiments},
	edition   = {2nd},
	publisher = {University of California Press},
	address   = {Berkeley, CA},
	year      = {1956},
	lccn      = {56-7310},
	url       = {https://archive.org/details/perceptionrepres0000brun}
}

@book{Suchman1987-XAI-PlansSituatedActions-CUP,
	author    = {Suchman, Lucy A.},
	title     = {Plans and Situated Actions: The Problem of Human-Machine Communication},
	publisher = {Cambridge University Press},
	address   = {Cambridge, UK},
	year      = {1987},
	isbn      = {9780521337397},
	url       = {https://www.cambridge.org/9780521337397}
}

@incollection{Bannon1991-XAI-HumanFactorsHumanActors-LawrenceErlbaum,
	author    = {Bannon, Liam J.},
	title     = {From Human Factors to Human Actors: The Role of Psychology and Human-Computer Interaction Studies in System Design},
	booktitle = {Design at Work: Cooperative Design of Computer Systems},
	editor    = {Greenbaum, Joan and Kyng, Morten},
	publisher = {Lawrence Erlbaum Associates},
	address   = {Hillsdale, NJ},
	year      = {1991},
	pages     = {25--44},
	isbn      = {9780805806113},
	doi       = {10.1201/9781315800349-4},
	url       = {https://doi.org/10.1201/9781315800349-4}
}

@book{Dourish2001-XAI-WhereTheActionIs-MITPress,
	author    = {Dourish, Paul},
	title     = {Where the Action Is: The Foundations of Embodied Interaction},
	publisher = {MIT Press},
	address   = {Cambridge, MA},
	year      = {2001},
	isbn      = {9780262041966},
	doi       = {10.7551/mitpress/7221.001.0001},
	url       = {https://doi.org/10.7551/mitpress/7221.001.0001}
}

@inproceedings{Ehsan2020-XAI-ReflectiveSociotechnicalApproach-Springer,
	author    = {Ehsan, Upol and Riedl, Mark O.},
	title     = {Human-Centered Explainable AI: Towards a Reflective Sociotechnical Approach},
	booktitle = {HCI International 2020 -- Late Breaking Papers: Multimodality and Intelligence},
	editor    = {Stephanidis, Constantine and Kurosu, Masaaki and Degen, Helmut and Reinerman-Jones, Lauren},
	series    = {Lecture Notes in Computer Science},
	volume    = {12424},
	publisher = {Springer},
	address   = {Cham},
	year      = {2020},
	pages     = {449--466},
	doi       = {10.1007/978-3-030-60117-1\_33},
	url       = {https://doi.org/10.1007/978-3-030-60117-1\_33}
}

@article{Ehsan2023-XAI-SociotechnicalGap-ACM,
	author    = {Ehsan, Upol and Saha, Koustuv and De Choudhury, Munmun and Riedl, Mark O.},
	title     = {Charting the Sociotechnical Gap in Explainable AI: A Framework to Address the Gap in XAI},
	journal   = {Proceedings of the ACM on Human-Computer Interaction},
	volume    = {7},
	number    = {CSCW1},
	articleno = {34},
	pages     = {1--32},
	year      = {2023},
	publisher = {ACM},
	doi       = {10.1145/3579467},
	url       = {https://doi.org/10.1145/3579467}
}

@article{Pirolli1999-XAI-InformationForaging-PsychReview,
	author   = {Pirolli, Peter and Card, Stuart K.},
	title    = {{Information Foraging}},
	journal  = {Psychological Review},
	year     = {1999},
	month    = {Oct},
	volume   = {106},
	number   = {4},
	pages    = {643--675},
	doi      = {10.1037/0033-295X.106.4.643},
}

@article{Endsley1995-XAI-SituationAwareness-HF,
	author   = {Endsley, Mica R.},
	title    = {{Toward a Theory of Situation Awareness in Dynamic Systems}},
	journal  = {Human Factors},
	year     = {1995},
	month    = {Mar},
	volume   = {37},
	number   = {1},
	pages    = {32--64},
	doi      = {10.1518/001872095779049543},
}

@incollection{Hart1988-XAI-NASATLX-ElsevierNorthHolland,
	author    = {Hart, Sandra G. and Staveland, Lowell E.},
	title     = {Development of {NASA-TLX} (Task Load Index): Results of Empirical and Theoretical Research},
	booktitle = {Human Mental Workload},
	editor    = {Hancock, Peter A. and Meshkati, Najmedin},
	series    = {Advances in Psychology},
	volume    = {52},
	publisher = {North-Holland, Elsevier},
	address   = {Amsterdam},
	year      = {1988},
	pages     = {139--183},
	doi       = {10.1016/S0166-4115(08)62386-9},
	url       = {https://doi.org/10.1016/S0166-4115(08)62386-9}
}

@article{Lee2004-XAI-TrustAutomation-HumanFactors,
	author    = {Lee, John D. and See, Katrina A.},
	title     = {Trust in Automation: Designing for Appropriate Reliance},
	journal   = {Human Factors: The Journal of the Human Factors and Ergonomics Society},
	volume    = {46},
	number    = {1},
	pages     = {50--80},
	year      = {2004},
	publisher = {SAGE Publications},
	doi       = {10.1518/hfes.46.1.50_30392},
	url       = {https://doi.org/10.1518/hfes.46.1.50_30392}
}

@article{Parasuraman2010-XAI-ComplacencyBiasAutomation-HumanFactors,
	author    = {Parasuraman, Raja and Manzey, Dietrich H.},
	title     = {Complacency and Bias in Human Use of Automation: An Attentional Integration},
	journal   = {Human Factors: The Journal of the Human Factors and Ergonomics Society},
	volume    = {52},
	number    = {3},
	pages     = {381--410},
	year      = {2010},
	publisher = {SAGE Publications},
	doi       = {10.1177/0018720810376055},
	url       = {https://doi.org/10.1177/0018720810376055}
}

@article{Kim2024-HumanCenteredEvaluationXAI-Frontiers,
	author       = {Kim, Jenia and Maathuis, Henry and Sent, Danielle},
	title        = {{Human-Centered Evaluation of Explainable AI Applications: A Systematic Review}},
	journal      = {{Frontiers in Artificial Intelligence}},
	year         = {2024},
	month        = {Jun},
	volume       = {7},
	pages        = {1456486},
	doi          = {10.3389/frai.2024.1456486},
	issn         = {2624-8212},
	publisher    = {Frontiers Media S.A.},
	address      = {Lausanne, Switzerland},
	howpublished = {\url{https://www.frontiersin.org/articles/10.3389/frai.2024.1456486}},
}

@article{Bayer2022-RoleDomainExpertiseTrust-JDS,
	author       = {Bayer, Sarah and Gimpel, Henner and Markgraf, Moritz},
	title        = {{The Role of Domain Expertise in Trusting and Following Explainable AI Decision Support Systems}},
	journal      = {{Journal of Decision Systems}},
	volume       = {32},
	number       = {1},
	pages        = {110--138},
	year         = {2022},
	month        = {Dec},
	publisher    = {Taylor \& Francis},
	address      = {Abingdon, United Kingdom},
	doi          = {10.1080/12460125.2021.1958505},
	howpublished =  {\url{https://doi.org/10.1080/12460125.2021.1958505}},
}

@inproceedings{BranleyBell2020-TrustUnderstandingXAI-HCI,
	author       = {Branley{-}Bell, Dawn and Whitworth, Rebecca and Coventry, Lynne},
	title        = {{User Trust and Understanding of Explainable AI: Exploring Algorithm Visualisations and User Biases}},
	booktitle    = {{Human-Computer Interaction: Human Values and Quality of Life}},
	editor       = {Kurosu, Masaaki},
	pages        = {382--399},
	year         = {2020},
	month        = {Jul},
	publisher    = {Springer International Publishing},
	address      = {Cham, Switzerland},
	series       = {Lecture Notes in Computer Science},
	volume       = {12183},
	doi          = {10.1007/978-3-030-49065-2\_27},
	isbn         = {978-3-030-49065-2},
	howpublished =  {\url{https://doi.org/10.1007/978-3-030-49065-2\_27}},
}

@inproceedings{Bunde2021-AIAssistedHateSpeechDetection-HICSS,
	author       = {Bunde, Enrico},
	title        = {{AI-Assisted and Explainable Hate Speech Detection for Social Media Moderators: A Design Science Approach}},
	booktitle    = {{Proceedings of the 54th Hawaii International Conference on System Sciences (HICSS)}},
	year         = {2021},
	month        = {Jan},
	pages        = {1--10},
	publisher    = {University of Hawai‘i at Mānoa},
	doi          = {10.24251/HICSS.2021.293},
	howpublished =  {\url{https://hdl.handle.net/10125/70865}},
}

@incollection{Faulhaber2021-EffectExplanationsTrust-MuC,
	author       = {Faulhaber, Anja K. and Ni, Ina and Schmidt, Ludger},
	title        = {{The Effect of Explanations on Trust in an Assistance System for Public Transport Users and the Role of the Propensity to Trust}},
	booktitle    = {{Proceedings of Mensch und Computer 2021}},
	pages        = {303--310},
	year         = {2021},
	month        = {Sep},
	publisher    = {Association for Computing Machinery},
	address      = {New York, NY, USA},
	doi          = {10.1145/3473856.3473886},
	howpublished =  {\url{https://doi.org/10.1145/3473856.3473886}},
	isbn         = {978-1-4503-8645-6},
	series       = {MuC '21},
	location     = {Ingolstadt, Germany},
}

@article{Fu2022-GPT2SP-TSE,
	author       = {Fu, Michael and Tantithamthavorn, Chakkrit},
	title        = {{GPT2SP: A Transformer-Based Agile Story Point Estimation Approach}},
	journal      = {{IEEE Transactions on Software Engineering}},
	volume       = {49},
	number       = {2},
	pages        = {611--625},
	year         = {2023},
	month        = {Feb},
	publisher    = {IEEE},
	address      = {Piscataway, NJ, USA},
	doi          = {10.1109/TSE.2022.3158252},
	issn         = {0098-5589},
	howpublished =  {\url{https://doi.org/10.1109/TSE.2022.3158252}},
}

@inproceedings{Kartikeya2022-TrustTransparencyXAI-SAI,
	author       = {Kartikeya, Arnav},
	title        = {{Examining Correlation Between Trust and Transparency with Explainable Artificial Intelligence}},
	booktitle    = {{Intelligent Computing}},
	editor       = {Arai, Kohei},
	pages        = {353--358},
	year         = {2022},
	month        = {Jul},
	publisher    = {Springer International Publishing},
	address      = {Cham, Switzerland},
	doi          = {10.1007/978-3-031-10464-0\_23},
	isbn         = {978-3-031-10464-0},
	howpublished =  {\url{https://doi.org/10.1007/978-3-031-10464-0\_23}},
}

@inproceedings{Kuehnlenz2023-SituationalExplanationsTrust-RO-MAN,
	author       = {K{\"u}hnlenz, Kolja and K{\"u}hnlenz, Barbara},
	title        = {{Study on the Impact of Situational Explanations and Prior Information Given to Users on Trust and Perceived Intelligence in Autonomous Driving in a Video-Based 2x2 Design}},
	booktitle    = {{Proceedings of the 2023 32nd IEEE International Conference on Robot and Human Interactive Communication (RO-MAN)}},
	pages        = {1509--1513},
	year         = {2023},
	month        = {Aug},
	publisher    = {IEEE},
	address      = {Piscataway, NJ, USA},
	doi          = {10.1109/RO-MAN57019.2023.10309319},
}

@article{Lundberg2022-TrustworthyIVNIDS-XAI-IEEEAccess,
	author       = {Lundberg, Hampus and Mowla, Nishat I. and Abedin, Sarder Fakhrul and Thar, Kyi and Mahmood, Aamir and Gidlund, Mikael and Raza, Shahid},
	title        = {{Experimental Analysis of Trustworthy In-Vehicle Intrusion Detection System Using eXplainable Artificial Intelligence (XAI)}},
	journal      = {{IEEE Access}},
	volume       = {10},
	pages        = {102831--102841},
	year         = {2022},
	month        = {Sep},
	publisher    = {IEEE},
	address      = {Piscataway, NJ, USA},
	doi          = {10.1109/ACCESS.2022.3208573},
	issn         = {2169-3536},
	howpublished =  {\url{https://doi.org/10.1109/ACCESS.2022.3208573}},
}

@inproceedings{Okumura2023-MIPCE-CounterfactualExplanations-ICANN,
	author       = {Okumura, Hiroyuki and Nagao, Tomoharu},
	title        = {{MIPCE: Generating Multiple Patches Counterfactual-Changing Explanations for Time Series Classification}},
	booktitle    = {{Artificial Neural Networks and Machine Learning -- ICANN 2023}},
	editor       = {Iliadis, Lazaros and Papaleonidas, Antonios and Angelov, Plamen and Jayne, Chrisina},
	pages        = {231--242},
	year         = {2023},
	month        = {Sep},
	publisher    = {Springer Nature Switzerland},
	address      = {Cham, Switzerland},
	doi          = {10.1007/978-3-031-44223-0\_19},
	isbn         = {978-3-031-44223-0},
	howpublished =  {\url{https://doi.org/10.1007/978-3-031-44223-0\_19}},
}

@inproceedings{Reeder2023-EvaluatingXAIUserGender-HCII,
	author       = {Reeder, Samuel and Jensen, Joshua and Ball, Robert},
	title        = {{Evaluating Explainable AI (XAI) in Terms of User Gender and Educational Background}},
	booktitle    = {{Artificial Intelligence in HCI}},
	editor       = {Degen, Helmut and Ntoa, Stavroula},
	pages        = {286--304},
	year         = {2023},
	month        = {Jul},
	publisher    = {Springer Nature Switzerland},
	address      = {Cham, Switzerland},
	series       = {Lecture Notes in Computer Science},
	doi          = {10.1007/978-3-031-35891-3\_18},
	isbn         = {978-3-031-35891-3},
	howpublished =  {\url{https://doi.org/10.1007/978-3-031-35891-3\_18}},
}

@article{Selten2023-JustLikeIThought-PAR,
	author       = {Selten, Friso and Robeer, Marcel and Grimmelikhuijsen, Stephan},
	title        = {{``Just Like I Thought'': Street-Level Bureaucrats Trust AI Recommendations if They Confirm Their Professional Judgment}},
	journal      = {{Public Administration Review}},
	volume       = {83},
	number       = {2},
	pages        = {263--278},
	year         = {2023},
	month        = {Mar},
	publisher    = {Wiley},
	address      = {Hoboken, NJ, USA},
	doi          = {10.1111/puar.13602},
	howpublished =  {\url{https://onlinelibrary.wiley.com/doi/10.1111/puar.13602}},
}

@article{Upasane2023-Type2FuzzyXAI-TAAI,
	author       = {Upasane, Shreyas J. and Hagras, Hani and Anisi, Mohammad Hossein and Savill, Stuart and Taylor, Ian and Manousakis, Kostas},
	title        = {{A Type-2 Fuzzy-Based Explainable AI System for Predictive Maintenance Within the Water Pumping Industry}},
	journal      = {{IEEE Transactions on Artificial Intelligence}},
	volume       = {5},
	number       = {2},
	pages        = {490--504},
	year         = {2024},
	month        = {Apr},
	publisher    = {IEEE},
	address      = {Piscataway, NJ, USA},
	doi          = {10.1109/TAI.2023.3279808},
}

@inproceedings{Wang2021-AreExplanationsHelpful-IUI,
	author       = {Wang, Xinru and Yin, Ming},
	title        = {{Are Explanations Helpful? A Comparative Study of the Effects of Explanations in AI-Assisted Decision-Making}},
	booktitle    = {{Proceedings of the 26th International Conference on Intelligent User Interfaces}},
	series       = {IUI '21},
	pages        = {318--328},
	year         = {2021},
	month        = {Apr},
	publisher    = {Association for Computing Machinery},
	address      = {New York, NY, USA},
	doi          = {10.1145/3397481.3450650},
	howpublished =  {\url{https://doi.org/10.1145/3397481.3450650}},
	isbn         = {978-1-4503-8017-1},
}

@article{Wysocki2023-CommunicationGapHealthcare-AI,
	author       = {Wysocki, Oskar and Davies, Jessica Katharine and Vigo, Markel and Armstrong, Anne Caroline and Landers, D{\'o}nal and Lee, Rebecca and Freitas, Andr{\'e}},
	title        = {{Assessing the Communication Gap Between AI Models and Healthcare Professionals: Explainability, Utility, and Trust in AI-Driven Clinical Decision-Making}},
	journal      = {{Artificial Intelligence}},
	volume       = {316},
	pages        = {103839},
	year         = {2023},
	month        = {Jan},
	doi          = {10.1016/j.artint.2022.103839},
	howpublished =  {\url{https://www.sciencedirect.com/science/article/pii/S0004370222001795}},
	issn         = {0004-3702},
}

@article{LaGatta2021-CASTLE-ESWA,
	author       = {La Gatta, Valerio and Moscato, Vincenzo and Postiglione, Marco and Sperl{\`i}, Giancarlo},
	title        = {{CASTLE: Cluster-Aided Space Transformation for Local Explanations}},
	journal      = {{Expert Systems with Applications}},
	volume       = {179},
	pages        = {115045},
	year         = {2021},
	month        = {Aug},
	doi          = {10.1016/j.eswa.2021.115045},
	howpublished =  {\url{https://www.sciencedirect.com/science/article/pii/S0957417421004863}},
	issn         = {0957-4174},
}

@article{LaGatta2021-PASTLE-PRL,
	author       = {La Gatta, Valerio and Moscato, Vincenzo and Postiglione, Marco and Sperl{\`i}, Giancarlo},
	title        = {{PASTLE: Pivot-Aided Space Transformation for Local Explanations}},
	journal      = {{Pattern Recognition Letters}},
	volume       = {149},
	pages        = {67--74},
	year         = {2021},
	doi          = {10.1016/j.patrec.2021.05.018},
	howpublished =  {\url{https://www.sciencedirect.com/science/article/pii/S0167865521002014}},
	issn         = {0167-8655},
}

@inproceedings{Nazaretsky2022-TeachersAI-LAK,
	author       = {Nazaretsky, Tanya and Bar, Carmel and Walter, Michal and Alexandron, Giora},
	title        = {{Empowering Teachers with AI: Co-Designing a Learning Analytics Tool for Personalized Instruction in the Science Classroom}},
	booktitle    = {{Proceedings of the 12th International Learning Analytics and Knowledge Conference (LAK22)}},
	pages        = {1--12},
	year         = {2022},
	publisher    = {Association for Computing Machinery},
	address      = {New York, NY, USA},
	doi          = {10.1145/3506860.3506861},
	howpublished =  {\url{https://doi.org/10.1145/3506860.3506861}},
	isbn         = {9781450395731},
}

@article{Brdnik2023-TrustSatisfactionXAI-LearningAnalytics,
	author       = {Brdnik, Sa{\v{s}}a and Podgorelec, Vili and {\v{S}}umak, Bo{\v{s}}tjan},
	title        = {{Assessing Perceived Trust and Satisfaction with Multiple Explanation Techniques in XAI-Enhanced Learning Analytics}},
	journal      = {{Electronics}},
	volume       = {12},
	number       = {12},
	pages        = {2594},
	year         = {2023},
	doi          = {10.3390/electronics12122594},
	howpublished =  {\url{https://www.mdpi.com/2079-9292/12/12/2594}},
	issn         = {2079-9292},
	publisher    = {MDPI},
	address      = {Basel, Switzerland},
}

@inproceedings{Forster2021-CapturingUsersReality-HICSS,
	author       = {F{\"o}rster, Maximilian and H{\"u}hn, Philipp and Klier, Mathias and Kluge, Kilian},
	title        = {{Capturing Users’ Reality: A Novel Approach to Generate Coherent Counterfactual Explanations}},
	booktitle    = {Proceedings of the 54th Hawaii International Conference on System Sciences (HICSS 2021)},
	year         = {2021},
	month        = {Jan},
	pages        = {1274--1283},
	doi          = {10.24251/hicss.2021.155},
	publisher    = {University of Hawai'i at M{\=a}noa},
	address      = {Wailea, HI, USA},
	howpublished = {\url{https://api.semanticscholar.org/CorpusID:232412682}},
}

@article{Kim2023-ExplainResultsToUsers-TFSC,
	author       = {Kim, Doha and Song, Yeosol and Kim, Songyie and Lee, Sewang and Wu, Yanqin and Shin, Jungwoo and Lee, Daeho},
	title        = {{How Should the Results of Artificial Intelligence Be Explained to Users? Research on Consumer Preferences in User-Centered Explainable Artificial Intelligence}},
	journal      = {{Technological Forecasting and Social Change}},
	volume       = {188},
	pages        = {122343},
	year         = {2023},
	month        = {Mar},
	publisher    = {Elsevier},
	address      = {Amsterdam, Netherlands},
	doi          = {10.1016/j.techfore.2023.122343},
	howpublished =  {\url{https://doi.org/10.1016/j.techfore.2023.122343}},
	issn         = {0040-1625},
}

@article{Meas2022-ExplainabilityTransparencyAHU-Sensors,
	author       = {Meas, Molika and Machlev, Ram and Kose, Ahmet and Tepljakov, Aleksei and Loo, Lauri and Levron, Yoash and Petlenkov, Eduard and Belikov, Juri},
	title        = {{Explainability and Transparency of Classifiers for Air-Handling Unit Faults Using Explainable Artificial Intelligence (XAI)}},
	journal      = {{Sensors}},
	volume       = {22},
	number       = {17},
	pages        = {6338},
	year         = {2022},
	month        = {Aug},
	publisher    = {MDPI},
	address      = {Basel, Switzerland},
	doi          = {10.3390/s22176338},
	howpublished =  {\url{https://doi.org/10.3390/s22176338}},
	pmid         = {36080795},
	pmcid        = {PMC9460735},
	issn         = {1424-8220},
}

@article{Nagy2024-InterpretableDropoutPrediction-IJAIED,
	author       = {Nagy, Marcell and Molontay, Roland},
	title        = {{Interpretable Dropout Prediction: Towards XAI-Based Personalized Intervention}},
	journal      = {{International Journal of Artificial Intelligence in Education}},
	volume       = {34},
	number       = {2},
	pages        = {274--300},
	year         = {2024},
	month        = {Jun},
	publisher    = {Springer},
	address      = {Cham, Switzerland},
	doi          = {10.1007/s40593-023-00331-8},
	howpublished =  {\url{https://doi.org/10.1007/s40593-023-00331-8}},
	issn         = {1560-4306},
}

@inproceedings{Polley2021-TowardsTrustworthinessXAI-SIGIR,
	author       = {Polley, Sayantan and Koparde, Rashmi Raju and Gowri, Akshaya Bindu and Perera, Maneendra and Nuernberger, Andreas},
	title        = {{Towards Trustworthiness in the Context of Explainable Search}},
	booktitle    = {{Proceedings of the 44th International ACM SIGIR Conference on Research and Development in Information Retrieval ({SIGIR} '21)}},
	pages        = {2580--2584},
	year         = {2021},
	month        = {Jul},
	publisher    = {Association for Computing Machinery},
	address      = {New York, NY, USA},
	doi          = {10.1145/3404835.3462799},
	howpublished =  {\url{https://doi.org/10.1145/3404835.3462799}},
	isbn         = {978-1-4503-8037-9},
}

@inproceedings{Scheers2021-ExplainableAdvisingDashboard-TEL,
	author       = {Scheers, Hanne and De Laet, Tinne},
	editor       = {De Laet, Tinne and Klemke, Roland and Alario-Hoyos, Carlos and Hilliger, Isabel and Ortega-Arranz, Alejandro},
	title        = {{Interactive and Explainable Advising Dashboard Opens the Black Box of Student Success Prediction}},
	booktitle    = {{Technology-Enhanced Learning for a Free, Safe, and Sustainable World}},
	pages        = {52--66},
	year         = {2021},
	publisher    = {Springer International Publishing},
	address      = {Cham},
	isbn         = {978-3-030-86436-1},
	doi          = {10.1007/978-3-030-86436-1\_5},
	howpublished =  {\url{https://doi.org/10.1007/978-3-030-86436-1\_5}},
}

@inproceedings{SchulzeWeddige2022-EffectsXAIVisualizations-ArtsIT,
	author       = {Schulze-Weddige, Sophia and Zylowski, Thorsten},
	title        = {{User Study on the Effects of Explainable AI Visualizations on Non-Experts}},
	booktitle    = {{ArtsIT, Interactivity and Game Creation}},
	pages        = {457--467},
	year         = {2022},
	month        = {Nov},
	editor       = {W{\"o}lfel, Matthias and Bernhardt, Johannes and Thiel, Sonja},
	publisher    = {Springer International Publishing},
	address      = {Cham, Switzerland},
	doi          = {10.1007/978-3-030-95531-1\_31},
	howpublished =  {\url{https://doi.org/10.1007/978-3-030-95531-1\_31}},
	isbn         = {978-3-030-95531-1},
}

@inproceedings{Swamy2023-TrustingExplainersXAI-LAK,
	author       = {Swamy, Vinitra and Du, Sijia and Marras, Mirko and Kaser, Tanja},
	title        = {{Trusting the Explainers: Teacher Validation of Explainable Artificial Intelligence for Course Design}},
	booktitle    = {{Proceedings of the 13th International Learning Analytics and Knowledge Conference ({LAK} 2023)}},
	pages        = {345--356},
	year         = {2023},
	month        = {Mar},
	publisher    = {Association for Computing Machinery},
	address      = {New York, NY, USA},
	doi          = {10.1145/3576050.3576147},
	howpublished =  {\url{https://doi.org/10.1145/3576050.3576147}},
	isbn         = {978-1-4503-9865-7},
}

@article{VanderWaa2020-InterpretableConfidenceMeasures-IJHCS,
	author       = {van der Waa, Jasper and Schoonderwoerd, Tjeerd and van Diggelen, Jurriaan and Neerincx, Mark},
	title        = {{Interpretable Confidence Measures for Decision Support Systems}},
	journal      = {{International Journal of Human-Computer Studies}},
	volume       = {144},
	pages        = {102493},
	year         = {2020},
	month        = {Oct},
	publisher    = {Elsevier},
	address      = {Amsterdam, Netherlands},
	doi          = {10.1016/j.ijhcs.2020.102493},
	howpublished =  {\url{https://doi.org/10.1016/j.ijhcs.2020.102493}},
	issn         = {1071-5819},
}

@article{Zlahtic2023-AgileMLDataCanyons-Medicine-AS,
	author       = {{\v{Z}}lahti{\'c}, Bojan and Zav{\v{r}}snik, Jernej and Bla{\v{z}}un Vo{\v{s}}ner, Helena and Kokol, Peter and {\v{S}}uran, David and Zav{\v{r}}snik, Tadej},
	title        = {{Agile Machine Learning Model Development Using Data Canyons in Medicine: A Step Towards Explainable Artificial Intelligence and Flexible Expert-Based Model Improvement}},
	journal      = {{Applied Sciences}},
	volume       = {13},
	number       = {14},
	pages        = {8329},
	year         = {2023},
	month        = {Jul},
	publisher    = {MDPI},
	address      = {Basel, Switzerland},
	doi          = {10.3390/app13148329},
	howpublished =  {\url{https://doi.org/10.3390/app13148329}},
	issn         = {2076-3417},
}

@inproceedings{Xu2023-DialogueExplanationsRuleBasedAI-EXTRAAMAS,
	author       = {Xu, Yifan and Collenette, Joe and Dennis, Louise and Dixon, Clare},
	title        = {{Dialogue Explanations for Rule-Based AI Systems}},
	booktitle    = {{Explainable and Transparent AI and Multi-Agent Systems}},
	pages        = {59--77},
	year         = {2023},
	month        = {May},
	editor       = {Calvaresi, Davide and Najjar, Amro and Omicini, Andrea and Aydogan, Reyhan and Carli, Rachele and Ciatto, Giovanni and Mualla, Yazan and Fr{\"a}mling, Kary},
	publisher    = {Springer Nature Switzerland},
	address      = {Cham, Switzerland},
	doi          = {10.1007/978-3-031-40878-6\_4},
	howpublished = {\url{https://doi.org/10.1007/978-3-031-40878-6\_4}},
	isbn         = {978-3-031-40878-6},
}

@article{Alufaisan2021-DoesXAIImproveDecisionMaking-AAAI,
	author       = {Alufaisan, Yasmeen and Marusich, Laura R. and Bakdash, Jonathan Z. and Zhou, Yan and Kantarcioglu, Murat},
	title        = {{Does Explainable Artificial Intelligence Improve Human Decision-Making?}},
	journal      = {{Proceedings of the AAAI Conference on Artificial Intelligence}},
	volume       = {35},
	number       = {8},
	pages        = {6618--6626},
	year         = {2021},
	month        = {May},
	publisher    = {AAAI Press},
	address      = {Palo Alto, CA, USA},
	doi          = {10.1609/aaai.v35i8.16819},
	howpublished =  {\url{https://doi.org/10.1609/aaai.v35i8.16819}},
}

@inproceedings{Cau2023-SupportingHighUncertaintyDecisionsXAI-IUI,
	author       = {Cau, Federico Maria and Hauptmann, Hanna and Spano, Lucio Davide and Tintarev, Nava},
	title        = {{Supporting High-Uncertainty Decisions Through AI and Logic-Style Explanations}},
	booktitle    = {{Proceedings of the 28th International Conference on Intelligent User Interfaces (IUI ’23)}},
	pages        = {251--263},
	year         = {2023},
	month        = {Mar},
	publisher    = {Association for Computing Machinery},
	address      = {New York, NY, USA},
	doi          = {10.1145/3581641.3584080},
	howpublished =  {\url{https://doi.org/10.1145/3581641.3584080}},
	isbn         = {979-8-4007-0106-1},
}

@article{Conati2021-PersonalizedXAI-ITS-AI,
	author       = {Conati, Cristina and Barral, Oswald and Putnam, Vanessa and Rieger, Lea},
	title        = {{Toward Personalized XAI: A Case Study in Intelligent Tutoring Systems}},
	journal      = {{Artificial Intelligence}},
	volume       = {298},
	pages        = {103503},
	year         = {2021},
	month        = {Jul},
	publisher    = {Elsevier},
	address      = {Amsterdam, Netherlands},
	doi          = {10.1016/j.artint.2021.103503},
	howpublished =  {\url{https://doi.org/10.1016/j.artint.2021.103503}},
	issn         = {0004-3702},
}

@article{Ghai2021-ExplainableActiveLearning-CSCW,
	author       = {Ghai, Bhavya and Liao, Q. Vera and Zhang, Yunfeng and Bellamy, Rachel and Mueller, Klaus},
	title        = {{Explainable Active Learning (XAL): Toward AI Explanations as Interfaces for Machine Teachers}},
	journal      = {{Proceedings of the ACM on Human-Computer Interaction}},
	volume       = {4},
	number       = {CSCW3},
	pages        = {235:1--235:28},
	year         = {2021},
	month        = {Jan},
	publisher    = {Association for Computing Machinery},
	address      = {New York, NY, USA},
	doi          = {10.1145/3432934},
	howpublished =  {\url{https://doi.org/10.1145/3432934}},
}

@article{Naiseh2023-ExplanationClassesTrustCalibration-IJHCS,
	author       = {Naiseh, Mohammad and Al-Thani, Dena and Jiang, Nan and Ali, Raian},
	title        = {{How the Different Explanation Classes Impact Trust Calibration: The Case of Clinical Decision Support Systems}},
	journal      = {{International Journal of Human-Computer Studies}},
	volume       = {169},
	pages        = {102941},
	year         = {2023},
	month        = {Feb},
	publisher    = {Elsevier},
	address      = {Amsterdam, Netherlands},
	doi          = {10.1016/j.ijhcs.2022.102941},
	howpublished =  {\url{https://doi.org/10.1016/j.ijhcs.2022.102941}},
	issn         = {1071-5819},
}

@article{Wang2023-NestedModelUserCentricXAI-TVCG,
	author       = {Wang, Qianwen and Huang, Kexin and Chandak, Payal and Zitnik, Marinka and Gehlenborg, Nils},
	title        = {{Extending the Nested Model for User-Centric XAI: A Design Study on GNN-Based Drug Repurposing}},
	journal      = {{IEEE Transactions on Visualization and Computer Graphics}},
	volume       = {29},
	number       = {1},
	pages        = {1266--1276},
	year         = {2023},
	month        = {Jan},
	publisher    = {IEEE},
	address      = {Piscataway, NJ, USA},
	doi          = {10.1109/TVCG.2022.3209435},
	howpublished =  {\url{https://doi.org/10.1109/TVCG.2022.3209435}},
	issn         = {1077-2626},
}

@inproceedings{Eriksson2022-XAISOC-IEEEBigData,
	author       = {Eriksson, H{\aa}kon Svee and Gr{\o}v, Gudmund},
	title        = {{Towards XAI in the SOC -- A User-Centric Study of Explainable Alerts with SHAP and LIME}},
	booktitle    = {{2022 IEEE International Conference on Big Data (Big Data)}},
	pages        = {2595--2600},
	year         = {2022},
	publisher    = {IEEE},
	address      = {Osaka, Japan},
	doi          = {10.1109/BigData55660.2022.10020248},
}

@inproceedings{Maltbie2021-XAIToolsPublicSector-FSE,
	author       = {Maltbie, Nicholas and Niu, Nan and Van Doren, Matthew and Johnson, Reese},
	title        = {{XAI Tools in the Public Sector: A Case Study on Predicting Combined Sewer Overflows}},
	booktitle    = {{Proceedings of the 29th ACM Joint Meeting on European Software Engineering Conference and Symposium on the Foundations of Software Engineering (ESEC/FSE 2021)}},
	pages        = {1032--1044},
	year         = {2021},
	month        = {Aug},
	publisher    = {Association for Computing Machinery},
	address      = {New York, NY, USA},
	doi          = {10.1145/3468264.3468547},
	howpublished =  {\url{https://doi.org/10.1145/3468264.3468547}},
	isbn         = {9781450385626},
}

@inproceedings{Wang2022-InterpretableDirectedDiversity-CHI,
	author       = {Wang, Yunlong and Venkatesh, Priyadarshini and Lim, Brian Y.},
	title        = {{Interpretable Directed Diversity: Leveraging Model Explanations for Iterative Crowd Ideation}},
	booktitle    = {{Proceedings of the 2022 CHI Conference on Human Factors in Computing Systems (CHI '22)}},
	pages        = {183:1--183:28},
	year         = {2022},
	month        = {Apr},
	publisher    = {Association for Computing Machinery},
	address      = {New York, NY, USA},
	doi          = {10.1145/3491102.3517551},
	howpublished =  {\url{https://doi.org/10.1145/3491102.3517551}},
	isbn         = {9781450391573},
}

@inproceedings{Abdul2020-COGAM-CognitiveLoadXAI-CHI,
	author       = {Abdul, Ashraf and von der Weth, Christian and Kankanhalli, Mohan and Lim, Brian Y.},
	title        = {{COGAM: Measuring and Moderating Cognitive Load in Machine Learning Model Explanations}},
	booktitle    = {{Proceedings of the 2020 CHI Conference on Human Factors in Computing Systems (CHI '20)}},
	pages        = {1--14},
	year         = {2020},
	month        = {Apr},
	publisher    = {Association for Computing Machinery},
	address      = {New York, NY, USA},
	doi          = {10.1145/3313831.3376615},
	howpublished =  {\url{https://doi.org/10.1145/3313831.3376615}},
	isbn         = {9781450367080},
}

@inproceedings{Adhikari2019-LEAFAGE-FUZZIEEE,
	author       = {Adhikari, Ajaya and Tax, David M. J. and Satta, Riccardo and Faeth, Matthias},
	title        = {{LEAFAGE: Example-Based and Feature-Importance-Based Explanations for Black-Box ML Models}},
	booktitle    = {{Proceedings of the 2019 IEEE International Conference on Fuzzy Systems (FUZZ-IEEE)}},
	pages        = {1--7},
	year         = {2019},
	month        = {Jun},
	publisher    = {IEEE},
	address      = {Piscataway, NJ, USA},
	doi          = {10.1109/FUZZ-IEEE.2019.8858846},
}

@inproceedings{Aechtner2022-UserPerceptionXAI-FUZZIEEE,
	author       = {Aechtner, Jonathan and Cabrera, Lena and Katwal, Dennis and Onghena, Pierre and Valenzuela, Diego Penroz and Wilbik, Anna},
	title        = {{Comparing User Perception of Explanations Developed with XAI Methods}},
	booktitle    = {{Proceedings of the 2022 IEEE International Conference on Fuzzy Systems (FUZZ-IEEE)}},
	pages        = {1--7},
	year         = {2022},
	month        = {Jul},
	publisher    = {IEEE},
	address      = {Piscataway, NJ, USA},
	doi          = {10.1109/FUZZ-IEEE55066.2022.9882743},
}

@article{Anjara2023-ExplainableCDSSThinkAloud-PLOSONE,
	author       = {Anjara, Sabrina G. and Janik, Adrianna and Dunford-Stenger, Amy and Mc Kenzie, Kenneth and Collazo-Lorduy, Ana and Torrente, Maria and Costabello, Luca and Provencio, Mariano},
	title        = {{Examining Explainable Clinical Decision Support Systems with Think Aloud Protocols}},
	journal      = {{PLoS One}},
	volume       = {18},
	number       = {9},
	pages        = {e0291443},
	year         = {2023},
	month        = {Sep},
	publisher    = {Public Library of Science},
	address      = {San Francisco, CA, USA},
	doi          = {10.1371/journal.pone.0291443},
	howpublished =  {\url{https://doi.org/10.1371/journal.pone.0291443}},
	issn         = {1932-6203},
	pmid         = {37708135},
	pmcid        = {PMC10501571},
}

@article{Avetisyan2022-ExplanationsSituationAwareness-TRF,
	author       = {Avetisyan, Lilit and Ayoub, Jackie and Zhou, Feng},
	title        = {{Investigating Explanations in Conditional and Highly Automated Driving: The Effects of Situation Awareness and Modality}},
	journal      = {{Transportation Research Part F: Traffic Psychology and Behaviour}},
	volume       = {89},
	pages        = {456--466},
	year         = {2022},
	month        = {Sep},
	publisher    = {Elsevier},
	address      = {Amsterdam, Netherlands},
	doi          = {10.1016/j.trf.2022.07.010},
	howpublished =  {\url{https://doi.org/10.1016/j.trf.2022.07.010}},
	issn         = {1369-8478},
}

@inproceedings{BenDavid2021-XAIFinancialAdvisors-AIES,
	author       = {Ben David, Daniel and Resheff, Yehezkel S. and Tron, Talia},
	title        = {{Explainable AI and Adoption of Financial Algorithmic Advisors: An Experimental Study}},
	booktitle    = {{Proceedings of the 2021 AAAI/ACM Conference on AI, Ethics, and Society (AIES '21)}},
	pages        = {390--400},
	year         = {2021},
	month        = {Jul},
	publisher    = {Association for Computing Machinery},
	address      = {New York, NY, USA},
	doi          = {10.1145/3461702.3462565},
	howpublished =  {\url{https://doi.org/10.1145/3461702.3462565}},
	isbn         = {9781450384735},
}

@inproceedings{Bertrand2023-FeatureBasedExplanationsFinance-FAccT,
	author       = {Bertrand, Astrid and Eagan, James R. and Maxwell, Winston},
	title        = {{Questioning the Ability of Feature-Based Explanations to Empower Non-Experts in Robo-Advised Financial Decision-Making}},
	booktitle    = {{Proceedings of the 2023 ACM Conference on Fairness, Accountability, and Transparency (FAccT '23)}},
	pages        = {943--958},
	year         = {2023},
	month        = {Jun},
	publisher    = {Association for Computing Machinery},
	address      = {New York, NY, USA},
	doi          = {10.1145/3593013.3594053},
	howpublished =  {\url{https://doi.org/10.1145/3593013.3594053}},
	isbn         = {9798400701924},
}

@inproceedings{Bhattacharya2023-DirectiveExplanationsDiabetes-IUI,
	author       = {Bhattacharya, Aditya and Ooge, Jeroen and Stiglic, Gregor and Verbert, Katrien},
	title        = {{Directive Explanations for Monitoring the Risk of Diabetes Onset: Introducing Directive Data-Centric Explanations and Combinations to Support What-If Explorations}},
	booktitle    = {{Proceedings of the 28th International Conference on Intelligent User Interfaces (IUI '23)}},
	pages        = {204--219},
	year         = {2023},
	month        = {Mar},
	publisher    = {Association for Computing Machinery},
	address      = {New York, NY, USA},
	doi          = {10.1145/3581641.3584075},
	howpublished =  {\url{https://doi.org/10.1145/3581641.3584075}},
	isbn         = {9798400701061},
}

@article{Chien2022-XFlag-FakeNewsDetection-IJHCI,
	author       = {Chien, Shih-Yi and Yang, Cheng-Jun and Yu, Fang},
	title        = {{XFlag: Explainable Fake News Detection Model on Social Media}},
	journal      = {{International Journal of Human–Computer Interaction}},
	volume       = {38},
	number       = {18--20},
	pages        = {1808--1827},
	year         = {2022},
	month        = {Dec},
	publisher    = {Taylor \& Francis},
	address      = {Abingdon, United Kingdom},
	doi          = {10.1080/10447318.2022.2062113},
	howpublished = {\url{https://doi.org/10.1080/10447318.2022.2062113}},
}

@article{Conijn2023-ExplanationsEssayScoring-JLA,
	author       = {Conijn, Rianne and Kahr, Patricia and Snijders, Chris},
	title        = {{The Effects of Explanations in Automated Essay Scoring Systems on Student Trust and Motivation}},
	journal      = {{Journal of Learning Analytics}},
	volume       = {10},
	number       = {1},
	pages        = {37--53},
	year         = {2023},
	month        = {Mar},
	publisher    = {Society for Learning Analytics Research},
	address      = {Edmonton, Canada},
	doi          = {10.18608/jla.2023.7801},
	howpublished =  {\url{https://learning-analytics.info/index.php/JLA/article/view/7801}},
}

@article{Das2023-ExplainableActivityRecognition-TIIS,
	author       = {Das, Devleena and Nishimura, Yasutaka and Vivek, Rajan P. and Takeda, Naoto and Fish, Sean T. and Pl{\"o}tz, Thomas and Chernova, Sonia},
	title        = {{Explainable Activity Recognition for Smart Home Systems}},
	journal      = {{ACM Transactions on Interactive Intelligent Systems}},
	volume       = {13},
	number       = {2},
	pages        = {7:1--7:39},
	year         = {2023},
	month        = {May},
	publisher    = {Association for Computing Machinery},
	address      = {New York, NY, USA},
	doi          = {10.1145/3561533},
	howpublished =  {\url{https://doi.org/10.1145/3561533}},
	issn         = {2160-6455},
}

@inproceedings{Deo2021-UserCentricExplainabilityFintech-HCII,
	author       = {Deo, Sahil and Sontakke, Neha},
	title        = {{User-Centric Explainability in Fintech Applications}},
	booktitle    = {{HCI International 2021 -- Posters}},
	editor       = {Stephanidis, Constantine and Antona, Margherita and Ntoa, Stavroula},
	pages        = {481--488},
	year         = {2021},
	month        = {Jul},
	publisher    = {Springer International Publishing},
	address      = {Cham, Switzerland},
	isbn         = {978-3-030-78642-7},
	doi          = {10.1007/978-3-030-78642-7\_64},
	howpublished =  {\url{https://doi.org/10.1007/978-3-030-78642-7\_64}},
}

@article{Fernandes2023-ExplainableWeightManagement-JMIR,
	author       = {Fernandes, Glenn J. and Choi, Arthur and Schauer, Jacob Michael and Pfammatter, Angela F. and Spring, Bonnie J. and Darwiche, Adnan and Alshurafa, Nabil I.},
	title        = {{An Explainable Artificial Intelligence Software Tool for Weight Management Experts (PRIMO): Mixed Methods Study}},
	journal      = {{Journal of Medical Internet Research}},
	volume       = {25},
	pages        = {e42047},
	year         = {2023},
	month        = {Sep},
	publisher    = {JMIR Publications},
	address      = {Toronto, Canada},
	doi          = {10.2196/42047},
	howpublished =  {\url{https://doi.org/10.2196/42047}},
	issn         = {1438-8871},
	pmid         = {37672333},
	pmcid        = {PMC10512114},
}

@inproceedings{Guo2022-BuildingTrustIML-IUI,
	author       = {Guo, Lijie and Daly, Elizabeth M. and Alkan, Oznur and Mattetti, Massimiliano and Cornec, Owen and Knijnenburg, Bart},
	title        = {{Building Trust in Interactive Machine Learning via User Contributed Interpretable Rules}},
	booktitle    = {{Proceedings of the 27th International Conference on Intelligent User Interfaces}},
	pages        = {537--548},
	year         = {2022},
	month        = {Mar},
	publisher    = {Association for Computing Machinery},
	address      = {New York, NY, USA},
	doi          = {10.1145/3490099.3511111},
	howpublished =  {\url{https://doi.org/10.1145/3490099.3511111}},
	isbn         = {978-1-4503-9144-3},
}

@article{HernandezBocanegra2023-ExplainingRecommendationsDialogModel-TIIS,
	author       = {Hern{\'a}ndez-Bocanegra, Diana C. and Ziegler, J{\"u}rgen},
	title        = {{Explaining Recommendations through Conversations: Dialog Model and the Effects of Interface Type and Degree of Interactivity}},
	journal      = {{ACM Transactions on Interactive Intelligent Systems}},
	volume       = {13},
	number       = {2},
	pages        = {1--47},
	year         = {2023},
	month        = {Jan},
	publisher    = {Association for Computing Machinery},
	address      = {New York, NY, USA},
	doi          = {10.1145/3579541},
	howpublished =  {\url{https://doi.org/10.1145/3579541}},
}

@inproceedings{Jang2023-TowardInterpretableML-PAKDD,
	author       = {Jang, Jisoo and Kim, Mina and Bui, Tien{-}Cuong and Li, Wen{-}Syan},
	title        = {{Toward Interpretable Machine Learning: Constructing Polynomial Models Based on Feature Interaction Trees}},
	booktitle    = {{Advances in Knowledge Discovery and Data Mining}},
	editor       = {Kashima, Hisashi and Ide, Tsuyoshi and Peng, Wen{-}Chih},
	pages        = {159--170},
	year         = {2023},
	month        = {May},
	publisher    = {Springer Nature Switzerland},
	address      = {Cham},
	isbn         = {978-3-031-33377-4},
	doi          = {10.1007/978-3-031-33377-4\_13},
	howpublished =  {\url{https://doi.org/10.1007/978-3-031-33377-4\_13}},
}

@article{Khodabandehloo2021-HealthXAI-FGCS,
	author       = {Khodabandehloo, Elham and Riboni, Daniele and Alimohammadi, Abbas},
	title        = {{HealthXAI: Collaborative and Explainable AI for Supporting Early Diagnosis of Cognitive Decline}},
	journal      = {{Future Generation Computer Systems}},
	volume       = {116},
	pages        = {168--189},
	year         = {2021},
	issn         = {0167-739X},
	doi          = {10.1016/j.future.2020.10.030},
	howpublished =  {\url{https://www.sciencedirect.com/science/article/pii/S0167739X20330144}},
}

@inproceedings{Larasati2022-XAI-BreastCancer-ICECET,
	author       = {Larasati, Retno},
	title        = {{Explainable AI for Breast Cancer Diagnosis: Application and User’s Understandability Perception}},
	booktitle    = {{2022 International Conference on Electrical, Computer and Energy Technologies (ICECET)}},
	pages        = {1--6},
	year         = {2022},
	publisher    = {IEEE},
	address      = {Prague, Czech Republic},
	doi          = {10.1109/ICECET55527.2022.9872950},
}

@article{Moradi2021-XAI-CIE-Elsevier,
	author       = {Moradi, Milad and Samwald, Matthias},
	title        = {{Post-hoc Explanation of Black-Box Classifiers Using Confident Itemsets}},
	journal      = {{Expert Systems with Applications}},
	volume       = {165},
	pages        = {113941},
	year         = {2021},
	month        = {May},
	issn         = {0957-4174},
	doi          = {10.1016/j.eswa.2020.113941},
	howpublished =  {\url{https://www.sciencedirect.com/science/article/pii/S0957417420307302}},
	publisher    = {Elsevier},
	address      = {Amsterdam, Netherlands},
}

@article{Neves2021-XAI-HeartbeatExplainability-Elsevier,
	author       = {Neves, In{\^e}s and Folgado, Duarte and Santos, Sara and Barandas, Mar{\'i}lia and Campagner, Andrea and Ronzio, Luca and Cabitza, Federico and Gamboa, Hugo},
	title        = {{Interpretable Heartbeat Classification Using Local Model-Agnostic Explanations on ECGs}},
	journal      = {{Computers in Biology and Medicine}},
	volume       = {133},
	pages        = {104393},
	year         = {2021},
	month        = {May},
	issn         = {0010-4825},
	doi          = {10.1016/j.compbiomed.2021.104393},
	howpublished =  {\url{https://www.sciencedirect.com/science/article/pii/S0010482521001876}},
	publisher    = {Elsevier},
	address      = {Amsterdam, Netherlands},
}

@inproceedings{Ooge2022-XAI-ElearningTrust-ACM,
	author       = {Ooge, Jeroen and Kat{\=o}, Shotallo and Verbert, Katrien},
	title        = {{Explaining Recommendations in E-Learning: Effects on Adolescents' Trust}},
	booktitle    = {{Proceedings of the 27th International Conference on Intelligent User Interfaces}},
	year         = {2022},
	month        = {Mar},
	pages        = {93--105},
	numpages     = {13},
	isbn         = {9781450391443},
	publisher    = {Association for Computing Machinery},
	address      = {New York, NY, USA},
	doi          = {10.1145/3490099.3511140},
	howpublished =  {\url{https://doi.org/10.1145/3490099.3511140}},
	location     = {Helsinki, Finland},
	series       = {IUI '22},
}

@inproceedings{Panigutti2022-XAI-AdviceTakingTrust-ACM,
	author       = {Panigutti, Cecilia and Beretta, Andrea and Giannotti, Fosca and Pedreschi, Dino},
	title        = {{Understanding the Impact of Explanations on Advice-Taking: A User Study for AI-Based Clinical Decision Support Systems}},
	booktitle    = {{Proceedings of the 2022 CHI Conference on Human Factors in Computing Systems}},
	year         = {2022},
	month        = {Apr},
	pages        = {1--9},
	articleno    = {568},
	numpages     = {9},
	isbn         = {9781450391573},
	doi          = {10.1145/3491102.3502104},
	publisher    = {Association for Computing Machinery},
	address      = {New York, NY, USA},
	howpublished = {\url{https://doi.org/10.1145/3491102.3502104}},
	location     = {New Orleans, LA, USA},
	series       = {CHI '22},
}

@inproceedings{Schellingerhout2022-XAI-CareerPathPredictions-CEUR,
	author       = {Schellingerhout, Roan and Medentsiy, Volodymyr and Marx, Maarten},
	editor       = {Kaya, Mesut and Bogers, Toine and Graus, David and Mesbah, Sepideh and Johnson, Chris and Guti{\'{e}}rrez, Francisco},
	title        = {{Explainable Career Path Predictions Using Neural Models}},
	booktitle    = {{Proceedings of the 2nd Workshop on Recommender Systems for Human Resources (RecSys-in-HR 2022) Co-located with the 16th ACM Conference on Recommender Systems (RecSys 2022)}},
	series       = {{CEUR Workshop Proceedings}},
	volume       = {3218},
	year         = {2022},
	month        = {Sep},
	pages        = {1--8},
	publisher    = {CEUR-WS.org},
	address      = {Aachen, Germany},
	howpublished =  {\url{https://ceur-ws.org/Vol-3218/RecSysHR2022-paper\_7.pdf}},
}

@article{Veldhuis2022-XAI-ForensicDNAReCo-Elsevier,
	author       = {Veldhuis, Marthe S. and Ari{\"e}ns, Simone and Ypma, Rolf J. F. and Abeel, Thomas and Benschop, Corina C. G.},
	title        = {{Explainable Artificial Intelligence in Forensics: Realistic Explanations for Number of Contributor Predictions of DNA Profiles}},
	journal      = {{Forensic Science International: Genetics}},
	volume       = {56},
	pages        = {102632},
	year         = {2022},
	month        = {Feb},
	issn         = {1872-4973},
	doi          = {10.1016/j.fsigen.2021.102632},
	howpublished =  {\url{https://www.sciencedirect.com/science/article/pii/S187249732100168X}},
	publisher    = {Elsevier},
	address      = {Amsterdam, Netherlands},
}

@misc{Warren2022-XAI-FeaturesExplainability-ArXiv,
	author       = {Warren, Greta and Keane, Mark T. and Byrne, Ruth M. J.},
	title        = {{Features of Explainability: How Users Understand Counterfactual and Causal Explanations for Categorical and Continuous Features in XAI}},
	year         = {2022},
	month        = {Apr},
	eprint       = {2204.10152},
	archivePrefix= {arXiv},
	primaryClass = {cs.HC},
	howpublished =  {\url{https://arxiv.org/abs/2204.10152}},
	publisher    = {arXiv, Cornell University},
	address      = {Ithaca, NY, USA},
}

@article{Weitz2021-XAI-VirtualAgentsTrust-Springer,
	author       = {Weitz, Katharina and Schiller, Dominik and Schlagowski, Ruben and Huber, Tobias and Andr{\'e}, Elisabeth},
	title        = {{“Let Me Explain!”: Exploring the Potential of Virtual Agents in Explainable AI Interaction Design}},
	journal      = {{Journal on Multimodal User Interfaces}},
	volume       = {15},
	number       = {2},
	pages        = {87--98},
	year         = {2021},
	month        = {Jun},
	issn         = {1783-8738},
	doi          = {10.1007/s12193-020-00332-0},
	howpublished = {\url{https://doi.org/10.1007/s12193-020-00332-0}},
	publisher    = {Springer},
	address      = {Cham, Switzerland},
}

@article{Zoeller2023-XAI-XAutoML-ACM,
	author       = {Z\"{o}ller, Marc-Andr\'{e} and Titov, Waldemar and Schlegel, Thomas and Huber, Marco F.},
	title        = {{XAutoML: A Visual Analytics Tool for Understanding and Validating Automated Machine Learning}},
	journal      = {{ACM Transactions on Interactive Intelligent Systems}},
	volume       = {13},
	number       = {4},
	pages        = {28:1--28:39},
	year         = {2023},
	month        = {Dec},
	issn         = {2160-6455},
	doi          = {10.1145/3625240},
	howpublished =  {\url{https://doi.org/10.1145/3625240}},
	publisher    = {Association for Computing Machinery},
	address      = {New York, NY, USA},
}

@inproceedings{Bansal2021-XAI-TeamPerformance-ACM,
	author       = {Bansal, Gagan and Wu, Tongshuang and Zhou, Joyce and Fok, Raymond and Nushi, Besmira and Kamar, Ece and Ribeiro, Marco Tulio and Weld, Daniel},
	title        = {{Does the Whole Exceed Its Parts? The Effect of AI Explanations on Complementary Team Performance}},
	booktitle    = {{Proceedings of the 2021 CHI Conference on Human Factors in Computing Systems}},
	year         = {2021},
	month        = {May},
	publisher    = {Association for Computing Machinery},
	address      = {New York, NY, USA},
	pages        = {81:1--81:16},
	isbn         = {9781450380966},
	doi          = {10.1145/3411764.3445717},
	howpublished =  {\url{https://doi.org/10.1145/3411764.3445717}},
}

@inproceedings{Bucinca2020-XAI-ProxyTasksEvaluation-ACM,
	author       = {Bu\c{c}inca, Zana and Lin, Phoebe and Gajos, Krzysztof Z. and Glassman, Elena L.},
	title        = {{Proxy Tasks and Subjective Measures Can Be Misleading in Evaluating Explainable AI Systems}},
	booktitle    = {{Proceedings of the 25th International Conference on Intelligent User Interfaces}},
	year         = {2020},
	month        = {Mar},
	publisher    = {Association for Computing Machinery},
	address      = {New York, NY, USA},
	pages        = {454--464},
	isbn         = {9781450371186},
	doi          = {10.1145/3377325.3377498},
	howpublished =  {\url{https://doi.org/10.1145/3377325.3377498}},
}

@article{Confalonieri2021-XAI-OntologyGlobalExplanations-Elsevier,
	author       = {Confalonieri, Roberto and Weyde, Tillman and Besold, Tarek R. and {Moscoso del Prado Martín}, Ferm{\'i}n},
	title        = {{Using Ontologies to Enhance Human Understandability of Global Post-hoc Explanations of Black-box Models}},
	journal      = {{Artificial Intelligence}},
	volume       = {296},
	pages        = {103471},
	year         = {2021},
	month        = {May},
	issn         = {0004-3702},
	doi          = {10.1016/j.artint.2021.103471},
	howpublished =  {\url{https://www.sciencedirect.com/science/article/pii/S0004370221000229}},
	publisher    = {Elsevier},
	address      = {Amsterdam, Netherlands},
}

@inproceedings{Ibrahim2023-XAI-FactualCounterfactualExplanations-IFAAMAS,
	author       = {Ibrahim, Lujain and Ghassemi, Mohammad M. and Alhanai, Tuka},
	title        = {{Do Explanations Improve the Quality of AI-assisted Human Decisions? An Algorithm-in-the-Loop Analysis of Factual \& Counterfactual Explanations}},
	booktitle    = {{Proceedings of the 2023 International Conference on Autonomous Agents and Multiagent Systems}},
	year         = {2023},
	month        = {May},
	publisher    = {International Foundation for Autonomous Agents and Multiagent Systems},
	address      = {Richland, SC, USA},
	pages        = {326--334},
	isbn         = {9781450394321},
}

@inproceedings{Jmoona2023-XAI-FlightDelayPrediction-Springer,
	author       = {Jmoona, Waleed and Ahmed, Mobyen Uddin and Islam, Mir Riyanul and Barua, Shaibal and Begum, Shahina and Ferreira, Ana and Cavagnetto, Nicola},
	editor       = {Maglogiannis, Ilias and Iliadis, Lazaros and MacIntyre, John and Dominguez, Manuel},
	title        = {{Explaining the Unexplainable: Role of XAI for Flight Take-Off Time Delay Prediction}},
	booktitle    = {{Artificial Intelligence Applications and Innovations}},
	year         = {2023},
	month        = {Jun},
	publisher    = {Springer Nature Switzerland},
	address      = {Cham, Switzerland},
	pages        = {81--93},
	isbn         = {978-3-031-34107-6},
}

@article{Raab2023-XAI-XAI4EEG-Springer,
	author       = {Raab, Dominik and Theissler, Andreas and Spiliopoulou, Myra},
	title        = {{XAI4EEG: Spectral and Spatio-Temporal Explanation of Deep Learning-Based Seizure Detection in EEG Time Series}},
	journal      = {{Neural Computing and Applications}},
	volume       = {35},
	number       = {14},
	pages        = {10051--10068},
	year         = {2023},
	month        = {May},
	doi          = {10.1007/s00521-022-07809-x},
	howpublished =  {\url{https://doi.org/10.1007/s00521-022-07809-x}},
	issn         = {1433-3058},
	publisher    = {Springer Nature},
	address      = {Cham, Switzerland},
}

@article{Schrills2023-XAI-TraceabilityAID-ACM,
	author       = {Schrills, Tim and Franke, Thomas},
	title        = {{How Do Users Experience Traceability of AI Systems? Examining Subjective Information Processing Awareness in Automated Insulin Delivery (AID) Systems}},
	journal      = {{ACM Transactions on Interactive Intelligent Systems}},
	volume       = {13},
	number       = {4},
	pages        = {25:1--25:34},
	year         = {2023},
	month        = {Dec},
	doi          = {10.1145/3588594},
	howpublished = {\url{https://doi.org/10.1145/3588594}},
	issn         = {2160-6455},
	publisher    = {Association for Computing Machinery},
	address      = {New York, NY, USA},
}

@article{Halpern2005-XAI-CausesExplanationsII-OUP,
	author    = {Halpern, Joseph Y. and Pearl, Judea},
	title     = {{Causes and Explanations: A Structural-Model Approach. Part II: Explanations}},
	journal   = {British Journal for the Philosophy of Science},
	volume    = {56},
	number    = {4},
	year      = {2005},
	month     = {Dec},
	pages     = {889--911},
	doi       = {10.1093/bjps/axi148},
	publisher = {Oxford University Press},
	address   = {Oxford, UK},
	howpublished =  {\url{https://www.cs.cornell.edu/home/halpern/papers/actcaus1.pdf}},
}

@book{Pearl2009-XAI-Causality-Cambridge,
	author    = {Pearl, Judea},
	title     = {{Causality: Models, Reasoning, and Inference}},
	edition   = {2},
	year      = {2009},
	month     = {Sep},
	publisher = {Cambridge University Press},
	address   = {Cambridge, UK},
	isbn      = {978-0-521-89560-6},
}

@article{Wachter2018-XAI-CounterfactualExplanations-HarvardJLT,
	author    = {Wachter, Sandra and Mittelstadt, Brent and Russell, Chris},
	title     = {{Counterfactual Explanations Without Opening the Black Box: Automated Decisions and the GDPR}},
	journal   = {Harvard Journal of Law \& Technology},
	volume    = {31},
	number    = {2},
	year      = {2018},
	month     = {Apr},
	pages     = {841--887},
	publisher = {Harvard Law School},
	address   = {Cambridge, MA, USA},
	howpublished =  {\url{https://jolt.law.harvard.edu/assets/articlePDFs/v31/Counterfactual-Explanations-without-Opening-the-Black-Box-Sandra-Wachter-et-al.pdf}},
}

@article{Hullermeier2021-XAI-AleatoricEpistemicUncertainty-Springer,
	author    = {H{\"u}llermeier, Eyke and Waegeman, Willem},
	title     = {{Aleatoric and epistemic uncertainty in machine learning: an introduction to concepts and methods}},
	journal   = {Machine Learning},
	volume    = {110},
	number    = {3},
	pages     = {457--506},
	year      = {2021},
	month     = {Mar},
	doi       = {10.1007/s10994-021-05946-3},
	howpublished =  {\url{https://link.springer.com/article/10.1007/s10994-021-05946-3}},
	publisher = {Springer},
	address   = {Berlin, Germany},
}

@article{Jiang2022-XAI-UserEpistemicUncertainty-Elsevier,
	author    = {Jiang, Jinglu and Kahai, Surinder and Yang, Ming},
	title     = {{Who needs explanation and when? Juggling explainable AI and user epistemic uncertainty}},
	journal   = {International Journal of Human-Computer Studies},
	volume    = {165},
	pages     = {102839},
	year      = {2022},
	month     = {Sep},
	doi       = {10.1016/j.ijhcs.2022.102839},
	howpublished =  {\url{https://www.sciencedirect.com/science/article/abs/pii/S1071581922000660}},
	publisher = {Elsevier},
	address   = {Amsterdam, The Netherlands},
}

\newpage




\appendix

\section{Analysis of human-centered evaluations for explainable AI}
\label{app:Analysis}

The 73 publications reviewed by Kim et al. \cite{Kim2024-HumanCenteredEvaluationXAI-Frontiers} were coded on four criteria: context of use, user role, user goals and tasks, and explanation-content elicitation. Each criterion was rated on a three-level scale (yes, partial, no) using the following decision rules.

\begin{itemize}
	\item \textbf{Context of use.} Yes: the context is described explicitly with social, virtual, or organizational conditions. Partial: the context is described implicitly or only partially, as part of the research narrative. No: the context is not described.
	\item \textbf{User role.} Yes: the role is specified with typical role properties, such as role name, age range, gender mix, education, work environment, tools, goals, tasks, and limitations. Partial: the role is described with some of these characteristics, or only implicitly. No: the role is kept vague or not identified.
	\item \textbf{User goals and tasks.} Yes: user goals and tasks are listed explicitly. Partial: some user goals or tasks are listed, or only an application-specific task is identified. No: user goals and tasks are not mentioned.
	\item \textbf{Explanation-content elicitation.} Yes: explanation content is elicited from target users without providing predefined options, with the intent of informing AI system design. Partial: predefined explanation options are offered to the target users as input to system design. No: no explanation content is elicited as input to system design.
\end{itemize}

The results of the analysis are reported in Tables~\ref{tab:AnalysisXAIPublicationsPart1}, \ref{tab:AnalysisXAIPublicationsPart2}, and~\ref{tab:AnalysisXAIPublicationsPart3}. Six publications could not be accessed and are listed with dashes in their result columns.\footnote{The inaccessible publications are: Bayer et al. (2022), Scheers et al. (2021), Chien et al. (2022), Deo \& Sontakke (2021), Jang et al. (2023), and Eriksson et al. (2022).}

\begin{table}[!h]
	\scriptsize
	\caption{Secondary analysis of 73 publications, focusing on human-centered evaluations of explainable AI, reviewed by Kim et al \cite{Kim2024-HumanCenteredEvaluationXAI-Frontiers}, part 1}
	\label{tab:AnalysisXAIPublicationsPart1}
	\begin{tabular}{>{\raggedright\arraybackslash}m{4cm} 
			>{\centering\arraybackslash}m{1.75cm} 
			>{\centering\arraybackslash}m{1.75cm} 
			>{\centering\arraybackslash}m{1.75cm} 
			>{\centering\arraybackslash}m{1.75cm} 
			>{\raggedright\arraybackslash}m{3cm}} 
		\hline
		Paper key   & Context of use defined & User role defined & User goals and user tasks defined & Explanation content elicited & Application domain \\
		\hline 
		Bayer et al. (2022)
		\cite{Bayer2022-RoleDomainExpertiseTrust-JDS}
		& -
		& -           
		& -  
		& - 
		& -\\
		
		Branley-Bell et al. (2020)
		\cite{BranleyBell2020-TrustUnderstandingXAI-HCI}
		& No
		& No            
		& No  
		& No
		& Healthcare \\
		
		Bunde (2021)
		\cite{Bunde2021-AIAssistedHateSpeechDetection-HICSS}   
		& No
		& No           
		& No  
		& No
		& Social media \\
		
		Faulhaber et al. (2021)
		\cite{Faulhaber2021-EffectExplanationsTrust-MuC}   
		& No
		& No           
		& No  
		& No
		& Public transportation \\
		
		Fu \& Tantithamthavorn (2022)
		\cite{Fu2022-GPT2SP-TSE}    
		& No           
		& No  
		& No
		& No
		& Software development \\
		
		Kartikeya (2022)
		\cite{Kartikeya2022-TrustTransparencyXAI-SAI}            
		& No           
		& No  
		& No
		& No
		& Recommendation service \\
		
		K{\"u}hnlenz \& K{\"u}hnlenz (2023)
		\cite{Kuehnlenz2023-SituationalExplanationsTrust-RO-MAN}            
		& No           
		& No  
		& No
		& No
		& Autonomous driving \\
		
		Lundberg et al. (2022)
		\cite{Lundberg2022-TrustworthyIVNIDS-XAI-IEEEAccess}        
		& No           
		& No  
		& No
		& No
		& Cybersecurity for automotive \\
		
		Okumura \& Nagao (2023)
		\cite{Okumura2023-MIPCE-CounterfactualExplanations-ICANN}       
		& No           
		& No  
		& No
		& No
		& Not defined \\
		
		Reeder et al. (2023)
		\cite{Reeder2023-EvaluatingXAIUserGender-HCII}        
		& No           
		& No  
		& No
		& No
		& Education \\
		
		Selten et al. (2023)
		\cite{Selten2023-JustLikeIThought-PAR}
		& Partial           
		& Partial  
		& Partial
		& Partial
		& Police work \\
		
		Upasane et al. (2023)
		\cite{Upasane2023-Type2FuzzyXAI-TAAI}
		& No           
		& No  
		& No
		& No
		& Water pumping \\
		
		Wang \& Yin (2021)
		\cite{Wang2021-AreExplanationsHelpful-IUI}
		& No           
		& No  
		& No
		& No
		& Decision making \\
		
		\hline
		
		Wysocki et al. (2023)
		\cite{Wysocki2023-CommunicationGapHealthcare-AI}            
		& No           
		& No  
		& No
		& No
		& Healthcare \\
		
		La Gatta et al. (2021)
		\cite{LaGatta2021-CASTLE-ESWA}    
		& No           
		& No  
		& No
		& No
		& Diverse \\
		
		La Gatta et al. (2021)
		\cite{LaGatta2021-PASTLE-PRL}    
		& No           
		& No  
		& No
		& No
		& Diverse \\
		
		Nazaretsky et al. (2022)
		\cite{Nazaretsky2022-TeachersAI-LAK}        
		& No           
		& Partial  
		& No
		& \cellcolor{gray!25} Yes
		& Education \\
		
		Brdnik et al. (2023)
		\cite{Brdnik2023-TrustSatisfactionXAI-LearningAnalytics}        
		& Partial           
		& Partial  
		& No
		& No
		& Education \\
		
		F{\"o}rster et al. (2021)
		\cite{Forster2021-CapturingUsersReality-HICSS}  
		& No           
		& No  
		& No
		& No
		& Not defined \\
		
		Kim et al. (2023)
		\cite{Kim2023-ExplainResultsToUsers-TFSC}  
		& No           
		& No  
		& No
		& No
		& Not defined \\
		
		Meas et al. (2022)
		\cite{Meas2022-ExplainabilityTransparencyAHU-Sensors}  
		& No           
		& No  
		& No
		& No
		& Building technology \\
		
		Nagy \& Molontay (2024)
		\cite{Nagy2024-InterpretableDropoutPrediction-IJAIED}  
		& No           
		& No  
		& No
		& No
		& Education \\
		
		Polley et al. (2021)
		\cite{Polley2021-TowardsTrustworthinessXAI-SIGIR}  
		& No           
		& No  
		& No
		& No
		& Book retrieval \\
		
		Scheers et al. (2021)
		\cite{Scheers2021-ExplainableAdvisingDashboard-TEL}  
		& -           
		& -  
		& -
		& -
		& - \\
		
		Schulze-Weddige \& Zylowski (2022)
		\cite{SchulzeWeddige2022-EffectsXAIVisualizations-ArtsIT}  
		& No           
		& No  
		& No
		& No
		& Sales \\
		
		Swamy et al. (2023)
		\cite{Swamy2023-TrustingExplainersXAI-LAK}  
		& Partial           
		& Partial  
		& No
		& No
		& Education \\
		
		van der Waa et al. (2020)
		\cite{VanderWaa2020-InterpretableConfidenceMeasures-IJHCS}  
		& No           
		& No  
		& No
		& No
		& Autonomous ship driving \\
		
		\hline

	\end{tabular}
\end{table}

\begin{table}[!h]
	\scriptsize
	\caption{Secondary analysis of 73 publications, focusing on human-centered evaluations of explainable AI, reviewed by Kim et al \cite{Kim2024-HumanCenteredEvaluationXAI-Frontiers}, part 2}
	\label{tab:AnalysisXAIPublicationsPart2}
	\begin{tabular}{>{\raggedright\arraybackslash}m{4cm} 
			>{\centering\arraybackslash}m{1.75cm} 
			>{\centering\arraybackslash}m{1.75cm} 
			>{\centering\arraybackslash}m{1.75cm} 
			>{\centering\arraybackslash}m{1.75cm} 
			>{\raggedright\arraybackslash}m{3cm}} 
		\hline
		Paper key    & Context of use defined & User role defined & User goals and user tasks defined & Explanation content elicited & Application domain \\
		
		\hline 
		
		{\v{Z}}lahti{\'c} et al. (2023)
		\cite{Zlahtic2023-AgileMLDataCanyons-Medicine-AS}  
		& No           
		& No  
		& No
		& No
		& Healthcare \\
		
		Xu et al. (2023)
		\cite{Xu2023-DialogueExplanationsRuleBasedAI-EXTRAAMAS}  
		& No           
		& No  
		& No
		& No
		& Not defined \\
		
		Alufaisan et al. (2021)
		\cite{Alufaisan2021-DoesXAIImproveDecisionMaking-AAAI}  
		& No           
		& No  
		& No
		& No
		& Decision making \\
		
		Cau et al. (2023)
		\cite{Cau2023-SupportingHighUncertaintyDecisionsXAI-IUI}  
		& Partial           
		& Partial  
		& Partial
		& No
		& Stock trading \\
		
		Conati et al. (2021)
		\cite{Conati2021-PersonalizedXAI-ITS-AI}  
		& Partial           
		& Partial  
		& Partial
		& Partial
		& Education \\
		
		Ghai et al. (2021)
		\cite{Ghai2021-ExplainableActiveLearning-CSCW}  
		& No           
		& No  
		& No
		& No
		& Decision making \\
		
		Naiseh et al. (2023)
		\cite{Naiseh2023-ExplanationClassesTrustCalibration-IJHCS}  
		& \cellcolor{gray!25} Yes           
		& Partial 
		& \cellcolor{gray!25} Yes
		& No
		& Healthcare \\
		
		Wang et al. (2023)
		\cite{Wang2023-NestedModelUserCentricXAI-TVCG}  
		& No           
		& No  
		& No
		& No
		& Pharmaceutical \\
		
		Eriksson et al. (2022)
		\cite{Eriksson2022-XAISOC-IEEEBigData}  
		& -           
		& -  
		& -
		& -
		& - \\
		
		Maltbie et al. (2021)
		\cite{Maltbie2021-XAIToolsPublicSector-FSE}  
		& \cellcolor{gray!25} Yes           
		& \cellcolor{gray!25} Yes  
		& \cellcolor{gray!25} Yes
		& Partial
		& Utility \\
		
		Wang et al. (2022)
		\cite{Wang2022-InterpretableDirectedDiversity-CHI}  
		& No           
		& No  
		& No
		& No
		& Ideation \\
		
		Abdul et al. (2020)
		\cite{Abdul2020-COGAM-CognitiveLoadXAI-CHI}  
		& No           
		& No  
		& No
		& No
		& Housing prices \\
		
		Adhikari et al. (2019)
		\cite{Adhikari2019-LEAFAGE-FUZZIEEE}  
		& No           
		& No  
		& No
		& No
		& Housing \\
		
		Aechtner et al. (2022)
		\cite{Aechtner2022-UserPerceptionXAI-FUZZIEEE}  
		& Partial           
		& \cellcolor{gray!25} Yes
		& \cellcolor{gray!25} Yes
		& No
		& Education \\
		
		Anjara et al. (2023)
		\cite{Anjara2023-ExplainableCDSSThinkAloud-PLOSONE}  
		& \cellcolor{gray!25} Yes           
		& \cellcolor{gray!25} Yes  
		& \cellcolor{gray!25} Yes
		& \cellcolor{gray!25} Yes
		& Healthcare \\
		
		Avetisyan et al. (2022)
		\cite{Avetisyan2022-ExplanationsSituationAwareness-TRF}  
		& No           
		& No  
		& No
		& No
		& Automated driving \\
		
		\hline
		
		Ben David et al. (2021)
		\cite{BenDavid2021-XAIFinancialAdvisors-AIES}  
		& No           
		& No  
		& No
		& No
		& Decision making \\
		
		Bertrand et al. (2023)
		\cite{Bertrand2023-FeatureBasedExplanationsFinance-FAccT}  
		& Partial           
		& Partial
		& \cellcolor{gray!25} Yes
		& No
		& Decision making \\
		
		Bhattacharya et al. (2023)
		\cite{Bhattacharya2023-DirectiveExplanationsDiabetes-IUI}  
		& No           
		& \cellcolor{gray!25} Yes  
		& \cellcolor{gray!25} Yes
		& No
		& Healthcare \\
		
		Chien et al. (2022)
		\cite{Chien2022-XFlag-FakeNewsDetection-IJHCI}  
		& -           
		& -  
		& -
		& -
		& - \\
		
		Conijin et al. (2023)
		\cite{Conijn2023-ExplanationsEssayScoring-JLA}  
		& Partial           
		& \cellcolor{gray!25} Yes  
		& \cellcolor{gray!25} Yes
		& No
		& Education \\
		
		Das et al. (2023)
		\cite{Das2023-ExplainableActivityRecognition-TIIS}  
		& Partial           
		& \cellcolor{gray!25} Yes
		& \cellcolor{gray!25} Yes
		& No
		& Smart Home \\
		
		Deo \& Sontakke (2021)
		\cite{Deo2021-UserCentricExplainabilityFintech-HCII}  
		& -           
		& -  
		& -
		& -
		& - \\
		
		Fernandes et al. (2023)
		\cite{Fernandes2023-ExplainableWeightManagement-JMIR}  
		& Partial           
		& \cellcolor{gray!25} Yes
		& \cellcolor{gray!25} Yes
		& No
		& Healthcare \\
		
		Guo et al. (2022)
		\cite{Guo2022-BuildingTrustIML-IUI}  
		& No           
		& No  
		& Partial
		& No
		& Decision making \\
		
		Hern{\'a}ndez-Bocanegra \& Ziegler (2023)
		\cite{HernandezBocanegra2023-ExplainingRecommendationsDialogModel-TIIS}  
		& No           
		& No  
		& No
		& No
		& Decision making \\
		
		Jang et al. (2023)
		\cite{Jang2023-TowardInterpretableML-PAKDD}  
		& -           
		& -  
		& -
		& -
		& - \\
		
		Khodabandehloo et al. (2021)
		\cite{Khodabandehloo2021-HealthXAI-FGCS}  
		& No           
		& Partial
		& Partial
		& \cellcolor{gray!25} Yes
		& Healthcare \\
		
		Lasarati (2022)
		\cite{Larasati2022-XAI-BreastCancer-ICECET}  
		& No           
		& No  
		& No
		& No
		& Healthcare \\
		
		Moradi \& Samwald (2021)
		\cite{Moradi2021-XAI-CIE-Elsevier}  
		& No           
		& No  
		& No
		& No
		& Not defined \\
		
		Neves et al. (2021)
		\cite{Neves2021-XAI-HeartbeatExplainability-Elsevier}  
		& No           
		& Partial  
		& Partial
		& No
		& Healthcare \\
		
		Ooge et al. (2022)
		\cite{Ooge2022-XAI-ElearningTrust-ACM}  
		& No           
		& Partial  
		& Partial
		& No
		& Education \\
		
		\hline
	\end{tabular}
\end{table}

\begin{table}[!h]
	\scriptsize
	\caption{Secondary analysis of 73 publications, focusing on human-centered evaluations of explainable AI, reviewed by Kim et al \cite{Kim2024-HumanCenteredEvaluationXAI-Frontiers}, part 3}
	\label{tab:AnalysisXAIPublicationsPart3}
	\begin{tabular}{>{\raggedright\arraybackslash}m{4cm} 
			>{\centering\arraybackslash}m{1.75cm} 
			>{\centering\arraybackslash}m{1.75cm} 
			>{\centering\arraybackslash}m{1.75cm} 
			>{\centering\arraybackslash}m{1.75cm} 
			>{\raggedright\arraybackslash}m{3cm}} 
		\hline
		Paper key    & Context of use defined & User role defined & User goals and user tasks defined & Explanation content elicited & Application domain \\
		
		\hline 
		
		Panigutti et al. (2022)
		\cite{Panigutti2022-XAI-AdviceTakingTrust-ACM}  
		& No           
		& Partial  
		& Partial
		& No
		& Healthcare \\
		
		Panigutti et al. (2023)
		\cite{Panigutti2023-XAI-CoDesignClinicalXAI-ACM}  
		& No           
		& Partial  
		& Partial
		& No
		& Healthcare \\
		
		Schellingerhout et al. (2022)
		\cite{Schellingerhout2022-XAI-CareerPathPredictions-CEUR}  
		& No           
		& Partial  
		& Partial
		& No
		& Career \\
		
		Veldhuis et al. (2022)
		\cite{Veldhuis2022-XAI-ForensicDNAReCo-Elsevier}  
		& No           
		& Partial  
		& Partial
		& No
		& Genetics \\
		
		Warren et al. (2022)
		\cite{Warren2022-XAI-FeaturesExplainability-ArXiv}  
		& No           
		& Partial  
		& Partial
		& No
		& Driving \\
		
		Weitz et al. (2021)
		\cite{Weitz2021-XAI-VirtualAgentsTrust-Springer}  
		& No           
		& No  
		& Partial
		& No
		& Not defined \\
		
		Z{\"o}ller et al. (2023)
		\cite{Zoeller2023-XAI-XAutoML-ACM}  
		& No
		& Partial
		& Partial
		& Partial
		& Healthcare\\
		
		\hline
		
		Bansal et al. (2021)
		\cite{Bansal2021-XAI-TeamPerformance-ACM}  
		& No           
		& No  
		& No
		& No
		& Decision making \\
		Bu\c{c}inca et al. (2020)
		\cite{Bucinca2020-XAI-ProxyTasksEvaluation-ACM}  
		& No           
		& No  
		& Partial
		& No
		& Decision making \\
		Confalonieri et al. (2021)
		\cite{Confalonieri2021-XAI-OntologyGlobalExplanations-Elsevier}  
		& No           
		& No  
		& No
		& No
		& Decision making \\
		Ibrahim et al. (2022)
		\cite{Ibrahim2023-XAI-FactualCounterfactualExplanations-IFAAMAS}  
		& No           
		& No  
		& Partial
		& No
		& Legal \\
		Jmoona et al. (2023)
		\cite{Jmoona2023-XAI-FlightDelayPrediction-Springer}  
		& No           
		& Yes  
		& Yes
		& No
		& Air Traffic Control \\
		Raab et al. (2023)
		\cite{Raab2023-XAI-XAI4EEG-Springer}  
		& No           
		& No  
		& Partial
		& No
		& Healthcare \\
		Schrills \& Franke (2023)
		\cite{Schrills2023-XAI-TraceabilityAID-ACM}  
		& No           
		& Partial  
		& Partial
		& No
		& Healthcare \\
		\hline
	\end{tabular}
\end{table}

\clearpage


\section{Overview about the user studies}
\label{app:Userstudies}


\subsection*{User Study 1: Facility Manager}

The first study \cite{Degen2021-inproceedings} focused on building management systems and the user role of facility manager. Commercial buildings, including campuses and large residential buildings, are typically equipped with several control systems, such as fire alarm, power, heating and cooling, ventilation, and security systems. Facility managers use a building management system (BMS) to observe and operate these systems. The BMS reports building status and issues to the facility manager, and continuously measures building control properties against defined setpoints to keep operation within specified ranges. The AI-generated outcome is a suggested action in response to a detected incident, and the explanations justify why the action was suggested.


\subsection*{User Study 2: Energy Engineers}

The second study \cite{Degen2022-inproceedings} also focuses on buildings and investigates the facility manager's counterpart, the role of energy engineer. An energy engineer supervises buildings for one or more customers who are either tenants or owners, and collaborates with the facility managers whom those customers have employed for daily building operations. The AI-generated outcome is a suggested action in response to a detected incident, and the explanations justify the suggested action. Although facility managers and energy engineers have similar user goals, their roles differ: the facility manager is responsible for a specific building, while the energy engineer is external and is paid by tenants or owners to provide building services.


\subsection*{User Study 3: Model Managers}

The third user study \cite{Degen2023-inproceedings} investigates the explanation needs of the model manager role in the manufacturing domain. Manufacturing involves multiple processing steps, each of which can introduce errors; quality assurance steps are therefore used to verify that parts comply with defined quality criteria. Quality assurance today relies on standard technology and manual inspection, and the intent is to replace some of the costly manual inspections with AI systems. Because AI systems can degrade over time, a model manager monitors one or more AI systems and addresses detected deviations in a timely manner. The AI-generated outcome is a suggested action in response to a detected deviation, and the explanations justify the suggested action.


\subsection*{User Study 4: System Tester}

The fourth user study \cite{Degen2024-inproceedings} investigates the explanation needs of a system tester working with a futuristic AI system that plans and conducts tests of AI-based software systems. A system tester ensures the quality of the larger software system, which includes one or more AI components. This involves developing a test strategy that defines the scope, methodologies, environments, and execution plan; designing test cases tailored to AI components, including scenarios and edge cases; preparing test inputs that adequately exercise the AI models; and analyzing the test results produced under the test environment. Test results are determined by the system under test executing within the test environment. Failing results may originate from a defect in the test case, the test environment, or the system under test; defects are reported and must be fixed before the next execution cycle. The futuristic AI system generates three outcomes: the test strategy, the test cases, and the test results, the latter including identified defects and responsive actions. Accordingly, three sets of explanations are needed, one for each outcome.


\subsection*{User Study 5: Data Scientist}

The fifth user study \cite{Degen2025-misc,Degen2025-XAI-ExplainDataScientists-Springer-InProceedings} investigates the explanation needs of a data scientist. Data scientists collect, analyze, and interpret large datasets to extract actionable insights, and they train, test, and validate machine learning models for their intended use. As in the fourth user study, this study considers a futuristic AI system that generates, deploys, and maintains machine learning models. The data scientist supplies the system with inputs expressing context, constraints, and objectives. The AI system then proceeds in three steps. It generates one or more combinations of machine learning model, model parameters, and dataset that address the provided inputs, with the data scientist selecting the best combination. It deploys the selected combination to the production environment. It monitors the deployed model's performance, detects quality deviations, and, when a deviation is detected, generates a suggested action. Explanation needs are identified for the first and third steps, which produce two AI-generated outcomes: a suggested model-parameter-dataset combination and a suggested action to address a detected deviation. Both outcomes require explanation content that justifies the outcome.


\subsection*{User Study 6: Head of Cybersecurity and IT Specialty Lead}

The sixth study (not published) investigates the explanation needs of two user roles in hospital cybersecurity: the Head of Cybersecurity and the IT Specialty Lead. The Head of Cybersecurity identifies, assesses, and prioritizes cybersecurity risks, reviews vulnerabilities and remediation options, and coordinates responses across IT, biomedical engineering, compliance, and clinical stakeholders. The IT Specialty Lead maintains and supports technical systems within a specialty area, including uptime, patches, incidents, and integrations, aligning technical decisions with clinical workflow and safety requirements. The AI-generated outcome is a suggested remediation action that is technically effective and operationally feasible given hospital constraints such as maintenance windows and device dependencies. The explanations justify the suggested remediation by identifying the prioritized vulnerabilities the remediation addresses, explaining why those vulnerabilities are locally exploitable and clinically consequential, describing the tradeoffs the remediation implies, reporting the system's confidence and the evidence behind it, and indicating whether the recommended action is deployable given vendor dependencies and clinical workflow impact.


\subsection*{Characteristics of the six user studies}

\begin{table}[!ht]
	\small
	\caption{Characteristics of the six user studies that provided explanation content for the hybrid inductive-deductive content analysis, including per-study meaning-unit counts and material-type breakdown}
	\label{tab:UserStudies:ExplanationContentVariations}
	\begin{scriptsize}
		\begin{tabular}
			{>{\raggedright\arraybackslash}m{1.5cm}
				>{\raggedright\arraybackslash}m{1.5cm}
				>{\raggedright\arraybackslash}m{1.5cm}
				>{\raggedright\arraybackslash}m{1.5cm}
				>{\raggedright\arraybackslash}m{1.5cm}
				>{\raggedright\arraybackslash}m{1.5cm}
				>{\raggedright\arraybackslash}m{1.5cm}}
			
			\hline
			
			Variation dimension &
			User study 1 &
			User study 2 &
			User study 3 &
			User study 4 &
			User study 5 &
			User study 6\\
			
			\hline
			
			Application domain &
			Building technology &
			Building technology &
			Manufacturing &
			AI software development &
			AI software development &
			Cybersecurity vulnerability management in hospitals\\
			
			\hline

			Application type &
			Anomaly detection and resolution &
			Anomaly detection and resolution &
			Anomaly detection and resolution &
			Software engineering artifact generation &
			Software engineering artifacts generation &
			Anomaly detection and resolution \\
			
			\hline
			
			User role &
			Facility manager &
			Energy engineer &
			Model manager &
			System tester &
			Data scientist &
			Cybersecurity expert\\
			
			\hline
			
			Lifecycle phase &
			Operation &
			Operation &
			Maintenance &
			Development &
			Development &
			Maintenance \\
			
			\hline
			
			Number of study participants &
			1 domain expert; 4 end users (one interview each, 60--90 minutes per interview) &
			1 domain expert; 11 end users (one interview each, 60--90 minutes per interview) &
			1 domain expert; 10 end users (one interview each, 60--90 minutes per interview) &
			1 domain expert; 12 end users (one interview each, 90--120 minutes per interview) &
			2 domain experts; 12 end users (two interviews per end user, about 90 minutes per interview) &
			Elicitation with 1 hospital IT expert and 2 cybersecurity experts; evaluation with cybersecurity experts and IT experts ongoing\\
			
			\hline
			
			Material type &
			User interface &
			User interface &
			Diagrammatic + User interface &
			Diagrammatic &
			Diagrammatic + User interface &
			Diagrammatic \\
			
			\hline
			
		\end{tabular}
		\begin{minipage}{\linewidth}
			\vspace{2pt}
			\footnotesize\textit{Note.} Study~5 is the largest single contributor at approximately 65\% of the corpus total of 325 units, reflecting UI granularity effects discussed in the Results section. Study 6 is in progress.
		\end{minipage}
	\end{scriptsize}
\end{table}

\clearpage


\section{Detailed Results}
\label{app:Results}

\subsection{Step~1: Corpus definition and variation design}
\label{sec:Results:Step1}

This subsection reports the size and basic characteristics of the explanation-content corpus derived from user studies~1 to~6 (see Appendix~\ref{app:Userstudies}). In total, 325 meaning units were extracted from multimodal study materials, including user-interface representations and diagrammatic mental models (Study~4). Table~\ref{tab:Result:MeaningUnits} summarizes the number of meaning units per user study and in total. The same table also reports the distribution of the stratified purposive reliability-validation subsample across user studies; subsample composition and sampling rationale are described in Step~9. The reliability subsample comprised 82 meaning units, corresponding to 25.2\% of the coded corpus.

\begin{table}[!ht]
	\caption{Distribution of all meaning units and of the stratified purposive reliability-validation subsample across user studies. Subsample composition and sampling rationale are described in Step~9.}
	\label{tab:Result:MeaningUnits}
	\begin{scriptsize}
		\begin{tabular}
			{>{\raggedright\arraybackslash}m{2.2cm}
				>{\centering\arraybackslash}m{1.15cm}
				>{\centering\arraybackslash}m{1.15cm}
				>{\centering\arraybackslash}m{1.15cm}
				>{\centering\arraybackslash}m{1.15cm}
				>{\centering\arraybackslash}m{1.15cm}
				>{\centering\arraybackslash}m{1.15cm} |
				>{\centering\arraybackslash}m{1.15cm}
			}
			\hline
			& Study~1 & Study~2 & Study~3 & Study~4 & Study~5 & Study~6 & Total \\
			\hline
			All meaning units ($n$) 
			& \cellcolor{gray!25} 12 
			& \cellcolor{gray!25} 23 
			& \cellcolor{gray!25} 18 
			& \cellcolor{gray!25} 47 
			& \cellcolor{gray!25} 211 
			& \cellcolor{gray!25} 14 
			& \cellcolor{gray!25} 325 \\
			
			All meaning units (\%) 
			& \cellcolor{gray!25} 3.7 
			& \cellcolor{gray!25} 7.1 
			& \cellcolor{gray!25} 5.5 
			& \cellcolor{gray!25} 14.5 
			& \cellcolor{gray!25} 64.9 
			& \cellcolor{gray!25} 4.3 
			& \cellcolor{gray!25} 100.0 \\
			
			\hline
			
			Reliability subsample ($n$) 
			& \cellcolor{gray!25} 8 
			& \cellcolor{gray!25} 6 
			& \cellcolor{gray!25} 6 
			& \cellcolor{gray!25} 12 
			& \cellcolor{gray!25} 45 
			& \cellcolor{gray!25} 5 
			& \cellcolor{gray!25} 82 \\
			
			Reliability subsample (\%) 
			& \cellcolor{gray!25} 9.8 
			& \cellcolor{gray!25} 7.3 
			& \cellcolor{gray!25} 7.3 
			& \cellcolor{gray!25} 14.6 
			& \cellcolor{gray!25} 54.9 
			& \cellcolor{gray!25} 6.1 
			& \cellcolor{gray!25} 100.0 \\
			
			\hline
			
		\end{tabular}
	\end{scriptsize}
\end{table}

The distribution of meaning units across user studies reflects two structural sources of variation that should be distinguished from sampling bias. First, elicitation format affects the granularity at which conceptual content is represented: a user interface yields one meaning unit per individually addressable UI element, whereas a mental model yields one meaning unit per conceptual node or contiguous statement. Because Study~5 involved the design of a detailed user interface, conceptual content that would constitute a single meaning unit in other studies, such as a causal factor with associated ranking properties and evidence or an individual actual element with its attributes, is represented as multiple meaning units in Study~5, one per rendered UI element. This format effect accounts for the higher absolute frequency of Input, What-if, and Individual-element codes in Study~5 relative to Studies~1 to~4 and~6. The same conceptual content, when elicited through a mental model in Study~4, yields fewer meaning units for equivalent conceptual coverage because the mental-model format represents each concept as a single node rather than decomposing it into individually addressable interface elements.

Second, the applicability of certain explanation types is constrained by system architecture rather than elicitation format. What-if reasoning over inputs presupposes that inputs are user-modifiable, which is the case in Studies~4 and~5 but not in the remaining studies where inputs are machine-generated from sensor readings, process data, or system logs. Because users cannot vary machine-generated inputs directly, counterfactual What-if reasoning over inputs is not a practically actionable explanation type in those contexts. The absence of What-if content in Studies~1,~2,~3, and~6 therefore reflects a genuine property of those explanation spaces rather than incomplete elicitation. Table~\ref{tab:Result:WhatIfDistribution} summarizes the relationship between input type and What-if code frequency across studies.

\begin{table}[!ht]
	\caption{Relationship between input type and What-if code frequency across user studies.}
	\label{tab:Result:WhatIfDistribution}
	\begin{scriptsize}
		\begin{tabular}
			{>{\raggedright\arraybackslash}m{2.5cm}
				>{\centering\arraybackslash}m{1.35cm}
				>{\centering\arraybackslash}m{1.35cm}
				>{\centering\arraybackslash}m{1.35cm}
				>{\centering\arraybackslash}m{1.35cm}
				>{\centering\arraybackslash}m{1.35cm}
				>{\centering\arraybackslash}m{1.35cm}
			}
			
			\hline
			
			& Study~1 & Study~2 & Study~3 & Study~4 & Study~5 & Study~6 \\
			
			\hline
			
			Input type 
			& Machine-generated 
			& Machine-generated 
			& Machine-generated 
			& User-provided 
			& User-provided 
			& Machine-generated \\
			
			Primary elicitation format 
			& User interface 
			& User interface 
			& User interface 
			& Mental model 
			& User interface 
			& Mental model \\
			
			What-if units 
			& 0 
			& 0 
			& 0 
			& 18 
			& 54 
			& 0 \\
			
			\hline
			
		\end{tabular}
	\end{scriptsize}
\end{table}

The scale imbalance created by Study~5 warrants explicit consideration of its implications for blueprint generalizability. The domain-of-origin distribution reported in Step~4 provides independent evidence of cross-domain coverage. AI-domain content (37.2\%), application-domain content (32.0\%), and system-domain content (21.5\%) all appear in at least four of the six studies, indicating that the explanatory distinctions that drove category formation are not specific to the UI-design context of Study~5. The Discussion section considers the implications of this analysis for the scope conditions under which the blueprint can be applied.

To assess whether the blueprint's structure and code frequencies are driven by Study~5, Table~\ref{tab:Sensitivity} reports the code frequency distribution for two sub-corpora: Study~5 alone ($n = 211$) and Studies~1--4 and~6 combined ($n = 114$). Frequencies are reported as percentages of each sub-corpus total.

\begin{table}[ht!]
	\scriptsize
	\caption{Code frequency distribution for Study~5 alone and Studies~1--4 and~6 combined, reported as percentages of each sub-corpus total. Full corpus percentages are included for reference.}
	\label{tab:Sensitivity}
	\begin{tabular}
		{>{\raggedright\arraybackslash}m{4.5cm}
			>{\centering\arraybackslash}m{2.5cm}
			>{\centering\arraybackslash}m{2.5cm}
			>{\centering\arraybackslash}m{2.5cm}
		}
		
		\hline
		
		Post-validation code &
		Study 5 only &
		Studies 1--4 + 6 &
		Full corpus \\
		
		& (\%, $n=211$) &
		(\%, $n=114$) &
		(\%, $n=325$) \\
		
		\hline
		
		\multicolumn{4}{l}{\textit{Rule-based group}} \\
		
		\hline
		
		PoVC 1.1 Rule base                            & 0.0  & 0.0  & 0.0  \\
		PoVC 1.2 Applied rule                         & 0.0  & 1.8  & 0.6  \\
		
		\hline
		\multicolumn{4}{l}{\textit{Causal group}} \\
		\hline
		
		PoVC 2.1 Context                              & 6.2  & 5.3  & 5.8  \\
		PoVC 2.2 Input                                & 26.1 & 23.7 & 25.2 \\
		PoVC 2.3 Causal factor                        & 0.9  & 3.5  & 1.8  \\
		PoVC 2.4 Outcome                              & 3.3  & 8.8  & 5.2  \\
		PoVC 2.5 Future state                         & 0.5  & 0.9  & 0.6  \\
		PoVC 2.6 Causal mechanism                     & 0.0  & 0.9  & 0.3  \\
		PoVC 2.7 What-if forward                      & 25.6 & 15.8 & 22.2 \\
		PoVC 2.8 What-if backward                     & 0.0  & 0.0  & 0.0  \\
		
		\hline
		\multicolumn{4}{l}{\textit{Epistemic group (actual)}} \\
		\hline
		
		PoVC 3.1 Set of ranked elements (actual)      & 3.8  & 8.8  & 5.5  \\
		PoVC 3.2 Individual element (actual)          & 30.3 & 30.7 & 30.2 \\
		
		\hline
		\multicolumn{4}{l}{\textit{Epistemic group (similar)}} \\
		\hline
		
		PoVC 4.1 Set of similar elements (similar)    & 1.9  & 0.0  & 1.2  \\
		PoVC 4.2 Individual similar element (similar) & 0.9  & 0.0  & 0.6  \\
		
		\hline
		
		Not assigned                                  & 0.5  & 0.9  & 0.6  \\
		
		\hline
		
	\end{tabular}
	
	\smallskip
	\begin{minipage}{\linewidth}
		\footnotesize
		Note 1: Per-study meaning-unit counts are reported in Appendix~\ref{app:Userstudies}, verified against the segmented corpus records. Study~5 is the largest single contributor at approximately 65\% of the corpus total of 325 units, reflecting UI granularity effects discussed in the Results section.\\
		Note 2: The zero frequencies for epistemic-similar codes (PoVC 4.1 and PoVC 4.2) in the Studies 1--4+6 column reflect the absence of historical case bases in those system architectures, not a claim that these codes are unimportant or inapplicable. The available sub-corpora show partial structural convergence, with strongest support for causal and epistemic-actual categories and weaker support for rule-based and epistemic-similar categories.
	\end{minipage}
\end{table}

Three patterns of cross-sub-corpus stability can be distinguished in the frequency distribution. First, codes whose relative frequency is comparable across both sub-corpora, specifically Context (6.2\% vs.\ 5.3\%), Individual element (30.3\% vs.\ 30.7\%), and Set of ranked elements (3.8\% vs.\ 8.8\%), indicate structural features of the blueprint that are not specific to the UI-design context of Study~5. 
Second, codes whose frequency is substantially higher in Study~5 but which remain present in the remainder sub-corpus, specifically What-if forward (25.6\% vs.\ 15.8\%) and Input (26.1\% vs.\ 23.7\%), reflect the architecture-driven scoping criterion identified in Finding~3: What-if content arises when inputs are user-modifiable, and Study~4 in the remainder sub-corpus also involves user-modifiable inputs, accounting for the non-zero remainder frequency. The frequency asymmetry is therefore partially but not entirely a Study~5 effect. 
Third, codes concentrated in Study 5 with zero representation in the remainder sub-corpus, specifically the epistemic (similar) group (PoVC 4.1: 1.9\% vs. 0.0\%; PoVC 4.2: 0.9\% vs. 0.0\%), reflect the system architecture of Study 5, which involved an AI system for data scientists with explicit access to a historical case base. This concentration is a corpus property rather than evidence that similar-outcome epistemic content is less important in other industrial contexts; it indicates only that the studied systems in Studies 1--4 and 6 did not expose historical case bases as an explanation modality. The category structure, as distinct from code frequencies, generalizes across sub-corpora. To clarify, the claim that the category structure generalizes means that when similar-outcome epistemic content does appear in a system (for example, in a case-based reasoning system), its constituent elements can be reliably coded using PoRC 4.1 and PoRC 4.2 according to the definitions and decision rules in the codebook. The claim is about the definitional distinctness and coding reliability of the categories, not about the frequency with which such content occurs across industrial domains.

Study~6 differed from Studies~1 to~5 in that it used a survey-based elicitation and validation format rather than interview-based elicitation.\footnote{The content validation study for Study~6 (importance ratings for user groups) is ongoing and will be reported separately; it does not affect the meaning unit extraction or coding reported here.} It was retained because it also produced concrete explanation-content statements that could be segmented into meaning units using the same unit-of-analysis rules as the other studies.

\clearpage


\subsection{Step~2: Meaning unit definition and coding hierarchy setup}
\label{sec:Results:Step2}

Meaning units were defined as distinct segments that expressed explanatory meaning, large enough to preserve contextual meaning but small enough to be coded consistently. Meaning units were fixed before coding so that validity and reliability evaluations referred to identical units of analysis.

Segmentation was applied to all six user studies without exclusion of any study material. The multimodal character of the corpus required application of distinct segmentation rules for three material types: user interface elements, diagrammatic mental model nodes and relations, and interview statements. No segmentation conflicts arose between material types because the unit-of-analysis rules were defined per material type rather than as a single cross-modal rule. A small number of meaning units required boundary judgements during segmentation, specifically in cases where a single UI element displayed content spanning two candidate explanation-content elements. In all such cases the unit was retained as a single meaning unit and the boundary judgement was documented in the coding log, with the final code assignment resolving the ambiguity at the coding stage rather than the segmentation stage.

It is also possible for a unit to pass the initial segmentation screen (expressing language that is explanatory in character) but to fail code assignment during open coding because its content refers to meta-level interface properties rather than to explanation content itself. Such units are retained in the corpus total to preserve the integrity of the meaning-unit count and included in all table totals and percentage denominators for transparency; they are reported in all tables under the label Not assigned. Two such units were identified during open coding in Step~3 and are reported there.


\subsection{Step~3: Open coding and normalization of preliminary codes}
\label{sec:Results:Step3}

This subsection reports the outcomes of open coding and code consolidation for user studies~1 to~6. Across 325 meaning units, open coding resulted in an initial pool of preliminary codes that captured the functional meaning of each unit. Preliminary codes were data-near descriptive labels formulated to reflect what information the content element conveys and what role it serves in an explanation, following open-coding procedures in qualitative content analysis.

Preliminary codes were then consolidated into a normalized set of pre-validation codes to reduce superficial variation while preserving functional distinctions that affect coding decisions. Consolidation actions included merging semantically equivalent codes, splitting codes that conflated distinct functions, and refining labels to reduce ambiguity and improve downstream reuse in the draft codebook. Consolidation decisions were guided by whether alternative labels would lead to different inclusion or exclusion decisions during coding.

In the pre-validation coding frame developed in Steps~3 to~7, each meaning unit received one pre-validation code assignment that captured its explanation-content type. The resulting pre-validation codes covered rule-based explanation content, causal explanation content, and epistemic explanation content. The causal perspective comprised context, input, causal factor, causal mechanism, outcome, future state, and what-if. The epistemic perspective distinguished sets of ranked elements from individual elements and distinguished content referring to the actual outcome from content referring to similar outcomes. The post-validation coding system is reported after Step~8 and in the revised codebook.

As anticipated in Step~2, two meaning units (one from Study~2, which described a display format preference, and one from Study~5, which described a navigation interaction) passed the initial segmentation screen but could not be assigned to any pre-validation code during open coding because their content referred to meta-level interface properties rather than explanation content. Both units are retained in the corpus total of 325 and appear in all tables under the label Not assigned; they are included in all table totals and percentage denominators for transparency, and their frequency is reported in the Not assigned row of each table. Their Not assigned status was confirmed during post-validation recoding in Step~9 and they remain Not assigned in the final dataset.

Beyond the two Not assigned units, open coding also produced boundary-level findings that informed the draft codebook. Four code boundaries were identified as potentially ambiguous and flagged for decision-rule development: Applied rule vs Causal factor, Causal factor vs Causal mechanism, Set of ranked elements vs Individual element, and Input vs What-if forward. These boundaries were subsequently addressed in the codebook construction reported in Step~6 and informed the subsample design for the reliability evaluation in Step~9.

PC~2.1 Context was carried forward without modification because its definition was unambiguous, no alternative formulations were identified during open coding, and no boundary conflicts with adjacent codes arose during the consolidation review. Table~\ref{tab:Result:ConsolidationPreliminaryCodes} documents how all other preliminary codes were consolidated into the pre-validation code set. Table~\ref{tab:Result:PreValidationCodes} reports the harmonized distribution of the pre-validation code set across user studies and in the corpus as a whole. All percentages in Table~\ref{tab:Result:PreValidationCodes} are computed against the full corpus of 325 meaning units, including the two Not assigned units, which are retained in the denominator for transparency. The full change log documenting the trigger, rationale, and outcome of each consolidation decision from preliminary codes to pre-validation codes is provided in Appendix~\ref{app:ChangeLogPreliminaryToPrevalidation}.

\begin{table}[!ht]
	\small
	\caption{Revision of the preliminary codes to pre-validation codes}
	\label{tab:Result:ConsolidationPreliminaryCodes}
	\begin{scriptsize}
		\begin{tabular}
			{>{\raggedright\arraybackslash}m{4.0cm}
				>{\raggedright\arraybackslash}m{3.5cm}
				>{\raggedright\arraybackslash}m{5.0cm}}
			\hline
			Preliminary code & Consolidation strategy & Pre-validation code \\
			\hline
			PC~1.1 Rule & Wording modified & PrVC~1.1 Rule attribute \\
			\hline
			PC~2.1 Context & -- & PrVC~2.1 Context \\
			PC~2.2 Input & Code split & PrVC~2.2 Input, PrVC~2.7 What-if \\
			PC~2.3 Causal chain & Code removed & -- \\
			PC~2.4 Causal factor & Renumbered & PrVC~2.3 Causal factor \\
			PC~2.5 Causal mechanism & Renumbered & PrVC~2.4 Causal mechanism \\
			PC~2.6 Outcome & Renumbered & PrVC~2.5 Outcome \\
			PC~2.7 Future state & Renumbered & PrVC~2.6 Future state \\
			\hline
			PC~3.1 Ranked elements (actual) & Wording modified & PrVC~3.1 Set of ranked elements with ranking mechanism (actual) \\
			PC~3.2 Element ranking property (actual) & Code merged & PrVC~3.2 Individual element with ranking properties, evidence, and/or attributes (actual) \\
			PC~3.3 Element attributes (actual) & Code merged & PrVC~3.2 Individual element with ranking properties, evidence, and/or attributes (actual) \\
			PC~3.4 Element evidence (actual) & Code merged & PrVC~3.2 Individual element with ranking properties, evidence, and/or attributes (actual) \\
			PC~3.5 Ranked elements (similar) & Wording modified; renumbered to group 4 & PrVC~4.1 Set of similar elements with filter mechanism (similar) \\
			PC~3.6 Element ranking property (similar) & Code merged; renumbered to group 4 & PrVC~4.2 Individual, similar element with filter properties, evidence, and/or attributes (similar) \\
			PC~3.7 Element attributes (similar) & Code merged; renumbered to group 4 & PrVC~4.2 Individual, similar element with filter properties, evidence, and/or attributes (similar) \\
			PC~3.8 Element evidence (similar) & Code merged; renumbered to group 4 & PrVC~4.2 Individual, similar element with filter properties, evidence, and/or attributes (similar) \\
			\hline
		\end{tabular}
	\end{scriptsize}
\end{table}

PC~2.2 Input was split into PrVC~2.2 Input and PrVC~2.7 What-if because user-entered input can also be used to define hypothetical scenarios that belong to counterfactual reasoning. PC~2.3 Causal chain was removed because it functioned as a higher-level grouping construct and was therefore better represented as a category-level element. PC~3.1 Ranked elements (actual) was renamed to PrVC~3.1 Set of ranked elements with ranking mechanism (actual) to make explicit that the code refers to multiple ranked elements. PC~3.5 Ranked elements (similar) was renamed to PrVC~4.1 Set of similar elements with filter mechanism (similar) to emphasize the set character and the applied filter of similar elements.

PC~3.2 Element ranking property (actual), PC~3.3 Element attributes (actual), and PC~3.4 Element evidence (actual) were merged into PrVC~3.2 Individual element with ranking properties, evidence, and/or attributes (actual) to simplify coding at the pre-validation code level and to shift finer distinctions into codebook definitions and examples. The same consolidation logic was applied to the similar-outcome perspective by merging PC~3.6, PC~3.7, and PC~3.8 into PrVC~4.2 Individual, similar element with filter properties, evidence, and/or attributes (similar). The definitions of the pre-validation codes are documented in Appendix~\ref{app:DraftCodebook}.

\begin{table}[!ht]
	\small
	\caption{Assignment of meaning units to the pre-validation code set ($N = 325$). All percentages are computed against the full corpus of 325 meaning units.}
	\label{tab:Result:PreValidationCodes}
	\begin{scriptsize}
		\begin{tabular}
			{>{\raggedright\arraybackslash}m{2.55cm}
				>{\centering\arraybackslash}m{0.95cm}
				>{\centering\arraybackslash}m{0.95cm}
				>{\centering\arraybackslash}m{0.95cm}
				>{\centering\arraybackslash}m{0.95cm}
				>{\centering\arraybackslash}m{0.95cm}
				>{\centering\arraybackslash}m{0.95cm}|
				>{\centering\arraybackslash}m{1.0cm}
				>{\centering\arraybackslash}m{1.0cm}}
			\hline
			Pre-validation code & Study~1 & Study~2 & Study~3 & Study~4 & Study~5 & Study~6 & Total & \% \\
			\hline
			PrVC~1.1 Rule attribute & 0 & 0 & 0 & 0 & 0 & \cellcolor{gray!25} 2 & \cellcolor{gray!25} 2 & \cellcolor{gray!25} 0.6 \\
			\hline
			PrVC~2.1 Context & \cellcolor{gray!25} 3 & \cellcolor{gray!25} 2 & 0 & 0 & \cellcolor{gray!25} 13 & \cellcolor{gray!25} 1 & \cellcolor{gray!25} 19 & \cellcolor{gray!25} 5.8 \\
			PrVC~2.2 Input & \cellcolor{gray!25} 1 & \cellcolor{gray!25} 2 & \cellcolor{gray!25} 2 & \cellcolor{gray!25} 21 & \cellcolor{gray!25} 55 & \cellcolor{gray!25} 1 & \cellcolor{gray!25} 82 & \cellcolor{gray!25} 25.2 \\
			PrVC~2.3 Causal factor & \cellcolor{gray!25} 1 & \cellcolor{gray!25} 1 & \cellcolor{gray!25} 1 & 0 & \cellcolor{gray!25} 2 & \cellcolor{gray!25} 1 & \cellcolor{gray!25} 6 & \cellcolor{gray!25} 1.8 \\
			PrVC~2.4 Causal mechanism & 0 & 0 & \cellcolor{gray!25} 1 & 0 & 0 & 0 & \cellcolor{gray!25} 1 & \cellcolor{gray!25} 0.3 \\
			PrVC~2.5 Outcome & \cellcolor{gray!25} 1 & \cellcolor{gray!25} 1 & \cellcolor{gray!25} 3 & \cellcolor{gray!25} 4 & \cellcolor{gray!25} 7 & \cellcolor{gray!25} 1 & \cellcolor{gray!25} 17 & \cellcolor{gray!25} 5.2 \\
			PrVC~2.6 Future state & 0 & 0 & 0 & 0 & \cellcolor{gray!25} 1 & \cellcolor{gray!25} 1 & \cellcolor{gray!25} 2 & \cellcolor{gray!25} 0.6 \\
			PrVC~2.7 What-if & 0 & 0 & 0 & \cellcolor{gray!25} 18 & \cellcolor{gray!25} 54 & 0 & \cellcolor{gray!25} 72 & \cellcolor{gray!25} 22.2 \\
			\hline
			PrVC~3.1 Set of ranked elements with ranking mechanism (actual) & \cellcolor{gray!25} 1 & \cellcolor{gray!25} 3 & \cellcolor{gray!25} 3 & 0 & \cellcolor{gray!25} 8 & \cellcolor{gray!25} 3 & \cellcolor{gray!25} 18 & \cellcolor{gray!25} 5.5 \\
			PrVC~3.2 Individual element with ranking properties, evidence, and/or attributes (actual) & \cellcolor{gray!25} 5 & \cellcolor{gray!25} 13 & \cellcolor{gray!25} 8 & \cellcolor{gray!25} 4 & \cellcolor{gray!25} 64 & \cellcolor{gray!25} 4 & \cellcolor{gray!25} 98 & \cellcolor{gray!25} 30.2 \\
			\hline
			PrVC~4.1 Set of similar elements with filter mechanism (similar) & 0 & 0 & 0 & 0 & \cellcolor{gray!25} 4 & 0 & \cellcolor{gray!25} 4 & \cellcolor{gray!25} 1.2 \\
			PrVC~4.2 Individual, similar element with filter properties, evidence, and/or attributes (similar) & 0 & 0 & 0 & 0 & \cellcolor{gray!25} 2 & 0 & \cellcolor{gray!25} 2 & \cellcolor{gray!25} 0.6 \\
			\hline
			Not assigned & 0 & \cellcolor{gray!25} 1 & 0 & 0 & \cellcolor{gray!25} 1 & 0 & \cellcolor{gray!25} 2 & \cellcolor{gray!25} 0.6 \\
			\hline
			Total & \cellcolor{gray!25} 12 & \cellcolor{gray!25} 23 & \cellcolor{gray!25} 18 & \cellcolor{gray!25} 47 & \cellcolor{gray!25} 211 & \cellcolor{gray!25} 14 & \cellcolor{gray!25} 325 & \cellcolor{gray!25} 99.8 \\
			\hline
		\end{tabular}
		\par\smallskip
		\footnotesize $^{a}$Total does not sum to 100.0 due to rounding.
	\end{scriptsize}
\end{table}

Across all user studies, PrVC~3.2 Individual element with ranking properties, evidence, and/or attributes (actual) was the most frequent pre-validation code (98, 30.2\%), followed by PrVC~2.2 Input (82, 25.2\%) and PrVC~2.7 What-if (72, 22.2\%). All percentages are computed against the full corpus of 325 meaning units, including the two Not assigned units, which are retained in the denominator for transparency. The concentration of PrVC~3.2 assignments in Study~5 reflects the same format effect described in Step~1. In Study~5, a detailed user interface was designed in which each selected causal element, such as a causal factor or outcome, was rendered as a set of individual UI elements, each displaying a distinct ranking property, evidence value, or attribute. Under the segmentation rules defined in Step~2, each such UI element constitutes a separate meaning unit. The same conceptual content, when elicited through a mental model or interview in other studies, would typically be represented as a single node or statement, yielding one meaning unit where the UI yields several. The higher absolute frequency of PrVC~3.2 in Study~5 therefore reflects greater granularity of representation rather than greater conceptual prevalence of epistemic explanation content in that domain. These three pre-validation codes jointly captured most explanation content in the corpus and therefore drove the largest share of codebook examples and boundary clarifications.


\subsection{Step~4: Domain-of-origin and explanation-type coding}
\label{sec:Results:Step4}

Each meaning unit was assigned a domain of origin to indicate which domain the explanation content belongs to and to support interpretation of what data sources and system components are needed to produce the content in an implemented system. The domain-of-origin annotation distinguishes the AI domain, the system domain, the application domain, and mixed origin. The mapping is depicted in Table~\ref{tab:Result:DomainOfOrigin}. The explanation-type annotation assigned to each meaning unit is reported in Step~7.

The two meaning units recorded as Not assigned in Step~3 were not assigned a domain of origin because they do not express explanation content and therefore have no identifiable information-source domain. They are retained in the table total of 325 for consistency with all other tables; their domain-of-origin cells are empty in the dataset and their frequency is reported in the Not assigned row. All percentages in Table~\ref{tab:Result:DomainOfOrigin} are computed against the full corpus of 325 meaning units.

\begin{table}[!ht]
	\small
	\caption{Domain-of-origin distribution across user studies (pre-validation code set, $N = 325$). All percentages are computed against the full corpus of 325 meaning units.}
	\label{tab:Result:DomainOfOrigin}
	\begin{scriptsize}
		\begin{tabular}
			{>{\raggedright\arraybackslash}m{2.2cm}
				>{\centering\arraybackslash}m{0.95cm}
				>{\centering\arraybackslash}m{0.95cm}
				>{\centering\arraybackslash}m{0.95cm}
				>{\centering\arraybackslash}m{0.95cm}
				>{\centering\arraybackslash}m{0.95cm}
				>{\centering\arraybackslash}m{0.95cm} |
				>{\centering\arraybackslash}m{1.0cm}
				>{\centering\arraybackslash}m{1.0cm}}
			\hline
			Domain & Study~1 & Study~2 & Study~3 & Study~4 & Study~5 & Study~6 & Total & \% \\
			\hline
			AI domain & \cellcolor{gray!25} 1 & \cellcolor{gray!25} 1 & \cellcolor{gray!25} 3 & \cellcolor{gray!25} 12 & \cellcolor{gray!25} 104 & 0 & \cellcolor{gray!25} 121 & \cellcolor{gray!25} 37.2 \\
			Application domain & \cellcolor{gray!25} 11 & \cellcolor{gray!25} 21 & \cellcolor{gray!25} 12 & \cellcolor{gray!25} 4 & \cellcolor{gray!25} 54 & \cellcolor{gray!25} 2 & \cellcolor{gray!25} 104 & \cellcolor{gray!25} 32.0 \\
			System domain & 0 & 0 & 0 & \cellcolor{gray!25} 31 & \cellcolor{gray!25} 27 & \cellcolor{gray!25} 12 & \cellcolor{gray!25} 70 & \cellcolor{gray!25} 21.5 \\
			Mixed domains & 0 & 0 & \cellcolor{gray!25} 3 & 0 & \cellcolor{gray!25} 25 & 0 & \cellcolor{gray!25} 28 & \cellcolor{gray!25} 8.6 \\
			Not assigned & 0 & \cellcolor{gray!25} 1 & 0 & 0 & \cellcolor{gray!25} 1 & 0 & \cellcolor{gray!25} 2 & \cellcolor{gray!25} 0.6 \\
			\hline
			Total & \cellcolor{gray!25} 12 & \cellcolor{gray!25} 23 & \cellcolor{gray!25} 18 & \cellcolor{gray!25} 47 & \cellcolor{gray!25} 211 & \cellcolor{gray!25} 14 & \cellcolor{gray!25} 325 & \cellcolor{gray!25} 99.9 \\
			\hline
		\end{tabular}
		\par\smallskip
		\footnotesize $^{a}$Total does not sum to 100.0 due to rounding.
	\end{scriptsize}
\end{table}

The domain-of-origin distribution shows that AI-domain content (37.2\%) and application-domain content (32.0\%) together account for approximately 69\% of the corpus. System-domain content (21.5\%) is concentrated in Studies~4 and~6, reflecting the prominence of system-level inputs and configuration data in those deployment contexts. Mixed-domain content (8.6\%) is concentrated in Studies~3 and~5, where single explanation elements drew on more than one information source simultaneously. These patterns indicate that an implemented system generating the full range of explanation content represented in the corpus would require access to at least three distinct data sources: the AI model's internal outputs (AI domain), the surrounding system's operational data (system domain), and application-domain knowledge (application domain). The 8.6\% of mixed-domain units confirms that some explanation elements draw on more than one of these sources simultaneously, implying that data integration across sources is required for a complete explanation interface rather than independent access to each source in isolation. The domain-of-origin annotation also directly supports the data-source rationale for the four-category structure derived in Step~5, where each category maps to a qualitatively different information source required to generate its content in an implemented system.


\subsection{Step~5: Abstraction to categories (researcher-led)}
\label{sec:Results:Step5}

Based on abstraction from the pre-validation codes, four categories were derived by the researcher: C1 Rule-based explanation group, C2 Causal explanation group, C3 Epistemic explanation group (actual), and C4 Epistemic explanation group (similar). The four-category structure rests on two principled distinctions. First, each category corresponds to a qualitatively different data source required to generate the content in an implemented system: rule representations for C1, causal or feature-attribution outputs for C2, current-case probabilistic or ranking outputs for C3, and a historical case base for C4. This data-source distinction is directly supported by the domain-of-origin annotation reported in Step~4, which showed that different explanation-content types draw systematically from different information domains. Second, the split of epistemic content into C3 and C4 reflects a fundamental difference in reference object: C3 refers to the current case being processed, while C4 refers to historical cases that are analogous but distinct. This distinction was identified by the researcher from the data: meaning units in the actual and similar groups appeared in different interface locations, served different user reasoning tasks, and required different system data sources. The identification of this distinction from the data, before it was assessed against the theoretical typology in Step~7, supports its data-driven rather than theoretically imposed character.

Alternative structures were considered and rejected on functional grounds. Splitting the causal group into antecedent and consequent subgroups was explored but not retained because all causal codes represent elements of a single explanatory chain designed to be read sequentially, and splitting the group would have disrupted this sequential logic without improving coding consistency. Merging C3 and C4 into a single epistemic group was explored but not retained because the actual and similar reference objects require different data sources and answer different user questions, making the distinction practically important for requirements specification.

The two meaning units recorded as Not assigned in Step~3 carry no pre-validation code and were therefore excluded from the abstraction step; they do not map to any of the four categories.

The pre-validation abstraction is depicted in Table~\ref{tab:Result:AbstractedCategories}.

\begin{table}[!ht]
	\small
	\caption{Abstraction of pre-validation codes to explanation-content categories}
	\label{tab:Result:AbstractedCategories}
	\begin{scriptsize}
		\begin{tabular}
			{>{\raggedright\arraybackslash}m{4.00cm}|
				>{\raggedright\arraybackslash}m{7.00cm}}
			\hline
			Pre-validation code & Abstracted category \\
			\hline
			PrVC~1.1 Rule attribute & C1 Rule-based explanation group \\
			\hline
			PrVC~2.1 Context & \multirow{7}{7cm}{C2 Causal explanation group} \\
			PrVC~2.2 Input & \\
			PrVC~2.3 Causal factor & \\
			PrVC~2.4 Causal mechanism & \\
			PrVC~2.5 Outcome & \\
			PrVC~2.6 Future state & \\
			PrVC~2.7 What-if & \\
			\hline
			PrVC~3.1 Set of ranked elements with ranking mechanism (actual) & \multirow{5}{7cm}{C3 Epistemic explanation group (actual)} \\
			PrVC~3.2 Individual element with ranking properties, evidence, and/or attributes (actual) & \\
			\hline
			PrVC~4.1 Set of similar elements with filter mechanism (similar) & \multirow{5}{7cm}{C4 Epistemic explanation group (similar)} \\
			PrVC~4.2 Individual, similar element with filter properties, evidence, and/or attributes (similar) & \\
			\hline
		\end{tabular}
	\end{scriptsize}
\end{table}


\subsection{Step~6: Codebook construction and iterative frame refinement}
\label{sec:Results:Step6}

The pre-validation coding frame and draft codebook were constructed to capture the stabilized definitions and boundaries of the pre-validation codes and categories, together with inclusion and exclusion criteria, evidence indicators, and examples (see \ref{app:CodingFrame} and \ref{app:DraftCodebook}). At the time of construction, the draft codebook contained 12 pre-validation codes organized into four categories, with inclusion and exclusion criteria, at least two positive examples and one negative example per code, and decision rules for the four boundaries identified as potentially ambiguous during open coding in Step~3: Applied rule vs Causal factor, Causal factor vs Causal mechanism, Set of ranked elements vs Individual element, and Input vs What-if forward. This documentation provided the reference artifact for the subsequent content-validity evaluation. After Step~8, the codebook was revised, and the revised post-validation version became the basis for recoding, reliability evaluation, and the final explanation-content blueprint.


\subsection{Step~7: Theory-informed coverage assessment}
\label{sec:Results:Step7}

To assess whether the pre-validation codes and categories covered established explanation types, the pre-validation code set was mapped to explanation types as a coverage check. The mapping is shown in Table~\ref{tab:Result:MappingToExplanationTypes}.

\begin{table}[!ht]
	\small
	\caption{Mapping of pre-validation codes to explanation types}
	\label{tab:Result:MappingToExplanationTypes}
	\begin{scriptsize}
		\begin{tabular}
			{>{\raggedright\arraybackslash}m{6.00cm}|
				>{\raggedright\arraybackslash}m{7.00cm}}
			\hline
			Pre-validation code & Explanation type \\
			\hline
			PrVC~1.1 Rule attribute & ET1 Rule-based explanations \\
			\hline
			PrVC~2.1 Context & \multirow{7}{7cm}{ET2 Causal explanations} \\
			PrVC~2.2 Input & \\
			PrVC~2.3 Causal factor & \\
			PrVC~2.4 Causal mechanism & \\
			PrVC~2.5 Outcome & \\
			PrVC~2.6 Future state & \\
			PrVC~2.7 What-if & \\
			\hline
			PrVC~2.7 What-if & ET3 Counterfactual explanations \\
			\hline
			PrVC~3.1 Set of ranked elements with ranking mechanism (actual) & \multirow{3}{7cm}{ET4 Contrastive explanations} \\
			PrVC~4.1 Set of similar elements with filter mechanism (similar) & \\
			\hline
			PrVC~3.2 Individual element with ranking properties, evidence, and/or attributes (actual) & ET5 Epistemic explanations \\
			\hline
			PrVC~4.2 Individual, similar element with filter properties, evidence, and/or attributes (similar) & ET6 Example-based explanations \\
			\hline
		\end{tabular}
	\end{scriptsize}
\end{table}

The assignment of PrVC~2.7 What-if to both ET2 Causal and ET3 Counterfactual reflects a theoretical relationship rather than an ambiguity in the coding system. Counterfactual reasoning is a subclass of causal reasoning in which the antecedent condition is hypothetical rather than observed, meaning What-if content inherently spans both explanation types depending on the direction of reasoning. This dual assignment was one of the findings motivating the post-validation split of What-if into two distinct codes: PoVC~2.7 What-if forward, which follows the natural direction of the causal chain from input to outcome and maps primarily to causal reasoning, and PoVC~2.8 What-if backward, which reasons from a hypothetical outcome back to required input changes and maps primarily to counterfactual reasoning. The post-validation split therefore resolves the dual assignment at the code level while preserving the theoretical relationship between causal and counterfactual explanation types at the category level.

The coverage assessment produced findings of three analytically distinct types, which should be differentiated rather than treated as equivalent evidence of discriminative power. The first finding is a genuine prospective correction with a direct design consequence. Rule-based explanations were identified as theoretically important but entirely absent from the material available in Studies~1 to~5. This gap directly motivated the inclusion of Study~6 in the corpus before the final analysis, which was selected specifically because it provided explanation-content material covering rule-based explanations. This finding constitutes the primary evidence that the coverage assessment retained corrective power: it produced a concrete remedial decision (the addition of a sixth study) that the category development process alone would not have generated. The second finding is a representational observation requiring interpretive caution. Contrastive explanations (18 units, 5.5\%) and example-based explanations (6 units, 1.8\%) were present in the corpus but weakly represented relative to causal (39.1\%) and epistemic (30.2\%) types. This does not mean that contrastive and example-based content are less important for explanation design: their low frequency likely reflects properties of the studied domains and system architectures (specifically, that the systems in Studies~1 to~4 and~6 did not expose ranked comparison sets or historical case bases as primary explanation modalities) rather than a general principle about their relevance. The blueprint retains these types with their corresponding codes (PoVC~3.1, PoVC~4.1, PoVC~4.2) because they are applicable in systems with different architectures: PoVC~3.1 contrastive content would arise in systems that present ranked alternatives alongside a primary recommendation, such as a clinical decision support tool displaying competing diagnoses ordered by probability; PoVC~4.1 and PoVC~4.2 example-based content would arise in case-based reasoning systems or diagnostic tools that retrieve and display historical cases analogous to the current input, such as a radiology AI that surfaces prior imaging studies from similar patients. Both architectures are well-documented in the XAI literature, and the absence of these system types from Studies~1 to~6 is a corpus property rather than a theoretical argument against their inclusion; this finding therefore highlights a limitation of the empirical coverage rather than a gap in the theoretical blueprint. The third finding is a theoretical clarification rather than a data discovery. The What-if code did not map cleanly to a single explanation type but spanned both causal and counterfactual reasoning. This reflects a well-established theoretical relationship: counterfactual reasoning is a subclass of causal reasoning in which the antecedent condition is hypothetical rather than observed. The dual assignment was resolvable by consulting the explanation-type literature rather than by inspecting the corpus data. Its analytic value lies in motivating the post-validation split of What-if into forward and backward codes at the conceptual level, not in demonstrating that the category development process produced an unexpected result.

In addition, each meaning unit in the pre-validation dataset was assigned an explanation-type label to quantify the distribution of explanation types across user studies. Table~\ref{tab:Result:ExplanationTypes} reports the explanation-type distribution. Because PrVC~2.7 What-if spans both ET2 Causal and ET3 Counterfactual in the coverage mapping (Table~\ref{tab:Result:MappingToExplanationTypes}), each What-if meaning unit is assigned to its primary explanation type in this table: What-if units are counted under Counterfactual to avoid double-counting, and the Causal row reflects only meaning units assigned exclusively to ET2. Input meaning units (PrVC~2.2, $n = 82$) are counted under Causal because they represent factual values of observed inputs, not hypothetical variations. The two Not assigned units are included in all totals and percentages. The Used explanation types row counts explanation types with at least one assigned meaning unit, excluding Not assigned units.

\begin{table}[!ht]
	\small
	\caption{Explanation-type distribution across user studies (pre-validation code set, $N = 325$). All percentages are computed against the full corpus of 325 meaning units.}
	\label{tab:Result:ExplanationTypes}
	\begin{scriptsize}
		\begin{tabular}
			{>{\raggedright\arraybackslash}m{2.3cm}
				>{\centering\arraybackslash}m{0.95cm}
				>{\centering\arraybackslash}m{0.95cm}
				>{\centering\arraybackslash}m{0.95cm}
				>{\centering\arraybackslash}m{0.95cm}
				>{\centering\arraybackslash}m{0.95cm}
				>{\centering\arraybackslash}m{0.95cm} |
				>{\centering\arraybackslash}m{1.0cm}
				>{\centering\arraybackslash}m{1.0cm}}
			\hline
			Explanation type & Study~1 & Study~2 & Study~3 & Study~4 & Study~5 & Study~6 & Total & \% \\
			\hline
			Causal & \cellcolor{gray!25} 6 & \cellcolor{gray!25} 6 & \cellcolor{gray!25} 7 & \cellcolor{gray!25} 25 & \cellcolor{gray!25} 78 & \cellcolor{gray!25} 5 & \cellcolor{gray!25} 127 & \cellcolor{gray!25} 39.1 \\
			Contrastive & \cellcolor{gray!25} 1 & \cellcolor{gray!25} 3 & \cellcolor{gray!25} 3 & 0 & \cellcolor{gray!25} 8 & \cellcolor{gray!25} 3 & \cellcolor{gray!25} 18 & \cellcolor{gray!25} 5.5 \\
			Counterfactual & 0 & 0 & 0 & \cellcolor{gray!25} 18 & \cellcolor{gray!25} 54 & 0 & \cellcolor{gray!25} 72 & \cellcolor{gray!25} 22.2 \\
			Epistemic & \cellcolor{gray!25} 5 & \cellcolor{gray!25} 13 & \cellcolor{gray!25} 8 & \cellcolor{gray!25} 4 & \cellcolor{gray!25} 64 & \cellcolor{gray!25} 4 & \cellcolor{gray!25} 98 & \cellcolor{gray!25} 30.2 \\
			Example-based & 0 & 0 & 0 & 0 & \cellcolor{gray!25} 6 & 0 & \cellcolor{gray!25} 6 & \cellcolor{gray!25} 1.8 \\
			Rule-based & 0 & 0 & 0 & 0 & 0 & \cellcolor{gray!25} 2 & \cellcolor{gray!25} 2 & \cellcolor{gray!25} 0.6 \\
			Not assigned & 0 & \cellcolor{gray!25} 1 & 0 & 0 & \cellcolor{gray!25} 1 & 0 & \cellcolor{gray!25} 2 & \cellcolor{gray!25} 0.6 \\
			\hline
			Total & \cellcolor{gray!25} 12 & \cellcolor{gray!25} 23 & \cellcolor{gray!25} 18 & \cellcolor{gray!25} 47 & \cellcolor{gray!25} 211 & \cellcolor{gray!25} 14 & \cellcolor{gray!25} 325 & \cellcolor{gray!25} 99.9 \\
			Used explanation types & \cellcolor{gray!25} 3 & \cellcolor{gray!25} 3 & \cellcolor{gray!25} 3 & \cellcolor{gray!25} 3 & \cellcolor{gray!25} 5 & \cellcolor{gray!25} 4 & \cellcolor{gray!25} -- & \cellcolor{gray!25} -- \\
			\hline
		\end{tabular}
		\par\smallskip
		\footnotesize $^{a}$Total does not sum to 100.0 due to rounding.
	\end{scriptsize}
\end{table}

All user studies include explanation content that can be assigned to at least three explanation types. Causal explanations were most frequent (127, 39.1\%), followed by epistemic explanations (98, 30.2\%), counterfactual explanations (72, 22.2\%), contrastive explanations (18, 5.5\%), example-based explanations (6, 1.8\%), and rule-based explanations (2, 0.6\%). The retention of the C4 epistemic group (similar) despite its small empirical footprint was motivated by the coverage assessment finding that example-based explanations are applicable in systems with case-based reasoning components even though the present corpus could not substantiate them as a frequent type. The explanation-type distribution provides additional evidence that the pre-validation category system already supported multiple explanation styles that are common in XAI design and evaluation.

The pre-validation explanation-content model comprised four groups organized identically to the post-validation model in terms of top-level structure: a rule-based group positioned above and optional relative to a causal group, followed by two epistemic groups distinguishing actual from similar outcomes. The post-validation model (shown in Figure~\ref{fig:XAIModel-Framework-PostValidation} after Step~8, which reflects revisions from the pre-validation structure) differs from the pre-validation model in three respects that follow directly from the expert review outcomes. First, the rule-based group was split from one code (PrVC~1.1 Rule attribute) into two codes (PoVC~1.1 Rule base, PoVC~1.2 Applied rule), reflecting the expert panel's finding that the distinction between a queryable rule repository and a specific applied rule serves different user information needs. Second, the causal group was extended from seven to eight codes by splitting PrVC~2.7 What-if into PoVC~2.7 What-if forward and PoVC~2.8 What-if backward, reflecting expert feedback on reasoning direction and the Step~7 dual-assignment finding. Third, decision rules were added to all four epistemic codes to resolve the set-versus-element boundary ambiguity that generated the highest comment frequency across the panel. The overall four-group structure, the sequential arrangement of the causal chain, and the parallel structure of the two epistemic groups were unchanged by the expert review, providing evidence of structural stability across the validation procedure.


\subsection{Step~8: Expert content validity and codebook revision}
\label{sec:Results:Step8}


\subsubsection*{Overview of the expert panel and procedure}

The expert panel comprised 11 participants, including seven experts from industry, three from academia, and one from government. In terms of AI-specific professional experience, five experts reported one year or less, four reported two to five years, one reported six to ten years, and one reported more than ten years. The panel included four female and seven male experts. Participants were based in three countries: the United States (seven), Germany (two), and Greece (two).

The panel was designed to cover two complementary expertise dimensions. Experts with substantial AI and XAI experience assessed the technical coherence of code definitions and boundaries. Experts with more limited AI experience but substantial application domain expertise (including cybersecurity, building technology, and related industrial domains covered by the corpus studies) assessed the relevance and understandability of explanation-content categories from the practitioner perspective. Because the blueprint is intended for use by practitioners who include both AI specialists and domain experts, this mixed-expertise composition was a deliberate design choice that assessed the codebook simultaneously from both perspectives. Subgroup-level I-CVI values were not computed separately for AI-experienced and domain-expert panel members because the panel size of 11 does not support reliable subgroup-level content-validity index computation with the recommended minimum of six raters per subgroup. The aggregate I-CVI values reported in Table~\ref{tab:ExpertValidationResults} therefore reflect the combined judgment of both subgroups. The qualitative comment analysis, reported below, was examined for any obvious patterns in the concerns raised by AI-experienced versus domain-expert panelists. However, because the panel size of 11 precludes reliable subgroup analysis (minimum recommended n=6 per subgroup), no systematic comparison is reported. The study does not claim that AI-experts and domain-experts agreed or differed systematically. Future validation studies with larger panels (minimum 12 experts with 6 per subgroup) should examine whether I-CVI values differ systematically between expertise types.

Experts provided asynchronous independent ratings. All experts rated all pre-validation codes. In addition to ratings, experts provided qualitative comments that motivated ratings and proposed revisions, missing content, and boundary clarifications.


\subsubsection*{Quantitative expert agreement indices}

Item-level expert agreement indices were computed for relevance, boundary clarity, and understandability. As noted in the Methodology section, the I-CVI and S-CVI/Ave statistics used here originate in psychometric scale development; they are applied in this study as practical agreement indices for codebook adequacy, where the proportion of experts endorsing a category definition provides a direct criterion for whether that definition requires revision before use as a reliable coding instrument, regardless of whether a latent construct is involved. For each pre-validation code and criterion, I-CVI was computed as the proportion of experts providing an acceptable rating, defined as 3 or 4 on the corresponding four-point scale. Scale-level indices were summarized as S-CVI-Ave by averaging I-CVI values across pre-validation codes for each criterion. With $N=11$ experts, the predefined item-level acceptability criterion was I-CVI~$\geq 0.82$, operationalized as at least $9/11$ experts rating the item at or above the threshold of 3. S-CVI-Ave values of at least 0.90 were treated as indicating excellent scale-level agreement. An expert-panel size of 11 exceeds common recommendations to use six or more experts when permitting limited disagreement and falls within typical ranges reported for expert codebook evaluation panels.

Table~\ref{tab:ExpertValidationResults} reports mean ratings and I-CVI values for each pre-validation code. At the scale level, expert agreement was excellent for relevance (S-CVI-Ave = 0.93), boundary clarity (S-CVI-Ave = 0.92), and understandability (S-CVI-Ave = 0.94). At the item level, all pre-validation codes met or exceeded the predefined acceptability criterion for all three criteria (I-CVI \(\geq 0.82\)). The lowest I-CVI values were 0.82 and occurred for relevance of PrVC 2.4, PrVC 2.7, and PrVC 3.1, for boundary clarity of PrVC 1.1, PrVC 2.7, and PrVC 3.1, and for understandability of PrVC 4.1. These patterns indicate broad acceptance of the structure while also motivating targeted clarification for content types with more dispersed ratings.

\begin{table}[ht]
	\scriptsize
	\setlength{\tabcolsep}{4pt}
	\renewcommand{\arraystretch}{1.15}
	\caption{Expert ratings ($N=11$) and item content validity (I-CVI) for each pre-validation code. Mean ratings are computed on the original 1 to 4 scale. I-CVI is the proportion of experts rating at or above the threshold of 3.}
	\label{tab:ExpertValidationResults}
	\begin{tabular}
		{>{\raggedright\arraybackslash}m{4.6cm}|
		>{\centering\arraybackslash}m{1.1cm}
		>{\centering\arraybackslash}m{1.1cm}|
		>{\centering\arraybackslash}m{1.1cm}
		>{\centering\arraybackslash}m{1.1cm}|
		>{\centering\arraybackslash}m{1.1cm}
		>{\centering\arraybackslash}m{1.1cm}}
		
		\hline
		
		Pre-validation code &
		\multicolumn{2}{c|}{Relevance} &
		\multicolumn{2}{c|}{Boundary clarity} &
		\multicolumn{2}{c}{Understandability} \\
		& Mean & I-CVI & Mean & I-CVI & Mean & I-CVI \\
		
		\hline
		
		PrVC~1.1 Rule attribute & 4.00 & 1.00 & 3.55 & 0.82 & 3.55 & 0.91 \\
		
		\hline
		
		PrVC~2.1 Context & 4.00 & 1.00 & 3.55 & 1.00 & 3.36 & 0.91 \\
		PrVC~2.2 Input & 4.00 & 1.00 & 3.82 & 1.00 & 3.73 & 1.00 \\
		PrVC~2.3 Causal factor & 4.00 & 1.00 & 3.45 & 0.91 & 3.64 & 0.91 \\
		PrVC~2.4 Causal mechanism & 3.64 & 0.82 & 3.45 & 0.91 & 3.73 & 0.91 \\
		PrVC~2.5 Outcome & 3.91 & 1.00 & 3.73 & 1.00 & 3.73 & 1.00 \\
		PrVC~2.6 Future state & 3.82 & 0.91 & 3.82 & 0.91 & 3.82 & 1.00 \\
		PrVC~2.7 What-if & 3.45 & 0.82 & 3.36 & 0.82 & 3.64 & 0.91 \\
		
		\hline
		
		PrVC~3.1 Set of ranked elements with ranking mechanism (actual) & 3.64 & 0.82 & 3.55 & 0.82 & 3.64 & 0.91 \\
		PrVC~3.2 Individual element with ranking properties, evidence, and/or attributes (actual) & 3.82 & 1.00 & 3.55 & 0.91 & 3.73 & 1.00 \\
		
		\hline
		
		PrVC~4.1 Set of similar elements with filter mechanism (similar) & 3.73 & 0.91 & 3.55 & 0.91 & 3.64 & 0.82 \\
		PrVC~4.2 Individual, similar element with filter properties, evidence, and/or attributes (similar) & 3.73 & 0.91 & 3.64 & 1.00 & 3.82 & 1.00 \\
		
		\hline
		
	\end{tabular}
	
	\footnotesize
	\textit{Scale-level indices:} S-CVI-Ave (relevance) = 0.93, S-CVI-Ave (boundary clarity) = 0.92, S-CVI-Ave (understandability) = 0.94.\\
	\textit{Item-level acceptability criterion:} With $N=11$ experts, an item is considered acceptable when I-CVI $\geq 0.82$ (at least $9/11$ experts rate the item at or above the threshold of 3).
\end{table}

At the scale level, expert agreement was excellent for relevance (S-CVI-Ave~=~0.93), boundary clarity (S-CVI-Ave~=~0.92), and understandability (S-CVI-Ave~=~0.94). At the item level, all pre-validation codes met or exceeded the predefined acceptability criterion for all three criteria (I-CVI $\geq 0.82$). The lowest I-CVI values were 0.82 and occurred for relevance of PrVC~2.4, PrVC~2.7, and PrVC~3.1, for boundary clarity of PrVC~1.1, PrVC~2.7, and PrVC~3.1, and for understandability of PrVC~4.1. These patterns indicate broad acceptance of the structure while also motivating targeted clarification for content types with more dispersed ratings.


\subsection{Subgroup patterns in expert ratings}

Although the panel size (n=5 with \(\leq\)1 year AI experience, n=6 with >1 year AI experience) precludes reliable statistical subgroup comparison (minimum recommended n=6 per subgroup for I-CVI computation \citep{Polit2006-XAI-ContentValidityIndex-Wiley}), descriptive patterns were examined to assess whether any code showed substantially different acceptance across expertise levels. Table \ref{tab:Subgroup-cvi} reports subgroup I-CVI values for all codes and criteria.

\begin{table}[htbp]
	\scriptsize
	\caption{Subgroup I-CVI comparison (Low AI experience: n=5, P1, P3, P6, P8, P11; High AI experience: n=6, P2, P4, P5, P7, P9, P10)}
	\label{tab:Subgroup-cvi}
	\begin{tabular}{lcccccc}
		
		\hline
		
		\multirow{2}{*}{Code} & \multicolumn{2}{c}{Relevance} & \multicolumn{2}{c}{Boundary} & \multicolumn{2}{c}{Understandability} \\
		\cline{2-7}
		& Low & High & Low & High & Low & High \\
		
		\hline
		
		PrVC 1.1 Rule attribute & 1.00 & 1.00 & 0.80 & 0.83 & 0.80 & 1.00 \\
		PrVC 2.1 Context & 1.00 & 1.00 & 1.00 & 1.00 & 0.80 & 1.00 \\
		PrVC 2.2 Input & 1.00 & 1.00 & 1.00 & 1.00 & 1.00 & 1.00 \\
		PrVC 2.3 Causal factor & 1.00 & 1.00 & 0.80 & 1.00 & 0.80 & 1.00 \\
		PrVC 2.4 Causal mechanism & 0.80 & 0.83 & 0.80 & 1.00 & 0.80 & 1.00 \\
		PrVC 2.5 Outcome & 1.00 & 1.00 & 1.00 & 1.00 & 1.00 & 1.00 \\
		PrVC 2.6 Future state & 0.80 & 1.00 & 0.80 & 1.00 & 1.00 & 1.00 \\
		PrVC 2.7 What-if & 0.80 & 0.83 & 0.80 & 0.83 & 0.80 & 1.00 \\
		PrVC 3.1 Set ranked (actual) & 0.80 & 0.83 & 0.80 & 0.83 & 0.80 & 1.00 \\
		PrVC 3.2 Individual (actual) & 1.00 & 1.00 & 0.80 & 1.00 & 1.00 & 1.00 \\
		PrVC 4.1 Set similar & 0.80 & 1.00 & 0.80 & 1.00 & 0.60 & 1.00 \\
		PrVC 4.2 Individual similar & 0.80 & 1.00 & 1.00 & 1.00 & 1.00 & 1.00 \\
		
		\hline
		
	\end{tabular}
\end{table}

Three patterns from this descriptive analysis warrant mention. First, all codes met or exceeded I-CVI \(\geq\) 0.80 in the high-AI-experience subgroup for all criteria. In the low-AI-experience subgroup, all codes met this threshold except PrVC 4.1 (Set of similar elements with filter mechanism, similar), which had understandability I-CVI = 0.60 (3 of 5 experts rated \(\geq\)3). This suggests that similar-case reasoning content may require additional plain-language guidance for practitioners without AI backgrounds.

Second, for every code and criterion where subgroups differed, the low-AI-experience subgroup had equal or lower I-CVI than the high-AI-experience subgroup. No code was rated systematically higher by low-AI experts. This directional consistency—while not statistically significant—suggests that AI experience may modestly facilitate codebook comprehension, particularly for abstract constructs like similarity mechanisms and future-state predictions.

Third, the qualitative themes raised by each subgroup showed substantial overlap (see comment analysis below). The set-versus-individual distinction was the most frequently raised theme in both subgroups (low-AI: 4/5; high-AI: 4/6), indicating that this boundary ambiguity is not an artifact of differential expertise but a genuine codebook design challenge. The rule-content separation theme was raised by slightly more high-AI experts (4/6 vs. 3/5), consistent with their greater familiarity with rule-based system architectures.

These descriptive observations should be interpreted cautiously. They do not constitute statistical evidence of subgroup differences, and future validation studies with larger panels (minimum 6 per subgroup) are needed to test whether AI experience moderates codebook usability. For the current instrument, the primary implication is that PrVC 4.1 (now PoRC 4.1) benefits from additional plain-language examples and explicit guidance on similarity interpretation, which have been incorporated into the post-reliability codebook (Appendix L).

\clearpage


\subsubsection*{Qualitative findings and codebook revisions}

The expert review in Step~8 evaluated the pre-validation coding system reported in Steps~3 to~7 and resulted in a revised post-validation codebook. Table~\ref{tab:Result:ExpertCommentThemes} summarizes the number of expert panelists raising each major revision theme and the corresponding codebook action taken.

\begin{table}[!ht]
	\scriptsize
	\caption{Number of expert panelists ($N = 11$) raising each major revision theme in qualitative comments and corresponding codebook action}
	\label{tab:Result:ExpertCommentThemes}
	\begin{tabular}
		{>{\raggedright\arraybackslash}m{7.5cm}
			>{\centering\arraybackslash}m{1.5cm}
			>{\raggedright\arraybackslash}m{3.5cm}}
		
		\hline
		
		Revision theme & Experts ($n$) & Codebook action \\
		
		\hline
		
		Separation of rule-related content from context and input; distinction between rule repository and case-specific rule application & 7 & Split PrVC~1.1 into PoVC~1.1 Rule base and PoVC~1.2 Applied rule \\
		Differentiation of causal factor and causal mechanism; clarification of mechanism scope and causal item connections & 6 & Boundary refined; examples revised; decision rules added \\
		Differentiation of What-if by reasoning direction (forward vs.\ backward) & 5 & Split PrVC~2.7 into PoVC~2.7 What-if forward and PoVC~2.8 What-if backward \\
		Clarification of set-versus-element distinction for ranked recommendations and similar-case reasoning; guidance on optional ranked sets; admissible similarity evidence & 8 & Decision rules added to PoVC~3.1, PoVC~3.2, PoVC~4.1, PoVC~4.2 \\
		
		\hline
		
	\end{tabular}
\end{table}

The distribution of expert comment counts across themes reveals a consistent pattern that is analytically significant in two respects. First, boundary ambiguity at the epistemic level (eight experts) and at the rule-content level (seven experts) generated more widespread concern than boundary ambiguity within the causal chain (six experts for causal factor vs mechanism) or at the level of reasoning direction (five experts for What-if). This ordering is consistent with the quantitative I-CVI results (see Table~\ref{tab:ExpertValidationResults}), where the lowest boundary-clarity scores occurred for PrVC 1.1 (I-CVI = 0.82), PrVC 2.7 (I-CVI = 0.82), and PrVC 3.1 (I-CVI = 0.82) (all codes that were subsequently split or had decision rules added). The convergence between quantitative I-CVI patterns and qualitative comment frequencies provides mutual validation for the revision decisions. Second, the four themes collectively span all four code groups (rule-based, causal, epistemic actual, epistemic similar), confirming that the expert review produced actionable feedback across the full breadth of the coding system rather than concentrating on a single group.

Examination of rating patterns by expertise level revealed a systematic difference between AI-experienced (more than one year of experience with AI systems) and AI-novice panelists (one year or less of experience with AI systems). AI-experienced experts (n=6) provided 48 comments and 15 low ratings (scores of 2 or below), whereas AI-novice experts (n=5) provided 23 comments and only 1 low rating. AI-experienced experts proposed revision actions in 67\% of their comments, compared to 52\% for AI-novice experts. This pattern indicates that the codebook was more readily acceptable to domain practitioners without deep AI backgrounds, while AI experts identified more boundary ambiguities requiring clarification. However, due to the small subgroup sizes (n=6 and n=5), these observed differences are descriptive only and should not be interpreted as statistically significant or definitive. Future studies with larger panels are needed to determine whether these patterns replicate.

Based on the qualitative comments, six major revision themes were identified, each leading to concrete codebook actions. Separation of rule-related content into rule repository versus case-specific rule application (7 experts) motivated the split of PrVC 1.1 into PoVC 1.1 Rule base and PoVC 1.2 Applied rule. Differentiation of causal factor and causal mechanism (6 experts) led to boundary refinement and added decision rules. Differentiation of What-if by reasoning direction (6 experts) motivated the split of PrVC 2.7 into PoVC 2.7 What-if forward and PoVC 2.8 What-if backward. Clarification of the set-versus-element distinction for ranked recommendations and similar-case reasoning (8 experts) led to decision rules added to PoVC 3.1, PoVC 3.2, PoVC 4.1, and PoVC 4.2. The need for an integrated causal chain example (6 experts) resulted in a new example added to the codebook. Operational definition of similarity for epistemic-similar codes (6 experts) led to refined definitions and additional decision rules.

An overview of the changes from the pre-validation code set to the post-validation code set is shown in Table~\ref{tab:Result:RevisionPostValidationCodesPart1}. The full comment matrix is provided in \ref{app:FullCommentMatrix}. The full change log documenting the trigger, rationale, and outcome of each revision from pre-validation codes to post-validation codes is provided in Appendix~\ref{app:ChangeLogPreliminaryToPrevalidation}.

\begin{table}[!ht]
	\scriptsize
	\caption{Revision of pre-validation codes into post-validation codes, Part 1}
	\label{tab:Result:RevisionPostValidationCodesPart1}
	\begin{tabular}
		{>{\raggedright\arraybackslash}m{4.2cm}
			>{\raggedright\arraybackslash}m{5.0cm}
			>{\raggedright\arraybackslash}m{3.7cm}}
		
		\hline
		
		Pre-validation code & Revision strategy & Post-validation code \\
		
		\hline
		
		PrVC~1.1 Rule attribute & Split into Rule base and Applied rule; definitions added and revised & PoVC~1.1 Rule base, PoVC~1.2 Applied rule \\
		
		\hline
		
		PrVC~2.1 Context & Decision rule introduced to separate Context (PoVC~2.1) from Input (PoVC~2.2) & PoVC~2.1 Context \\
		PrVC~2.2 Input & Decision rule introduced to separate Input (PoVC~2.2) from Context (PoVC~2.1) & PoVC~2.2 Input \\
		PrVC~2.3 Causal factor & Boundary refined; examples revised to correct misalignment with definition; decision rules introduced to separate Causal factor (PoVC~2.3) from Applied rule (PoVC~1.2) and from Causal mechanism (PoVC~2.6) & PoVC~2.3 Causal factor \\
		
		PrVC~2.4 Causal mechanism & Definition refined; decision rule introduced to separate Causal mechanism (PoVC~2.6) from Causal factor (PoVC~2.3) & PoVC~2.6 Causal mechanism \\
		PrVC~2.5 Outcome & Decision rule introduced to separate Outcome (PoVC~2.4) from Future state (PoVC~2.5) & PoVC~2.4 Outcome \\
		PrVC~2.6 Future state & Decision rule introduced to separate Future state (PoVC~2.5) from Outcome (PoVC~2.4) & PoVC~2.5 Future state \\
		
		PrVC~2.7 What-if & Split into What-if forward and What-if backward, motivated by expert feedback on reasoning direction and confirmed by Step~7 dual-assignment finding & PoVC~2.7 What-if forward, PoVC~2.8 What-if backward \\
		
		\hline
	
		PrVC~3.1 Set of ranked elements with ranking mechanism (actual) & Definition refined; decision rules introduced & PoVC~3.1 Set of ranked elements with ranking mechanism (actual) \\
		PrVC~3.2 Individual element with ranking properties, evidence, and/or attributes (actual) & Definition refined; decision rules introduced & PoVC~3.2 Individual element with ranking properties, evidence, and/or attributes (actual) \\
		
		\hline
		
		PrVC~4.1 Set of similar elements with filter mechanism (similar) & Title reworded and definition refined; decision rules introduced & PoVC~4.1 Set of similar elements with similarity mechanism and set success-score information (similar) \\
		PrVC~4.2 Individual, similar element with filter properties, evidence, and/or attributes (similar) & Title reworded and definition refined; decision rules introduced & PoVC~4.2 Individual, similar element with similarity property, success score, evidence, and/or attribute (similar) \\
		
		\hline
		
	\end{tabular}
\end{table}

\subsubsection*{Explanation content model for local, post-hoc explanations (post-validation, reflecting revisions from the pre-validation structure described in Step~7)}

\begin{figure}[h!]
	\small
	\includegraphics[width=1.0\linewidth]{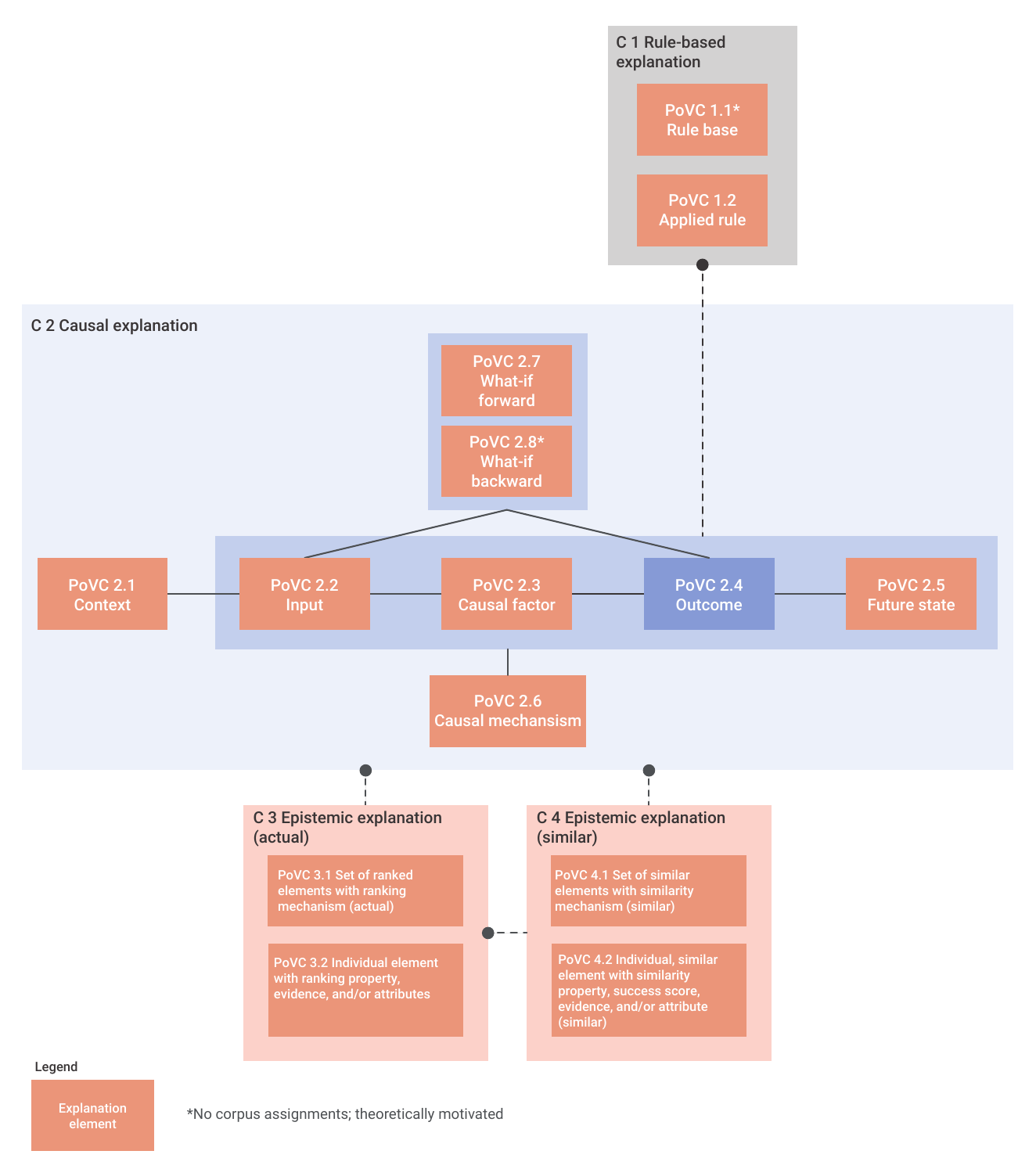}
	\caption{Explanation content model for local, post-hoc explanations (post-validation, reflecting revisions from the pre-validation structure described in Step~7). PoVC 1.1 Rule base and PoVC 2.8 What-if backward are not represented in the corpus and theoretically motivated.}
	\label{fig:XAIModel-Framework-PostValidation}
\end{figure}

The post-validation explanation-content model is shown in Figure~\ref{fig:XAIModel-Framework-PostValidation}. The model comprises four groups. The rule-based group is positioned above the causal group and is marked as optional, reflecting that rule-based content is applicable only in systems with explicit decision logic. The causal group contains eight codes arranged to reflect the explanatory chain from context through input, causal factor, causal mechanism, outcome, and future state, with PoVC~2.7 What-if forward and PoVC~2.8 What-if backward positioned as branches representing hypothetical reasoning over the causal chain. The epistemic group for actual outcomes contains two codes distinguishing the set of ranked elements from individual elements. The epistemic group for similar outcomes contains two parallel codes distinguishing the set of similar cases from individual similar cases. Rule-based content can serve to justify or derive elements of the causal chain, and the epistemic groups represent alternative perspectives on the causal outcome rather than sequential extensions of it. This post-validation model represents the finalized explanation-content blueprint that was subsequently used for recoding, reliability evaluation, and final dataset construction.

Table~\ref{tab:Result:PostReliabilityCode} provides a condensed summary of the final post-reliability codebook (PoRC) for quick reference. Full definitions, inclusion and exclusion criteria, decision rules, and examples are provided in~\ref{app:PostReliabilityCodebook}.

\begin{table}[h]
	\scriptsize
	\caption{Post-reliability explanation content codes (PoRC) - Summary}
	\label{tab:Result:PostReliabilityCode}
	\begin{tabular}
		{>{\raggedright\arraybackslash}m{2.5cm}
		>{\raggedright\arraybackslash}m{2.5cm}
		>{\raggedright\arraybackslash}m{8.0cm}}
		
		\hline
		
		\textbf{Code} & \textbf{Label} & \textbf{Brief definition} \\
		
		\hline
		
		PoRC 1.1* & Rule base & Repository or set of available rules for case assessment \\
		PoRC 1.2 & Applied rule & Rule evaluated and applied to current case \\
		
		\hline
		
		PoRC 2.1 & Context & Surrounding situation that helps interpret the case \\
		PoRC 2.2 & Input & Data provided to AI system to generate outcome \\
		PoRC 2.3 & Causal factor & Condition believed to have influenced the outcome \\
		PoRC 2.4 & Outcome & Result produced for current case \\
		PoRC 2.5 & Future state & Predicted state after outcome has been applied \\
		PoRC 2.6 & Causal mechanism & Description of causal relationship within causal chain \\
		PoRC 2.7 & What-if forward & Hypothetical input change and expected effect on outcome \\
		PoRC 2.8* & What-if backward & Desired outcome and required input change to achieve it \\
		
		\hline
		
		PoRC 3.1 & Set of ranked elements (actual) & Multiple candidate elements in ranked order \\
		PoRC 3.2 & Individual element (actual) & One candidate element with properties and evidence \\
		
		\hline
		
		PoRC 4.1 & Set of similar elements (similar) & Multiple similar past cases with success-score information \\
		PoRC 4.2 & Individual similar element (similar) & One similar past case with similarity property and score \\
		
		\hline
		
	\end{tabular}
	\footnotesize \textit{Note:} Codes marked with an asterisk (*) are theoretically proposed extensions without empirical corpus assignments (see Limitation~4 in Section~\ref{sec:Limitations}). Full definitions, inclusion and exclusion criteria, decision rules, and examples are provided in the supplementary materials (see~\ref{app:PostReliabilityCodebook}).
\end{table}

\clearpage


\subsection{Step~9: Corpus recoding after validity revisions and preparation for reliability evaluation}
\label{sec:Results:Step9}

After completing Step~8, the corpus was reviewed to identify all meaning units affected by substantial revisions to the pre-validation coding system. All affected meaning units were recoded by the ground-truth coder using the revised, version-controlled post-validation codebook to ensure consistent interpretation across the dataset.

The ground-truth coder split all units previously assigned to PrVC~1.1 Rule attribute into PoVC~1.1 Rule base and PoVC~1.2 Applied rule using the revised definitions from Step~8; all 2 units (Study~6) were assigned to PoVC~1.2 Applied rule and none to PoVC~1.1 Rule base. The ground-truth coder reviewed all 72 units previously assigned to PrVC~2.7 What-if against the revised What-if forward and What-if backward definitions; all 72 units were assigned to PoVC~2.7 What-if forward and none to PoVC~2.8 What-if backward, reflecting that the available corpus material exclusively represented forward-direction hypothetical reasoning.

Boundary revisions to Context, Input, Outcome, Future state, Causal factor, and Causal mechanism triggered review of all meaning units assigned to these codes. For each boundary-revised code, the ground-truth coder re-read every meaning unit assigned to that code against the revised definition and decision rules from Step~8, applying the new boundary criteria to determine whether the existing assignment remained correct. In all cases the existing assignment was confirmed; no units were reassigned. All per-code counts for these codes are therefore identical in the pre-validation and post-validation tables.

After recoding, the revised codebook was frozen for the pending reliability evaluation. To ensure the independence and transparency of the reliability procedure, the following sequence was strictly observed and is documented here in a single location for clarity. First, all recoding motivated by the expert panel revisions in Step~8 was completed using the revised post-validation codebook. Second, the post-validation codebook was frozen; no further changes to code definitions, decision rules, boundaries, or examples were permitted after this point. Third, the frozen codebook, the pre-specified reliability subsample composition, the equivalence rule for PoVC~2.2 and PoVC~2.7, the acceptance criteria, and the stopping rules were pre-registered on OSF on March~30, 2026, prior to any independent coding activity; the registration is currently under embargo and will become publicly available upon publication of this article (https://osf.io/up2sh). Fourth, independent coding of the reliability subsample was conducted on April~3, 2026, after pre-registration was complete. This sequence ensures that the codebook, subsample, and decision rules were fully specified and publicly committed before any reliability data were collected, precluding post-hoc adjustment of criteria in response to observed coding outcomes.

\subsubsection*{Post-validation coding system: empirical and theoretical codes}

The post-validation coding system comprises fourteen codes organized into four groups: rule-based, causal, epistemic (actual), and epistemic (similar). Twelve of these codes have empirical corpus assignments derived from the analysis of Studies~1 to~6. Two codes (PoVC~1.1 Rule base and PoVC~2.8 What-if backward) have no corpus assignments because the system architectures required to instantiate them were not present in the studied corpus.

The blueprint is therefore hybrid in nature: twelve codes are empirically grounded; two are theoretically proposed. PoVC~1.1 Rule base is applicable in systems that expose a queryable rule repository (for example, a system in which users can browse or search the full set of rules governing a class of decisions, as opposed to seeing only the specific rules applied to the current case, which is PoVC~1.2 Applied rule). PoVC~2.8 What-if backward is applicable when users reason from a hypothetical outcome back to required input changes (for example, "what would the inputs need to be for the system to recommend option X instead?"). Both codes are grounded in the XAI literature and address system architectures that are practically relevant in deployed AI systems. Future empirical work with systems that expose rule repositories or support backward-direction counterfactual reasoning should test whether these codes receive corpus assignments and whether their definitions require refinement.

Table~\ref{tab:Result:PostValidationCodes} reports the distribution of the post-validation code set across user studies and in the corpus as a whole following recoding. Only the rows for the two split codes differ from Table~\ref{tab:Result:PreValidationCodes}: PrVC~1.1 Rule attribute is replaced by PoVC~1.1 Rule base (0 units) and PoVC~1.2 Applied rule (2 units, Study~6); PrVC~2.7 What-if is replaced by PoVC~2.7 What-if forward (72 units) and PoVC~2.8 What-if backward (0 units). All other counts are unchanged.

\begin{table}[!ht]
	\scriptsize
	\caption{Assignment of meaning units to the post-validation code set ($N = 325$). All percentages are computed against the full corpus of 325 meaning units. Only the rows for the two split codes differ from Table~\ref{tab:Result:PreValidationCodes}.}
	\label{tab:Result:PostValidationCodes}
	\begin{tabular}
		{>{\raggedright\arraybackslash}m{2.55cm}
			>{\centering\arraybackslash}m{0.95cm}
			>{\centering\arraybackslash}m{0.95cm}
			>{\centering\arraybackslash}m{0.95cm}
			>{\centering\arraybackslash}m{0.95cm}
			>{\centering\arraybackslash}m{0.95cm}
			>{\centering\arraybackslash}m{0.95cm}|
			>{\centering\arraybackslash}m{1.0cm}
			>{\centering\arraybackslash}m{1.0cm}}
		
		\hline
		
		Post-validation code & Study~1 & Study~2 & Study~3 & Study~4 & Study~5 & Study~6 & Total & \% \\
		
		\hline
		
		PoVC~1.1 Rule base* & 0 & 0 & 0 & 0 & 0 & 0 & 0 & 0.0 \\
		PoVC~1.2 Applied rule & 0 & 0 & 0 & 0 & 0 & \cellcolor{gray!25} 2 & \cellcolor{gray!25} 2 & \cellcolor{gray!25} 0.6 \\
		
		\hline
		
		PoVC~2.1 Context & \cellcolor{gray!25} 3 & \cellcolor{gray!25} 2 & 0 & 0 & \cellcolor{gray!25} 13 & \cellcolor{gray!25} 1 & \cellcolor{gray!25} 19 & \cellcolor{gray!25} 5.8 \\
		PoVC~2.2 Input & \cellcolor{gray!25} 1 & \cellcolor{gray!25} 2 & \cellcolor{gray!25} 2 & \cellcolor{gray!25} 21 & \cellcolor{gray!25} 55 & \cellcolor{gray!25} 1 & \cellcolor{gray!25} 82 & \cellcolor{gray!25} 25.2 \\
		PoVC~2.3 Causal factor & \cellcolor{gray!25} 1 & \cellcolor{gray!25} 1 & \cellcolor{gray!25} 1 & 0 & \cellcolor{gray!25} 2 & \cellcolor{gray!25} 1 & \cellcolor{gray!25} 6 & \cellcolor{gray!25} 1.8 \\
		PoVC~2.4 Outcome & \cellcolor{gray!25} 1 & \cellcolor{gray!25} 1 & \cellcolor{gray!25} 3 & \cellcolor{gray!25} 4 & \cellcolor{gray!25} 7 & \cellcolor{gray!25} 1 & \cellcolor{gray!25} 17 & \cellcolor{gray!25} 5.2 \\
		PoVC~2.5 Future state & 0 & 0 & 0 & 0 & \cellcolor{gray!25} 1 & \cellcolor{gray!25} 1 & \cellcolor{gray!25} 2 & \cellcolor{gray!25} 0.6 \\
		PoVC~2.6 Causal mechanism & 0 & 0 & \cellcolor{gray!25} 1 & 0 & 0 & 0 & \cellcolor{gray!25} 1 & \cellcolor{gray!25} 0.3 \\
		PoVC~2.7 What-if forward & 0 & 0 & 0 & \cellcolor{gray!25} 18 & \cellcolor{gray!25} 54 & 0 & \cellcolor{gray!25} 72 & \cellcolor{gray!25} 22.2 \\
		PoVC~2.8 What-if backward* & 0 & 0 & 0 & 0 & 0 & 0 & 0 & 0.0 \\
		
		\hline
		
		PoVC~3.1 Set of ranked elements with ranking mechanism (actual) & \cellcolor{gray!25} 1 & \cellcolor{gray!25} 3 & \cellcolor{gray!25} 3 & 0 & \cellcolor{gray!25} 8 & \cellcolor{gray!25} 3 & \cellcolor{gray!25} 18 & \cellcolor{gray!25} 5.5 \\
		PoVC~3.2 Individual element with ranking properties, evidence, and/or attributes (actual) & \cellcolor{gray!25} 5 & \cellcolor{gray!25} 13 & \cellcolor{gray!25} 8 & \cellcolor{gray!25} 4 & \cellcolor{gray!25} 64 & \cellcolor{gray!25} 4 & \cellcolor{gray!25} 98 & \cellcolor{gray!25} 30.2 \\
		
		\hline
		
		PoVC~4.1 Set of similar elements with similarity mechanism and set success-score information (similar) & 0 & 0 & 0 & 0 & \cellcolor{gray!25} 4 & 0 & \cellcolor{gray!25} 4 & \cellcolor{gray!25} 1.2 \\
		PoVC~4.2 Individual, similar element with similarity property, success score, evidence, and/or attribute (similar) & 0 & 0 & 0 & 0 & \cellcolor{gray!25} 2 & 0 & \cellcolor{gray!25} 2 & \cellcolor{gray!25} 0.6 \\
		
		\hline
		
		Not assigned & 0 & \cellcolor{gray!25} 1 & 0 & 0 & \cellcolor{gray!25} 1 & 0 & \cellcolor{gray!25} 2 & \cellcolor{gray!25} 0.6 \\
		
		\hline
		
		Total & \cellcolor{gray!25} 12 & \cellcolor{gray!25} 23 & \cellcolor{gray!25} 18 & \cellcolor{gray!25} 47 & \cellcolor{gray!25} 211 & \cellcolor{gray!25} 14 & \cellcolor{gray!25} 325 & \cellcolor{gray!25} 99.8$^{a}$ \\
		
		\hline
		
	\end{tabular}
	\footnotesize $^{a}$Total does not sum to 100.0 due to rounding.\\
	\footnotesize $^{*}$The post-validation codes PoVC~1.1 Rule base and PoVC~2.8 What-if backward do not have a corpus assignment and are theoretically motivated.
\end{table}

The reliability subsample of 82 meaning units (25.2\% of the corpus) was constructed using stratified purposive sampling after recoding was complete and the post-validation codebook was frozen. Study~5 remained dominant in the subsample because it dominates the corpus overall. The subsample over-represents Studies~1 and~6 relative to their corpus proportions: Study~1 constitutes 3.7\% of the corpus (12/325) but 9.8\% of the subsample (8/82), and Study~6 constitutes 4.3\% of the corpus (14/325) but 6.1\% of the subsample (5/82). These deviations reflect representational balancing across explanation-content types rather than over- or under-sampling of particular studies, and were necessary to ensure adequate coverage of codes that are rare in the corpus overall (PoVC~1.2 Applied rule, PoVC~2.6 Causal mechanism) but concentrated in these studies.

The subsample was designed to include coding-difficulty cases at three borderline code boundaries (PoVC~1.2 Applied rule versus PoVC~2.3 Causal factor, PoVC~3.1 Set of ranked elements versus PoVC~3.2 Individual element, and PoVC~2.3 Causal factor versus PoVC~2.6 Causal mechanism) and one dual-valid code pair (PoVC~2.2 Input and PoVC~2.7 What-if forward), for which both codes constitute correct assignments for the same class of meaning units. The three pre-identified boundaries were selected as the primary loci of expected coding difficulty based on open-coding observations and expert comment patterns. Disagreements at other boundaries remained possible and, if observed, would be addressed through consensus resolution and decision-rule addition in the post-reliability codebook. In total, 32 units are genuinely borderline cases at the three code boundaries, and a further 32 units are dual-valid cases; together these 64 units (78\% of the subsample) represent coding-difficulty cases. Because the subsample deliberately overrepresents such cases, the resulting reliability statistics should be interpreted as conservative lower-bound estimates relative to expected full-corpus agreement.

Of the 32 borderline units, the 6 units assigned to PoVC~2.3 Causal factor are shared across two boundaries: 3 units were selected to represent the Applied rule versus Causal factor boundary and 3 units were selected to represent the Causal factor versus Causal mechanism boundary. The single PoVC~2.6 Causal mechanism unit represents the Causal factor versus Causal mechanism boundary on the mechanism side. The 2 PoVC~1.2 Applied rule units represent the Applied rule versus Causal factor boundary on the rule side. The 4 PoVC~3.1 and 19 PoVC~3.2 units all represent the Set versus Individual element boundary.

Table~\ref{tab:Result:SubsampleComposition} documents the composition of the reliability subsample. Borderline units require disambiguation between two codes. Dual-valid units are those for which both PoVC~2.2 Input and PoVC~2.7 What-if forward are correct assignments; disagreements between these codes are treated as agreements in reliability computation. The 6 borderline Causal factor units are split across two boundaries: 3 at the Applied rule vs Causal factor boundary and 3 at the Causal factor vs Causal mechanism boundary.

\begin{table}[!ht]
	\caption{Composition of the reliability subsample ($N = 82$).}
	\label{tab:Result:SubsampleComposition}
	\begin{scriptsize}
		\begin{tabular}
			{>{\raggedright\arraybackslash}m{4.5cm}
				>{\centering\arraybackslash}m{1.5cm}
				>{\centering\arraybackslash}m{1.5cm}
				>{\centering\arraybackslash}m{1.5cm}
				>{\raggedright\arraybackslash}m{3.0cm}}
			\hline
			Post-validation code & $n$ & Borderline & Dual-valid & Criterion \\
			\hline
			PoVC~1.1 Rule base & 0 & 0 & 0 & No corpus meaning units \\
			PoVC~1.2 Applied rule & 2 & 2 & 0 & Applied rule vs Causal factor \\
			PoVC~2.1 Context & 4 & 0 & 0 & -- \\
			PoVC~2.2 Input & 17 & 0 & 17 & Input or What-if forward both correct \\
			PoVC~2.3 Causal factor & 6 & 6 & 0 & 3 at Applied rule vs Causal factor; 3 at Causal factor vs Causal mechanism \\
			PoVC~2.4 Outcome & 4 & 0 & 0 & -- \\
			PoVC~2.5 Future state & 2 & 0 & 0 & -- \\
			PoVC~2.6 Causal mechanism & 1 & 1 & 0 & Causal factor vs Causal mechanism \\
			PoVC~2.7 What-if forward & 15 & 0 & 15 & Input or What-if forward both correct \\
			PoVC~2.8 What-if backward & 0 & 0 & 0 & No corpus meaning units \\
			PoVC~3.1 Set of ranked elements (actual) & 4 & 4 & 0 & Set vs Individual element \\
			PoVC~3.2 Individual element (actual) & 19 & 19 & 0 & Set vs Individual element \\
			PoVC~4.1 Set of similar elements (similar) & 4 & 0 & 0 & -- \\
			PoVC~4.2 Individual similar element (similar) & 2 & 0 & 0 & -- \\
			Not assigned & 2 & 0 & 0 & -- \\
			\hline
			Total & \cellcolor{gray!25} 82 & \cellcolor{gray!25} 32 & \cellcolor{gray!25} 32 & \\
			\hline
		\end{tabular}
	\end{scriptsize}
\end{table}

\clearpage


\subsection{Step~10: Reliability evaluation}
\label{sec:Results:Step10}

The reliability subsample was coded using the frozen post-validation codebook, in which codes carry the prefix PoVC. Following consensus resolution of disagreements and the addition of decision rules at three boundaries, the codebook was updated to its post-reliability version, in which all codes carry the prefix PoRC. The post-reliability codebook retains the same code numbering as the post-validation codebook; only the prefix changes from PoVC to PoRC. References to PoRC codes in the remainder of this subsection therefore refer to the post-reliability versions of the corresponding PoVC codes, with identical numbering.

Independent reliability coding was completed in accordance with the pre-registered procedure (OSF, pre-registered March~30, 2026), codebook frozen prior to pre-registration and independent coding conducted on April~3, 2026, on the stratified purposive subsample of 82 meaning units (25.2\% of the corpus), comprising 32 genuinely borderline units at three code boundaries and 32 dual-valid units for which both PoVC~2.2 Input and PoVC~2.7 What-if forward constitute correct assignments (64 of 82 units, 78\% coding-difficulty cases in total). Because PoVC~2.2 and PoVC~2.7 are dual-valid, assignments to one code by the ground-truth coder and to the other by the second coder were not counted as disagreements, and reliability statistics were computed on the adjusted coding matrix in which such pairs are resolved as agreements. Under this equivalence rule, percent agreement was 93.9\% (77 of 82 unique units). Krippendorff's alpha was $\alpha = 0.920$ and Cohen's kappa was $\kappa = 0.920$, both substantially exceeding the threshold of 0.800 required for reliable data for drawing conclusions. The two statistics produced identical values to three decimal places, which reflects the absence of the systematic disagreement patterns that typically cause kappa and alpha to diverge; the conditions under which the two statistics differ (specifically, systematic disagreement patterns that inflate or deflate kappa relative to alpha) were not present in the reliability data.

Table~\ref{tab:Result:ReliabilityAgreement} reports inter-rater agreement and disagreement by code boundary. Agreements include dual-valid PoVC~2.2/PoVC~2.7 pairs resolved as agreements under the pre-registered equivalence rule. Because the 6 borderline Causal factor units appear in two boundary rows (3 in the Applied rule vs Causal factor row and 3 in the Causal factor vs Causal mechanism row), boundary-row counts sum to 88 rather than 82, and boundary-row agreement counts sum to 83 rather than 77; the unique-unit count is 82 and the unique-agreement count is 77, and it is these unique counts that are used to compute percent agreement (77/82 = 93.9\%). Code labels use the PoVC prefix reflecting the post-validation codebook version in use during reliability coding; corresponding post-reliability codes carry the PoRC prefix.

\begin{table}[!ht]
	\small
	\caption{Inter-rater agreement and disagreement by code boundary in the reliability subsample ($N = 82$ unique units).}
	\label{tab:Result:ReliabilityAgreement}
	\begin{scriptsize}
		\begin{tabular}
			{>{\raggedright\arraybackslash}m{6.5cm}
				>{\centering\arraybackslash}m{1.8cm}
				>{\centering\arraybackslash}m{1.8cm}
				>{\centering\arraybackslash}m{1.8cm}}
			\hline
			Code boundary or group & Boundary-row counts$^{*}$ & Agreements & Disagreements \\
			\hline
			Non-borderline, non-dual-valid units & 18 & 18 & 0 \\
			Dual-valid PoVC~2.2 / PoVC~2.7 (equivalence rule applied) & 32 & 32 & 0 \\
			PoVC~1.2 Applied rule vs PoVC~2.3 Causal factor & 5 & 5 & 0 \\
			PoVC~2.3 Causal factor vs PoVC~2.6 Causal mechanism & 4 & 4 & 0 \\
			PoVC~3.1 Set vs PoVC~3.2 Individual element & 23 & 21 & 2 \\
			PoVC~2.6 Causal mechanism vs adjacent codes & 2 & 0 & 2 \\
			PoVC~3.2 Individual element vs PoVC~2.2 Input & 4 & 3 & 1 \\
			\hline
			\multicolumn{4}{p{11cm}}{Boundary-row counts sum to 88; boundary-row agreement counts sum to 83. Unique units: 82; unique agreements: 77; percent agreement: 93.9\%.} \\
			\hline
		\end{tabular}
		$^{*}$This row includes the single PoVC~2.6 unit from the ground-truth dataset and one unit assigned Not assigned by the ground-truth coder but PoVC~2.6 by the second coder; the boundary row therefore includes units assigned to PoVC~2.6 by either coder.
	\end{scriptsize}
\end{table}

The five disagreements occurred across three boundaries. Two disagreements occurred at the boundary between PoVC~3.1 Set of ranked elements and PoVC~3.2 Individual element, where the ground-truth coder assigned PoVC~3.1 and the second coder assigned PoVC~3.2 in both cases. Two further disagreements occurred at the boundary between PoVC~2.6 Causal mechanism and adjacent codes: in one case the ground-truth coder assigned Not assigned while the second coder assigned PoVC~2.6, and in another the ground-truth coder assigned PoVC~2.6 while the second coder assigned PoVC~3.2. One disagreement occurred at the boundary between PoVC~3.2 Individual element and PoVC~2.2 Input, where the ground-truth coder assigned PoVC~3.2 and the second coder assigned PoVC~2.2. Not all of these boundaries were among the three pre-identified borderline boundaries, which likely contributed to the disagreements. The affected boundaries were clarified in the codebook by adding explicit decision rules at the PoRC~2.2 vs.\ PoRC~3.2, PoRC~2.6 vs.\ PoRC~3.2, and PoRC~3.1 vs.\ PoRC~3.2 boundaries, producing the post-reliability codebook in which all codes carry the prefix PoRC. No structural changes to the code system were required. The five disagreements were resolved by consensus and the full corpus dataset was finalized using the frozen post-reliability codebook. The change log documenting the trigger, rationale, and outcome of each revision from the post-validation codebook to the post-reliability codebook is provided in Appendix~\ref{app:ChangeLogPostvalidationToPostReliability}.

\clearpage


\section{Coding frame (pre-validation)}
\label{app:CodingFrame}


\subsection*{Level 1: Category}

\begin{itemize}
	\item C1 Rule-based explanation
	\item C2 Causal explanation
	\item C3 Epistemic explanation (actual)
	\item C4 Epistemic explanation (similar)
\end{itemize}


\subsection*{Domains}

\begin{itemize}
	\item D1 AI domain
	\item D2 System domain
	\item D3 Application domain
	\item D4 Mixed domain
\end{itemize}


\subsection*{Explanation types}

\begin{itemize}
	\item ET1 Causal
	\item ET2 Epistemic
	\item ET3 Counterfactual
	\item ET4 Contrastive
	\item ET5 Example-based
	\item ET6 Rule-based
\end{itemize}

\clearpage


\section{Domain-of-origin types}
\label{app:DomainTypes}

\subsubsection*{DT1 AI domain}
\begin{itemize}
	\item \textbf{Definition} The domain perspective focused on model-level concepts and evidence that describe how the ML component behaves as an AI system, including learning, inference, uncertainty, performance, and data-related properties.
	
	\item \textbf{Include}
	\begin{itemize}
		\item Model performance metrics, error patterns, and calibration or uncertainty information.
		\item Training and inference details such as architecture, hyperparameters, feature selection, and version lineage.
		\item Data properties used for learning and evaluation such as label distribution, dataset split, modality, and drift indicators.
	\end{itemize}
	
	\item \textbf{Exclude}
	\begin{itemize}
		\item End-user task outcomes and operational business impact unless explicitly linked to AI metrics as objectives.
		\item System integration, infrastructure, latency budgets, and reliability concerns unless they are framed as model-inference properties.
		\item Facility, process, or asset-specific operational context not directly tied to model behavior.
	\end{itemize}
	
	\item \textbf{Typical evidence}
	\begin{itemize}
		\item Manufacturing: Confusion matrix for defect classes, dataset version history, feature importances, misclassification examples.
		\item Building technology: F1 scores for fault classes, calibration curves, drift reports across buildings, training-dataset coverage.
	\end{itemize}
	
	\item \textbf{Examples}
	\begin{itemize}
		\item Manufacturing: The model F1 improved after adding more examples of reflective defects and adjusting the decision threshold.
		\item Building technology: The classifier is under-calibrated for rare chiller faults and shows increased false negatives in winter data.
	\end{itemize}
\end{itemize}

\subsection*{DT2 System domain}
\begin{itemize}
	\item \textbf{Definition} The domain perspective focused on the engineered system that hosts and integrates the ML component, emphasizing architecture, interfaces, infrastructure, operational constraints, reliability, security, and maintainability.
	
	\item \textbf{Include}
	\begin{itemize}
		\item System architecture, integration interfaces, data pipelines, and deployment environments.
		\item Operational properties such as latency, availability, throughput, resource usage, monitoring, and rollback mechanisms.
		\item Safety, security, governance, and compliance constraints for operating the system.
	\end{itemize}
	
	\item \textbf{Exclude}
	\begin{itemize}
		\item Model-internal learning details unless required to explain system behavior such as inference resource use.
		\item Purely application outcomes such as scrap reduction or comfort improvement without system-level considerations.
		\item Domain context that does not affect system design or operation.
	\end{itemize}
	
	\item \textbf{Typical evidence}
	\begin{itemize}
		\item Manufacturing: Edge gateway specifications, OPC UA interface definitions, SCADA integration logs, container-version manifests.
		\item Building technology: BACnet integration setup, gateway CPU and memory reports, security-review records, monitoring dashboards.
	\end{itemize}
	
	\item \textbf{Examples}
	\begin{itemize}
		\item Manufacturing: The edge runtime meets the p95 inference-latency target under peak throughput and supports safe rollback.
		\item Building technology: The analytics service must throttle polling to avoid overloading BACnet networks and controllers.
	\end{itemize}
\end{itemize}

\subsection*{DT3 Application domain}
\begin{itemize}
	\item \textbf{Definition} The domain perspective focused on real-world work, processes, and outcomes in the user and business context where the system is applied, including tasks, decisions, impacts, and operational objectives.
	
	\item \textbf{Include}
	\begin{itemize}
		\item User goals, tasks, workflows, and decision points such as diagnosis, approval, dispatch, and mitigation selection.
		\item Operational outcome objectives such as downtime reduction, scrap reduction, energy savings, safety, comfort, and compliance.
		\item Contextual factors such as facility conditions, production-line settings, occupancy, and maintenance practices.
	\end{itemize}
	
	\item \textbf{Exclude}
	\begin{itemize}
		\item Model-performance metrics unless they are translated into application outcomes or user-relevant success criteria.
		\item Infrastructure and system-integration details unless they affect application operations and user tasks directly.
		\item Purely technical model-lineage information without application relevance.
	\end{itemize}
	
	\item \textbf{Typical evidence}
	\begin{itemize}
		\item Manufacturing: OEE and downtime reports, scrap rates, maintenance work orders, operator notes, process-change records.
		\item Building technology: Comfort complaint logs, energy-consumption trends, service-dispatch records, occupancy schedules.
	\end{itemize}
	
	\item \textbf{Examples}
	\begin{itemize}
		\item Manufacturing: The recommended maintenance action prevents unplanned line stoppages and reduces scrap for a packaging line.
		\item Building technology: Fault prioritization reduces comfort complaints while lowering unnecessary service dispatches.
	\end{itemize}
\end{itemize}

\subsection*{DT4 Mixed domain}
\begin{itemize}
	\item \textbf{Definition} A classification used when an explanation concept or item simultaneously expresses meaningful content from more than one domain perspective, such that it cannot be assigned to a single domain without losing essential meaning.
	
	\item \textbf{Include}
	\begin{itemize}
		\item Objectives or scores that combine application outcomes with AI and system measures.
		\item Evidence that jointly links model behavior, system behavior, and application impact.
		\item Items that describe tradeoffs across domains, such as quality versus latency versus cost or safety versus false alarms.
	\end{itemize}
	
	\item \textbf{Exclude}
	\begin{itemize}
		\item Items that mention another domain only as background context without substantive linkage.
		\item Cases where one domain is clearly primary and others are incidental.
		\item Pure aggregation labels without a clear cross-domain interpretation.
	\end{itemize}
	
	\item \textbf{Typical evidence}
	\begin{itemize}
		\item Manufacturing: A ranked model list where ranking combines defect F1, edge latency, and expected downtime reduction.
		\item Building technology: A recommendation ROI computed from predicted fault probability, gateway resource limits, and energy savings.
	\end{itemize}
	
	\item \textbf{Examples}
	\begin{itemize}
		\item Manufacturing: Model selection chooses a slightly less accurate model because it meets edge latency and improves throughput impact.
		\item Building technology: A responsive action is prioritized using predicted fault risk, comfort impact, and implementation cost on site.
	\end{itemize}
\end{itemize}

\clearpage


\section{Change log from preliminary to pre-validation code}
\label{app:ChangeLogPreliminaryToPrevalidation}

\begin{table}[!ht]
	\small
	\caption{Change log from preliminary codes to pre-validation codes, part 1}
	\label{tab:Result:ChangeLogPreliminaryToPrevaliationPart1}
	\begin{scriptsize}
		\begin{tabular}
			{>{\raggedright\arraybackslash}m{4.0cm}
				>{\raggedright\arraybackslash}m{3.0cm}
				>{\raggedright\arraybackslash}m{3.0cm}
				>{\raggedright\arraybackslash}m{4.0cm}}
			\hline
			
			Preliminary code & Trigger & Change rationale & Pre-validation code \\
			
			\hline
			
			PC~1.1 Rule &
			Difficulty assigning mentions of rule parts to the code &
			The code name ``Rule'' is too broad for rule parts, such as conditions &
			PrVC~1.1 Rule attribute \\
			
			\hline
			
			PC~2.1 Context &
			-- &
			-- &
			PrVC~2.1 Context \\
			
			PC~2.2 Input &
			Meaning units reflecting user inputs that influence the outcome can also be used for what-if analysis &
			No code existed for what-if content; a dedicated code is therefore introduced &
			PrVC~2.2 Input, PrVC~2.7 What-if \\
			
			PC~2.3 Causal chain &
			It is challenging to map a meaning unit reflecting a causal chain (a group of causal elements) to a single code that is comparable to other codes &
			Causal chain reflects a group of causal explanation content, being equivalent to the category ``Causal explanation,'' which is on a different abstraction level than each element of the causal chain; to rectify this, the causal chain was removed as a code &
			Code removed \\
			
			PC~2.4 Causal factor &
			Removal of PC~2.3 Causal chain &
			Code renumbered; no substantive change to the code definition &
			PrVC~2.3 Causal factor \\
			
			PC~2.5 Causal mechanism &
			Removal of PC~2.3 Causal chain &
			Code renumbered; no substantive change to the code definition &
			PrVC~2.4 Causal mechanism \\
			
			PC~2.6 Outcome &
			Removal of PC~2.3 Causal chain &
			Code renumbered; no substantive change to the code definition &
			PrVC~2.5 Outcome \\
			
			PC~2.7 Future state &
			Removal of PC~2.3 Causal chain &
			Code renumbered; no substantive change to the code definition &
			PrVC~2.6 Future state \\
			
			\hline
			
		\end{tabular}
	\end{scriptsize}
\end{table}

\begin{table}[!ht]
	\small
	\caption{Change log from preliminary codes to pre-validation codes, part 2}
	\label{tab:Result:ChangeLogPreliminaryToPrevaliationPart2}
	\begin{scriptsize}
		\begin{tabular}
			{>{\raggedright\arraybackslash}m{4.0cm}
				>{\raggedright\arraybackslash}m{3.0cm}
				>{\raggedright\arraybackslash}m{3.0cm}
				>{\raggedright\arraybackslash}m{4.0cm}}
			\hline
			
			Preliminary code & Trigger & Change rationale & Pre-validation code \\
			
			\hline
			
			PC~3.1 Ranked elements (actual) &
			Assigning meaning units to this code is challenging because the code name does not clearly articulate that it refers to a set of elements, and it is not clear that the set includes the ranking mechanism &
			Clarify the scope of the code by rewording it &
			PrVC~3.1 Set of ranked elements with ranking mechanism (actual) \\
			
			PC~3.2 Element ranking property (actual) &
			The element (actual) code consists of three subcodes --- properties, attributes, and evidence --- and there is a risk that three subcodes would not pass reliability validation due to excessive complexity &
			Simplify the code schema by merging the three subcodes element properties, element attributes, and element evidence into one code element (actual); no substantive content is lost, only the internal subcode distinction is collapsed &
			PrVC~3.2 Individual element with ranking properties, evidence, and/or attributes (actual) \\
			
			PC~3.3 Element attributes (actual) &
			The element (actual) code consists of three subcodes --- properties, attributes, and evidence --- and there is a risk that three subcodes would not pass reliability validation due to excessive complexity &
			Simplify the code schema by merging the three subcodes element properties, element attributes, and element evidence into one code element (actual); no substantive content is lost, only the internal subcode distinction is collapsed &
			PrVC~3.2 Individual element with ranking properties, evidence, and/or attributes (actual) \\
			
			PC~3.4 Element evidence (actual) &
			The element (actual) code consists of three subcodes --- properties, attributes, and evidence --- and there is a risk that three subcodes would not pass reliability validation due to excessive complexity &
			Simplify the code schema by merging the three subcodes element properties, element attributes, and element evidence into one code element (actual); no substantive content is lost, only the internal subcode distinction is collapsed &
			PrVC~3.2 Individual element with ranking properties, evidence, and/or attributes (actual) \\
			
			\hline
			
		\end{tabular}
	\end{scriptsize}
\end{table}

\begin{table}[!ht]
	\small
	\caption{Change log from preliminary codes to pre-validation codes, part 3}
	\label{tab:Result:ChangeLogPreliminaryToPrevaliationPart3}
	\begin{scriptsize}
		\begin{tabular}
			{>{\raggedright\arraybackslash}m{4.0cm}
				>{\raggedright\arraybackslash}m{3.0cm}
				>{\raggedright\arraybackslash}m{3.0cm}
				>{\raggedright\arraybackslash}m{4.0cm}}
			\hline
			
			Preliminary code & Trigger & Change rationale & Pre-validation code \\
			
			\hline
			
			PC~3.5 Ranked elements (similar) &
			Assigning meaning units to this code is challenging because the code name does not clearly articulate that it refers to a set of elements, and it is not clear that the set includes the filter mechanism that selects elements for the set &
			Clarify the scope of the code by rewording it; to separate this set from the set of elements (actual), a different code number is used &
			PrVC~4.1 Set of similar elements with filter mechanism (similar) \\
			
			PC~3.6 Element ranking property (similar) &
			The element (similar) code consists of three subcodes --- properties, attributes, and evidence --- and there is a risk that three subcodes would not pass reliability validation due to excessive complexity &
			Simplify the code schema by merging the three subcodes element properties, element attributes, and element evidence into one code element (similar); no substantive content is lost, only the internal subcode distinction is collapsed &
			PrVC~4.2 Individual, similar element with filter properties, evidence, and/or attributes (similar) \\
			
			PC~3.7 Element attributes (similar) &
			The element (similar) code consists of three subcodes --- properties, attributes, and evidence --- and there is a risk that three subcodes would not pass reliability validation due to excessive complexity &
			Simplify the code schema by merging the three subcodes element properties, element attributes, and element evidence into one code element (similar); no substantive content is lost, only the internal subcode distinction is collapsed &
			PrVC~4.2 Individual, similar element with filter properties, evidence, and/or attributes (similar) \\
			
			PC~3.8 Element evidence (similar) &
			The element (similar) code consists of three subcodes --- properties, attributes, and evidence --- and there is a risk that three subcodes would not pass reliability validation due to excessive complexity &
			Simplify the code schema by merging the three subcodes element properties, element attributes, and element evidence into one code element (similar); no substantive content is lost, only the internal subcode distinction is collapsed &
			PrVC~4.2 Individual, similar element with filter properties, evidence, and/or attributes (similar) \\
			
			\hline
			
		\end{tabular}
	\end{scriptsize}
\end{table}

\clearpage


\section{Explanation types}
\label{app:ExplanationTypes}

\subsubsection*{ET1 Rule-based explanation}
\begin{itemize}
	\item \textbf{Definition} A rule-based explanation justifies an outcome by referring to the rule base and by showing which applied rule, or which applied rules, support the predicted or recommended outcome for the current case. The explanation makes explicit which conditions of the applied rule are satisfied and how these conditions support the outcome. \cite{VanderWaa2021-XAI-RuleExampleComparison-Elsevier}
	
	\item \textbf{Include} References to the rule base, applied rules for the current case, if--then rules, decision criteria, threshold-based conditions, conjunctions of feature constraints, and local rule sets that indicate why the current instance is assigned to the present outcome. \cite{VanderWaa2021-XAI-RuleExampleComparison-Elsevier}
	
	\item \textbf{Exclude} Explanations that primarily identify causal mechanisms, explanations that compare the actual outcome with an explicit alternative, explanations that explore hypothetical changes to inputs, and explanations that justify the outcome through retrieved prior cases rather than applied rules. \cite{VanderWaa2021-XAI-RuleExampleComparison-Elsevier}
	
	\item \textbf{Typical evidence} A displayed rule base, one or more applied rules highlighted for the current case, rule lists, decision sets, threshold statements, anchored conditions, or local surrogate rules that show which rule conditions were satisfied and how they support the outcome. \cite{VanderWaa2021-XAI-RuleExampleComparison-Elsevier}
	
	\item \textbf{Examples}
	\begin{itemize}
		\item Manufacturing: ``The outcome is classified as surface defect based on the applied rule: if edge contrast exceeds threshold, defect length is above 3~mm, and the detected shape is elongated, then classify as surface defect.''
		\item Building technology: ``The system diagnoses a heating-valve fault based on the applied rule: if valve command is high, room temperature remains below setpoint, and measured supply temperature does not increase, then diagnose heating-valve fault.''
	\end{itemize}
\end{itemize}

\subsubsection*{ET2 Causal explanation}
\begin{itemize}
	\item \textbf{Definition} A causal explanation identifies one or more causes of the actual outcome and characterizes how they bring about that outcome. It may also specify the mechanism that links cause to effect and the intervention-relevant dependencies that support the causal claim. \cite{Halpern2005-XAI-CausesExplanationsII-OUP,Pearl2009-XAI-Causality-Cambridge}
	
	\item \textbf{Include} Cause-and-effect statements, causal factors, causal relationships, causal chains, causal mechanisms, and interventional claims of the form ``if X had not occurred, Y would not have occurred.''
	
	\item \textbf{Exclude} Pure comparisons between alternatives without a causal account, hypothetical scenario exploration without a claim about the actual cause, and confidence or uncertainty statements that do not explain why the outcome occurred.
	
	\item \textbf{Typical evidence} Causal graphs or structural causal models, intervention or ablation studies, time-ordered logs that support precedence and linkage, controlled experiments, and root-cause analysis reports that connect conditions to outcomes with stated mechanisms.
	
	\item \textbf{Examples}
	\begin{itemize}
		\item Manufacturing: ``The defect spike was caused by lens contamination after a coolant leak, which reduced image contrast and increased false rejects.''
		\item Building technology: ``The comfort violation was caused by a stuck heating valve, which limited hot-water flow and prevented supply-air temperature from reaching setpoint.''
	\end{itemize}
\end{itemize}

\subsubsection*{ET3 Counterfactual explanation}
\begin{itemize}
	\item \textbf{Definition} A what-if explanation describes how the predicted or recommended outcome would change under one or more specified hypothetical changes to inputs, operating conditions, or assumptions. It supports scenario exploration and sensitivity assessment, but does not require the changes to be minimal, feasible, or action oriented. \cite{Wachter2018-XAI-CounterfactualExplanations-HarvardJLT}
	
	\item \textbf{Include} A specified change to one or more inputs, parameters, or context conditions; the resulting change in outcome, score, or recommendation under that scenario; sensitivity statements that show how much a variable must change to alter the outcome; and multiple scenario comparisons such as best case, nominal, and worst-case settings.
	
	\item \textbf{Exclude} Counterfactual recourse statements that require minimal and feasible changes to achieve a specific alternative outcome, pure contrastive explanations that compare outcomes without changing inputs, example-based or case-based explanations that justify an outcome by retrieved instances, and causal narratives that explain why the actual outcome happened without exploring alternative settings.
	
	\item \textbf{Typical evidence} Parameter-sweep results, sensitivity curves, interactive sliders with updated predictions and confidence values, scenario tables that list changed inputs and resulting outputs, simulation or digital-twin outputs for modified operating conditions, and logs that record scenario assumptions and computed outcomes.
	
	\item \textbf{Examples}
	\begin{itemize}
		\item Manufacturing: ``If conveyor speed is reduced by 10\% while illumination stays constant, the predicted reject rate decreases from 8\% to 5\%.''
		\item Building technology: ``If outdoor-air temperature drops by 5~$^\circ$C under the same setpoint, the predicted heating shortfall risk increases from 0.30 to 0.48.''
	\end{itemize}
\end{itemize}

\subsubsection*{ET4 Contrastive explanation}
\begin{itemize}
	\item \textbf{Definition} A contrastive explanation answers a question of the form ``Why P rather than Q'' by identifying the foil Q and highlighting the difference makers that favor the actual outcome P over that alternative. \cite{Miller2019-XAI-ExplanationAI-Elsevier}
	
	\item \textbf{Include} An explicit or contextually recoverable foil, comparative factors that distinguish P from Q, explanations that identify why one candidate was selected over another from a set, and explanations framed around ``rather than'' comparisons.
	
	\item \textbf{Exclude} Explanations that only state the actual cause without reference to an alternative, hypothetical scenario exploration based on changed inputs, and confidence statements that do not compare alternatives.
	
	\item \textbf{Typical evidence} Side-by-side comparison tables, candidate lists with explicit selection rationale, feature or factor deltas between P and Q, decision logs that record why an alternative was not chosen, and competing-hypotheses lists with eliminations.
	
	\item \textbf{Examples}
	\begin{itemize}
		\item Manufacturing: ``Why classify as scratch rather than stain: elongated edges and consistent directionality match scratch patterns, while stain candidates lack edge structure.''
		\item Building technology: ``Why diagnose valve stuck rather than sensor drift: command-response mismatch is present, while the parallel reference sensor agrees with the reported temperature.''
	\end{itemize}
\end{itemize}

\subsubsection*{ET5 Epistemic explanation (actual outcome)}
\begin{itemize}
	\item \textbf{Definition} An epistemic explanation for the actual outcome communicates the system's uncertainty about the current output when that uncertainty is due to incomplete knowledge, such as sparse training data, missing inputs, weak evidence, or model uncertainty. \cite{Hullermeier2021-XAI-AleatoricEpistemicUncertainty-Springer,Jiang2022-XAI-UserEpistemicUncertainty-Elsevier}
	
	\item \textbf{Include}
	\begin{itemize}
		\item Confidence or uncertainty indicators for the current prediction, such as confidence scores or predictive uncertainty estimates. \cite{Jiang2022-XAI-UserEpistemicUncertainty-Elsevier}
		\item Statements about knowledge limitations that affect reliability, such as insufficient evidence, missing input information, or sparse data coverage. \cite{Hullermeier2021-XAI-AleatoricEpistemicUncertainty-Springer}
		\item Explanatory qualifiers that calibrate reliance on the current output, such as low confidence due to limited historical support. \cite{Hullermeier2021-XAI-AleatoricEpistemicUncertainty-Springer}
	\end{itemize}
	
	\item \textbf{Exclude}
	\begin{itemize}
		\item Causal or rule-based rationales that explain why the outcome occurs. \cite{VanderWaa2021-XAI-RuleExampleComparison-Elsevier}
		\item What-if, contrastive, or counterfactual content that focuses on alternative outcomes or changed conditions. \cite{VanderWaa2021-XAI-RuleExampleComparison-Elsevier}
		\item Pure aleatoric-randomness statements without a knowledge-limitation component. \cite{Hullermeier2021-XAI-AleatoricEpistemicUncertainty-Springer}
	\end{itemize}
	
	\item \textbf{Typical evidence}
	\begin{itemize}
		\item A displayed confidence score or uncertainty estimate tied to the current output. \cite{Jiang2022-XAI-UserEpistemicUncertainty-Elsevier}
		\item Indicators of incomplete information affecting certainty, such as missing inputs or weak signals. \cite{Hullermeier2021-XAI-AleatoricEpistemicUncertainty-Springer}
		\item Reliability cues that help users decide whether to rely on the current recommendation. \cite{Jiang2022-XAI-UserEpistemicUncertainty-Elsevier}
	\end{itemize}
	
	\item \textbf{Examples}
	\begin{itemize}
		\item Predictive maintenance: ``Bearing-failure risk is high, but confidence is moderate because this motor type has limited historical data in the training set.''
		\item Visual quality inspection: ``Defect class is identified as scratch with low confidence due to poor illumination and partial occlusion in the image.''
		\item Process control: ``Yield drop is predicted with elevated uncertainty because several upstream sensor channels are missing or out of calibration.''
	\end{itemize}
\end{itemize}

\subsubsection*{ET6 Example-based explanation}
\begin{itemize}
	\item \textbf{Definition} An example-based explanation justifies an outcome by referring to one or more specific prior instances that are similar to the current case and that illustrate why the present outcome is plausible, typical, or supported. The explanation uses representative or nearest cases as the explanatory basis rather than rules or causal mechanisms. \cite{VanderWaa2021-XAI-RuleExampleComparison-Elsevier}
	
	\item \textbf{Include} Retrieved similar cases, nearest-neighbor examples, precedents, representative instances, and panels of comparable examples that show similar inputs and associated outcomes. \cite{VanderWaa2021-XAI-RuleExampleComparison-Elsevier}
	
	\item \textbf{Exclude} Pure confidence or uncertainty indicators without reference to specific instances, rule lists or threshold criteria that are not grounded in examples, contrastive explanations centered on ``why P rather than Q,'' and hypothetical scenario exploration under changed inputs. \cite{VanderWaa2021-XAI-RuleExampleComparison-Elsevier}
	
	\item \textbf{Typical evidence} Top similar-case lists, nearest-neighbor displays, example panels with similarity scores, retrieved precedents with outcomes, and side-by-side instance comparisons that show how the current case resembles known examples.
	
	\item \textbf{Examples}
	\begin{itemize}
		\item Manufacturing: ``This image is classified as scratch because it closely matches three prior scratch cases with the same elongated-edge pattern and contrast profile.''
		\item Building technology: ``This incident is flagged as a stuck valve because it matches earlier cases in which valve command stayed high while room temperature and supply response remained low.''
	\end{itemize}
\end{itemize}

\clearpage


\section{Codebook (draft)}
\label{app:DraftCodebook}

\subsection*{Explanation categories}

\subsubsection*{C1 Rule-based explanation}
\begin{itemize}
	\item \textbf{Definition} A rule-based explanation is an explanation that justifies an outcome by stating one or more explicit if-then rules or decision conditions that map input conditions to an outcome. The explanation communicates the decision logic in a symbolic, human-readable form, such as thresholds, Boolean combinations, or ordered rules, indicating which rule(s) applied in the current case and why this led to the outcome.
	
	\item \textbf{Include}
	\begin{itemize}
		\item If-then rules that specify conditions under which a prediction, classification, recommendation, or alert is produced.
		\item Decision rules expressed as thresholds, Boolean logic, scoring rules, decision tables, or rule lists, including ordered rules with precedence.
		\item Explanations that highlight the rule(s) triggered in the current case and the relevant input values that satisfied the rule conditions.
		\item Rule-based policies or compliance checks when they are used as the decision logic and presented to justify an outcome.
	\end{itemize}
	
	\item \textbf{Exclude}
	\begin{itemize}
		\item Post-hoc feature-attribution lists that rank features by importance without expressing explicit decision conditions.
		\item Example-based explanations that justify an outcome by citing similar cases rather than explicit rules.
		\item Causal narratives that describe mechanisms or causal chains without specifying decision conditions.
		\item Purely procedural descriptions of how the model was trained or what data it uses, without stating the rule(s) that led to the specific outcome.
		\item Uninterpretable model internals, for example weights, unless translated into explicit decision rules.
	\end{itemize}
	
	\item \textbf{Typical evidence}
	\begin{itemize}
		\item Manufacturing: ``If vibration RMS exceeds 7~mm/s for longer than 10~minutes and temperature exceeds 80$^\circ$C, then predict bearing fault.''
		\item Building technology: ``If CO$_2$ $>$ 1000~ppm and damper position $<$ 20\% during occupancy, then raise indoor-air-quality alert.''
		\item Cybersecurity: ``If software version matches a known vulnerable pattern and the asset is internet-facing, then flag the CVE as critical and recommend patching.''
	\end{itemize}
	
	\item \textbf{Examples}
	\begin{itemize}
		\item Manufacturing: ``The system predicts an impending bearing fault because the rule `vibration RMS $>$ 7~mm/s for 10~minutes AND temperature $>$ 80$^\circ$C' is satisfied (vibration = 8.2~mm/s, temperature = 83$^\circ$C).''
		\item Building technology: ``An indoor-air-quality alert is raised because CO$_2$ exceeded 1000~ppm while the outside-air damper remained below 20\% during occupied hours.''
		\item Cybersecurity: ``The vulnerability is marked as critical because the rule `Log4j version 2.14.1 AND internet exposure = true' is satisfied, triggering the remediation recommendation to upgrade.''
	\end{itemize}
\end{itemize}

\subsubsection*{C2 Causal explanation}
\begin{itemize}
	\item \textbf{Definition} A causal explanation is an explanation that justifies an outcome by stating one or more causes and describing how those causes produced the outcome. It expresses a cause-to-effect relationship that goes beyond correlation, linking initiating conditions and intermediate factors or mechanisms to the observed outcome in a way that supports understanding, diagnosis, or intervention.
	
	\item \textbf{Include}
	\begin{itemize}
		\item Statements that identify one or more causal factors or drivers of the outcome, for example root cause or contributing cause.
		\item Explanations that describe a causal mechanism or causal chain connecting factors to the outcome, for example intermediate steps or propagation.
		\item Explanations that indicate directionality and causal influence, for example led to, resulted in, due to, because, including multi-cause accounts.
		\item Causal accounts that connect evidence to an outcome to support actions such as troubleshooting, mitigation, or prevention.
	\end{itemize}
	
	\item \textbf{Exclude}
	\begin{itemize}
		\item Pure statistical-association or correlation statements without a causal claim.
		\item Rule-based decision-logic explanations that only state decision conditions without asserting real-world causation.
		\item Counterfactual or recourse explanations that focus on how inputs must change to obtain a different outcome without explaining why the actual outcome occurred.
		\item Example-based explanations that justify an outcome by citing similar cases rather than causal reasoning.
		\item Descriptions of the data pipeline or model-training process that do not explain why this specific outcome occurred.
	\end{itemize}
	
	\item \textbf{Typical evidence}
	\begin{itemize}
		\item Manufacturing: Failure-analysis narrative linking component wear to process deviation and defect formation, supported by sensor trends.
		\item Building technology: Fault-diagnosis narrative linking actuator malfunction to airflow changes and comfort violations, supported by control logs.
		\item Cybersecurity: Attack-analysis narrative linking initial compromise to privilege escalation and impact, supported by authentication logs and alerts.
	\end{itemize}
	
	\item \textbf{Examples}
	\begin{itemize}
		\item Manufacturing: ``Nozzle wear reduced fill precision, which caused spills that contaminated the inspection optics, leading to increased false rejects.''
		\item Building technology: ``A stuck damper reduced outside-air intake, which raised CO$_2$ levels during occupancy and triggered the air-quality alert.''
		\item Cybersecurity: ``A phishing email led to credential theft, enabling unauthorized VPN access and lateral movement, which resulted in ransomware deployment and service disruption.''
	\end{itemize}
\end{itemize}

\subsubsection*{C3 Epistemic explanation (actual)}
\begin{itemize}
	\item \textbf{Definition} An epistemic explanation (actual) is an explanation that justifies the actual outcome by communicating the system's uncertainty, confidence, or evidence strength regarding that outcome. It characterizes how strongly the system believes the actual outcome holds, based on probabilistic reasoning, likelihood, prediction confidence, posterior belief, or related uncertainty measures, and may indicate how reliable the outcome is expected to be.
	
	\item \textbf{Include}
	\begin{itemize}
		\item Confidence, probability, or likelihood statements about the actual outcome, for example ``spam probability = 0.92'' or ``fault likelihood high''.
		\item Uncertainty information such as prediction intervals, credible intervals, error bars, entropy, margin to decision boundary, or ambiguity indicators tied to the actual outcome.
		\item Evidence-strength or reliability cues for the actual outcome, for example calibration statements, model-certainty categories, or quality scores, when presented as a justification for the outcome.
		\item References to alternative outcomes only to contextualize uncertainty, for example ``spam 0.62 vs.\ not spam 0.38,'' while keeping the actual outcome as the explained result.
	\end{itemize}
	
	\item \textbf{Exclude}
	\begin{itemize}
		\item Explanations that describe why the outcome occurred in causal terms.
		\item Rule-based or procedural explanations that state decision conditions without uncertainty information.
		\item Contrastive explanations whose primary purpose is to compare the actual outcome with a specific alternative outcome.
		\item Counterfactual or recourse explanations that describe what would need to change to obtain a different outcome.
		\item Pure data-quality or provenance statements unless explicitly linked to confidence in the actual outcome.
	\end{itemize}
	
	\item \textbf{Typical evidence}
	\begin{itemize}
		\item Manufacturing: Predicted defect probability with a confidence score and a note about uncertainty due to low image quality.
		\item Building technology: Fault-likelihood score with a confidence band and an ambiguity warning due to missing sensor readings.
		\item Cybersecurity: Risk score for an identified vulnerability with probability of exploitation and a confidence level based on observed exposure and asset telemetry.
	\end{itemize}
	
	\item \textbf{Examples}
	\begin{itemize}
		\item Manufacturing: ``The system predicts a bearing fault with 0.87 probability; uncertainty is moderate because recent vibration data contains gaps.''
		\item Building technology: ``The economizer fault is detected with high confidence (confidence = 0.91), supported by consistent deviations across three sensors.''
		\item Cybersecurity: ``The system flags CVE-2021-44228 as present with 0.95 confidence; confidence is lower for exploitability because external exposure could not be fully confirmed.''
	\end{itemize}
\end{itemize}

\subsubsection*{C4 Epistemic explanation (similar)}
\begin{itemize}
	\item \textbf{Definition} An epistemic explanation (similar) is an explanation that justifies the current outcome by referencing historic cases that are similar to the current case and using those similarities as evidence for an element in the causal chain. It communicates the system's belief or confidence in the outcome by showing that comparable prior instances produced the same or highly similar outcomes, often including similarity scores, match criteria, and how frequently the outcome occurred among the retrieved cases.
	
	\item \textbf{Include}
	\begin{itemize}
		\item References to one or more historic cases deemed similar to the current case, used as evidence for the current outcome.
		\item Similarity information such as nearest neighbors, match scores, distances, feature overlap, or retrieved-case sets, when tied to confidence in the current outcome.
		\item Frequency or distribution summaries over similar cases, for example ``8 of the 10 most similar cases were defects of type X''.
		\item Case-based or example-based probability statements derived from the similar-case set, for example ``based on similar cases, likelihood of fault is high''.
	\end{itemize}
	
	\item \textbf{Exclude}
	\begin{itemize}
		\item Purely illustrative examples shown for explanation without being used as evidence for the current outcome.
		\item Counterfactual examples that show how the outcome would change under modified inputs.
		\item Contrastive examples whose main purpose is to compare the current case to a foil case with a different outcome.
		\item Causal narratives explaining why the outcome occurred, if they do not rely on similar cases as evidence.
		\item Rule-based decision conditions that do not involve retrieval of similar historic cases.
	\end{itemize}
	
	\item \textbf{Typical evidence}
	\begin{itemize}
		\item Manufacturing: Retrieval of past production runs with similar vibration signatures, showing that the same fault pattern preceded bearing failure.
		\item Building technology: Similar past days or zones with comparable temperature and valve behavior, where the same fault label was confirmed.
		\item Cybersecurity: Prior incidents with similar alert sequences or vulnerability configurations, showing how often they led to exploitation or high-risk classification.
	\end{itemize}
	
	\item \textbf{Examples}
	\begin{itemize}
		\item Manufacturing: ``The system predicts bearing wear because the current spectrum matches 9 prior cases (similarity $\geq$ 0.92), and 8 of those cases were confirmed as outer-race wear.''
		\item Building technology: ``This condition is classified as an economizer fault because 12 similar historical episodes showed the same damper and temperature pattern, and 10 were resolved by damper repair.''
		\item Cybersecurity: ``The alert is rated high risk because it matches recent incident patterns in 6 similar cases (same CVE, exposure, and log sequence), and 5 of those cases resulted in confirmed compromise.''
	\end{itemize}
\end{itemize}

\subsection*{Pre-validation codes}

\subsubsection*{PrVC~1.1 Rule attribute}
\begin{itemize}
	\item \textbf{Definition} Conditions that specify when a rule applies, typically expressed in the form IF <condition(s)> THEN <conclusion>. The <conclusion> part includes one or multiple causal explanation elements.
	
	\item \textbf{Include}
	\begin{itemize}
		\item Explicit antecedent conditions stated as predicates, thresholds, ranges, or membership checks, including combined conditions using AND or OR.
		\item Condition sets that reference measurable values, states, events, or categorical properties used to determine whether the rule applies.
		\item Condition evaluations shown as met or not met, including which specific predicates passed or failed.
	\end{itemize}
	
	\item \textbf{Exclude}
	\begin{itemize}
		\item Pure outcome statements without an explicit condition part.
		\item Probabilistic statements, feature weights, or scores that do not define a testable antecedent.
		\item General policies or guidance phrased without operational conditions that can be checked against a case.
	\end{itemize}
	
	\item \textbf{Typical evidence}
	\begin{itemize}
		\item Manufacturing: IF vibration RMS exceeds 7.5~mm/s AND bearing temperature exceeds 85$^\circ$C AND the condition persists for at least 10~minutes.
		\item Building technology: IF supply-air temperature deviates by more than 3$^\circ$C from setpoint AND the valve command remains at 100\% for at least 15~minutes AND airflow remains below the minimum expected value.
		\item Cybersecurity: IF the software bill of materials indicates \texttt{log4j-core} version 2.0--2.14.1 AND an outbound LDAP connection attempt is observed in logs AND the service is reachable from the internet.
	\end{itemize}
	
	\item \textbf{Examples}
	\begin{itemize}
		\item Manufacturing: IF vibration RMS exceeds 7.5~mm/s AND bearing temperature exceeds 85$^\circ$C THEN the likely causal factor is bearing wear and the recommended outcome is to schedule bearing replacement.
		\item Building technology: IF supply-air temperature deviates by more than 3$^\circ$C from setpoint AND the valve command remains at 100\% for at least 15~minutes THEN the likely causal factor is a stuck heating valve and the recommended outcome is to dispatch maintenance to inspect and repair the actuator.
		\item Cybersecurity: IF \texttt{log4j-core} version 2.0--2.14.1 is present AND exploitation indicators, e.g.\ JNDI lookup patterns, appear in request logs AND the service is internet-facing THEN the causal factor is exposure to CVE-2021-44228 and the recommended outcome is to patch to Log4j 2.17.1 or later and apply mitigations until patching is complete.
	\end{itemize}
\end{itemize}

\subsubsection*{PrVC~2.1 Context}
\begin{itemize}
	\item \textbf{Definition} Situation or circumstances under which an outcome was generated. It captures descriptive information that helps interpret the outcome and the entire causal chain, such as the operational environment, setting, time and location, involved assets or process segment, and relevant constraints or states of the surrounding system.
	
	\item \textbf{Include}
	\begin{itemize}
		\item Environmental and operational setting information, for example plant area, building zone, production line, cybersecurity network segment, or operating mode.
		\item Time, location, and asset or process identifiers that help situate the outcome, for example timestamp, site, machine ID, or host name.
		\item Boundary conditions and constraints that affect interpretation or action, for example planned maintenance window, safety restrictions, policy scope, or access limitations.
	\end{itemize}
	
	\item \textbf{Exclude}
	\begin{itemize}
		\item Measurements, signals, or records that directly trigger or determine the outcome.
		\item Causal statements that explain why the outcome occurred.
		\item Recommended actions or decisions.
	\end{itemize}
	
	\item \textbf{Typical evidence}
	\begin{itemize}
		\item Manufacturing: ``Line 3, Station 12, night shift, product variant B, humidity control active.''
		\item Building technology: ``Building A, 3rd floor west wing, weekday occupancy schedule, AHU-2 operating in economizer mode.''
		\item Cybersecurity: ``Production subnet DMZ, internet-facing service, asset criticality high, change-freeze period in effect.''
	\end{itemize}
	
	\item \textbf{Examples}
	\begin{itemize}
		\item Manufacturing: ``Incident occurred on Press~\#4 in Plant~2 during the start-up phase after a tooling change.''
		\item Building technology: ``Temperature anomaly detected in Zone~3B at 08:15 on a cold day while the building was in pre-occupancy warm-up.''
		\item Cybersecurity: ``Vulnerability detected on an internet-facing application server in the production environment during a scheduled weekly scan.''
	\end{itemize}
\end{itemize}

\subsubsection*{PrVC~2.2 Input}
\begin{itemize}
	\item \textbf{Definition} Input is the data provided to the AI system to generate an outcome. It captures the case-specific values, signals, records, and/or user-supplied entries that are used by the system's model or decision logic when producing a prediction, classification, recommendation, or alert.
	
	\item \textbf{Include}
	\begin{itemize}
		\item Sensor readings, time series, measurements, images, audio, or log files ingested by the AI component.
		\item Structured feature values derived from raw data, for example aggregated statistics, extracted descriptors, or embeddings, when they are used by the system to compute the outcome.
		\item User-provided data used by the system to generate the outcome, for example form entries, parameter settings, uploaded files, or prompt text.
	\end{itemize}
	
	\item \textbf{Exclude}
	\begin{itemize}
		\item Background circumstances that frame the case but are not used to compute the outcome.
		\item The identified vulnerability, cause, or explanatory factor inferred from the input.
		\item Recommended actions or decisions produced by the system.
		\item Post-outcome verification data collected after an intervention.
	\end{itemize}
	
	\item \textbf{Typical evidence}
	\begin{itemize}
		\item Manufacturing: Camera image of a product, sensor stream from a station, or a feature vector computed from vibration data used to predict a bearing fault.
		\item Building technology: Temperature and humidity readings, occupancy signals, and setpoints used to predict an energy anomaly.
		\item Cybersecurity: Vulnerability scan results, software inventory with version strings, and configuration snapshots used to identify a CVE.
	\end{itemize}
	
	\item \textbf{Examples}
	\begin{itemize}
		\item Manufacturing: ``Input includes the last 30 seconds of vibration data from Motor M2 and the extracted spectral peaks used to predict an impending failure.''
		\item Building technology: ``Input includes zone temperature, supply-air temperature, valve position, and occupancy status used to infer an AHU fault.''
		\item Cybersecurity: ``Input includes the detected Log4j library version and an internet-exposure flag from the scan report used to identify CVE-2021-44228.''
	\end{itemize}
\end{itemize}

\subsubsection*{PrVC~2.3 Causal factor}
\begin{itemize}
	\item \textbf{Definition} A causal factor is an explanation element that represents a causal connection between the input and the outcome. It captures what is believed to have influenced the outcome.
	
	\item \textbf{Include}
	\begin{itemize}
		\item Stated causes, drivers, or contributing conditions that explain why the outcome occurred, for example ``insufficient illumination caused misclassification''.
		\item Identified issues, faults, vulnerabilities, or anomalies that connect input evidence to the system's recommendation or decision, for example ``CVE-2021-44228 present'' as the vulnerability driving remediation.
		\item Contributing factors that are part of a causal chain, single factor or multiple factors, and that can be referenced as targets for intervention.
	\end{itemize}
	
	\item \textbf{Exclude}
	\begin{itemize}
		\item Raw observations, measurements, or records used to compute or detect the outcome.
		\item Background circumstances that frame the case but are not used as causes in the explanation.
		\item Descriptions of how the factor leads to the outcome in terms of process or mechanism.
	\end{itemize}
	
	\item \textbf{Typical evidence}
	\begin{itemize}
		\item Manufacturing: ``Excessive vibration at 120~Hz indicates bearing wear'' used as the driver for a maintenance recommendation.
		\item Building technology: ``Damper stuck at 10\%'' identified as the fault driving an air-quality or comfort alert.
		\item Cybersecurity: ``CVE-2021-44228 present'' identified as the vulnerability driving the recommendation to patch.
	\end{itemize}
	
	\item \textbf{Examples}
	\begin{itemize}
		\item Manufacturing: ``The most likely cause of the predicted failure is bearing wear driven by sustained high vibration at the outer-race frequency.''
		\item Building technology: ``The alert is attributed to a stuck outside-air damper that reduces ventilation during occupancy.''
		\item Cybersecurity: ``The high-risk rating is driven by the presence of CVE-2021-44228 on an internet-facing application server.''
	\end{itemize}
\end{itemize}

\subsubsection*{PrVC~2.4 Causal mechanism}
\begin{itemize}
	\item \textbf{Definition} A causal mechanism is an explanation element that describes the causal relationship between two elements of the causal chain.
	
	\item \textbf{Include}
	\begin{itemize}
		\item Process descriptions that explain how a causal factor produces the outcome, including physical, organizational, computational, or behavioral processes.
		\item Intermediate links that connect adjacent causal-chain elements, for example input $\rightarrow$ causal factor, causal factor $\rightarrow$ outcome, or outcome $\rightarrow$ future state, by naming the causal principle in between.
		\item Domain-specific mechanism statements that describe a functional relationship.
	\end{itemize}
	
	\item \textbf{Exclude}
	\begin{itemize}
		\item Causal-factor statements that name a cause but do not describe a pathway.
		\item Pure statistical-association statements that do not express a how-or-why relationship unless explicitly framed as a causal pathway.
		\item Raw observations used to infer the relationship, such as sensor readings, logs, or measurements.
		\item Recommended actions, decisions, or controls.
	\end{itemize}
	
	\item \textbf{Typical evidence}
	\begin{itemize}
		\item Manufacturing: Text explaining how vibration at a specific frequency indicates bearing wear, which increases friction and leads to overheating.
		\item Building technology: Text explaining how a stuck damper reduces outside-air intake, which increases CO$_2$ levels and triggers an indoor-air-quality alert.
		\item Cybersecurity: Text explaining how a vulnerable component enables remote-code execution, which supports privilege escalation and increases compromise risk, motivating remediation.
	\end{itemize}
	
	\item \textbf{Examples}
	\begin{itemize}
		\item Manufacturing: ``Contamination on the lens reduces image contrast, causing the edge detector to miss boundaries, which increases false defect classifications.''
		\item Building technology: ``If the heating valve remains open, heat continues to enter the zone despite the setpoint, which raises zone temperature and triggers an overheating alert.''
		\item Cybersecurity: ``A vulnerable logging library can be exploited via crafted input that triggers message lookups, enabling remote-code execution, which increases the likelihood of system compromise and drives the recommendation to patch.''
	\end{itemize}
\end{itemize}

\subsubsection*{PrVC~2.5 Outcome (actual)}
\begin{itemize}
	\item \textbf{Definition} Outcome (actual) is the observed result of an AI system for the current case. It is the entity that is explained.
	
	\item \textbf{Include}
	\begin{itemize}
		\item Predicted or classified labels and values, for example ``defect detected,'' ``energy anomaly,'' or ``high risk''.
		\item Recommended actions or interventions, for example ``inspect sensor,'' ``replace component,'' or ``apply patch''.
		\item System decisions and triggers, for example ``raise alarm,'' ``block request,'' or ``schedule maintenance''.
		\item Rankings or selections when the system chooses one item among alternatives.
	\end{itemize}
	
	\item \textbf{Exclude}
	\begin{itemize}
		\item The evidence or data used to produce the result.
		\item Causes, vulnerabilities, or contributing conditions used to justify the result.
		\item Alternative outcomes not selected for the current case.
		\item Post-intervention verification results or expected post-state.
	\end{itemize}
	
	\item \textbf{Typical evidence}
	\begin{itemize}
		\item Manufacturing: ``Defect type = surface scratch'' or ``Quality check failed'' generated by vision inspection.
		\item Building technology: ``Fault detected: AHU economizer malfunction'' or ``Energy anomaly score = high.''
		\item Cybersecurity: ``Vulnerability detected: CVE-2021-44228'' or ``Recommended remediation: upgrade Log4j to 2.17.1.''
	\end{itemize}
	
	\item \textbf{Examples}
	\begin{itemize}
		\item Manufacturing: ``The system flags Unit 1837 as defective due to a suspected surface scratch and recommends manual inspection.''
		\item Building technology: ``The system reports an overheating fault for Zone~3B and recommends checking the reheat valve.''
		\item Cybersecurity: ``The system identifies CVE-2021-44228 on Server A and recommends applying the vendor patch within 48 hours.''
	\end{itemize}
\end{itemize}

\subsubsection*{PrVC~2.6 Future state}
\begin{itemize}
	\item \textbf{Definition} Future state is the predicted state that is expected to occur after suggested decisions, represented as outcomes, have been applied.
	
	\item \textbf{Include}
	\begin{itemize}
		\item Predicted post-intervention system states, for example ``fault cleared,'' ``risk reduced,'' ``performance restored,'' or ``vulnerability removed''.
		\item Expected verification or confirmation states following an action, for example ``rescan shows no finding'' or ``sensor reading returns to normal range''.
		\item Anticipated side effects or tradeoffs if explicitly stated as part of the post-action state, for example ``service restart required'' or ``temporary downtime expected''.
		\item Follow-up milestones or conditions that define success after implementing the outcome, for example ``compliance restored'' or ``incident closed''.
	\end{itemize}
	
	\item \textbf{Exclude}
	\begin{itemize}
		\item The action itself or the recommended decision.
		\item The causes, vulnerabilities, or contributing conditions that motivate the action.
		\item The data used to generate the initial outcome and the background situation.
		\item Hypothetical alternatives that are not tied to applying the suggested outcome.
	\end{itemize}
	
	\item \textbf{Typical evidence}
	\begin{itemize}
		\item Manufacturing: ``After recalibration, defect rate returns to baseline and the line resumes normal throughput.''
		\item Building technology: ``After valve replacement, zone temperature stabilizes within setpoint band and energy consumption decreases.''
		\item Cybersecurity: ``After patching, a validation scan shows CVE-2021-44228 is no longer present and exposure risk is downgraded.''
	\end{itemize}
	
	\item \textbf{Examples}
	\begin{itemize}
		\item Manufacturing: ``If the operator cleans the lens as recommended, the inspection confidence is expected to increase and false rejects should drop.''
		\item Building technology: ``If the reheat valve is repaired, the system expects the overheating alert to clear and comfort complaints to decrease.''
		\item Cybersecurity: ``If the patch is applied and the service is restarted, the next scheduled scan is expected to report the vulnerability as resolved.''
	\end{itemize}
\end{itemize}

\subsubsection*{PrVC~2.7 What-if}
\begin{itemize}
	\item \textbf{Definition} A what-if element is a specified change to an input or the outcome itself, used to explore directional dependencies between inputs and outcomes. In a forward what-if analysis, inputs are varied to observe how the outcome changes. In a reverse what-if analysis, an alternative outcome is specified to explore which input changes would be required to reach that outcome.
	
	\item \textbf{Include}
	\begin{itemize}
		\item Parameter changes and hypothetical input values, for example ``set temperature to 21$^\circ$C,'' ``increase illumination by 15\%,'' or ``assume network exposure is removed''.
		\item Scenario toggles and constraint changes that the system treats as inputs for recomputation.
		\item Outputs that explicitly report how the outcome changes under the specified modification.
	\end{itemize}
	
	\item \textbf{Exclude}
	\begin{itemize}
		\item Explanations of why the current outcome occurred without varying anything.
		\item Minimal recourse-oriented changes required to achieve a specific alternative outcome.
		\item Comparisons between the actual outcome and a specific alternative outcome without a simulated change.
	\end{itemize}
	
	\item \textbf{Typical evidence}
	\begin{itemize}
		\item Manufacturing: Slider or input field that changes a process parameter and shows predicted defect probability.
		\item Building technology: Scenario tool that changes occupancy or setpoints and shows predicted energy use or comfort impact.
		\item Cybersecurity: Simulator that changes asset exposure, compensating controls, or patch availability and shows updated risk or remediation priority.
	\end{itemize}
	
	\item \textbf{Examples}
	\begin{itemize}
		\item Manufacturing: ``What if illumination is increased by 10\%? The predicted false reject rate decreases from 6\% to 2\%.''
		\item Building technology: ``What if the supply-air temperature setpoint is raised by 1$^\circ$C? Predicted heating energy decreases by 8\% with a small comfort penalty.''
		\item Cybersecurity: ``What if the internet-facing service is restricted to internal access? The risk score for the CVE drops and the remediation priority changes from urgent to high.''
	\end{itemize}
\end{itemize}

\subsubsection*{PrVC~3.1 Set of ranked elements with ranking mechanism (actual)}
\begin{itemize}
	\item \textbf{Definition} A set of ranked elements with ranking properties (actual) is an explanation element that presents multiple candidate elements as potential elements of the causal chain, that is, causal factor, outcome, or future state, together with the properties used to rank them and the ranking mechanism that determines the ordering. The top-ranked candidate, typically the one with the highest confidence or utility score, is selected for the causal chain.
	
	\item \textbf{Include}
	\begin{itemize}
		\item Ranked lists of alternative outcome candidates for the current case.
		\item Candidate-level ranking properties such as confidence, probability, utility, risk reduction, cost, or effort.
		\item A stated ranking mechanism or selection rule that explains why the top candidate is chosen.
	\end{itemize}
	
	\item \textbf{Exclude}
	\begin{itemize}
		\item Single outcomes without alternative candidates or ordering.
		\item Lists of inputs or observations used to produce the outcome.
		\item Lists of candidate causes or vulnerabilities rather than outcome candidates.
	\end{itemize}
	
	\item \textbf{Typical evidence}
	\begin{itemize}
		\item Manufacturing: Ranked corrective actions with confidence scores and an indication of the selected action.
		\item Building technology: Ranked fault resolutions with expected savings and feasibility scores and a selection rule.
		\item Cybersecurity: Ranked remediations with priority scores derived from severity and exposure.
	\end{itemize}
	
	\item \textbf{Examples}
	\begin{itemize}
		\item Manufacturing: ``(1) Recalibrate lighting 0.82, (2) Clean lens 0.74, (3) Replace camera 0.41'' and the system recommends option (1).
		\item Building technology: ``Top resolution selected because it has the highest weighted score from savings and feasibility.''
		\item Cybersecurity: ``Upgrade library selected because it has the highest risk-reduction priority score.''
	\end{itemize}
\end{itemize}

\subsubsection*{PrVC~3.2 Individual element with ranking properties, evidence, and/or attributes (actual)}
\begin{itemize}
	\item \textbf{Definition} An individual element with describing information. This information can include a ranking-property value, supporting evidence, and other attributes.
	
	\item \textbf{Include}
	\begin{itemize}
		\item The selected element for the current case, for example the chosen remediation, chosen action, or chosen finding, with its ranking-property value, such as confidence or priority.
		\item Evidence that supports selection, such as sensor readings, log excerpts, scan findings, similarity references, or rule matches.
		\item Attributes that characterize the selected element, such as severity, location, affected asset, time, scope, constraints, or expected impact.
	\end{itemize}
	
	\item \textbf{Exclude}
	\begin{itemize}
		\item Sets of multiple candidates shown as a ranked list.
		\item The overall system-outcome statement without any justification detail.
		\item The raw data stream or record used to compute the selection when it is not presented as justification.
	\end{itemize}
	
	\item \textbf{Typical evidence}
	\begin{itemize}
		\item Manufacturing: Selected root cause ``Camera lens contamination'' with confidence 0.78, supported by a blur metric and a recent maintenance record.
		\item Building technology: Selected fault ``Stuck valve'' with a priority score, supported by a valve-position trend and zone-temperature deviation.
		\item Cybersecurity: Selected remediation ``Upgrade OpenSSL'' with an urgency score, supported by CVE-match evidence and exposure status.
	\end{itemize}
	
	\item \textbf{Examples}
	\begin{itemize}
		\item Manufacturing: ``Selected action: Clean lens (confidence 0.74); evidence: increasing blur score; attributes: Station 3, Camera C2.''
		\item Building technology: ``Selected diagnosis: valve-actuator failure (priority 92); evidence: command issued but position unchanged; attributes: AHU-4, Zone 2B.''
		\item Cybersecurity: ``Selected remediation: patch within 48h (priority 95); evidence: CVE-2021-44228 detected in inventory; attributes: internet-facing service, high severity.''
	\end{itemize}
\end{itemize}

\subsubsection*{PrVC~4.1 Set of similar elements with filter mechanism (similar)}
\begin{itemize}
	\item \textbf{Definition} A set of similar elements with set properties presents multiple similar cases, for example historic cases that are similar to an actual element, PrVC~3.2 Individual element. It is used to support an actual-element selection.
	
	\item \textbf{Include}
	\begin{itemize}
		\item A collection of similar cases, assets, incidents, or instances explicitly linked to the current case through similarity.
		\item Set-level properties that summarize the group, such as number of similar cases, similarity range, outcome-consistency rate, or distribution of outcomes.
		\item Set-selection properties that define how the set was formed, such as top-$k$ neighbors, similarity threshold, time window, or filtering criteria.
	\end{itemize}
	
	\item \textbf{Exclude}
	\begin{itemize}
		\item One representative similar case selected and described in detail.
		\item A ranked list of outcome candidates for the current case.
		\item General confidence statements without an explicit set of similar cases.
	\end{itemize}
	
	\item \textbf{Typical evidence}
	\begin{itemize}
		\item Manufacturing: ``Top 10 similar incidents'' with average similarity score, proportion resolved by a specific action, and time-to-resolution statistics.
		\item Building technology: ``Similar zones'' set with count, common symptom profile, and percentage that required the same repair.
		\item Cybersecurity: ``Similar assets'' set with number of hosts affected by the same CVE, exposure categories, and remediation-success rate.
	\end{itemize}
	
	\item \textbf{Examples}
	\begin{itemize}
		\item Manufacturing: ``10 similar cases found (similarity 0.72--0.91); 8 of 10 were resolved by cleaning the lens; median recovery time 2 hours.''
		\item Building technology: ``7 similar fault episodes; 6 had the same root cause; 5 were resolved by actuator replacement within 24 hours.''
		\item Cybersecurity: ``12 similar hosts match the same CVE; 9 are internet-facing; patch-success rate in this set is 92\% based on prior rescans.''
	\end{itemize}
\end{itemize}

\subsubsection*{PrVC~4.2 Individual, similar element with filter properties, evidence, and/or attributes (similar)}
\begin{itemize}
	\item \textbf{Definition} An individual, similar element that describes one specific similar case, for example a historic case that is similar to an actual element, see PrVC~3.2. A similar element can be an individual element or it can be part of a similar-element set, see PrVC~4.1. It is used to support an actual-element selection.
	
	\item \textbf{Include}
	\begin{itemize}
		\item One referenced similar case or instance together with an element-level property, such as similarity score, distance, match strength, or outcome-consistency indicator.
		\item Evidence supporting similarity or outcome alignment, such as matching conditions, shared signals, log patterns, scan findings, or rule matches.
		\item Attributes that characterize similar elements, such as context, asset identifiers, timestamps, environment, severity, or operational conditions.
	\end{itemize}
	
	\item \textbf{Exclude}
	\begin{itemize}
		\item A group of multiple similar cases summarized with set-level statistics.
		\item A similar case shown without any similarity property, evidence, or attributes beyond a name or identifier.
		\item Candidate outcomes ranked for the current case rather than similar cases from the past.
	\end{itemize}
	
	\item \textbf{Typical evidence}
	\begin{itemize}
		\item Manufacturing: ``Similar incident ID 24-031'' with similarity 0.86, supported by a matching blur-metric trend and identical camera model.
		\item Building technology: ``Similar zone episode'' with a similarity score and evidence from overlapping temperature and valve-position patterns.
		\item Cybersecurity: ``Similar host finding'' with a match score, evidence of the same vulnerable package version, and the same exposure profile.
	\end{itemize}
	
	\item \textbf{Examples}
	\begin{itemize}
		\item Manufacturing: ``Similar case: Station 3, Camera C2, similarity 0.88; evidence: same blur spike and lighting conditions; attributes: occurred after the lens-cleaning interval was exceeded.''
		\item Building technology: ``Similar case: AHU-4 fault from last month, similarity 0.81; evidence: same command vs.\ position mismatch; attributes: same zone and outside-air temperature range.''
		\item Cybersecurity: ``Similar case: SRV-12, match 0.90; evidence: same package version flagged and same CVE signature; attributes: internet-facing service, high severity, patched successfully within 24h.''
	\end{itemize}
\end{itemize}

\clearpage


\section{Content-validity survey}
\label{app:Content-Validity-Round1}

For each pre-validation code, the definition, inclusion criteria, exclusion criteria, evidence indicators, and examples are displayed. Afterwards, the following questions were asked per pre-validation code:

\subsubsection*{Q1 How relevant is the pre-validation code?}

\textbf{Relevance} means that the pre-validation code is necessary for the explanation-content model. Removing it would leave the model incomplete with respect to its intended scope.

\begin{itemize}
	\item[$\square$] 1 - Not relevant
	\item[$\square$] 2 - Somewhat relevant, major revision needed
	\item[$\square$] 3 - Quite relevant, minor revision needed
	\item[$\square$] 4 - Very relevant
\end{itemize}

\vspace{0.5cm}

\subsubsection*{Q2 How clear are the boundaries to other pre-validation codes?}

\textbf{Boundary clarity} means that the inclusion and exclusion criteria define the pre-validation code's scope in a way that clearly separates it from neighboring pre-validation codes.

\begin{itemize}
	\item[$\square$] 1 - Not clear
	\item[$\square$] 2 - Somewhat clear, major revision needed
	\item[$\square$] 3 - Quite clear, minor revision needed
	\item[$\square$] 4 - Very clear
\end{itemize}

\vspace{0.5cm}

\subsubsection*{Q3 How understandable is the intended meaning from the label and definition?}

\textbf{Understandability} means that the intended meaning of a pre-validation code can be easily and correctly grasped from its label and definition, without needing additional explanation or interpretation.

\begin{itemize}
	\item[$\square$] 1 - Not understandable
	\item[$\square$] 2 - Somewhat understandable, major revision needed
	\item[$\square$] 3 - Quite understandable, minor revision needed
	\item[$\square$] 4 - Very understandable
\end{itemize}

\vspace{0.5cm}

\subsubsection*{Q4 What is the proposed action?}
\begin{itemize}
	\item[$\square$] Remove pre-validation code
	\item[$\square$] Add a new pre-validation code (add suggestion in comment area below)
	\item[$\square$] Move in hierarchy (specify new parents; add suggestions in comment area below)
	\item[$\square$] Merge with another pre-validation code (add suggestions in comment area below)
	\item[$\square$] Split into two or more codes (add suggestion in comment area below)
	\item[$\square$] Revise wording, e.g.\ label or definition (add suggestion in comment area below)
	\item[$\square$] Keep as is
\end{itemize}

\subsubsection*{Q5 Comments (free text)}

\clearpage


\section{Full comment matrix}
\label{app:FullCommentMatrix}

\begin{table}[ht]
	\scriptsize
	\setlength{\tabcolsep}{4pt}
	\renewcommand{\arraystretch}{1.2}
	\caption{Summarized expert comments per pre-validation code ($N=11$) (Part~1). Themes are paraphrased and clustered across participants. The support column reports how many experts mentioned the theme at least once.}
	\label{tab:FullCommentMatrixPart1}
	\begin{tabular}
		{>{\raggedright\arraybackslash}m{2.7cm}
			>{\centering\arraybackslash}m{1.5cm}
			>{\raggedright\arraybackslash}m{8.8cm}}
		
		\hline
		
		Pre-validation code & Support & Summarized comment themes (paraphrased) \\
		
		\hline
		
		PrVC~1.1 Rule attribute & 7/11 &
		(1) Clarify inclusion and exclusion so that explicit antecedent conditions are clearly separated from other information, especially when context variables trigger rules.
		(2) Add rule metadata to improve interpretability, such as rule origin, confidence, priority, and applicability scope.
		(3) Consider renaming the label to reduce confusion, for example \emph{Rule} instead of \emph{Rule attribute}.
		(4) Several experts perceived overlap between rules and causal explanation and requested clearer differentiation.
		(5) Some experts requested clearer separation between a repository of available rules and case-specific rule applicability. \\
		
		\hline
		
		PrVC~2.1 Context & 8/11 &
		(1) Boundary between context and input is ambiguous, especially when contextual variables are also used as model features.
		(2) Examples should be less open to interpretation and should distinguish background conditions from measured inputs more clearly.
		(3) Some experts suggested that context is foundational and should be integrated or more tightly connected to the rule-based or causal content.
		(4) Minor wording issues indicate that the label and inclusion and exclusion text could be made easier to follow without examples. \\
		
		\hline
		
		PrVC~2.2 Input & 7/11 &
		(1) Boundary with context should be clarified when the same variables can be interpreted as background or input.
		(2) Terminology should avoid confusion between input and evidence, since evidence is also used later in epistemic components.
		(3) Some experts requested clearer wording and examples, even though the core idea of raw data analyzed by the system was generally understood.
		(4) One expert noted that the level of detail may depend on task constraints and decision criteria. \\
		
		\hline
		
		PrVC~2.3 Causal factor & 6/11 &
		(1) The examples and the typical-evidence description were perceived as misaligned with the definition, with examples appearing to describe input rather than causal factors.
		(2) Clarify that causal factors may be inferred or attributed and not necessarily verified causal truth in data-driven systems.
		(3) Consider supporting plural causal factors and guidance on how multiple factors may be represented or quantified.
		(4) Some experts requested clearer differentiation from rule-based content and rule attributes.
		(5) Minor wording refinements were suggested to improve clarity of the definition. \\
		
		\hline
		
	\end{tabular}
\end{table}

\begin{table}[ht]
	\scriptsize
	\setlength{\tabcolsep}{4pt}
	\renewcommand{\arraystretch}{1.2}
	\caption{Summarized expert comments per pre-validation code ($N=11$) (Part~2). Themes are paraphrased and clustered across participants. The support column reports how many experts mentioned the theme at least once.}
	\label{tab:FullCommentMatrixPart2}
	\begin{tabular}
		{>{\raggedright\arraybackslash}m{2.7cm}
			>{\centering\arraybackslash}m{1.5cm}
			>{\raggedright\arraybackslash}m{8.8cm}}
		\hline
		
		Pre-validation code & Support & Summarized comment themes (paraphrased) \\
		
		\hline
		
		PrVC~2.4 Causal mechanism & 8/11 &
		(1) Clarify the exclusion criterion for simple factor statements by defining what counts as an acceptable minimal pathway versus no mechanism.
		(2) Specify whether the mechanism refers to domain-level mechanisms, system-internal computational processes, or both, particularly for data-driven models.
		(3) Provide an integrated example showing how PrVC~2.1 to PrVC~2.7 work together, since several experts perceived overlap between factor, mechanism, and outcome components.
		(4) Improve the figure-level representation, for example by treating the mechanism as a container for the chain from input to factor to outcome to future state.
		(5) Some experts suggested simplification or partial consolidation to make the framework more practical.
		(6) Some experts noted that the level of mechanism detail may depend on task goals and decision criteria. \\
		
		\hline
		
		PrVC~2.5 Outcome (actual) & 5/11 &
		(1) Generally seen as well defined and relevant.
		(2) Some experts requested guidance for cases with multiple AI components contributing to identification versus recommendation and how outcomes should be handled then.
		(3) Improve wording of outcome statements to be more direct and less conditional when appropriate.
		(4) Some experts suggested that exclusion statements may be too hard to make complete and should be minimized in favor of inclusion guidance. \\
		
		\hline
		
		PrVC~2.6 Future state & 5/11 &
		(1) Seen as useful to express expected remediation effect and contributes beyond outcome.
		(2) Some concern about relevance across all explanations, suggesting clearer scope conditions for when future state is required.
		(3) Some experts perceived placement issues within the causal group and requested tighter alignment with the causal mechanism and what-if content.
		(4) Some experts reported confusion between future state and ranked epistemic content, especially when future state is interpreted as a confidence estimate for remediation success.
		(5) Clarify whether anticipated side effects, tradeoffs, or required conditions belong in this content type. \\
		
		\hline
		
		PrVC~2.7 What-if & 6/11 &
		(1) Generally viewed as adding exploration of alternatives, but requires clearer integration with context and the causal chain.
		(2) Experts requested clearer coupling of what-if with the causal mechanism and better illustration in the framework figure.
		(3) Distinguish forward what-if from reverse what-if, since they support different reasoning goals such as hypothesis testing and goal seeking.
		(4) Some experts noted that the usefulness of what-if depends on decision criteria and constraints that shape the explored changes.
		(5) One expert suggested that the current label may still feel abstract and would benefit from more concrete examples. \\
		
		\hline
		
	\end{tabular}
\end{table}

\begin{table}[ht]
	\scriptsize
	\setlength{\tabcolsep}{4pt}
	\renewcommand{\arraystretch}{1.2}
	\caption{Summarized expert comments per pre-validation code ($N=11$) (Part~3). Themes are paraphrased and clustered across participants. The support column reports how many experts mentioned the theme at least once.}
	\label{tab:FullCommentMatrixPart3}
	\begin{tabular}
		{>{\raggedright\arraybackslash}m{2.7cm}
			>{\centering\arraybackslash}m{1.5cm}
			>{\raggedright\arraybackslash}m{8.8cm}}
		
		\hline
		
		Pre-validation code & Support & Summarized comment themes (paraphrased) \\
		
		\hline
		
		PrVC~3.1 Set of ranked elements with ranking mechanism (actual) & 6/11 &
		(1) Clarify how PrVC~3.1 differs from PrVC~3.2, especially whether ranking mechanism and alternative candidates are essential differentiators.
		(2) Some experts noted that many practical domains present only the top recommendation rather than an explicit ranked list, suggesting guidance on when ranked sets are optional.
		(3) Some experts suggested that the category may bundle multiple independent elements and could be split for usability.
		(4) Some experts reported confusion between ranked remediations and future state and questioned whether ranked sets are interpreted as case-level alternatives or as summaries over historical events.
		(5) One expert questioned whether ranked lists of outcomes might overlap with the outcome category and requested clearer differentiation. \\
		
		\hline
		
		PrVC~3.2 Individual element with ranking properties, evidence, and/or attributes (actual) & 6/11 &
		(1) Clarify distinction from PrVC~3.1.
		(2) Some experts suggested splitting the category because it currently mixes confidence, evidence, and other attributes.
		(3) Some experts noted that the value of the additional element-level detail depends on task goals and decision criteria.
		(4) One expert suggested that the epistemic items may need stronger connection to the causal explanation or clearer justification for being kept separate. \\
		
		\hline
		
		PrVC~4.1 Set of similar elements with filter mechanism (similar) & 6/11 &
		(1) Similarity and pattern matching were viewed as useful for justification, especially for patching decisions.
		(2) Some experts requested clearer guidance on what similarity means and whether users can explore similar cases.
		(3) Boundary with PrVC~4.2 should be clarified, and some experts questioned whether two epistemic groups are necessary or whether merging and restructuring would be more usable.
		(4) Some experts raised interpretability concerns for probability-like set summaries and noted that values such as 80\% require context to be meaningful.
		(5) One expert identified a minor wording issue in the label text. \\
		
		\hline
		
		PrVC~4.2 Individual, similar element with filter properties, evidence, and/or attribute (similar) & 6/11 &
		(1) Boundary with PrVC~4.1 should be clarified, particularly set versus individual value.
		(2) Some experts questioned whether separating set and individual is necessary for usability, even if it improves completeness of the taxonomy.
		(3) Minor wording and labeling suggestions were raised for clarity, including avoiding repeated use of the word set in the label.
		(4) Some experts asked for clearer guidance on how similarity is assessed and what dimensions define similarity. \\
		
		\hline
		
	\end{tabular}
\end{table}

\clearpage


\section{Pre-registration summary (OSF, registered 30 March 2026)}
\label{app:Preregistration}

This appendix summarizes the content of the OSF pre-registration (https://osf.io/up2sh) for the reliability evaluation reported in Step~9 and Step~10. The pre-registration was completed on 30 March 2026, prior to any independent coding activity. Independent coding was conducted on 3 April 2026. The following procedural elements were pre-specified.

\textit{Codebook version.} The frozen post-validation codebook, in which all codes carry the prefix PoVC, constituted the coding instrument for the reliability evaluation. No changes to code definitions, decision rules, boundaries, or examples were permitted after freezing.

\textit{Subsample composition.} The reliability subsample comprised 82 meaning units (25.2\% of the corpus), selected using stratified purposive sampling to ensure that all post-validation codes with corpus assignments were represented by at least two units, that borderline cases at three pre-identified code boundaries were included, and that dual-valid cases for the PoVC~2.2 and PoVC~2.7 pair were included in sufficient number to constitute a demanding reproducibility test.

\textit{Equivalence rule.} Assignments to PoVC~2.2 Input and PoVC~2.7 What-if forward by different coders on the same unit were pre-specified as equivalent and not counted as disagreements, because both assignments constitute correct descriptions of the same class of meaning units. No other code pairs were treated as equivalent.

\textit{Acceptance criteria.} Krippendorff's $\alpha \geq 0.800$ was pre-specified as the threshold for reliable data for drawing conclusions. Cohen's $\kappa$ was pre-specified as a supplementary statistic with no separate acceptance threshold.

\textit{Stopping rule.} If the acceptance criterion was not met, code definitions, decision rules, and examples were to be revised, affected corpus material recoded, and reliability testing repeated on a newly drawn subsample using the updated frozen codebook. The procedure was to continue until the acceptance criterion was met or until the codebook could not be further revised without structural changes to the category system.

\clearpage


\section{Change log from pre-content validation to post-content validation code}
\label{app:ChangeLogPrevalidationToPostvalidation}

\begin{table}[!ht]
	\scriptsize
	\caption{Change log from pre-validation to post-validation codes, part 1}
	\label{tab:Result:ChangeLogPrevalidationToPostvaliationPart1}
	\begin{scriptsize}
		\begin{tabular}
			{>{\raggedright\arraybackslash}m{4.0cm}
				>{\raggedright\arraybackslash}m{3.0cm}
				>{\raggedright\arraybackslash}m{3.0cm}
				>{\raggedright\arraybackslash}m{4.0cm}}
			\hline
			
			Pre-validation code & Trigger & Change rationale & Post-validation code \\
			
			\hline
			
			PrVC~1.1 Rule attribute &
			Some experts considered that rules should be part of the causal chain. They did not distinguish between the rule as a decision mechanism and the concluding element as part of the causal chain &
			The rule attribute needed to be separated into a set of available rules and an applied rule for the actual case; to separate between applied rules and causal elements, a decision rule was introduced. &
			PoVC~1.1 Rule base, PoVC~1.2 Applied rule \\
			
			\hline
			
			PrVC~2.1 Context &
			Some experts requested that the distinction between the context and the input is not clear enough. They correctly commented that aspects of the context could also be an input. &
			Both definitions (context and input) were revised and a decision rule was introduced to better distinguish between context and input. &
			PoVC~2.1 Context \\
			
			PrVC~2.2 Input &
			Some experts requested to improve the distinction between the context and the input. They correctly commented that aspects of the context could also be an input.  &
			Both definitions (context and input) were revised and a decision rule was introduced to better distinguish between context and input. &
			PoVC~2.2 Input \\
			
			PrVC~2.3 Causal factor &
			Some experts requested to improve the distinction between causal factor and rule-based explanations. They outlined that a rule should be part of the causal chain. A distinction is needed between rules and elements of the causal chain. Additionally, some experts noted that the provided examples appeared to describe input rather than causal factors, indicating a misalignment between the definition and the examples.  &
			The three definitions (applied rule, causal element, outcome) were revised and decision rules were introduced to better distinguish between rules, causal factors, and outcomes. &
			PoVC~2.3 Causal factor \\

			\hline
			
		\end{tabular}
	\end{scriptsize}
\end{table}

\begin{table}[!ht]
	\scriptsize
	\caption{Change log from pre-validation to post-validation codes, part 2}
	\label{tab:Result:ChangeLogPrevalidationToPostvaliationPart2}
	\begin{scriptsize}
		\begin{tabular}
			{>{\raggedright\arraybackslash}m{4.0cm}
				>{\raggedright\arraybackslash}m{3.0cm}
				>{\raggedright\arraybackslash}m{3.0cm}
				>{\raggedright\arraybackslash}m{4.0cm}}
			\hline
			
			Pre-validation code & Trigger & Change rationale & Post-validation code \\
			
			\hline
			
			PrVC~2.4 Outcome &
			Some experts requested a better distinction between the outcome and the future state. &
			The definitions of outcome and future state were revised; a decision rule was introduced to clarify the distinction between outcome and future state. &
			PoVC~2.4 Outcome \\
			
			PrVC~2.5 Future state &
			Some experts requested clarification when future states are applicable, and what the difference is between outcome and future states. &
			The definitions of outcome and future state were revised; a decision rule was introduced to clarify the distinction between outcome and future state. &
			PoVC~2.5 Future state \\
			
			PrVC~2.6 Causal mechanism &
			Some experts requested clarification to identify a causal mechanism and to which domains the causal mechanism belongs. Some experts requested clearer distinction between causal mechanism and other elements of the causal chain. &
			The definition of causal mechanisms was revised and a decision rule between causal mechanism and causal factor was introduced. &
			PoVC~2.6 Causal mechanism \\
			
			PrVC~2.7 What-if &
			Some experts provided comments to separate whether what-if applies to the input, assessing the impact on the outcome, or on the outcome, assessing the impact on the input.  &
			What-if was separated into What-if forward (modifying the input to assess the impact on the outcome) and What-if backward (modifying the outcome to assess the required change in the input); the split was independently confirmed by the Step~7 coverage assessment, which identified that PrVC~2.7 spanned both causal and counterfactual explanation types. &
			PoVC~2.7 What-if forward, PoVC~2.8 What-if backward \\
			
			\hline
			
		\end{tabular}
	\end{scriptsize}
\end{table}

\begin{table}[!ht]
	\scriptsize
	\caption{Change log from pre-validation to post-validation codes, part 3}
	\label{tab:Result:ChangeLogPrevalidationToPostvaliationPart3}
	\begin{scriptsize}
		\begin{tabular}
			{>{\raggedright\arraybackslash}m{4.0cm}
				>{\raggedright\arraybackslash}m{3.0cm}
				>{\raggedright\arraybackslash}m{3.0cm}
				>{\raggedright\arraybackslash}m{4.0cm}}
			\hline
			
			Pre-validation code & Trigger & Change rationale & Post-validation code \\
			
			\hline
			
			PrVC~3.1 Set of ranked elements with ranking mechanism (actual) &
			Some experts requested a better distinction between PrVC~3.1 Set of ranked elements and PrVC~3.2 Individual elements &
			The definition was revised and decision rules between PrVC~3.1 and PrVC~3.2 were introduced. &
			PoVC~3.1 Set of ranked elements with ranking mechanism (actual) \\
			
			PrVC~3.2 Individual element with ranking properties, evidence, and/or attributes (actual) &
			Some experts requested a better distinction between PrVC~3.1 Set of ranked elements and PrVC~3.2 Individual elements &
			The definition was revised and decision rules between PrVC~3.1 and PrVC~3.2 were introduced. &
			PoVC~3.2 Individual element with ranking properties, evidence, and/or attributes (actual) \\
			
			\hline
			
			PrVC~4.1 Set of similar elements with filter mechanism (similar) &
			Some experts requested clarification of what similarity means; some experts requested clarification between PrVC~4.1 and PrVC~4.2. Additionally, some experts raised interpretability concerns about probability-like set summaries, noting that values such as 80\% require context to be meaningful. &
			The name of the code and the definitions were revised; decision rules between PrVC~4.1 and PrVC~4.2 were introduced. &
			PoVC~4.1 Set of similar elements with similarity mechanism and set success-score information (similar) \\
			
			PrVC~4.2 Individual, similar element with filter properties, evidence, and/or attributes (similar) &
			Some experts requested clarification of what similarity means; some experts requested clarification between PrVC~4.1 and PrVC~4.2. Additionally, some experts raised interpretability concerns about probability-like set summaries, noting that values such as 80\% require context to be meaningful. &
			The name of the code and the definitions were revised; decision rules between PrVC~4.1 and PrVC~4.2 were introduced. &
			PoVC~4.2 Individual, similar element with similarity property, success score, evidence, and/or attribute (similar) \\
			
			\hline
			
		\end{tabular}
	\end{scriptsize}
\end{table}

\clearpage


\section{Codebook (post-expert review)}
\label{app:PostExpertReviewCodebook}


\subsection*{Explanation categories}

\subsubsection*{C~1 Rule-based explanation}
\begin{itemize}
	\item \textbf{Definition} A rule-based explanation is an explanation that justifies a case-specific element of the explanation by referring to documented decision rules. It distinguishes the \emph{rule base} as a set of available rules that encode general knowledge from the \emph{applied rule} set that captures which rules were evaluated and used in the current case. The explanation communicates decision conditions in a symbolic, human-readable form such as thresholds and Boolean combinations and reports the case-specific evaluation of these conditions as met, not met, or not decidable based on the available inputs and context. Applied rules can support or constrain elements of the causal explanation group, most commonly the outcome recommendation, ranked candidates, or admissibility constraints, while remaining outside the causal chain.
	
	\item \textbf{Include}
	\begin{itemize}
		\item A documented \emph{rule base} that provides uniquely identifiable rules with traceable provenance and scope.
		\item A case-specific \emph{applied rule} that reports which rule was evaluated for the case and how its conditions were assessed based on the available inputs and context.
		\item Explicit decision conditions expressed as thresholds, predicates, ranges, membership checks, Boolean logic, scoring rules, decision tables, or ordered rule lists with precedence.
		\item Condition-evaluation results that indicate which predicates were met, not met, or not decidable and which input evidence was used for the evaluation.
		\item Rule-based policies or compliance checks when they are presented as decision logic that supports or constrains a case-specific element such as an outcome recommendation or a ranked-candidate set.
	\end{itemize}
	
	\item \textbf{Exclude}
	\begin{itemize}
		\item Case-specific causal attributions that treat a rule as a cause without a separate causal-factor and mechanism description.
		\item Post-hoc feature-attribution lists that rank features by importance without expressing explicit decision conditions.
		\item Example-based explanations that justify an element by citing similar cases rather than by reporting an evaluated rule.
		\item Causal narratives that describe mechanisms or causal chains without specifying decision conditions and their evaluation.
		\item Procedural descriptions of training data or model development without reporting the rule(s) evaluated in the current case.
		\item Uninterpretable model internals such as weights unless translated into explicit decision rules.
	\end{itemize}
	
	\item \textbf{Typical evidence}
	\begin{itemize}
		\item Manufacturing rule base: ``If vibration RMS exceeds 7~mm/s for longer than 10~minutes and temperature exceeds 80$^\circ$C, then predict bearing fault.''
		\item Building technology rule base: ``If CO$_2$ $>$ 1000~ppm and damper position $<$ 20\% during occupancy, then raise indoor-air-quality alert.''
		\item Cybersecurity rule base: ``If software version matches a known vulnerable pattern and the asset is internet-facing, then flag the vulnerability as critical and recommend patching.''
	\end{itemize}
	
	\item \textbf{Examples}
	\begin{itemize}
		\item Manufacturing applied rule: ``Rule R-MFG-12 was evaluated and applied. Vibration RMS $>$ 7~mm/s for 10~minutes is met (8.2~mm/s). Temperature $>$ 80$^\circ$C is met (83$^\circ$C). The rule supports the predicted bearing-fault outcome and the recommendation to inspect the bearing.''
		\item Building technology applied rule: ``Rule R-BLD-07 was evaluated and applied. CO$_2$ $>$ 1000~ppm is met (1120~ppm). Damper position $<$ 20\% during occupancy is met (12\%). The rule supports raising an indoor-air-quality alert.''
		\item Cybersecurity applied rule: ``Rule R-CYB-21 was evaluated and applied. Log4Shell (CVE-2021-44228) is detected in inventory, which is met. Internet-facing exposure is met. The rule supports assigning criticality and the recommendation to patch within 48 hours.''
	\end{itemize}
\end{itemize}

\subsubsection*{C~2 Causal explanation}
\begin{itemize}
	\item \textbf{Definition} A causal explanation is an explanation that justifies an outcome by stating one or more causes and describing how those causes produced the outcome. It expresses a cause-to-effect relationship that goes beyond correlation, linking initiating conditions and intermediate factors or mechanisms to the observed outcome in a way that supports understanding, diagnosis, or intervention.
	
	\item \textbf{Include}
	\begin{itemize}
		\item Statements that identify one or more causal factors or drivers of the outcome, for example root cause or contributing cause.
		\item Explanations that describe a causal mechanism or causal chain connecting factors to the outcome, for example intermediate steps or propagation.
		\item Explanations that indicate directionality and causal influence, for example led to, resulted in, due to, because, including multi-cause accounts.
		\item Causal accounts that connect evidence to an outcome to support actions such as troubleshooting, mitigation, or prevention.
	\end{itemize}
	
	\item \textbf{Exclude}
	\begin{itemize}
		\item Pure statistical-association or correlation statements without a causal claim.
		\item Rule-based decision-logic explanations that only state decision conditions without asserting real-world causation.
		\item Counterfactual or recourse explanations that focus on how inputs must change to obtain a different outcome without explaining why the actual outcome occurred.
		\item Example-based explanations that justify an outcome by citing similar cases rather than causal reasoning.
		\item Descriptions of the data pipeline or model-training process that do not explain why this specific outcome occurred.
	\end{itemize}
	
	\item \textbf{Typical evidence}
	\begin{itemize}
		\item Manufacturing: Failure-analysis narrative linking component wear to process deviation and defect formation, supported by sensor trends.
		\item Building technology: Fault-diagnosis narrative linking actuator malfunction to airflow changes and comfort violations, supported by control logs.
		\item Cybersecurity: Attack-analysis narrative linking initial compromise to privilege escalation and impact, supported by authentication logs and alerts.
	\end{itemize}
	
	\item \textbf{Examples}
	\begin{itemize}
		\item Manufacturing: ``Nozzle wear reduced fill precision, which caused spills that contaminated the inspection optics, leading to increased false rejects.''
		\item Building technology: ``A stuck damper reduced outside-air intake, which raised CO$_2$ levels during occupancy and triggered the air-quality alert.''
		\item Cybersecurity: ``A phishing email led to credential theft, enabling unauthorized VPN access and lateral movement, which resulted in ransomware deployment and service disruption.''
	\end{itemize}
\end{itemize}

\subsubsection*{C~3 Epistemic explanation (actual)}
\begin{itemize}
	\item \textbf{Definition} An epistemic explanation (actual) is an explanation that justifies the actual outcome by communicating the system's uncertainty, confidence, or evidence strength regarding that outcome. It characterizes how strongly the system believes the actual outcome holds, based on probabilistic reasoning, likelihood, prediction confidence, posterior belief, or related uncertainty measures, and may indicate how reliable the outcome is expected to be.
	
	\item \textbf{Include}
	\begin{itemize}
		\item Confidence, probability, or likelihood statements about the actual outcome.
		\item Uncertainty information such as prediction intervals, credible intervals, error bars, entropy, margin to decision boundary, or ambiguity indicators tied to the actual outcome.
		\item Evidence-strength or reliability cues for the actual outcome, when presented as a justification for the outcome.
		\item References to alternative outcomes only to contextualize uncertainty, while keeping the actual outcome as the explained result.
	\end{itemize}
	
	\item \textbf{Exclude}
	\begin{itemize}
		\item Explanations that describe why the outcome occurred in causal terms.
		\item Rule-based or procedural explanations that state decision conditions without uncertainty information.
		\item Contrastive explanations whose primary purpose is to compare the actual outcome with a specific alternative outcome.
		\item Counterfactual or recourse explanations that describe what would need to change to obtain a different outcome.
		\item Pure data-quality or provenance statements unless explicitly linked to confidence in the actual outcome.
	\end{itemize}
	
	\item \textbf{Typical evidence}
	\begin{itemize}
		\item Manufacturing: Predicted defect probability with a confidence score and a note about uncertainty due to low image quality.
		\item Building technology: Fault-likelihood score with a confidence band and an ambiguity warning due to missing sensor readings.
		\item Cybersecurity: Risk score for an identified vulnerability with probability of exploitation and a confidence level based on observed exposure and asset telemetry.
	\end{itemize}
	
	\item \textbf{Examples}
	\begin{itemize}
		\item Manufacturing: ``The system predicts a bearing fault with 0.87 probability; uncertainty is moderate because recent vibration data contains gaps.''
		\item Building technology: ``The economizer fault is detected with high confidence (confidence = 0.91), supported by consistent deviations across three sensors.''
		\item Cybersecurity: ``The system flags CVE-2021-44228 as present with 0.95 confidence; confidence is lower for exploitability because external exposure could not be fully confirmed.''
	\end{itemize}
\end{itemize}

\subsubsection*{C~4 Epistemic explanation (similar)}
\begin{itemize}
	\item \textbf{Definition} An epistemic explanation (similar) is an explanation that justifies the current outcome by referencing historic cases that are similar to the current case and using those similarities as evidence for an element in the causal chain. It communicates the system's belief or confidence in the outcome by showing that comparable prior instances produced the same or highly similar outcomes, often including similarity scores, match criteria, and how frequently the outcome occurred among the retrieved cases.
	
	\item \textbf{Include}
	\begin{itemize}
		\item References to one or more historic cases deemed similar to the current case, used as evidence for the current outcome.
		\item Similarity information such as nearest neighbors, match scores, distances, feature overlap, or retrieved-case sets, when tied to confidence in the current outcome.
		\item Frequency or distribution summaries over similar cases.
		\item Case-based or example-based probability statements derived from the similar-case set.
	\end{itemize}
	
	\item \textbf{Exclude}
	\begin{itemize}
		\item Purely illustrative examples shown for explanation without being used as evidence for the current outcome.
		\item Counterfactual examples that show how the outcome would change under modified inputs.
		\item Contrastive examples whose main purpose is to compare the current case to a foil case with a different outcome.
		\item Causal narratives explaining why the outcome occurred, if they do not rely on similar cases as evidence.
		\item Rule-based decision conditions that do not involve retrieval of similar historic cases.
	\end{itemize}
	
	\item \textbf{Typical evidence}
	\begin{itemize}
		\item Manufacturing: Retrieval of past production runs with similar vibration signatures, showing that the same fault pattern preceded bearing failure.
		\item Building technology: Similar past days or zones with comparable temperature and valve behavior, where the same fault label was confirmed.
		\item Cybersecurity: Prior incidents with similar alert sequences or vulnerability configurations, showing how often they led to exploitation or high-risk classification.
	\end{itemize}
	
	\item \textbf{Examples}
	\begin{itemize}
		\item Manufacturing: ``The system predicts bearing wear because the current spectrum matches 9 prior cases (similarity $\geq$ 0.92), and 8 of those cases were confirmed as outer-race wear.''
		\item Building technology: ``This condition is classified as an economizer fault because 12 similar historical episodes showed the same damper and temperature pattern, and 10 were resolved by damper repair.''
		\item Cybersecurity: ``The alert is rated high risk because it matches recent incident patterns in 6 similar cases (same CVE, exposure, and log sequence), and 5 of those cases resulted in confirmed compromise.''
	\end{itemize}
\end{itemize}


\subsection*{Explanation content codes (post-content validation)}

\subsubsection*{Shared examples}

To demonstrate dependencies between explanation-content items, we use the following four shared domain examples consistently across the initial-code definitions. Note that the manufacturing, building-technology, and cybersecurity examples are drawn from the six corpus studies. The medical-device example (infusion pump, ICU setting) is a constructed illustration intended to broaden applicability and was not part of the corpus used to develop or validate the model.

\paragraph{Manufacturing}

A vision-based inspection system flags a product as defective on [\emph{PoVC~2.1 Context}: Press~\#4 in Plant~2, during start-up after a tooling change, operator Maria Chen, line operator, ext.~4582]. The system uses [\emph{PoVC~2.2 Input}: camera image, blur score, illumination value, edge-detection confidence, and maintenance record]. It identifies [\emph{PoVC~2.3 Causal factor}: lens contamination]. The explanation states [\emph{PoVC~2.6 Causal mechanism}: lens contamination reduces image contrast, which causes the edge detector to miss boundaries and increases false defect classifications]. The system produces [\emph{PoVC~2.4 Outcome}: product flagged as defective; manual inspection recommended]. It also states [\emph{PoVC~2.5 Future state}: if the lens is cleaned, false rejects are expected to decrease and inspection confidence is expected to improve]. A hypothetical variation is [\emph{PoVC~2.7 What-if forward}: if illumination is increased by 10\%, the predicted false reject rate decreases from 6\% to 2\%]. The available repository of rules is [\emph{PoVC~1.1 Rule base}: Rule~1: IF blur score exceeds threshold AND illumination is below lower bound THEN likely causal factor = lens contamination; Rule~2: IF vibration remains normal but current draw rises sharply THEN likely causal factor = motor overload]. The case-specific rule evaluation is [\emph{PoVC~1.2 Applied rule}: IF blur score exceeds threshold AND illumination is below lower bound THEN likely causal factor = lens contamination; condition status: met]. The ranked candidates are [\emph{PoVC~3.1 Set of ranked elements with ranking mechanism (actual)}: (1) Clean lens, confidence = 0.82; (2) Recalibrate lighting, confidence = 0.74; (3) Replace camera, confidence = 0.41]. One candidate element is [\emph{PoVC~3.2 Individual element with ranking properties, evidence, and/or attributes (actual)}: action = Clean lens; confidence = 0.82; evidence = rising blur score; attributes = Camera~C2, Station~3]. Similar past cases are [\emph{PoVC~4.1 Set of similar elements with similarity mechanism and set success-score information (similar)}: 10 similar cases selected by a similarity mechanism with a similarity threshold of 0.72; with respect to the selected actual element action = Clean lens, 8 cases are successful, 1 case is inconclusive, and 1 case is not successful]. One similar case is [\emph{PoVC~4.2 Individual, similar element with similarity property, success score, evidence, and/or attribute (similar)}: Station~3, Camera~C2; similarity score = 0.88; success score = successful relative to the selected actual element action = Clean lens; evidence = same blur spike and lighting conditions; attribute = same production station after a tooling change].

\paragraph{Building-technology}
A building-management system reports an overheating fault in [\emph{PoVC~2.1 Context}: Zone~3B, at 08:15, during pre-occupancy warm-up on a cold day, facility engineer David Kumar, ext.~2147]. The system uses [\emph{PoVC~2.2 Input}: zone temperature, supply-air temperature, valve command, valve position, and occupancy schedule]. It identifies [\emph{PoVC~2.3 Causal factor}: stuck reheat valve]. The explanation states [\emph{PoVC~2.6 Causal mechanism}: if the reheat valve remains open, heat continues to enter the zone despite the setpoint, which raises zone temperature and triggers an overheating alert]. The system produces [\emph{PoVC~2.4 Outcome}: overheating fault reported; valve inspection recommended]. It also states [\emph{PoVC~2.5 Future state}: if the valve is repaired, the overheating alert is expected to clear and comfort complaints are expected to decrease]. A hypothetical variation is [\emph{PoVC~2.7 What-if forward}: if the supply-air temperature setpoint is lowered by 1$^\circ$C, predicted overheating risk decreases]. The available repository of rules is [\emph{PoVC~1.1 Rule base}: Rule~A: IF zone temperature exceeds setpoint by more than 3$^\circ$C AND valve command remains at 100\% THEN likely causal factor = stuck reheat valve; Rule~B: IF zone temperature remains elevated while valve position does not follow the command signal THEN likely causal factor = actuator nonresponse]. The case-specific rule evaluation is [\emph{PoVC~1.2 Applied rule}: IF zone temperature exceeds setpoint by more than 3$^\circ$C AND valve command remains at 100\% THEN likely causal factor = stuck reheat valve; condition status: met]. The ranked candidates are [\emph{PoVC~3.1 Set of ranked elements with ranking mechanism (actual)}: (1) Inspect valve actuator, confidence = 0.86; (2) Recalibrate temperature sensor, confidence = 0.52; (3) Adjust occupancy schedule, confidence = 0.21]. One candidate element is [\emph{PoVC~3.2 Individual element with ranking properties, evidence, and/or attributes (actual)}: action = Inspect valve actuator; confidence = 0.86; evidence = command issued but valve position unchanged; attributes = AHU~4, Zone~3B]. Similar past cases are [\emph{PoVC~4.1 Set of similar elements with similarity mechanism and set success-score information (similar)}: 7 similar overheating episodes selected by a similarity mechanism with a defined similarity threshold; with respect to the selected actual element action = Inspect valve actuator, 5 cases are successful, 1 case is inconclusive, and 1 case is not successful]. One similar case is [\emph{PoVC~4.2 Individual, similar element with similarity property, success score, evidence, and/or attribute (similar)}: AHU~4 fault from last month; similarity score = 0.81; success score = successful relative to the selected actual element action = Inspect valve actuator; evidence = same command-versus-position mismatch; attribute = same zone and similar outside-air conditions].

\paragraph{Cybersecurity}
A vulnerability-management system detects CVE-2021-44228 on [\emph{PoVC~2.1 Context}: an internet-facing production server, during a scheduled weekly scan, high-criticality asset, change-freeze period, analyst Priya Lopez, cybersecurity analyst, ext.~7719]. The system uses [\emph{PoVC~2.2 Input}: software inventory, detected \texttt{log4j-core} version, exposure flag, scan findings, and configuration snapshot]. It identifies [\emph{PoVC~2.3 Causal factor}: presence of CVE-2021-44228 on an internet-facing asset]. The explanation states [\emph{PoVC~2.6 Causal mechanism}: a vulnerable logging library can be exploited through crafted input that triggers message lookups, enabling remote code execution and increasing compromise risk]. The system produces [\emph{PoVC~2.4 Outcome}: high-priority remediation recommendation to patch within 48 hours]. It also states [\emph{PoVC~2.5 Future state}: if the patch is applied and the service is restarted, the next validation scan is expected to report the vulnerability as resolved]. A hypothetical variation is [\emph{PoVC~2.7 What-if forward}: if internet exposure is removed, the risk score drops from urgent to high]. The available repository of rules is [\emph{PoVC~1.1 Rule base}: Rule~X: IF \texttt{log4j-core} version 2.0--2.14.1 is present AND the service is internet-facing THEN recommended outcome = patch or mitigate CVE-2021-44228; Rule~Y: IF exploitation indicators are present in logs AND the affected service is internet-facing THEN recommended outcome = immediate remediation]. The case-specific rule evaluation is [\emph{PoVC~1.2 Applied rule}: IF \texttt{log4j-core} version 2.0--2.14.1 is present AND the service is internet-facing THEN recommended outcome = patch or mitigate CVE-2021-44228; condition status: met]. The ranked candidates are [\emph{PoVC~3.1 Set of ranked elements with ranking mechanism (actual)}: (1) Patch immediately, confidence = 0.95; (2) Isolate service, confidence = 0.78; (3) Monitor only, confidence = 0.18]. One candidate element is [\emph{PoVC~3.2 Individual element with ranking properties, evidence, and/or attributes (actual)}: action = Patch within 48 hours; confidence = 0.95; evidence = vulnerable version detected on an internet-facing high-criticality asset; attributes = Server~A, production environment]. Similar past cases are [\emph{PoVC~4.1 Set of similar elements with similarity mechanism and set success-score information (similar)}: 12 similar hosts selected by a similarity mechanism with a defined similarity threshold based on the same CVE and exposure profile; with respect to the selected actual element action = Patch within 48 hours, 9 cases are successful, 2 cases are inconclusive, and 1 case is not successful]. One similar case is [\emph{PoVC~4.2 Individual, similar element with similarity property, success score, evidence, and/or attribute (similar)}: SRV-12; similarity score = 0.90; success score = successful relative to the selected actual element action = Patch within 48 hours; evidence = same package version and same exposure profile; attribute = same server role and criticality level].


\subsubsection*{PoVC~1.1 Rule base}

\begin{itemize}
	\item \textbf{Definition} A rule base is a repository or set of available rules that can be used to assess a case. Each rule consists of a condition part and a conclusion part, typically expressed as IF <condition(s)> THEN <conclusion>. The condition part specifies when the rule applies. The conclusion part typically refers to one or more elements of the causal explanation group, such as a causal factor, an outcome, or a future state. The rule base represents the available set of rules independent of whether any individual rule is confirmed for the current case.
	
	\item \textbf{Decision rule (PoVC~1.1 vs.\ PoVC~1.2):} If the explanation presents a general rule, decision logic, threshold condition, policy, or reusable if-then structure without tying it to the current case as satisfied, violated, or triggered, code it as \emph{PoVC~1.1 Rule base}. If the explanation instead shows how such a rule applies to the current case, for example by stating that its conditions are met, not met, or partially met and that this leads to the current result, code it as \emph{PoVC~1.2 Applied rule}.
	
	\item \textbf{Include}
	\begin{itemize}
		\item Sets, repositories, libraries, catalogs, or collections of explicit if-then rules, decision rules, thresholds, Boolean combinations, scoring rules, or rule lists that are available for case assessment.
		\item Sets of rules whose condition parts reference measurable values, states, events, asset properties, or categorical properties.
		\item Rule-base metadata, if shown, such as rule origin, priority, confidence, scope, policy source, version, or number of available rules.
	\end{itemize}
	
	\item \textbf{Exclude}
	\begin{itemize}
		\item A single rule shown in isolation rather than as part of a rule set or repository (if evaluated for the current case, it belongs to \emph{PoVC~1.2 Applied rule}).
		\item A case-specific statement that a rule has been confirmed as applying in the current case (this belongs to \emph{PoVC~1.2 Applied rule}).
		\item Pure outcome statements without an explicit condition part (these belong to \emph{PoVC~2.4 Outcome}).
		\item Probabilistic statements, feature weights, or scores that do not define a testable rule condition (these belong to \emph{PoVC~3.2 Individual element with ranking properties, evidence, and/or attributes (actual)} when presented for the current case).
		\item General guidance or policies that are not expressed as operational conditions that can be checked against a case (depending on their role, they may be coded as \emph{PoVC~2.1 Context} or remain uncoded).
	\end{itemize}
	
	\item \textbf{Typical evidence}
	\begin{itemize}
		\item A repository, library, catalog, or list that contains multiple available rules for case assessment.
		\item Repeated if-then structures or other explicit rule formulations shown as a set rather than as a single case-specific statement.
		\item Rule-base metadata such as rule identifiers, source, version, scope, priority, confidence, or grouping by domain topic.
		\item Grouping or navigation structures that organize available rules by asset type, fault class, vulnerability class, operating mode, or policy domain.
	\end{itemize}
	
	\item \textbf{Examples}
	\begin{itemize}
		\item Manufacturing: [Context] A vision-based inspection system flags a product as defective on Press~\#4 in Plant~2 during start-up after a tooling change. [Rule base] Rule~1: IF blur score exceeds threshold AND illumination is below the lower bound THEN [Causal factor] lens contamination. Rule~2: IF vibration remains normal but current draw rises sharply THEN [Causal factor] motor overload.
		
		\item Building technology: [Context] A building-management system reports an overheating fault in Zone~3B at 08:15 during pre-occupancy warm-up on a cold day. [Rule base] Rule~A: IF zone temperature exceeds setpoint by more than 3$^\circ$C AND valve command remains at 100\% THEN [Causal factor] stuck reheat valve. Rule~B: IF zone temperature remains elevated while valve position does not follow the command signal THEN [Causal factor] actuator nonresponse.
		
		\item Cybersecurity: [Context] A vulnerability-management system detects CVE-2021-44228 on an internet-facing production server during a scheduled weekly scan. [Rule base] Rule~X: IF \texttt{log4j-core} version 2.0--2.14.1 is present AND the service is internet-facing THEN [Outcome] patch or mitigate CVE-2021-44228. Rule~Y: IF exploitation indicators are present in logs AND the affected service is internet-facing THEN [Outcome] immediate remediation.
		
	\end{itemize}
\end{itemize}

\subsubsection*{PoVC~1.2 Applied rule}

\begin{itemize}
	\item \textbf{Definition} An applied rule is a rule from the rule base whose applicability is evaluated for the current case and whose conclusion is relevant for interpreting the case. Each rule consists of a condition part and a conclusion part, typically expressed as IF <condition(s)> THEN <conclusion>. For an applied rule, the condition part is evaluated as either \emph{met} or \emph{not decidable}. A status of \emph{not decidable} indicates unresolved applicability because required case evidence is missing. The conclusion part typically supports one or more elements of the causal explanation group, such as a causal factor, an outcome, or a future state. An applied rule supports case interpretation, but it is not itself the causal chain.
	
	\item \textbf{Decision rule (PoVC~1.2 vs.\ PoVC~1.1):} If the explanation states that a rule is satisfied, violated, triggered, or otherwise instantiated in the current case, code it as \emph{PoVC~1.2 Applied rule}. If the explanation only presents the reusable rule itself, including its general conditions, thresholds, or logic, without linking it to the current case, code it as \emph{PoVC~1.1 Rule base}.
	
	\item \textbf{Decision rule (PoVC~1.2 vs.\ PoVC~2.3 or PoVC~2.4):}
	If the explanation states that a rule is satisfied, violated, triggered, or otherwise instantiated in the current case, code it as \emph{PoVC~1.2 Applied rule}. An applied rule may motivate or justify a causal factor (\emph{PoVC~2.3}) or an outcome (\emph{PoVC~2.4}) within the causal chain, but the rule statement itself and the causal or outcome content it supports are distinct explanation-content elements and must be coded separately. Code \emph{PoVC~1.2} for the rule statement and \emph{PoVC~2.3} or \emph{PoVC~2.4} for the causal or outcome content that the rule motivates, if both are present in the same meaning unit or adjacent meaning units.
	
	\item \textbf{Include}
	\begin{itemize}
		\item Rules whose condition part is satisfied in the current case.
		\item Rules whose applicability cannot be fully decided because required case evidence is missing, but whose conclusion remains relevant to case interpretation.
		\item Explicit display of which predicates are met and which predicates remain unresolved.
		\item Rule conclusions that are linked to the current case because the rule conditions were evaluated for the case.
	\end{itemize}
	
	\item \textbf{Exclude}
	\begin{itemize}
		\item Rules listed without case-specific evaluation (\emph{PoVC~1.1 Rule base}).
		\item Rules whose conditions are established as not met in the current case (\emph{PoVC~1.1 Rule base}, not \emph{PoVC~1.2 Applied rule}).
		\item Causal claims that do not refer to a rule whose applicability was evaluated for the case (\emph{PoVC~2.3 Causal factor} or \emph{PoVC~2.6 Causal mechanism}, depending on whether the statement names a cause or explains a pathway).
		\item Pure outcome statements without an explicit rule condition part (\emph{PoVC~2.4 Outcome}).
	\end{itemize}
	
	\item \textbf{Typical evidence}
	\begin{itemize}
		\item A case-level rule evaluation that marks predicates as satisfied or unresolved.
		\item A rule-check view showing whether the available evidence is sufficient to establish that the rule applies.
		\item A display that links a rule conclusion to the current case while also indicating missing evidence when applicability remains unresolved.
		\item Diagnostic text that distinguishes between a rule being generally available and its applicability for the present case.
	\end{itemize}
	
	\item \textbf{Examples}
	\begin{itemize}
		\item Manufacturing: [Context] A vision-based inspection system flags a product as defective on Press~\#4 in Plant~2 during start-up after a tooling change. [Applied rule] Condition part: IF blur score exceeds threshold AND illumination is below the lower bound. Conclusion part: THEN [Causal factor] lens contamination. [Applied rule status] met.
		
		\item Building technology: [Context] A building-management system reports an overheating fault in Zone~3B at 08:15 during pre-occupancy warm-up on a cold day. [Applied rule] Condition part: IF zone temperature exceeds setpoint by more than 3$^\circ$C AND valve command remains at 100\%. Conclusion part: THEN [Causal factor] stuck reheat valve. [Applied rule status] met.
		
		\item Cybersecurity: [Context] A vulnerability-management system detects CVE-2021-44228 on an internet-facing production server during a scheduled weekly scan. [Applied rule] Condition part: IF \texttt{log4j-core} version 2.0--2.14.1 is present AND the service is internet-facing. Conclusion part: THEN [Outcome] patch or mitigate CVE-2021-44228. [Applied rule status] met.
		
		\item Cybersecurity (not decidable example): [Context] A vulnerability-management system evaluates an additional rule for the same case. [Applied rule] Condition part: IF exploitation indicators are present in logs AND the affected service is internet-facing. Conclusion part: THEN [Outcome] immediate remediation. [Applied rule status] not decidable.
		
	\end{itemize}
\end{itemize}


\subsubsection*{PoVC~2.1 Context}
\begin{itemize}
	\item \textbf{Definition} Context captures the surrounding situation in which the case occurs. It provides background information that helps interpret the case, the outcome, and the rest of the explanation. Context situates the case but does not, by itself, constitute the measured case evidence that the system uses to compute the outcome. Context can include time, location, operating mode, asset identity, environmental setting, organizational constraints, and other boundary conditions.
	
	\item \textbf{Decision rule (PoVC~2.1 vs.\ PoVC~2.2):} If the information primarily situates the case and helps interpret the outcome without being used as case evidence in the system's computation or decision logic, code it as \emph{PoVC~2.1 Context}. If the same type of information is explicitly used by the system's model, rule logic, or decision process to compute, rank, classify, recommend, or trigger the outcome, code it as \emph{PoVC~2.2 Input}.
	
	\item \textbf{Include}
	\begin{itemize}
		\item Environmental and operational setting information, such as plant area, building zone, production mode, network segment, maintenance mode, operator name, operator role, or operator contact information.
		\item Time, location, asset identifiers, process segment, or host identity that situate the case.
		\item Boundary conditions and constraints that affect interpretation or action, such as change-freeze periods, maintenance windows, policy scope, or access limitations.
	\end{itemize}
	
	\item \textbf{Exclude}
	\begin{itemize}
		\item Measurements, signals, records, or user entries that are explicitly used by the system to compute the outcome (\emph{PoVC~2.2 Input}).
		\item Causal statements that explain why the outcome occurred (\emph{PoVC~2.3 Causal factor} or \emph{PoVC~2.6 Causal mechanism}).
		\item Recommended actions or decisions (\emph{PoVC~2.4 Outcome}).
	\end{itemize}
	
	\item \textbf{Typical evidence}
	\begin{itemize}
		\item Environmental or organizational information that situates the case without functioning as computational input.
		\item Time, location, asset, host, or user identity information that helps interpret the case.
		\item Operational constraints or surrounding system conditions that affect how the case should be understood or acted upon.
	\end{itemize}
	
	\item \textbf{Examples}
	\begin{itemize}
		\item Manufacturing: A vision-based inspection system flags a product as defective on [Context] Press~\#4 in Plant~2, during start-up after a tooling change, operator Maria Chen, line operator, ext.~4582.

		\item Building technology: A building-management system reports an overheating fault in [Context] Zone~3B, at 08:15, during pre-occupancy warm-up on a cold day, facility engineer David Kumar, ext.~2147.

		\item Cybersecurity: A vulnerability-management system detects CVE-2021-44228 on [Context] an internet-facing production server, during a scheduled weekly scan, high-criticality asset, change-freeze period, analyst Priya Lopez, cybersecurity analyst, ext.~7719.
	\end{itemize}
	
\end{itemize}


\subsubsection*{PoVC~2.2 Input}
\begin{itemize}
	\item \textbf{Definition} Input is the data provided to the AI system to generate an outcome. It captures the case-specific values, signals, records, and/or user supplied entries that are used by the system's model or decision logic when producing a prediction, classification, recommendation, or alert.
	
	\item \textbf{Decision rule (PoVC~2.2 vs.\ PoVC~2.1):} If the information is explicitly used by the system's model, rule logic, or decision process to compute, rank, classify, recommend, or trigger the outcome, code it as \emph{PoVC~2.2 Input}. If the same type of information only situates the case and helps interpret it without being used as case evidence in the system's computation, code it as \emph{PoVC~2.1 Context}.
	
	\item \textbf{Include}
	\begin{itemize}
		\item Sensor readings, time series, measurements, images, audio, or log files ingested by the AI component.
		\item Structured feature values derived from raw data when they are used by the system to compute the outcome.
		\item Data that is used as input and that includes context information, for example environmental temperature, operating mode of a machine, or location of a part.
	\end{itemize}
	
	\item \textbf{Exclude}
	\begin{itemize}
		\item Background circumstances that frame the case but are not used to compute the outcome (\emph{PoVC~2.1 Context}).
		\item The identified vulnerability, cause, or explanatory factor inferred from the input (\emph{PoVC~2.3 Causal factor}).
		\item Recommended actions or decisions produced by the system (\emph{PoVC~2.4 Outcome}).
		\item Post-outcome verification data collected after an intervention (\emph{PoVC~2.5 Future state}).
	\end{itemize}
	
	\item \textbf{Typical evidence}
	\begin{itemize}
		\item Manufacturing: Camera image of a product, sensor stream from a station, or a feature vector computed from vibration data used to predict a bearing fault.
		\item Building technology: Temperature and humidity readings, occupancy signals, and setpoints used to predict an energy anomaly.
		\item Cybersecurity: Vulnerability scan results, software inventory with version strings, and configuration snapshots used to identify a CVE.
	\end{itemize}
	
	\item \textbf{Examples}
	\begin{itemize}
		\item Manufacturing: [Context] The vision-based inspection system uses [Input] camera image, blur score, illumination value, edge-detection confidence, and maintenance record.
		
		\item Building technology: [Context] The building-management system uses [Input] zone temperature, supply-air temperature, valve command, valve position, and occupancy schedule.

		\item Cybersecurity: [Context] The vulnerability-management system uses [Input] software inventory, detected 
		\texttt{log4j-core} version, exposure flag, scan findings, and configuration snapshot.
	\end{itemize}
\end{itemize}


\subsubsection*{PoVC~2.3 Causal factor}
\begin{itemize}
	\item \textbf{Definition} A causal factor is an explanation element that states what is believed to have influenced an outcome. A case may have no causal factor, one causal factor, or multiple causal factors. The presentation of one or more causal factors is a causal claim without evidence by itself. Support for this causal claim is provided by additional explanation content, in particular a causal mechanism that connects a causal factor to an outcome or to a future state.
	
	\item \textbf{Decision rule (PoVC~2.3 vs.\ PoVC~1.2):} If the explanation identifies a specific condition, variable, feature, or contributing element that played a role in producing the current outcome, code it as \emph{PoVC~2.3 Causal factor}, regardless of whether that factor originates from the AI domain, the system domain, or the application domain. The key distinction from \emph{PoVC~1.2 Applied rule} is that a causal factor describes what influenced the outcome, whereas an applied rule states that a formal decision rule was satisfied, violated, or triggered in the current case. If both a rule statement and the contributing condition it references are expressed as distinct content elements in the same or adjacent meaning units, code \emph{PoVC~1.2} for the rule statement and \emph{PoVC~2.3} for the contributing condition; if only one is expressed, assign the code that matches what is actually present.
	
	\item \textbf{Decision rule (PoVC~2.3 vs.\ PoVC~2.6):} If the statement names what is believed to have caused, driven, or contributed to the outcome, code it as \emph{PoVC~2.3 Causal factor}. If the statement instead explains how one element of the causal chain leads to another, code it as \emph{PoVC~2.6 Causal mechanism}.
	
	\item \textbf{Include}
	\begin{itemize}
		\item Stated causes, drivers, or contributing conditions that are presented as influencing the outcome.
		\item One or more identified issues, faults, vulnerabilities, or anomalies that are presented as reasons for the outcome.
		\item Contributing factors that are part of a broader causal chain and that may serve as targets for intervention.
	\end{itemize}
	
	\item \textbf{Exclude}
	\begin{itemize}
		\item Raw observations, measurements, signals, records, or user entries used to compute or infer the outcome (\emph{Input}).
		\item Background circumstances that situate the case but are not themselves presented as causes (\emph{Context}).
		\item Statements that explain how a factor leads to an outcome or future state (\emph{Causal mechanism}).
		\item Outcomes, recommendations, or decisions produced by the system (\emph{Outcome}).
	\end{itemize}
	
	\item \textbf{Typical evidence}
	\begin{itemize}
		\item Statements naming one or more suspected or inferred causes.
		\item Labels for faults, vulnerabilities, anomalies, or contributing conditions that are presented as influencing the outcome.
		\item Lists of candidate causes when the explanation claims that several factors may contribute to the same outcome.
	\end{itemize}
	
	\item \textbf{Examples}
	\begin{itemize}
		\item Manufacturing: [Context] In the shared manufacturing case, the system identifies [Causal factor] lens contamination.

		\item Building technology: [Context] In the shared building-technology case, the system identifies [Causal factor] stuck reheat valve.

		\item Cybersecurity: [Context] In the shared cybersecurity case, the system identifies [Causal factor] presence of CVE-2021-44228 on an internet-facing asset.
	\end{itemize}
\end{itemize}


\subsubsection*{PoVC~2.4 Outcome}
\begin{itemize}
	\item \textbf{Definition} An outcome is an explanation element that states the result produced for the current case. It can be a detected state, a classification, a prediction, an alert, a recommendation, a decision, or a selected action. The outcome is the result that the other explanation content helps interpret and justify.
	
	\item \textbf{Decision rule (PoVC~2.4 vs.\ PoVC~1.2):} If the explanation states the result, conclusion, recommendation, or action that the AI system produced for the current case, code it as \emph{PoVC~2.4 Outcome}, regardless of whether that result is expressed as a classification, a ranked recommendation, a risk score, or an actionable suggestion. The key distinction from \emph{PoVC~1.2 Applied rule} is that an outcome describes what the system concluded or recommended, whereas an applied rule states that a formal decision rule was satisfied, violated, or triggered in the current case and thereby justifies or motivates that conclusion. If both a rule statement and the outcome it motivates are expressed as distinct content elements in the same or adjacent meaning units, code \emph{PoVC~1.2} for the rule statement and \emph{PoVC~2.4} for the outcome the rule produces; if only one is expressed, assign the code that matches what is actually present.
	
	\item \textbf{Decision rule (PoVC~2.4 vs.\ PoVC~2.5):} If the statement expresses the result produced for the current case, code it as \emph{PoVC~2.4 Outcome}, even when that result is phrased as a future state, forecast, or expected condition. If the statement instead describes a post-intervention state that is expected after the outcome has been applied, code it as \emph{PoVC~2.5 Future state}.
	
	\item \textbf{Include}
	\begin{itemize}
		\item Stated results, such as classifications, predictions, detections, alerts, diagnoses, or risk assessments.
		\item Recommended actions, interventions, remediations, or decisions produced for the current case.
		\item Selected items or top-ranked actions when the system chooses one result for the current case.
	\end{itemize}
	
	\item \textbf{Exclude}
	\begin{itemize}
		\item Background circumstances that situate the case without being the result itself (\emph{Context}).
		\item Raw observations, signals, measurements, records, or user entries used by the system (\emph{Input}).
		\item Claimed causes or contributing conditions presented as influencing the result (\emph{Causal factor}).
		\item Statements that describe and support a causal relationship within the causal chain (\emph{Causal mechanism}).
		\item Predicted post-action states after an outcome has been applied (\emph{Future state}).
		\item Hypothetical variations of the case used to explore alternative results (\emph{PoVC~2.7 What-if forward} or \emph{PoVC~2.8 What-if backward}).
	\end{itemize}
	
	\item \textbf{Typical evidence}
	\begin{itemize}
		\item Statements naming the result for the current case.
		\item Alerts, recommendations, or selected actions presented as the system output.
		\item Labels, scores, or decisions shown as the case-specific result that requires explanation.
	\end{itemize}
	
	\item \textbf{Examples}
	\begin{itemize}
		\item Manufacturing: [Context] In the shared manufacturing case, the system produces [Outcome] product flagged as defective; manual inspection recommended.

		\item Building technology: [Context] In the shared building-technology case, the system produces [Outcome] overheating fault reported; valve inspection recommended.

		\item Cybersecurity: [Context] In the shared cybersecurity case, the system produces [Outcome] high-priority remediation recommendation to patch within 48 hours.
	\end{itemize}
\end{itemize}


\subsubsection*{PoVC~2.5 Future state}
\begin{itemize}
	\item \textbf{Definition} Future state is the predicted state that is expected to occur after suggested decisions, represented as outcomes, have been applied, and it is only applicable when the outcome is an actionable intervention that can plausibly change the state of the system or environment.
	
	\item \textbf{Decision rule (PoVC~2.5 vs.\ PoVC~2.4):} If the statement describes the state that is expected after the outcome has been applied, code it as \emph{PoVC~2.5 Future state}. If the statement itself is the result produced for the current case, including when that result is formulated as a forecast, predicted future condition, or expected state, code it as \emph{PoVC~2.4 Outcome}.
	
	\item \textbf{Include}
	\begin{itemize}
		\item Predicted post-intervention system states.
		\item Expected verification or confirmation states following an applied outcome.
		\item Anticipated side effects or tradeoffs explicitly stated as part of the post-action state.
		\item Follow-up success states or milestones expressed as achieved conditions after implementing the outcome.
	\end{itemize}
	
	\item \textbf{Exclude}
	\begin{itemize}
		\item The recommended action or decision itself, without the predicted post-intervention state change (\emph{outcomes}).
		\item The motivating causes, vulnerabilities, or contributing conditions (\emph{causal factors}).
		\item Case evidence and situational background (\emph{input} and \emph{context}).
		\item Hypothetical scenarios that vary conditions or actions without assuming the specific recommended outcome is applied (\emph{PoVC~2.7 What-if forward} or \emph{PoVC~2.8 What-if backward}).
	\end{itemize}
	
	\item \textbf{Typical evidence}
	\begin{itemize}
		\item Predicted post-intervention states after the recommended outcome has been applied.
		\item Expected verification or confirmation results.
		\item Anticipated side effects, tradeoffs, or follow-up conditions.
		\item Follow-up success conditions or milestones.
	\end{itemize}
	
	\item \textbf{Examples}
	\begin{itemize}
		\item Manufacturing: [Outcome] product flagged as defective; manual inspection recommended. [Future state] if the lens is cleaned, false rejects are expected to decrease and inspection confidence is expected to improve.

		\item Building technology: [Outcome] overheating fault reported; valve inspection recommended. [Future state] if the valve is repaired, the overheating alert is expected to clear and comfort complaints are expected to decrease.

		\item Cybersecurity: [Outcome] high-priority remediation recommendation to patch within 48 hours. [Future state] if the patch is applied and the service is restarted, the next validation scan is expected to report the vulnerability as resolved.
	\end{itemize}
\end{itemize}


\subsubsection*{PoVC~2.6 Causal mechanism}

\begin{itemize}
	\item \textbf{Definition} A causal mechanism is an explanation element that describes and supports a causal relationship within the causal chain. In particular, a causal mechanism can provide support for the causal relationship between an input and a causal factor, between two causal factors, between a causal factor and an outcome, and/or between an outcome and a future state.
	
	\item \textbf{Decision rule (PoVC~2.6 vs.\ PoVC~2.3):} If the statement explains how one element of the causal chain leads to another, code it as \emph{PoVC~2.6 Causal mechanism}. If the statement only names the suspected or inferred cause without describing the connecting process, pathway, or linkage, code it as \emph{PoVC~2.3 Causal factor}.
	
	\item \textbf{Include}
	\begin{itemize}
		\item Statements that explain how one element of the causal chain leads to another.
		\item Descriptions of physical, technical, organizational, biological, or computational processes that connect elements of the causal chain.
		\item Intermediate causal explanations that link input to causal factor, one causal factor to another, causal factor to outcome, or outcome to future state.
	\end{itemize}
	
	\item \textbf{Exclude}
	\begin{itemize}
		\item Statements that only name a cause without describing a causal relationship (\emph{Causal factor}).
		\item Raw observations, signals, measurements, records, or user entries used by the system (\emph{Input}).
		\item Background circumstances that situate the case without explaining a causal relationship (\emph{Context}).
		\item Decisions, alerts, recommendations, or classifications produced by the system (\emph{Outcome}).
	\end{itemize}
	
	\item \textbf{Typical evidence}
	\begin{itemize}
		\item Explanatory statements that describe how one element affects another.
		\item Descriptions of process dynamics, propagation effects, or system behaviors that connect adjacent elements in the causal chain.
		\item Statements that justify why a claimed causal factor is linked to an outcome or why an action is expected to lead to a future state.
	\end{itemize}
	
	\item \textbf{Examples}
	\begin{itemize}
		\item Manufacturing: [Causal mechanism] lens contamination reduces image contrast, which causes the edge detector to miss boundaries, supporting the causal relationship between [Causal factor] lens contamination and [Outcome] product flagged as defective.

		\item Building technology: [Causal mechanism] a stuck reheat valve continues to add heat despite the setpoint, which raises zone temperature, supporting the causal relationship between [Causal factor] stuck reheat valve and [Outcome] overheating fault reported.

		\item Cybersecurity: [Causal mechanism] a vulnerable logging library can be exploited through crafted input that triggers message lookups, enabling remote code execution, supporting the causal relationship between [Causal factor] presence of CVE-2021-44228 on an internet-facing asset and [Outcome] high-priority remediation recommendation.
	\end{itemize}
\end{itemize}


\subsubsection*{PoVC~2.7 What-if forward}
\begin{itemize}
	\item \textbf{Definition} A what-if forward explanation element describes a hypothetical change to an input and the expected effect of that change on the outcome for the current case. It emphasizes a directional relationship between input and outcome by starting from a changed input and projecting the resulting outcome.
	
	\item \textbf{Decision rule (PoVC~2.7 vs.\ PoVC~2.8):} If the explanation starts from a hypothetical change to one or more inputs and projects the resulting outcome for the current case, code it as \emph{PoVC~2.7 What-if forward}. If the explanation instead starts from a desired outcome and reasons backward to the input change required to achieve it, code it as \emph{PoVC~2.8 What-if backward}.
	
	\item \textbf{Include}
	\begin{itemize}
		\item Hypothetical changes to one or more input values used for the current case.
		\item Statements that project how the outcome would change if the input were changed.
		\item Forward-looking scenario variations that explicitly connect an altered input to an expected outcome.
	\end{itemize}
	
	\item \textbf{Exclude}
	\begin{itemize}
		\item Goal-oriented statements that start from a desired outcome and ask what input would need to change to achieve it (\emph{PoVC~2.8 What-if backward}).
		\item Statements that describe the current input-to-outcome relation without a hypothetical change.
		\item Statements that describe the expected state after implementing the recommended outcome without framing it as a hypothetical input variation (\emph{PoVC~2.5 Future state}).
	\end{itemize}
	
	\item \textbf{Typical evidence}
	\begin{itemize}
		\item Conditional statements of the form ``if this input changed, the outcome would change as follows.''
		\item Scenario exploration statements that vary one input and state the expected outcome.
		\item Sensitivity-style reasoning about how the outcome responds to an input change.
	\end{itemize}
	
	\item \textbf{Examples}
	\begin{itemize}
		\item Manufacturing: [What-if forward] if illumination is increased by 10\%, the predicted false reject rate decreases from 6\% to 2\%.

		\item Building technology: [What-if forward] if the supply-air temperature setpoint is lowered by 1$^\circ$C, predicted overheating risk decreases.

		\item Cybersecurity: [What-if forward] if internet exposure is removed, the risk score drops from urgent to high.
	\end{itemize}
\end{itemize}


\subsubsection*{PoVC~2.8 What-if backward}
\begin{itemize}
	\item \textbf{Definition} A what-if backward explanation element describes a desired outcome and the input change that would be required to achieve it for the current case. It emphasizes a directional relationship between input and outcome by starting from a target outcome and reasoning backward to the input that would need to change.
	
	\item \textbf{Decision rule (PoVC~2.8 vs.\ PoVC~2.7):} If the explanation starts from a desired outcome for the current case and identifies the input change needed to achieve it, code it as \emph{PoVC~2.8 What-if backward}. If the explanation instead starts from a changed input and projects the resulting outcome, code it as \emph{PoVC~2.7 What-if forward}.
	
	\item \textbf{Include}
	\begin{itemize}
		\item Goal-oriented hypothetical statements that begin with a desired outcome for the current case.
		\item Statements that identify what input would need to change to achieve the target outcome.
		\item Reverse reasoning that links a target outcome to a required input modification.
	\end{itemize}
	
	\item \textbf{Exclude}
	\begin{itemize}
		\item Forward scenario exploration that starts from a changed input and projects the resulting outcome (\emph{PoVC~2.7 What-if forward}).
		\item Statements that only describe the current outcome without reverse reasoning (\emph{PoVC~2.4 Outcome}).
		\item Statements that only describe the expected post-action state after the recommended outcome has been implemented (\emph{PoVC~2.5 Future state}).
		\item Ranked alternatives without an explicit target-outcome and required-input relation (\emph{PoVC~3.1 Set of ranked elements with ranking mechanism (actual)}).
	\end{itemize}
	
	\item \textbf{Typical evidence}
	\begin{itemize}
		\item Goal-oriented statements of the form ``to achieve this outcome, this input would need to change.''
		\item Reverse reasoning from a desired outcome to a required input modification.
		\item Recourse-style reasoning that identifies how the input must change to reach the target outcome.
	\end{itemize}
	
	\item \textbf{Examples}
	\begin{itemize}
		\item Manufacturing: [What-if backward] if the goal is to reduce the false reject rate to 2\%, illumination would need to be increased by about 10\%.

		\item Building technology: [What-if backward] if the goal is to clear the overheating alert, the supply-air temperature setpoint would need to be lowered by 1$^\circ$C.

		\item Cybersecurity: [What-if backward] if the goal is to reduce the risk score from urgent to high, internet exposure would need to be removed.
	\end{itemize}
\end{itemize}


\subsubsection*{PoVC~3.1 Set of ranked elements with ranking mechanism (actual)}
\begin{itemize}
	\item \textbf{Definition} A set of ranked elements with ranking mechanism (actual) is an explanation element that presents multiple candidate elements for the current case in ranked order.
	
	\item \textbf{Decision rule (PoVC~3.1 vs.\ PoVC~3.2):} If the explanation presents multiple candidate elements for the current case together with an ordering, selection logic, or set-level rule, code it as \emph{PoVC~3.1 Set of ranked elements with ranking mechanism (actual)}. If the explanation instead focuses on one candidate element and describes only that element's ranking properties, evidence, or attributes, code it as \emph{PoVC~3.2 Individual element with ranking properties, evidence, and/or attributes (actual)}.
	
	\item \textbf{Decision rule (PoVC~3.1 vs.\ PoVC~4.1):} If the set presents candidate elements for the current case, such as possible actions, decisions, recommendations, or outcomes to be selected or compared for this case, code it as \emph{PoVC~3.1}. If the set instead presents retrieved similar past cases used to support reasoning about the current case, code it as \emph{PoVC~4.1 Set of similar elements with similarity mechanism and set success-score information (similar)}.
	
	\item \textbf{Include}
	\begin{itemize}
		\item A set of candidate elements for the current case, such as alternative actions, recommendations, remediations, controls, or decisions.
		\item Rank positions and ranking properties used to order the candidate elements.
		\item A ranking mechanism or selection rule that explains how the ordering was produced.
		\item Set-level selection parameters, such as eligibility constraints, top-$k$ selection, thresholds, time windows, or robustness rules.
	\end{itemize}
	
	\item \textbf{Exclude}
	\begin{itemize}
		\item One single candidate element described individually (\emph{PoVC~3.2}).
		\item The final case result when no ranked candidate set is shown (\emph{PoVC~2.4 Outcome}).
		\item Candidate causes or vulnerabilities presented as causal claims rather than ranked recommendations (\emph{PoVC~2.3 Causal factor}).
	\end{itemize}
	
	\item \textbf{Typical evidence}
	\begin{itemize}
		\item A displayed ordered list of candidate elements for the current case.
		\item Rank positions, ranking properties, and a stated or inferable mechanism that explains the ordering.
	\end{itemize}
	
	\item \textbf{Examples}
	\begin{itemize}
		\item Manufacturing: [Set of ranked elements with ranking mechanism (actual)] (1) Clean lens, confidence = 0.82; (2) Recalibrate lighting, confidence = 0.74; (3) Replace camera, confidence = 0.41. Ordered by confidence; highest-confidence eligible action selected.

		\item Building technology: [Set of ranked elements with ranking mechanism (actual)] (1) Inspect valve actuator, confidence = 0.86; (2) Recalibrate temperature sensor, confidence = 0.52; (3) Adjust occupancy schedule, confidence = 0.21.

		\item Cybersecurity: [Set of ranked elements with ranking mechanism (actual)] (1) Patch immediately, confidence = 0.95; (2) Isolate service, confidence = 0.78; (3) Monitor only, confidence = 0.18.
	\end{itemize}
\end{itemize}


\subsubsection*{PoVC~3.2 Individual element with ranking properties, evidence, and/or attributes (actual)}
\begin{itemize}
	\item \textbf{Definition} An individual element with ranking properties, evidence, and/or attributes (actual) is an explanation element that describes one candidate element for the current case. The element can be selected or not selected. The description can include one or more ranking properties, supporting evidence, and descriptive attributes.
	
	\item \textbf{Decision rule (PoVC~3.2 vs.\ PoVC~3.1):} If the explanation focuses on one candidate element for the current case and describes that element's ranking properties, supporting evidence, and/or attributes, code it as \emph{PoVC~3.2}. If the explanation presents several candidate elements together and emphasizes their ordering, comparison, or set-level selection logic, code it as \emph{PoVC~3.1}.
	
	\item \textbf{Decision rule (PoVC~3.2 vs.\ PoVC~4.2):} If the element is a candidate element for the current case, code it as \emph{PoVC~3.2}. If the element instead describes one retrieved similar past case used to support reasoning about the current case, code it as \emph{PoVC~4.2 Individual, similar element with similarity property, success score, evidence, and/or attribute (similar)}.
	
	\item \textbf{Include}
	\begin{itemize}
		\item One candidate element for the current case, regardless of whether it was selected or not selected.
		\item One or more ranking properties, evidence supporting the assessment of that element, and attributes that characterize the element.
	\end{itemize}
	
	\item \textbf{Exclude}
	\begin{itemize}
		\item A ranked set of multiple candidate elements (\emph{PoVC~3.1}).
		\item The overall case result when no specific candidate element is described individually (\emph{PoVC~2.4 Outcome}).
		\item Similar past cases rather than candidate elements for the current case (\emph{PoVC~4.2}).
	\end{itemize}
	
	\item \textbf{Examples}
	\begin{itemize}
		\item Manufacturing: [Individual element (actual)] action = Clean lens; confidence = 0.82; evidence = rising blur score; attributes = Camera~C2, Station~3.

		\item Building technology: [Individual element (actual)] action = Inspect valve actuator; confidence = 0.86; evidence = command issued but valve position unchanged; attributes = AHU~4, Zone~3B.

		\item Cybersecurity: [Individual element (actual)] action = Patch within 48 hours; confidence = 0.95; evidence = vulnerable version detected on an internet-facing high-criticality asset; attributes = Server~A, production environment.
	\end{itemize}
\end{itemize}


\subsubsection*{PoVC~4.1 Set of similar elements with similarity mechanism and set success-score information (similar)}
\begin{itemize}
	\item \textbf{Definition} A set of similar elements with similarity mechanism and set success-score information (similar) is an explanation element that presents a set of similar cases for the current case, constructed through a similarity mechanism with a threshold or inclusion rule, together with set success-score information showing how many similar cases are successful, inconclusive, or not successful relative to the actual selected element.
	
	\item \textbf{Decision rule (PoVC~4.1 vs.\ PoVC~4.2):} If the explanation presents multiple similar cases together with a similarity mechanism and set-level success-score information, code it as \emph{PoVC~4.1}. If the explanation instead focuses on one specific similar case and describes that case's similarity property, success score, evidence, or attributes, code it as \emph{PoVC~4.2}.
	
	\item \textbf{Decision rule (PoVC~4.1 vs.\ PoVC~3.1):} If the set presents retrieved similar past cases used to support reasoning about the current case, code it as \emph{PoVC~4.1}. If the set instead presents candidate elements for the current case that are being ranked, selected, or compared, code it as \emph{PoVC~3.1}.
	
	\item \textbf{Include}
	\begin{itemize}
		\item A collection of similar cases linked to the current case through similarity, with a similarity mechanism and threshold.
		\item Set success-score information showing distribution across successful, inconclusive, and not successful relative to the actual selected element.
		\item Set-level summary values such as number of similar cases, similarity range, or aggregate confidence derived from the success-score distribution.
	\end{itemize}
	
	\item \textbf{Exclude}
	\begin{itemize}
		\item One individual similar case described separately (\emph{PoVC~4.2}).
		\item A ranked set of candidate elements for the current case (\emph{PoVC~3.1}).
		\item General confidence statements not derived from an explicit set of similar cases.
	\end{itemize}
	
	\item \textbf{Examples}
	\begin{itemize}
		\item Manufacturing: [Set of similar elements] 10 similar cases, similarity threshold 0.72; relative to action = Clean lens: 8 successful, 1 inconclusive, 1 not successful.

		\item Building technology: [Set of similar elements] 7 similar episodes, similarity threshold defined; relative to action = Inspect valve actuator: 5 successful, 1 inconclusive, 1 not successful.

		\item Cybersecurity: [Set of similar elements] 12 similar hosts, same CVE and exposure profile; relative to action = Patch within 48 hours: 9 successful, 2 inconclusive, 1 not successful.
	\end{itemize}
\end{itemize}


\subsubsection*{PoVC~4.2 Individual, similar element with similarity property, success score, evidence, and/or attribute (similar)}
\begin{itemize}
	\item \textbf{Definition} An individual, similar element with similarity property, success score, evidence, and/or attribute (similar) is an explanation element that describes one specific similar case relative to the current case, including an element-specific similarity property, a success score indicating whether it was successful, inconclusive, or not successful relative to the actual selected element, and optionally evidence and attributes.
	
	\item \textbf{Decision rule (PoVC~4.2 vs.\ PoVC~4.1):} If the explanation focuses on one specific similar case, code it as \emph{PoVC~4.2}. If the explanation presents several similar cases together with set-level information, code it as \emph{PoVC~4.1}.
	
	\item \textbf{Decision rule (PoVC~4.2 vs.\ PoVC~3.2):} If the element describes one retrieved similar past case, code it as \emph{PoVC~4.2}. If the element is a candidate element for the current case, code it as \emph{PoVC~3.2}.
	
	\item \textbf{Include}
	\begin{itemize}
		\item One referenced similar case with an element-specific similarity property such as similarity score or distance value.
		\item A success score indicating whether the similar case was successful, inconclusive, or not successful relative to the actual selected element.
		\item Evidence supporting the similarity relation and attributes characterizing the similar case.
	\end{itemize}
	
	\item \textbf{Exclude}
	\begin{itemize}
		\item A group of multiple similar cases summarized at set level (\emph{PoVC~4.1}).
		\item A candidate element for the current case rather than a similar past case (\emph{PoVC~3.2}).
		\item A similar case shown without an element-specific similarity property.
	\end{itemize}
	
	\item \textbf{Examples}
	\begin{itemize}
		\item Manufacturing: [Individual similar element] Station~3, Camera~C2; similarity score = 0.88; success score = successful relative to action = Clean lens; evidence = same blur spike and lighting conditions; attribute = same station after tooling change.

		\item Building technology: [Individual similar element] AHU~4 fault last month; similarity score = 0.81; success score = successful relative to action = Inspect valve actuator; evidence = same command-versus-position mismatch; attribute = same zone and outside-air conditions.

		\item Cybersecurity: [Individual similar element] SRV-12; similarity score = 0.90; success score = successful relative to action = Patch within 48 hours; evidence = same package version and exposure profile; attribute = same server role and criticality.
	\end{itemize}
\end{itemize}

\clearpage


\section{Change log from post-content validation to post-reliability validation codebook}
\label{app:ChangeLogPostvalidationToPostReliability}

\begin{table}[!ht]
	\small
	\caption{Change log from post-content validation to post-reliability validation codebook, part 1}
	\label{tab:Result:ChangeLogPostvalidationToPostReliabilityPart1}
	\begin{scriptsize}
		\begin{tabular}
			{>{\raggedright\arraybackslash}m{4.0cm}
				>{\raggedright\arraybackslash}m{3.0cm}
				>{\raggedright\arraybackslash}m{3.0cm}
				>{\raggedright\arraybackslash}m{4.0cm}}
			\hline
			
			Post-validation code (pre-reliability) &
			Trigger &
			Change rationale &
			Post-reliability code \\
			
			\hline
			
			PoVC~2.2 Input &
			One disagreement occurred at the boundary between
			PoVC~3.2 and PoVC~2.2, where the ground-truth coder
			assigned PoVC~3.2 and the second coder assigned
			PoVC~2.2. &
			The definitions of PoVC~2.2 and PoVC~3.2 were revised
			and a decision rule was introduced to clarify that input
			is data that precedes and drives the outcome, whereas a
			candidate element is part of the output space the system
			evaluates or selects from. An exclusion entry for
			candidate elements was added to PoVC~2.2. &
			PoRC~2.2 Input (definition and exclusions revised;
			decision rule PoRC~2.2 vs.\ PoRC~3.2 added) \\
			
			\hline
			
			PoVC~2.6 Causal mechanism &
			Two disagreements occurred at the boundary between
			PoVC~2.6 and adjacent codes: in one case the
			ground-truth coder assigned PoVC~Not assigned while the
			second coder assigned PoVC~2.6, and in another the
			ground-truth coder assigned PoVC~2.6 while the second
			coder assigned PoVC~3.2. &
			A decision rule was introduced to clarify the boundary
			between PoVC~2.6, PoVC~3.2, and uncoded content. The
			rule establishes that PoVC~2.6 requires a statement that
			explains \emph{how} one causal chain element produces or
			leads to another. Content that only names a cause belongs
			to PoVC~2.3, content that describes a candidate element
			belongs to PoVC~3.2, and content that provides
			background or procedural information without describing a
			causal pathway remains uncoded. An exclusion entry for
			candidate element descriptions was added to PoVC~2.6. &
			PoRC~2.6 Causal mechanism (exclusions revised; decision
			rule PoRC~2.6 vs.\ PoRC~3.2 and uncoded added) \\
			
			\hline
			
		\end{tabular}
	\end{scriptsize}
\end{table}

\begin{table}[!ht]
	\small
	\caption{Change log from post-content validation to post-reliability validation codebook, part 2}
	\label{tab:Result:ChangeLogPostvalidationToPostReliabilityPart2}
	\begin{scriptsize}
		\begin{tabular}
			{>{\raggedright\arraybackslash}m{4.0cm}
				>{\raggedright\arraybackslash}m{3.0cm}
				>{\raggedright\arraybackslash}m{3.0cm}
				>{\raggedright\arraybackslash}m{4.0cm}}
			\hline
			
			Post-validation code (pre-reliability) &
			Trigger &
			Change rationale &
			Post-reliability code \\
			
			\hline
			
			PoVC~3.1 Set of ranked elements with ranking mechanism
			(actual) &
			Two disagreements occurred at the boundary between
			PoVC~3.1 and PoVC~3.2, where the ground-truth coder
			assigned PoVC~3.1 and the second coder assigned
			PoVC~3.2. &
			The decision rule distinguishing PoVC~3.1 from PoVC~3.2
			was clarified to make the set-versus-individual
			distinction more explicit. The revised rule states that
			two or more candidate elements presented together with
			rank positions, comparative scores, or an ordering or
			selection mechanism constitute PoVC~3.1, whereas a
			single candidate element, even when carrying a rank
			position or confidence score, constitutes PoVC~3.2. A
			single element extracted or highlighted from a ranked
			list and described individually belongs to PoVC~3.2
			regardless of whether the surrounding ranked list is
			also present. &
			PoRC~3.1 Set of ranked elements with ranking mechanism
			(actual) (decision rule PoRC~3.1 vs.\ PoRC~3.2
			clarified) \\
			
			\hline
			
		\end{tabular}
	\end{scriptsize}
\end{table}

\begin{table}[!ht]
	\small
	\caption{Change log from post-content validation to post-reliability validation codebook, part 3}
	\label{tab:Result:ChangeLogPostvalidationToPostReliabilityPart3}
	\begin{scriptsize}
		\begin{tabular}
			{>{\raggedright\arraybackslash}m{4.0cm}
				>{\raggedright\arraybackslash}m{3.0cm}
				>{\raggedright\arraybackslash}m{3.0cm}
				>{\raggedright\arraybackslash}m{4.0cm}}
			\hline
			
			Post-validation code (pre-reliability) &
			Trigger &
			Change rationale &
			Post-reliability code \\
			
			\hline
			
			PoVC~3.2 Individual element with ranking properties,
			evidence, and/or attributes (actual) &
			One disagreement occurred at the boundary between
			PoVC~3.2 and PoVC~2.2, where the ground-truth coder
			assigned PoVC~3.2 and the second coder assigned
			PoVC~2.2. Two disagreements occurred at the boundary
			between PoVC~3.1 and PoVC~3.2, where the ground-truth
			coder assigned PoVC~3.1 and the second coder assigned
			PoVC~3.2. &
			A decision rule was introduced to clarify the boundary
			between PoVC~3.2 and PoVC~2.2, establishing that a
			candidate element described with an action label, rank
			position, confidence score, supporting evidence, or
			descriptive attributes belongs to PoVC~3.2 even when
			some of its attributes overlap with input variables. The
			decision rule distinguishing PoVC~3.2 from PoVC~3.1 was
			also clarified in alignment with the revision made to
			PoVC~3.1. An exclusion entry for raw input data was
			added to PoVC~3.2. &
			PoRC~3.2 Individual element with ranking properties,
			evidence, and/or attributes (actual) (exclusions
			revised; decision rule PoRC~3.2 vs.\ PoRC~2.2 added;
			decision rule PoRC~3.2 vs.\ PoRC~3.1 clarified) \\
			
			\hline
			
		\end{tabular}
	\end{scriptsize}
\end{table}

\clearpage


\section{Codebook (post-reliability validation)}
\label{app:PostReliabilityCodebook}


\subsection*{Explanation categories}

\subsubsection*{C~1 Rule-based explanation}
\begin{itemize}
	\item \textbf{Definition} A rule-based explanation is an explanation that justifies a case-specific element of the explanation by referring to documented decision rules. It distinguishes the \emph{rule base} as a set of available rules that encode general knowledge from the \emph{applied rule} set that captures which rules were evaluated and used in the current case. The explanation communicates decision conditions in a symbolic, human-readable form such as thresholds and Boolean combinations and reports the case-specific evaluation of these conditions as met, not met, or not decidable based on the available inputs and context. Applied rules can support or constrain elements of the causal explanation group, most commonly the outcome recommendation, ranked candidates, or admissibility constraints, while remaining outside the causal chain.
	
	\item \textbf{Include}
	\begin{itemize}
		\item A documented \emph{rule base} that provides uniquely identifiable rules with traceable provenance and scope.
		\item A case-specific \emph{applied rule} that reports which rule was evaluated for the case and how its conditions were assessed based on the available inputs and context.
		\item Explicit decision conditions expressed as thresholds, predicates, ranges, membership checks, Boolean logic, scoring rules, decision tables, or ordered rule lists with precedence.
		\item Condition-evaluation results that indicate which predicates were met, not met, or not decidable and which input evidence was used for the evaluation.
		\item Rule-based policies or compliance checks when they are presented as decision logic that supports or constrains a case-specific element such as an outcome recommendation or a ranked-candidate set.
	\end{itemize}
	
	\item \textbf{Exclude}
	\begin{itemize}
		\item Case-specific causal attributions that treat a rule as a cause without a separate causal-factor and mechanism description.
		\item Post-hoc feature-attribution lists that rank features by importance without expressing explicit decision conditions.
		\item Example-based explanations that justify an element by citing similar cases rather than by reporting an evaluated rule.
		\item Causal narratives that describe mechanisms or causal chains without specifying decision conditions and their evaluation.
		\item Procedural descriptions of training data or model development without reporting the rule(s) evaluated in the current case.
		\item Uninterpretable model internals such as weights unless translated into explicit decision rules.
	\end{itemize}
	
	\item \textbf{Typical evidence}
	\begin{itemize}
		\item Manufacturing rule base: ``If vibration RMS exceeds 7~mm/s for longer than 10~minutes and temperature exceeds 80$^\circ$C, then predict bearing fault.''
		\item Building technology rule base: ``If CO$_2$ $>$ 1000~ppm and damper position $<$ 20\% during occupancy, then raise indoor-air-quality alert.''
		\item Cybersecurity rule base: ``If software version matches a known vulnerable pattern and the asset is internet-facing, then flag the vulnerability as critical and recommend patching.''
	\end{itemize}
	
	\item \textbf{Examples}
	\begin{itemize}
		\item Manufacturing applied rule: ``Rule R-MFG-12 was evaluated and applied. Vibration RMS $>$ 7~mm/s for 10~minutes is met (8.2~mm/s). Temperature $>$ 80$^\circ$C is met (83$^\circ$C). The rule supports the predicted bearing-fault outcome and the recommendation to inspect the bearing.''
		\item Building technology applied rule: ``Rule R-BLD-07 was evaluated and applied. CO$_2$ $>$ 1000~ppm is met (1120~ppm). Damper position $<$ 20\% during occupancy is met (12\%). The rule supports raising an indoor-air-quality alert.''
		\item Cybersecurity applied rule: ``Rule R-CYB-21 was evaluated and applied. Log4Shell (CVE-2021-44228) is detected in inventory, which is met. Internet-facing exposure is met. The rule supports assigning criticality and the recommendation to patch within 48 hours.''
	\end{itemize}
\end{itemize}

\subsubsection*{C~2 Causal explanation}
\begin{itemize}
	\item \textbf{Definition} A causal explanation is an explanation that justifies an outcome by stating one or more causes and describing how those causes produced the outcome. It expresses a cause-to-effect relationship that goes beyond correlation, linking initiating conditions and intermediate factors or mechanisms to the observed outcome in a way that supports understanding, diagnosis, or intervention.
	
	\item \textbf{Include}
	\begin{itemize}
		\item Statements that identify one or more causal factors or drivers of the outcome, for example root cause or contributing cause.
		\item Explanations that describe a causal mechanism or causal chain connecting factors to the outcome, for example intermediate steps or propagation.
		\item Explanations that indicate directionality and causal influence, for example led to, resulted in, due to, because, including multi-cause accounts.
		\item Causal accounts that connect evidence to an outcome to support actions such as troubleshooting, mitigation, or prevention.
	\end{itemize}
	
	\item \textbf{Exclude}
	\begin{itemize}
		\item Pure statistical-association or correlation statements without a causal claim.
		\item Rule-based decision-logic explanations that only state decision conditions without asserting real-world causation.
		\item Counterfactual or recourse explanations that focus on how inputs must change to obtain a different outcome without explaining why the actual outcome occurred.
		\item Example-based explanations that justify an outcome by citing similar cases rather than causal reasoning.
		\item Descriptions of the data pipeline or model-training process that do not explain why this specific outcome occurred.
	\end{itemize}
	
	\item \textbf{Typical evidence}
	\begin{itemize}
		\item Manufacturing: Failure-analysis narrative linking component wear to process deviation and defect formation, supported by sensor trends.
		\item Building technology: Fault-diagnosis narrative linking actuator malfunction to airflow changes and comfort violations, supported by control logs.
		\item Cybersecurity: Attack-analysis narrative linking initial compromise to privilege escalation and impact, supported by authentication logs and alerts.
	\end{itemize}
	
	\item \textbf{Examples}
	\begin{itemize}
		\item Manufacturing: ``Nozzle wear reduced fill precision, which caused spills that contaminated the inspection optics, leading to increased false rejects.''
		\item Building technology: ``A stuck damper reduced outside-air intake, which raised CO$_2$ levels during occupancy and triggered the air-quality alert.''
		\item Cybersecurity: ``A phishing email led to credential theft, enabling unauthorized VPN access and lateral movement, which resulted in ransomware deployment and service disruption.''
	\end{itemize}
\end{itemize}

\subsubsection*{C~3 Epistemic explanation (actual)}
\begin{itemize}
	\item \textbf{Definition} An epistemic explanation (actual) is an explanation that justifies the actual outcome by communicating the system's uncertainty, confidence, or evidence strength regarding that outcome. It characterizes how strongly the system believes the actual outcome holds, based on probabilistic reasoning, likelihood, prediction confidence, posterior belief, or related uncertainty measures, and may indicate how reliable the outcome is expected to be.
	
	\item \textbf{Include}
	\begin{itemize}
		\item Confidence, probability, or likelihood statements about the actual outcome.
		\item Uncertainty information such as prediction intervals, credible intervals, error bars, entropy, margin to decision boundary, or ambiguity indicators tied to the actual outcome.
		\item Evidence-strength or reliability cues for the actual outcome, when presented as a justification for the outcome.
		\item References to alternative outcomes only to contextualize uncertainty, while keeping the actual outcome as the explained result.
	\end{itemize}
	
	\item \textbf{Exclude}
	\begin{itemize}
		\item Explanations that describe why the outcome occurred in causal terms.
		\item Rule-based or procedural explanations that state decision conditions without uncertainty information.
		\item Contrastive explanations whose primary purpose is to compare the actual outcome with a specific alternative outcome.
		\item Counterfactual or recourse explanations that describe what would need to change to obtain a different outcome.
		\item Pure data-quality or provenance statements unless explicitly linked to confidence in the actual outcome.
	\end{itemize}
	
	\item \textbf{Typical evidence}
	\begin{itemize}
		\item Manufacturing: Predicted defect probability with a confidence score and a note about uncertainty due to low image quality.
		\item Building technology: Fault-likelihood score with a confidence band and an ambiguity warning due to missing sensor readings.
		\item Cybersecurity: Risk score for an identified vulnerability with probability of exploitation and a confidence level based on observed exposure and asset telemetry.
	\end{itemize}
	
	\item \textbf{Examples}
	\begin{itemize}
		\item Manufacturing: ``The system predicts a bearing fault with 0.87 probability; uncertainty is moderate because recent vibration data contains gaps.''
		\item Building technology: ``The economizer fault is detected with high confidence (confidence = 0.91), supported by consistent deviations across three sensors.''
		\item Cybersecurity: ``The system flags CVE-2021-44228 as present with 0.95 confidence; confidence is lower for exploitability because external exposure could not be fully confirmed.''
	\end{itemize}
\end{itemize}

\subsubsection*{C~4 Epistemic explanation (similar)}
\begin{itemize}
	\item \textbf{Definition} An epistemic explanation (similar) is an explanation that justifies the current outcome by referencing historic cases that are similar to the current case and using those similarities as evidence for an element in the causal chain. It communicates the system's belief or confidence in the outcome by showing that comparable prior instances produced the same or highly similar outcomes, often including similarity scores, match criteria, and how frequently the outcome occurred among the retrieved cases.
	
	\item \textbf{Include}
	\begin{itemize}
		\item References to one or more historic cases deemed similar to the current case, used as evidence for the current outcome.
		\item Similarity information such as nearest neighbors, match scores, distances, feature overlap, or retrieved-case sets, when tied to confidence in the current outcome.
		\item Frequency or distribution summaries over similar cases.
		\item Case-based or example-based probability statements derived from the similar-case set.
	\end{itemize}
	
	\item \textbf{Exclude}
	\begin{itemize}
		\item Purely illustrative examples shown for explanation without being used as evidence for the current outcome.
		\item Counterfactual examples that show how the outcome would change under modified inputs.
		\item Contrastive examples whose main purpose is to compare the current case to a foil case with a different outcome.
		\item Causal narratives explaining why the outcome occurred, if they do not rely on similar cases as evidence.
		\item Rule-based decision conditions that do not involve retrieval of similar historic cases.
	\end{itemize}
	
	\item \textbf{Typical evidence}
	\begin{itemize}
		\item Manufacturing: Retrieval of past production runs with similar vibration signatures, showing that the same fault pattern preceded bearing failure.
		\item Building technology: Similar past days or zones with comparable temperature and valve behavior, where the same fault label was confirmed.
		\item Cybersecurity: Prior incidents with similar alert sequences or vulnerability configurations, showing how often they led to exploitation or high-risk classification.
	\end{itemize}
	
	\item \textbf{Examples}
	\begin{itemize}
		\item Manufacturing: ``The system predicts bearing wear because the current spectrum matches 9 prior cases (similarity $\geq$ 0.92), and 8 of those cases were confirmed as outer-race wear.''
		\item Building technology: ``This condition is classified as an economizer fault because 12 similar historical episodes showed the same damper and temperature pattern, and 10 were resolved by damper repair.''
		\item Cybersecurity: ``The alert is rated high risk because it matches recent incident patterns in 6 similar cases (same CVE, exposure, and log sequence), and 5 of those cases resulted in confirmed compromise.''
	\end{itemize}
\end{itemize}


\subsection*{Explanation content codes (post-reliability validation)}

\subsubsection*{Shared examples}

To demonstrate dependencies between explanation-content items, we use the following four shared domain examples consistently across the initial-code definitions. Note that the manufacturing, building-technology, and cybersecurity examples are drawn from the six corpus studies. The medical-device example (infusion pump, ICU setting) is a constructed illustration intended to broaden applicability and was not part of the corpus used to develop or validate the model.

\paragraph{Manufacturing}

A vision-based inspection system flags a product as defective on [\emph{PoRC~2.1 Context}: Press~\#4 in Plant~2, during start-up after a tooling change, operator Maria Chen, line operator, ext.~4582]. The system uses [\emph{PoRC~2.2 Input}: camera image, blur score, illumination value, edge-detection confidence, and maintenance record]. It identifies [\emph{PoRC~2.3 Causal factor}: lens contamination]. The explanation states [\emph{PoRC~2.6 Causal mechanism}: lens contamination reduces image contrast, which causes the edge detector to miss boundaries and increases false defect classifications]. The system produces [\emph{PoRC~2.4 Outcome}: product flagged as defective; manual inspection recommended]. It also states [\emph{PoRC~2.5 Future state}: if the lens is cleaned, false rejects are expected to decrease and inspection confidence is expected to improve]. A hypothetical variation is [\emph{PoRC~2.7 What-if forward}: if illumination is increased by 10\%, the predicted false reject rate decreases from 6\% to 2\%]. The available repository of rules is [\emph{PoRC~1.1 Rule base}: Rule~1: IF blur score exceeds threshold AND illumination is below lower bound THEN likely causal factor = lens contamination; Rule~2: IF vibration remains normal but current draw rises sharply THEN likely causal factor = motor overload]. The case-specific rule evaluation is [\emph{PoRC~1.2 Applied rule}: IF blur score exceeds threshold AND illumination is below lower bound THEN likely causal factor = lens contamination; condition status: met]. The ranked candidates are [\emph{PoRC~3.1 Set of ranked elements with ranking mechanism (actual)}: (1) Clean lens, confidence = 0.82; (2) Recalibrate lighting, confidence = 0.74; (3) Replace camera, confidence = 0.41]. One candidate element is [\emph{PoRC~3.2 Individual element with ranking properties, evidence, and/or attributes (actual)}: action = Clean lens; confidence = 0.82; evidence = rising blur score; attributes = Camera~C2, Station~3]. Similar past cases are [\emph{PoRC~4.1 Set of similar elements with similarity mechanism and set success-score information (similar)}: 10 similar cases selected by a similarity mechanism with a similarity threshold of 0.72; with respect to the selected actual element action = Clean lens, 8 cases are successful, 1 case is inconclusive, and 1 case is not successful]. One similar case is [\emph{PoRC~4.2 Individual, similar element with similarity property, success score, evidence, and/or attribute (similar)}: Station~3, Camera~C2; similarity score = 0.88; success score = successful relative to the selected actual element action = Clean lens; evidence = same blur spike and lighting conditions; attribute = same production station after a tooling change].

\paragraph{Building-technology}

A building-management system reports an overheating fault in [\emph{PoRC~2.1 Context}: Zone~3B, at 08:15, during pre-occupancy warm-up on a cold day, facility engineer David Kumar, ext.~2147]. The system uses [\emph{PoRC~2.2 Input}: zone temperature, supply-air temperature, valve command, valve position, and occupancy schedule]. It identifies [\emph{PoRC~2.3 Causal factor}: stuck reheat valve]. The explanation states [\emph{PoRC~2.6 Causal mechanism}: if the reheat valve remains open, heat continues to enter the zone despite the setpoint, which raises zone temperature and triggers an overheating alert]. The system produces [\emph{PoRC~2.4 Outcome}: overheating fault reported; valve inspection recommended]. It also states [\emph{PoRC~2.5 Future state}: if the valve is repaired, the overheating alert is expected to clear and comfort complaints are expected to decrease]. A hypothetical variation is [\emph{PoRC~2.7 What-if forward}: if the supply-air temperature setpoint is lowered by 1$^\circ$C, predicted overheating risk decreases]. The available repository of rules is [\emph{PoRC~1.1 Rule base}: Rule~A: IF zone temperature exceeds setpoint by more than 3$^\circ$C AND valve command remains at 100\% THEN likely causal factor = stuck reheat valve; Rule~B: IF zone temperature remains elevated while valve position does not follow the command signal THEN likely causal factor = actuator nonresponse]. The case-specific rule evaluation is [\emph{PoRC~1.2 Applied rule}: IF zone temperature exceeds setpoint by more than 3$^\circ$C AND valve command remains at 100\% THEN likely causal factor = stuck reheat valve; condition status: met]. The ranked candidates are [\emph{PoRC~3.1 Set of ranked elements with ranking mechanism (actual)}: (1) Inspect valve actuator, confidence = 0.86; (2) Recalibrate temperature sensor, confidence = 0.52; (3) Adjust occupancy schedule, confidence = 0.21]. One candidate element is [\emph{PoRC~3.2 Individual element with ranking properties, evidence, and/or attributes (actual)}: action = Inspect valve actuator; confidence = 0.86; evidence = command issued but valve position unchanged; attributes = AHU~4, Zone~3B]. Similar past cases are [\emph{PoRC~4.1 Set of similar elements with similarity mechanism and set success-score information (similar)}: 7 similar overheating episodes selected by a similarity mechanism with a defined similarity threshold; with respect to the selected actual element action = Inspect valve actuator, 5 cases are successful, 1 case is inconclusive, and 1 case is not successful]. One similar case is [\emph{PoRC~4.2 Individual, similar element with similarity property, success score, evidence, and/or attribute (similar)}: AHU~4 fault from last month; similarity score = 0.81; success score = successful relative to the selected actual element action = Inspect valve actuator; evidence = same command-versus-position mismatch; attribute = same zone and similar outside-air conditions].

\paragraph{Cybersecurity}

A vulnerability-management system detects CVE-2021-44228 on [\emph{PoRC~2.1 Context}: an internet-facing production server, during a scheduled weekly scan, high-criticality asset, change-freeze period, analyst Priya Lopez, cybersecurity analyst, ext.~7719]. The system uses [\emph{PoRC~2.2 Input}: software inventory, detected \texttt{log4j-core} version, exposure flag, scan findings, and configuration snapshot]. It identifies [\emph{PoRC~2.3 Causal factor}: presence of CVE-2021-44228 on an internet-facing asset]. The explanation states [\emph{PoRC~2.6 Causal mechanism}: a vulnerable logging library can be exploited through crafted input that triggers message lookups, enabling remote code execution and increasing compromise risk]. The system produces [\emph{PoRC~2.4 Outcome}: high-priority remediation recommendation to patch within 48 hours]. It also states [\emph{PoRC~2.5 Future state}: if the patch is applied and the service is restarted, the next validation scan is expected to report the vulnerability as resolved]. A hypothetical variation is [\emph{PoRC~2.7 What-if forward}: if internet exposure is removed, the risk score drops from urgent to high]. The available repository of rules is [\emph{PoRC~1.1 Rule base}: Rule~X: IF \texttt{log4j-core} version 2.0--2.14.1 is present AND the service is internet-facing THEN recommended outcome = patch or mitigate CVE-2021-44228; Rule~Y: IF exploitation indicators are present in logs AND the affected service is internet-facing THEN recommended outcome = immediate remediation]. The case-specific rule evaluation is [\emph{PoRC~1.2 Applied rule}: IF \texttt{log4j-core} version 2.0--2.14.1 is present AND the service is internet-facing THEN recommended outcome = patch or mitigate CVE-2021-44228; condition status: met]. The ranked candidates are [\emph{PoRC~3.1 Set of ranked elements with ranking mechanism (actual)}: (1) Patch immediately, confidence = 0.95; (2) Isolate service, confidence = 0.78; (3) Monitor only, confidence = 0.18]. One candidate element is [\emph{PoRC~3.2 Individual element with ranking properties, evidence, and/or attributes (actual)}: action = Patch within 48 hours; confidence = 0.95; evidence = vulnerable version detected on an internet-facing high-criticality asset; attributes = Server~A, production environment]. Similar past cases are [\emph{PoRC~4.1 Set of similar elements with similarity mechanism and set success-score information (similar)}: 12 similar hosts selected by a similarity mechanism with a defined similarity threshold based on the same CVE and exposure profile; with respect to the selected actual element action = Patch within 48 hours, 9 cases are successful, 2 cases are inconclusive, and 1 case is not successful]. One similar case is [\emph{PoRC~4.2 Individual, similar element with similarity property, success score, evidence, and/or attribute (similar)}: SRV-12; similarity score = 0.90; success score = successful relative to the selected actual element action = Patch within 48 hours; evidence = same package version and same exposure profile; attribute = same server role and criticality level].


\subsubsection*{PoRC~1.1 Rule base}

\begin{itemize}
	\item \textbf{Definition} A rule base is a repository or set of available rules that can be used to assess a case. Each rule consists of a condition part and a conclusion part, typically expressed as IF <condition(s)> THEN <conclusion>. The condition part specifies when the rule applies. The conclusion part typically refers to one or more elements of the causal explanation group, such as a causal factor, an outcome, or a future state. The rule base represents the available set of rules independent of whether any individual rule is confirmed for the current case.
	
	\item \textbf{Decision rule (PoRC~1.1 vs.\ PoRC~1.2):} If the explanation presents a general rule, decision logic, threshold condition, policy, or reusable if-then structure without tying it to the current case as satisfied, violated, or triggered, code it as \emph{PoRC~1.1 Rule base}. If the explanation instead shows how such a rule applies to the current case, for example by stating that its conditions are met, not met, or partially met and that this leads to the current result, code it as \emph{PoRC~1.2 Applied rule}.
	
	\item \textbf{Include}
	\begin{itemize}
		\item Sets, repositories, libraries, catalogs, or collections of explicit if-then rules, decision rules, thresholds, Boolean combinations, scoring rules, or rule lists that are available for case assessment.
		\item Sets of rules whose condition parts reference measurable values, states, events, asset properties, or categorical properties.
		\item Rule-base metadata, if shown, such as rule origin, priority, confidence, scope, policy source, version, or number of available rules.
	\end{itemize}
	
	\item \textbf{Exclude}
	\begin{itemize}
		\item A single rule shown in isolation rather than as part of a rule set or repository (if evaluated for the current case, it belongs to \emph{PoRC~1.2 Applied rule}).
		\item A case-specific statement that a rule has been confirmed as applying in the current case (this belongs to \emph{PoRC~1.2 Applied rule}).
		\item Pure outcome statements without an explicit condition part (these belong to \emph{PoRC~2.4 Outcome}).
		\item Probabilistic statements, feature weights, or scores that do not define a testable rule condition (these belong to \emph{PoRC~3.2 Individual element with ranking properties, evidence, and/or attributes (actual)} when presented for the current case).
		\item General guidance or policies that are not expressed as operational conditions that can be checked against a case (depending on their role, they may be coded as \emph{PoRC~2.1 Context} or remain uncoded).
	\end{itemize}
	
	\item \textbf{Typical evidence}
	\begin{itemize}
		\item A repository, library, catalog, or list that contains multiple available rules for case assessment.
		\item Repeated if-then structures or other explicit rule formulations shown as a set rather than as a single case-specific statement.
		\item Rule-base metadata such as rule identifiers, source, version, scope, priority, confidence, or grouping by domain topic.
		\item Grouping or navigation structures that organize available rules by asset type, fault class, vulnerability class, operating mode, or policy domain.
	\end{itemize}
	
	\item \textbf{Examples}
	\begin{itemize}
		\item Manufacturing: [Context] A vision-based inspection system flags a product as defective on Press~\#4 in Plant~2 during start-up after a tooling change. [Rule base] Rule~1: IF blur score exceeds threshold AND illumination is below the lower bound THEN [Causal factor] lens contamination. Rule~2: IF vibration remains normal but current draw rises sharply THEN [Causal factor] motor overload.

		\item Building technology: [Context] A building-management system reports an overheating fault in Zone~3B at 08:15 during pre-occupancy warm-up on a cold day. [Rule base] Rule~A: IF zone temperature exceeds setpoint by more than 3$^\circ$C AND valve command remains at 100\% THEN [Causal factor] stuck reheat valve. Rule~B: IF zone temperature remains elevated while valve position does not follow the command signal THEN [Causal factor] actuator nonresponse.

		\item Cybersecurity: [Context] A vulnerability-management system detects CVE-2021-44228 on an internet-facing production server during a scheduled weekly scan. [Rule base] Rule~X: IF \texttt{log4j-core} version 2.0--2.14.1 is present AND the service is internet-facing THEN [Outcome] patch or mitigate CVE-2021-44228. Rule~Y: IF exploitation indicators are present in logs AND the affected service is internet-facing THEN [Outcome] immediate remediation.
	\end{itemize}
\end{itemize}

\subsubsection*{PoRC~1.2 Applied rule}

\begin{itemize}
	\item \textbf{Definition} An applied rule is a rule from the rule base whose applicability is evaluated for the current case and whose conclusion is relevant for interpreting the case. Each rule consists of a condition part and a conclusion part, typically expressed as IF <condition(s)> THEN <conclusion>. For an applied rule, the condition part is evaluated as either \emph{met} or \emph{not decidable}. A status of \emph{not decidable} indicates unresolved applicability because required case evidence is missing. The conclusion part typically supports one or more elements of the causal explanation group, such as a causal factor, an outcome, or a future state. An applied rule supports case interpretation, but it is not itself the causal chain.
	
	\item \textbf{Decision rule (PoRC~1.2 vs.\ PoRC~1.1):} If the explanation states that a rule is satisfied, violated, triggered, or otherwise instantiated in the current case, code it as \emph{PoRC~1.2 Applied rule}. If the explanation only presents the reusable rule itself, including its general conditions, thresholds, or logic, without linking it to the current case, code it as \emph{PoRC~1.1 Rule base}.
	
	\item \textbf{Decision rule (PoRC~1.2 vs.\ PoRC~2.3 or PoRC~2.4):} If the explanation states that a rule is satisfied, violated, triggered, or otherwise instantiated in the current case, code it as \emph{PoRC~1.2 Applied rule}. An applied rule may motivate or justify a causal factor (\emph{PoRC~2.3}) or an outcome (\emph{PoRC~2.4}) within the causal chain, but the rule statement itself and the causal or outcome content it supports are distinct explanation-content elements and must be coded separately. Code \emph{PoRC~1.2} for the rule statement and \emph{PoRC~2.3} or \emph{PoRC~2.4} for the causal or outcome content that the rule motivates, if both are present in the same meaning unit or adjacent meaning units.
	
	\item \textbf{Include}
	\begin{itemize}
		\item Rules whose condition part is satisfied in the current case.
		\item Rules whose applicability cannot be fully decided because required case evidence is missing, but whose conclusion remains relevant to case interpretation.
		\item Explicit display of which predicates are met and which predicates remain unresolved.
		\item Rule conclusions that are linked to the current case because the rule conditions were evaluated for the case.
	\end{itemize}
	
	\item \textbf{Exclude}
	\begin{itemize}
		\item Rules listed without case-specific evaluation (\emph{PoRC~1.1 Rule base}).
		\item Rules whose conditions are established as not met in the current case (\emph{PoRC~1.1 Rule base}, not \emph{PoRC~1.2 Applied rule}).
		\item Causal claims that do not refer to a rule whose applicability was evaluated for the case (\emph{PoRC~2.3 Causal factor} or \emph{PoRC~2.6 Causal mechanism}, depending on whether the statement names a cause or explains a pathway).
		\item Pure outcome statements without an explicit rule condition part (\emph{PoRC~2.4 Outcome}).
	\end{itemize}
	
	\item \textbf{Typical evidence}
	\begin{itemize}
		\item A case-level rule evaluation that marks predicates as satisfied or unresolved.
		\item A rule-check view showing whether the available evidence is sufficient to establish that the rule applies.
		\item A display that links a rule conclusion to the current case while also indicating missing evidence when applicability remains unresolved.
		\item Diagnostic text that distinguishes between a rule being generally available and its applicability for the present case.
	\end{itemize}
	
	\item \textbf{Examples}
	\begin{itemize}
		\item Manufacturing: [Context] A vision-based inspection system flags a product as defective on Press~\#4 in Plant~2 during start-up after a tooling change. [Applied rule] Condition part: IF blur score exceeds threshold AND illumination is below the lower bound. Conclusion part: THEN [Causal factor] lens contamination. [Applied rule status] met.

		\item Building technology: [Context] A building-management system reports an overheating fault in Zone~3B at 08:15 during pre-occupancy warm-up on a cold day. [Applied rule] Condition part: IF zone temperature exceeds setpoint by more than 3$^\circ$C AND valve command remains at 100\%. Conclusion part: THEN [Causal factor] stuck reheat valve. [Applied rule status] met.

		\item Cybersecurity: [Context] A vulnerability-management system detects CVE-2021-44228 on an internet-facing production server during a scheduled weekly scan. [Applied rule] Condition part: IF \texttt{log4j-core} version 2.0--2.14.1 is present AND the service is internet-facing. Conclusion part: THEN [Outcome] patch or mitigate CVE-2021-44228. [Applied rule status] met.

		\item Cybersecurity (not decidable example): [Context] A vulnerability-management system evaluates an additional rule for the same case. [Applied rule] Condition part: IF exploitation indicators are present in logs AND the affected service is internet-facing. Conclusion part: THEN [Outcome] immediate remediation. [Applied rule status] not decidable.
	\end{itemize}
\end{itemize}


\subsubsection*{PoRC~2.1 Context}

\begin{itemize}
	\item \textbf{Definition} Context captures the surrounding situation in which the case occurs. It provides background information that helps interpret the case, the outcome, and the rest of the explanation. Context situates the case but does not, by itself, constitute the measured case evidence that the system uses to compute the outcome. Context can include time, location, operating mode, asset identity, environmental setting, organizational constraints, and other boundary conditions.
	
	\item \textbf{Decision rule (PoRC~2.1 vs.\ PoRC~2.2):} If the information primarily situates the case and helps interpret the outcome without being used as case evidence in the system's computation or decision logic, code it as \emph{PoRC~2.1 Context}. If the same type of information is explicitly used by the system's model, rule logic, or decision process to compute, rank, classify, recommend, or trigger the outcome, code it as \emph{PoRC~2.2 Input}.
	
	\item \textbf{Include}
	\begin{itemize}
		\item Environmental and operational setting information, such as plant area, building zone, production mode, network segment, maintenance mode, operator name, operator role, or operator contact information.
		\item Time, location, asset identifiers, process segment, or host identity that situate the case.
		\item Boundary conditions and constraints that affect interpretation or action, such as change-freeze periods, maintenance windows, policy scope, or access limitations.
	\end{itemize}
	
	\item \textbf{Exclude}
	\begin{itemize}
		\item Measurements, signals, records, or user entries that are explicitly used by the system to compute the outcome (\emph{PoRC~2.2 Input}).
		\item Causal statements that explain why the outcome occurred (\emph{PoRC~2.3 Causal factor} or \emph{PoRC~2.6 Causal mechanism}).
		\item Recommended actions or decisions (\emph{PoRC~2.4 Outcome}).
	\end{itemize}
	
	\item \textbf{Typical evidence}
	\begin{itemize}
		\item Environmental or organizational information that situates the case without functioning as computational input.
		\item Time, location, asset, host, or user identity information that helps interpret the case.
		\item Operational constraints or surrounding system conditions that affect how the case should be understood or acted upon.
	\end{itemize}
	
	\item \textbf{Examples}
	\begin{itemize}
		\item Manufacturing: A vision-based inspection system flags a product as defective on [Context] Press~\#4 in Plant~2, during start-up after a tooling change, operator Maria Chen, line operator, ext.~4582.

		\item Building technology: A building-management system reports an overheating fault in [Context] Zone~3B, at 08:15, during pre-occupancy warm-up on a cold day, facility engineer David Kumar, ext.~2147.

		\item Cybersecurity: A vulnerability-management system detects CVE-2021-44228 on [Context] an internet-facing production server, during a scheduled weekly scan, high-criticality asset, change-freeze period, analyst Priya Lopez, cybersecurity analyst, ext.~7719.

	\end{itemize}
\end{itemize}


\subsubsection*{PoRC~2.2 Input}

\begin{itemize}
	\item \textbf{Definition} Input is the data provided to the AI system to generate an outcome. It captures the case-specific values, signals, records, and/or user supplied entries that are used by the system's model or decision logic when producing a prediction, classification, recommendation, or alert.
	
	\item \textbf{Decision rule (PoRC~2.2 vs.\ PoRC~2.1):} If the information is explicitly used by the system's model, rule logic, or decision process to compute, rank, classify, recommend, or trigger the outcome, code it as \emph{PoRC~2.2 Input}. If the same type of information only situates the case and helps interpret it without being used as case evidence in the system's computation, code it as \emph{PoRC~2.1 Context}.
	
	\item \textbf{Decision rule (PoRC~2.2 vs.\ PoRC~3.2):} If the content describes raw data, signals, measurements, records, or feature values that are fed into the system to produce the outcome, code it as \emph{PoRC~2.2 Input}. If the content instead describes one candidate element for the current case, such as a specific action, recommendation, or decision option, together with its ranking properties, supporting evidence, or descriptive attributes, code it as \emph{PoRC~3.2 Individual element with ranking properties, evidence, and/or attributes (actual)}. The key distinction is directionality: input precedes and drives the outcome, whereas a candidate element is part of the output space that the system evaluates or selects from.
	
	\item \textbf{Include}
	\begin{itemize}
		\item Sensor readings, time series, measurements, images, audio, or log files ingested by the AI component.
		\item Structured feature values derived from raw data when they are used by the system to compute the outcome.
		\item Data that is used as input and that includes context information, for example environmental temperature, operating mode of a machine, or location of a part.
	\end{itemize}
	
	\item \textbf{Exclude}
	\begin{itemize}
		\item Background circumstances that frame the case but are not used to compute the outcome (\emph{PoRC~2.1 Context}).
		\item The identified vulnerability, cause, or explanatory factor inferred from the input (\emph{PoRC~2.3 Causal factor}).
		\item Recommended actions or decisions produced by the system (\emph{PoRC~2.4 Outcome}).
		\item Post-outcome verification data collected after an intervention (\emph{PoRC~2.5 Future state}).
		\item Candidate elements for the current case described with ranking properties, evidence, or attributes (\emph{PoRC~3.2 Individual element with ranking properties, evidence, and/or attributes (actual)}).
	\end{itemize}
	
	\item \textbf{Typical evidence}
	\begin{itemize}
		\item Manufacturing: Camera image of a product, sensor stream from a station, or a feature vector computed from vibration data used to predict a bearing fault.
		\item Building technology: Temperature and humidity readings, occupancy signals, and setpoints used to predict an energy anomaly.
		\item Cybersecurity: Vulnerability scan results, software inventory with version strings, and configuration snapshots used to identify a CVE.
	\end{itemize}
	
	\item \textbf{Examples}
	\begin{itemize}
		\item Manufacturing: [Context] The vision-based inspection system uses [Input] camera image, blur score, illumination value, edge-detection confidence, and maintenance record.

		\item Building technology: [Context] The building-management system uses [Input] zone temperature, supply-air temperature, valve command, valve position, and occupancy schedule.

		\item Cybersecurity: [Context] The vulnerability-management system uses [Input] software inventory, detected \texttt{log4j-core} version, exposure flag, scan findings, and configuration snapshot.
	\end{itemize}
\end{itemize}


\subsubsection*{PoRC~2.3 Causal factor}

\begin{itemize}
	\item \textbf{Definition} A causal factor is an explanation element that states what is believed to have influenced an outcome. A case may have no causal factor, one causal factor, or multiple causal factors. The presentation of one or more causal factors is a causal claim without evidence by itself. Support for this causal claim is provided by additional explanation content, in particular a causal mechanism that connects a causal factor to an outcome or to a future state.
	
	\item \textbf{Decision rule (PoRC~2.3 vs.\ PoRC~1.2):} If the explanation identifies a specific condition, variable, feature, or contributing element that played a role in producing the current outcome, code it as \emph{PoRC~2.3 Causal factor}, regardless of whether that factor originates from the AI domain, the system domain, or the application domain. The key distinction from \emph{PoRC~1.2 Applied rule} is that a causal factor describes what influenced the outcome, whereas an applied rule states that a formal decision rule was satisfied, violated, or triggered in the current case. If both a rule statement and the contributing condition it references are expressed as distinct content elements in the same or adjacent meaning units, code \emph{PoRC~1.2} for the rule statement and \emph{PoRC~2.3} for the contributing condition; if only one is expressed, assign the code that matches what is actually present.
	
	\item \textbf{Decision rule (PoRC~2.3 vs.\ PoRC~2.6):} If the statement names what is believed to have caused, driven, or contributed to the outcome, code it as \emph{PoRC~2.3 Causal factor}. If the statement instead explains how one element of the causal chain leads to another, code it as \emph{PoRC~2.6 Causal mechanism}.
	
	\item \textbf{Include}
	\begin{itemize}
		\item Stated causes, drivers, or contributing conditions that are presented as influencing the outcome.
		\item One or more identified issues, faults, vulnerabilities, or anomalies that are presented as reasons for the outcome.
		\item Contributing factors that are part of a broader causal chain and that may serve as targets for intervention.
	\end{itemize}
	
	\item \textbf{Exclude}
	\begin{itemize}
		\item Raw observations, measurements, signals, records, or user entries used to compute or infer the outcome (\emph{Input}).
		\item Background circumstances that situate the case but are not themselves presented as causes (\emph{Context}).
		\item Statements that explain how a factor leads to an outcome or future state (\emph{Causal mechanism}).
		\item Outcomes, recommendations, or decisions produced by the system (\emph{Outcome}).
	\end{itemize}
	
	\item \textbf{Typical evidence}
	\begin{itemize}
		\item Statements naming one or more suspected or inferred causes.
		\item Labels for faults, vulnerabilities, anomalies, or contributing conditions that are presented as influencing the outcome.
		\item Lists of candidate causes when the explanation claims that several factors may contribute to the same outcome.
	\end{itemize}
	
	\item \textbf{Examples}
	\begin{itemize}
		\item Manufacturing: [Context] In the shared manufacturing case, the system identifies [Causal factor] lens contamination.

		\item Building technology: [Context] In the shared building-technology case, the system identifies [Causal factor] stuck reheat valve.

		\item Cybersecurity: [Context] In the shared cybersecurity case, the system identifies [Causal factor] presence of CVE-2021-44228 on an internet-facing asset.
	\end{itemize}
\end{itemize}


\subsubsection*{PoRC~2.4 Outcome}

\begin{itemize}
	\item \textbf{Definition} An outcome is an explanation element that states the result produced for the current case. It can be a detected state, a classification, a prediction, an alert, a recommendation, a decision, or a selected action. The outcome is the result that the other explanation content helps interpret and justify.
	
	\item \textbf{Decision rule (PoRC~2.4 vs.\ PoRC~1.2):} If the explanation states the result, conclusion, recommendation, or action that the AI system produced for the current case, code it as \emph{PoRC~2.4 Outcome}, regardless of whether that result is expressed as a classification, a ranked recommendation, a risk score, or an actionable suggestion. The key distinction from \emph{PoRC~1.2 Applied rule} is that an outcome describes what the system concluded or recommended, whereas an applied rule states that a formal decision rule was satisfied, violated, or triggered in the current case and thereby justifies or motivates that conclusion. If both a rule statement and the outcome it motivates are expressed as distinct content elements in the same or adjacent meaning units, code \emph{PoRC~1.2} for the rule statement and \emph{PoRC~2.4} for the outcome the rule produces; if only one is expressed, assign the code that matches what is actually present.
	
	\item \textbf{Decision rule (PoRC~2.4 vs.\ PoRC~2.5):} If the statement expresses the result produced for the current case, code it as \emph{PoRC~2.4 Outcome}, even when that result is phrased as a future state, forecast, or expected condition. If the statement instead describes a post-intervention state that is expected after the outcome has been applied, code it as \emph{PoRC~2.5 Future state}.
	
	\item \textbf{Include}
	\begin{itemize}
		\item Stated results, such as classifications, predictions, detections, alerts, diagnoses, or risk assessments.
		\item Recommended actions, interventions, remediations, or decisions produced for the current case.
		\item Selected items or top-ranked actions when the system chooses one result for the current case.
	\end{itemize}
	
	\item \textbf{Exclude}
	\begin{itemize}
		\item Background circumstances that situate the case without being the result itself (\emph{Context}).
		\item Raw observations, signals, measurements, records, or user entries used by the system (\emph{Input}).
		\item Claimed causes or contributing conditions presented as influencing the result (\emph{Causal factor}).
		\item Statements that describe and support a causal relationship within the causal chain (\emph{Causal mechanism}).
		\item Predicted post-action states after an outcome has been applied (\emph{Future state}).
		\item Hypothetical variations of the case used to explore alternative results (\emph{PoRC~2.7 What-if forward} or \emph{PoRC~2.8 What-if backward}).
	\end{itemize}
	
	\item \textbf{Typical evidence}
	\begin{itemize}
		\item Statements naming the result for the current case.
		\item Alerts, recommendations, or selected actions presented as the system output.
		\item Labels, scores, or decisions shown as the case-specific result that requires explanation.
	\end{itemize}
	
	\item \textbf{Examples}
	\begin{itemize}
		\item Manufacturing: [Context] In the shared manufacturing case, the system produces [Outcome] product flagged as defective; manual inspection recommended.

		\item Building technology: [Context] In the shared building-technology case, the system produces [Outcome] overheating fault reported; valve inspection recommended.

		\item Cybersecurity: [Context] In the shared cybersecurity case, the system produces [Outcome] high-priority remediation recommendation to patch within 48 hours.
	\end{itemize}
\end{itemize}


\subsubsection*{PoRC~2.5 Future state}

\begin{itemize}
	\item \textbf{Definition} Future state is the predicted state that is expected to occur after suggested decisions, represented as outcomes, have been applied, and it is only applicable when the outcome is an actionable intervention that can plausibly change the state of the system or environment.
	
	\item \textbf{Decision rule (PoRC~2.5 vs.\ PoRC~2.4):} If the statement describes the state that is expected after the outcome has been applied, code it as \emph{PoRC~2.5 Future state}. If the statement itself is the result produced for the current case, including when that result is formulated as a forecast, predicted future condition, or expected state, code it as \emph{PoRC~2.4 Outcome}.
	
	\item \textbf{Include}
	\begin{itemize}
		\item Predicted post-intervention system states.
		\item Expected verification or confirmation states following an applied outcome.
		\item Anticipated side effects or tradeoffs explicitly stated as part of the post-action state.
		\item Follow-up success states or milestones expressed as achieved conditions after implementing the outcome.
	\end{itemize}
	
	\item \textbf{Exclude}
	\begin{itemize}
		\item The recommended action or decision itself, without the predicted post-intervention state change (\emph{outcomes}).
		\item The motivating causes, vulnerabilities, or contributing conditions (\emph{causal factors}).
		\item Case evidence and situational background (\emph{input} and \emph{context}).
		\item Hypothetical scenarios that vary conditions or actions without assuming the specific recommended outcome is applied (\emph{PoRC~2.7 What-if forward} or \emph{PoRC~2.8 What-if backward}).
	\end{itemize}
	
	\item \textbf{Typical evidence}
	\begin{itemize}
		\item Predicted post-intervention states after the recommended outcome has been applied.
		\item Expected verification or confirmation results.
		\item Anticipated side effects, tradeoffs, or follow-up conditions.
		\item Follow-up success conditions or milestones.
	\end{itemize}
	
	\item \textbf{Examples}
	\begin{itemize}
		\item Manufacturing: [Outcome] product flagged as defective; manual inspection recommended. [Future state] if the lens is cleaned, false rejects are expected to decrease and inspection confidence is expected to improve.

		\item Building technology: [Outcome] overheating fault reported; valve inspection recommended. [Future state] if the valve is repaired, the overheating alert is expected to clear and comfort complaints are expected to decrease.

		\item Cybersecurity: [Outcome] high-priority remediation recommendation to patch within 48 hours. [Future state] if the patch is applied and the service is restarted, the next validation scan is expected to report the vulnerability as resolved.
	\end{itemize}
\end{itemize}


\subsubsection*{PoRC~2.6 Causal mechanism}

\begin{itemize}
	\item \textbf{Definition} A causal mechanism is an explanation element that describes and supports a causal relationship within the causal chain. In particular, a causal mechanism can provide support for the causal relationship between an input and a causal factor, between two causal factors, between a causal factor and an outcome, and/or between an outcome and a future state.
	
	\item \textbf{Decision rule (PoRC~2.6 vs.\ PoRC~2.3):} If the statement explains how one element of the causal chain leads to another, code it as \emph{PoRC~2.6 Causal mechanism}. If the statement only names the suspected or inferred cause without describing the connecting process, pathway, or linkage, code it as \emph{PoRC~2.3 Causal factor}.
	
	\item \textbf{Decision rule (PoRC~2.6 vs.\ PoRC~3.2 and uncoded):} If the statement describes a process, pathway, or mechanism that connects two elements of the causal chain, such as how a factor produces an outcome or how an outcome leads to a future state, code it as \emph{PoRC~2.6 Causal mechanism}. If the statement instead describes one candidate element for the current case and its ranking properties, supporting evidence, or attributes, code it as \emph{PoRC~3.2 Individual element with ranking properties, evidence, and/or attributes (actual)}. If the statement provides background or procedural information without describing a causal pathway or a candidate element, leave it uncoded. The key test for \emph{PoRC~2.6} is whether the statement explains \emph{how} one causal chain element produces or leads to another; if it only names a cause or describes a candidate, it does not qualify.
	
	\item \textbf{Include}
	\begin{itemize}
		\item Statements that explain how one element of the causal chain leads to another.
		\item Descriptions of physical, technical, organizational, biological, or computational processes that connect elements of the causal chain.
		\item Intermediate causal explanations that link input to causal factor, one causal factor to another, causal factor to outcome, or outcome to future state.
	\end{itemize}
	
	\item \textbf{Exclude}
	\begin{itemize}
		\item Statements that only name a cause without describing a causal relationship (\emph{Causal factor}).
		\item Raw observations, signals, measurements, records, or user entries used by the system (\emph{Input}).
		\item Background circumstances that situate the case without explaining a causal relationship (\emph{Context}).
		\item Decisions, alerts, recommendations, or classifications produced by the system (\emph{Outcome}).
		\item Descriptions of one candidate element with ranking properties, evidence, or attributes (\emph{PoRC~3.2 Individual element with ranking properties, evidence, and/or attributes (actual)}).
	\end{itemize}
	
	\item \textbf{Typical evidence}
	\begin{itemize}
		\item Explanatory statements that describe how one element affects another.
		\item Descriptions of process dynamics, propagation effects, or system behaviors that connect adjacent elements in the causal chain.
		\item Statements that justify why a claimed causal factor is linked to an outcome or why an action is expected to lead to a future state.
	\end{itemize}
	
	\item \textbf{Examples}
	\begin{itemize}
		\item Manufacturing: [Causal mechanism] lens contamination reduces image contrast, which causes the edge detector to miss boundaries, supporting the causal relationship between [Causal factor] lens contamination and [Outcome] product flagged as defective.

		\item Building technology: [Causal mechanism] a stuck reheat valve continues to add heat despite the setpoint, which raises zone temperature, supporting the causal relationship between [Causal factor] stuck reheat valve and [Outcome] overheating fault reporte

		\item Cybersecurity: [Causal mechanism] a vulnerable logging library can be exploited through crafted input that triggers message lookups, enabling remote code execution, supporting the causal relationship between [Causal factor] presence of CVE-2021-44228 on an internet-facing asset and [Outcome] high-priority remediation recommendation.
	\end{itemize}
\end{itemize}


\subsubsection*{PoRC~2.7 What-if forward}

\begin{itemize}
	\item \textbf{Definition} A what-if forward explanation element describes a hypothetical change to an input and the expected effect of that change on the outcome for the current case. It emphasizes a directional relationship between input and outcome by starting from a changed input and projecting the resulting outcome.
	
	\item \textbf{Decision rule (PoRC~2.7 vs.\ PoRC~2.8):} If the explanation starts from a hypothetical change to one or more inputs and projects the resulting outcome for the current case, code it as \emph{PoRC~2.7 What-if forward}. If the explanation instead starts from a desired outcome and reasons backward to the input change required to achieve it, code it as \emph{PoRC~2.8 What-if backward}.
	
	\item \textbf{Include}
	\begin{itemize}
		\item Hypothetical changes to one or more input values used for the current case.
		\item Statements that project how the outcome would change if the input were changed.
		\item Forward-looking scenario variations that explicitly connect an altered input to an expected outcome.
	\end{itemize}
	
	\item \textbf{Exclude}
	\begin{itemize}
		\item Goal-oriented statements that start from a desired outcome and ask what input would need to change to achieve it (\emph{PoRC~2.8 What-if backward}).
		\item Statements that describe the current input-to-outcome relation without a hypothetical change.
		\item Statements that describe the expected state after implementing the recommended outcome without framing it as a hypothetical input variation (\emph{PoRC~2.5 Future state}).
	\end{itemize}
	
	\item \textbf{Typical evidence}
	\begin{itemize}
		\item Conditional statements of the form ``if this input changed, the outcome would change as follows.''
		\item Scenario exploration statements that vary one input and state the expected outcome.
		\item Sensitivity-style reasoning about how the outcome responds to an input change.
	\end{itemize}
	
	\item \textbf{Examples}
	\begin{itemize}
		\item Manufacturing: [What-if forward] if illumination is increased by 10\%, the predicted false reject rate decreases from 6\% to 2\%.

		\item Building technology: [What-if forward] if the supply-air temperature setpoint is lowered by 1$^\circ$C, predicted overheating risk decreases.

		\item Cybersecurity: [What-if forward] if internet exposure is removed, the risk score drops from urgent to high.
	\end{itemize}
\end{itemize}


\subsubsection*{PoRC~2.8 What-if backward}

\begin{itemize}
	\item \textbf{Definition} A what-if backward explanation element describes a desired outcome and the input change that would be required to achieve it for the current case. It emphasizes a directional relationship between input and outcome by starting from a target outcome and reasoning backward to the input that would need to change.
	
	\item \textbf{Decision rule (PoRC~2.8 vs.\ PoRC~2.7):} If the explanation starts from a desired outcome for the current case and identifies the input change needed to achieve it, code it as \emph{PoRC~2.8 What-if backward}. If the explanation instead starts from a changed input and projects the resulting outcome, code it as \emph{PoRC~2.7 What-if forward}.
	
	\item \textbf{Include}
	\begin{itemize}
		\item Goal-oriented hypothetical statements that begin with a desired outcome for the current case.
		\item Statements that identify what input would need to change to achieve the target outcome.
		\item Reverse reasoning that links a target outcome to a required input modification.
	\end{itemize}
	
	\item \textbf{Exclude}
	\begin{itemize}
		\item Forward scenario exploration that starts from a changed input and projects the resulting outcome (\emph{PoRC~2.7 What-if forward}).
		\item Statements that only describe the current outcome without reverse reasoning (\emph{PoRC~2.4 Outcome}).
		\item Statements that only describe the expected post-action state after the recommended outcome has been implemented (\emph{PoRC~2.5 Future state}).
		\item Ranked alternatives without an explicit target-outcome and required-input relation (\emph{PoRC~3.1 Set of ranked elements with ranking mechanism (actual)}).
	\end{itemize}
	
	\item \textbf{Typical evidence}
	\begin{itemize}
		\item Goal-oriented statements of the form ``to achieve this outcome, this input would need to change.''
		\item Reverse reasoning from a desired outcome to a required input modification.
		\item Recourse-style reasoning that identifies how the input must change to reach the target outcome.
	\end{itemize}
	
	\item \textbf{Examples}
	\begin{itemize}
		\item Manufacturing: [What-if backward] if the goal is to reduce the false reject rate to 2\%, illumination would need to be increased by about 10\%.

		\item Building technology: [What-if backward] if the goal is to clear the overheating alert, the supply-air temperature setpoint would need to be lowered by 1$^\circ$C.

		\item Cybersecurity: [What-if backward] if the goal is to reduce the risk score from urgent to high, internet exposure would need to be removed.
	\end{itemize}
\end{itemize}


\subsubsection*{PoRC~3.1 Set of ranked elements with ranking mechanism (actual)}

\begin{itemize}
	\item \textbf{Definition} A set of ranked elements with ranking mechanism (actual) is an explanation element that presents multiple candidate elements for the current case in ranked order.
	
	\item \textbf{Decision rule (PoRC~3.1 vs.\ PoRC~3.2):} If the explanation presents multiple candidate elements for the current case together with an ordering, selection logic, or set-level rule, code it as \emph{PoRC~3.1 Set of ranked elements with ranking mechanism (actual)}. If the explanation instead focuses on one candidate element and describes only that element's ranking properties, evidence, or attributes, code it as \emph{PoRC~3.2 Individual element with ranking properties, evidence, and/or attributes (actual)}.
	
	\item \textbf{Decision rule (PoRC~3.1 vs.\ PoRC~3.2, clarified):} The critical distinction is the number of candidate elements presented and whether a set-level ordering or selection mechanism is shown. If the content presents two or more candidate elements together with rank positions, comparative scores, or an explicit ordering or selection rule that applies across the set, code it as \emph{PoRC~3.1}. If the content presents only one candidate element, even if that element carries a rank position or confidence score, code it as \emph{PoRC~3.2}. A single element extracted or highlighted from a ranked list and described individually belongs to \emph{PoRC~3.2}, not \emph{PoRC~3.1}, regardless of whether the surrounding ranked list is also present.
	
	\item \textbf{Decision rule (PoRC~3.1 vs.\ PoRC~4.1):} If the set presents candidate elements for the current case, such as possible actions, decisions, recommendations, or outcomes to be selected or compared for this case, code it as \emph{PoRC~3.1}. If the set instead presents retrieved similar past cases used to support reasoning about the current case, code it as \emph{PoRC~4.1 Set of similar elements with similarity mechanism and set success-score information (similar)}.
	
	\item \textbf{Include}
	\begin{itemize}
		\item A set of candidate elements for the current case, such as alternative actions, recommendations, remediations, controls, or decisions.
		\item Rank positions and ranking properties used to order the candidate elements.
		\item A ranking mechanism or selection rule that explains how the ordering was produced.
		\item Set-level selection parameters, such as eligibility constraints, top-$k$ selection, thresholds, time windows, or robustness rules.
	\end{itemize}
	
	\item \textbf{Exclude}
	\begin{itemize}
		\item One single candidate element described individually (\emph{PoRC~3.2}).
		\item The final case result when no ranked candidate set is shown (\emph{PoRC~2.4 Outcome}).
		\item Candidate causes or vulnerabilities presented as causal claims rather than ranked recommendations (\emph{PoRC~2.3 Causal factor}).
	\end{itemize}
	
	\item \textbf{Typical evidence}
	\begin{itemize}
		\item A displayed ordered list of candidate elements for the current case.
		\item Rank positions, ranking properties, and a stated or inferable mechanism that explains the ordering.
	\end{itemize}
	
	\item \textbf{Examples}
	\begin{itemize}
		\item Manufacturing: [Set of ranked elements with ranking mechanism (actual)] (1) Clean lens, confidence = 0.82; (2) Recalibrate lighting, confidence = 0.74; (3) Replace camera, confidence = 0.41. Ordered by confidence; highest-confidence eligible action selected.

		\item Building technology: [Set of ranked elements with ranking mechanism (actual)] (1) Inspect valve actuator, confidence = 0.86; (2) Recalibrate temperature sensor, confidence = 0.52; (3) Adjust occupancy schedule, confidence = 0.21.

		\item Cybersecurity: [Set of ranked elements with ranking mechanism (actual)] (1) Patch immediately, confidence = 0.95; (2) Isolate service, confidence = 0.78; (3) Monitor only, confidence = 0.18.
	\end{itemize}
\end{itemize}


\subsubsection*{PoRC~3.2 Individual element with ranking properties, evidence, and/or attributes (actual)}

\begin{itemize}
	\item \textbf{Definition} An individual element with ranking properties, evidence, and/or attributes (actual) is an explanation element that describes one candidate element for the current case. The element can be selected or not selected. The description can include one or more ranking properties, supporting evidence, and descriptive attributes.
	
	\item \textbf{Decision rule (PoRC~3.2 vs.\ PoRC~3.1):} If the explanation focuses on one candidate element for the current case and describes that element's ranking properties, supporting evidence, and/or attributes, code it as \emph{PoRC~3.2}. If the explanation presents several candidate elements together and emphasizes their ordering, comparison, or set-level selection logic, code it as \emph{PoRC~3.1}.
	
	\item \textbf{Decision rule (PoRC~3.2 vs.\ PoRC~4.2):} If the element is a candidate element for the current case, code it as \emph{PoRC~3.2}. If the element instead describes one retrieved similar past case used to support reasoning about the current case, code it as \emph{PoRC~4.2 Individual, similar element with similarity property, success score, evidence, and/or attribute (similar)}.
	
	\item \textbf{Decision rule (PoRC~3.2 vs.\ PoRC~2.2):} If the content describes one candidate element for the current case, including its action label, rank position, confidence score, supporting evidence, or descriptive attributes, code it as \emph{PoRC~3.2}, even when some of its attributes overlap with input variables. If the content instead lists raw data, signals, measurements, records, or feature values that are fed into the system before the outcome is produced, code it as \emph{PoRC~2.2 Input}. The key distinction is directionality: input precedes and drives the outcome, whereas a candidate element is part of the output space that the system evaluates or selects from.
	
	\item \textbf{Include}
	\begin{itemize}
		\item One candidate element for the current case, regardless of whether it was selected or not selected.
		\item One or more ranking properties, evidence supporting the assessment of that element, and attributes that characterize the element.
	\end{itemize}
	
	\item \textbf{Exclude}
	\begin{itemize}
		\item A ranked set of multiple candidate elements (\emph{PoRC~3.1}).
		\item The overall case result when no specific candidate element is described individually (\emph{PoRC~2.4 Outcome}).
		\item Similar past cases rather than candidate elements for the current case (\emph{PoRC~4.2}).
		\item Raw input data, signals, or feature values that precede and drive the outcome (\emph{PoRC~2.2 Input}).
	\end{itemize}
	
	\item \textbf{Examples}
	\begin{itemize}
		\item Manufacturing: [Individual element (actual)] action = Clean lens; confidence = 0.82; evidence = rising blur score; attributes = Camera~C2, Station~3.

		\item Building technology: [Individual element (actual)] action = Inspect valve actuator; confidence = 0.86; evidence = command issued but valve position unchanged; attributes = AHU~4, Zone~3B.

		\item Cybersecurity: [Individual element (actual)] action = Patch within 48 hours; confidence = 0.95; evidence = vulnerable version detected on an internet-facing high-criticality asset; attributes = Server~A, production environment.
	\end{itemize}
\end{itemize}


\subsubsection*{PoRC~4.1 Set of similar elements with similarity mechanism and set success-score information (similar)}

\begin{itemize}
	\item \textbf{Definition} A set of similar elements with similarity mechanism and set success-score information (similar) is an explanation element that presents a set of similar cases for the current case, constructed through a similarity mechanism with a threshold or inclusion rule, together with set success-score information showing how many similar cases are successful, inconclusive, or not successful relative to the actual selected element.
	
	\item \textbf{Decision rule (PoRC~4.1 vs.\ PoRC~4.2):} If the explanation presents multiple similar cases together with a similarity mechanism and set-level success-score information, code it as \emph{PoRC~4.1}. If the explanation instead focuses on one specific similar case and describes that case's similarity property, success score, evidence, or attributes, code it as \emph{PoRC~4.2}.
	
	\item \textbf{Decision rule (PoRC~4.1 vs.\ PoRC~3.1):} If the set presents retrieved similar past cases used to support reasoning about the current case, code it as \emph{PoRC~4.1}. If the set instead presents candidate elements for the current case that are being ranked, selected, or compared, code it as \emph{PoRC~3.1}.
	
	\item \textbf{Include}
	\begin{itemize}
		\item A collection of similar cases linked to the current case through similarity, with a similarity mechanism and threshold.
		\item Set success-score information showing distribution across successful, inconclusive, and not successful relative to the actual selected element.
		\item Set-level summary values such as number of similar cases, similarity range, or aggregate confidence derived from the success-score distribution.
	\end{itemize}
	
	\item \textbf{Exclude}
	\begin{itemize}
		\item One individual similar case described separately (\emph{PoRC~4.2}).
		\item A ranked set of candidate elements for the current case (\emph{PoRC~3.1}).
		\item General confidence statements not derived from an explicit set of similar cases.
	\end{itemize}
	
	\item \textbf{Examples}
	\begin{itemize}
		\item Manufacturing: [Set of similar elements] 10 similar cases, similarity threshold 0.72; relative to action = Clean lens: 8 successful, 1 inconclusive, 1 not successful.

		\item Building technology: [Set of similar elements] 7 similar episodes, similarity threshold defined; relative to action = Inspect valve actuator: 5 successful, 1 inconclusive, 1 not successful.

		\item Cybersecurity: [Set of similar elements] 12 similar hosts, same CVE and exposure profile; relative to action = Patch within 48 hours: 9 successful, 2 inconclusive, 1 not successful.
	\end{itemize}
\end{itemize}


\subsubsection*{PoRC~4.2 Individual, similar element with similarity property, success score, evidence, and/or attribute (similar)}

\begin{itemize}
	\item \textbf{Definition} An individual, similar element with similarity property, success score, evidence, and/or attribute (similar) is an explanation element that describes one specific similar case relative to the current case, including an element-specific similarity property, a success score indicating whether it was successful, inconclusive, or not successful relative to the actual selected element, and optionally evidence and attributes.
	
	\item \textbf{Decision rule (PoRC~4.2 vs.\ PoRC~4.1):} If the explanation focuses on one specific similar case, code it as \emph{PoRC~4.2}. If the explanation presents several similar cases together with set-level information, code it as \emph{PoRC~4.1}.
	
	\item \textbf{Decision rule (PoRC~4.2 vs.\ PoRC~3.2):} If the element describes one retrieved similar past case, code it as \emph{PoRC~4.2}. If the element is a candidate element for the current case, code it as \emph{PoRC~3.2}.
	
	\item \textbf{Include}
	\begin{itemize}
		\item One referenced similar case with an element-specific similarity property such as similarity score or distance value.
		\item A success score indicating whether the similar case was successful, inconclusive, or not successful relative to the actual selected element.
		\item Evidence supporting the similarity relation and attributes characterizing the similar case.
	\end{itemize}
	
	\item \textbf{Exclude}
	\begin{itemize}
		\item A group of multiple similar cases summarized at set level (\emph{PoRC~4.1}).
		\item A candidate element for the current case rather than a similar past case (\emph{PoRC~3.2}).
		\item A similar case shown without an element-specific similarity property.
	\end{itemize}
	
	\item \textbf{Examples}
	\begin{itemize}
		\item Manufacturing: [Individual similar element] Station~3, Camera~C2; similarity score = 0.88; success score = successful relative to action = Clean lens; evidence = same blur spike and lighting conditions; attribute = same station after tooling change.

		\item Building technology: [Individual similar element] AHU~4 fault last month; similarity score = 0.81; success score = successful relative to action = Inspect valve actuator; evidence = same command-versus-position mismatch; attribute = same zone and outside-air conditions.

		\item Cybersecurity: [Individual similar element] SRV-12; similarity score = 0.90; success score = successful relative to action = Patch within 48 hours; evidence = same package version and exposure profile; attribute = same server role and criticality.
	\end{itemize}
\end{itemize}

\clearpage


\section{Annotated Explanation Content Model (Cheat Sheet)}
\label{app:AnnotatedExplanationContentModel}

\begin{figure}[h!]
	\centering
	\includegraphics[width=1.0\linewidth]{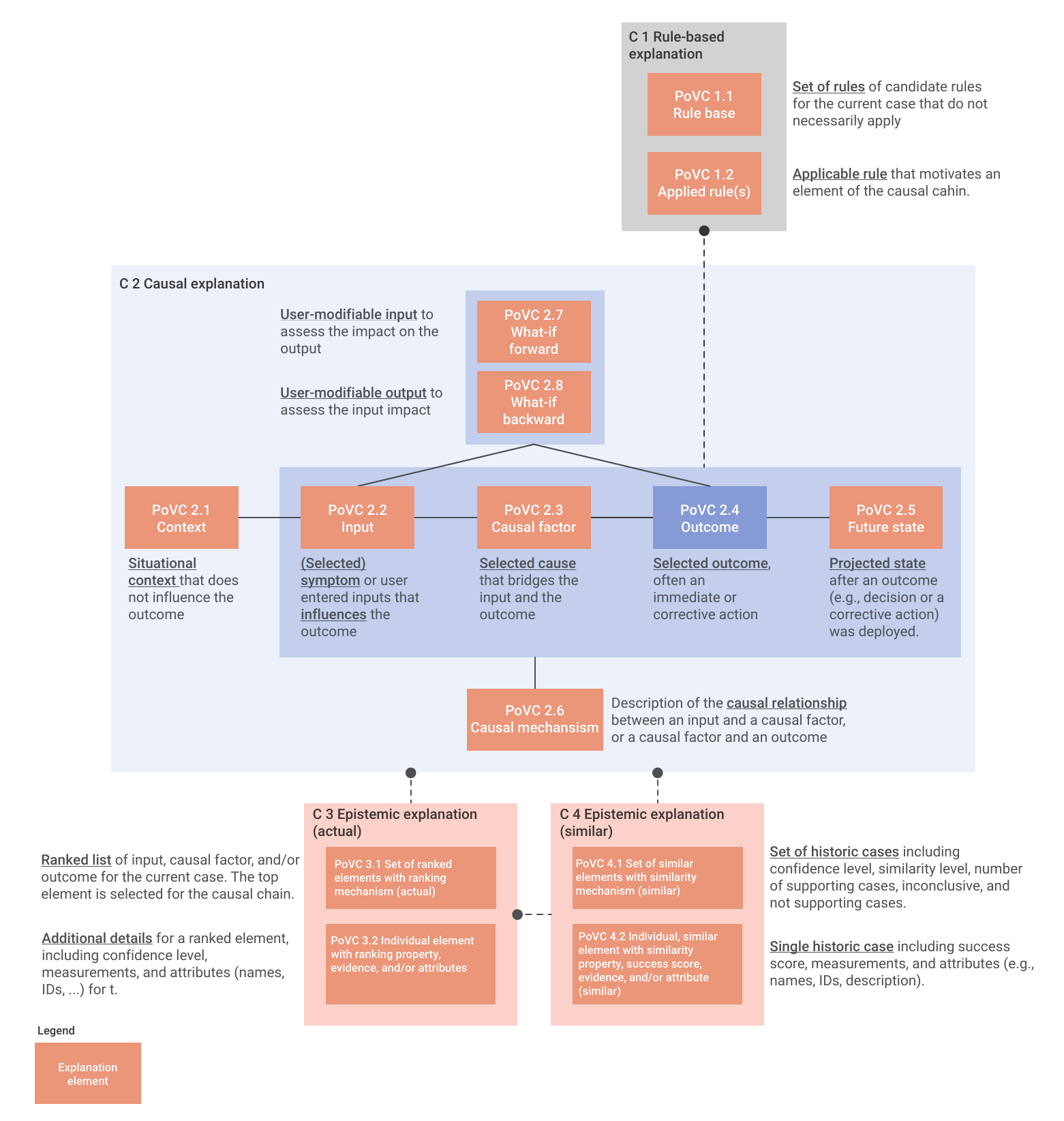}
	\caption{Explanation content model for local, post-hoc explanations with coding hints}
	\label{fig:XAIModel-Framework-PostValidation_Annotated}
\end{figure}

Figure~\ref{fig:XAIModel-Framework-PostValidation_Annotated} depicts the explanation content model with coding hints.

\section{Extended proposed use modes and illustrative application logic}
\label{app:UseModesExtended}

The post-reliability explanation-content blueprint supports three proposed use modes. These use modes follow directly from the code structure and reliability evidence, but none has been empirically validated with practitioners in the current study: the reliability evaluation demonstrates that independent researchers can apply the codebook consistently, not that domain practitioners can use it effectively as a requirements, audit, or comparative tool. The use modes described below should therefore be understood as design intentions and theoretical propositions that require empirical validation in future work. The worked examples are author-constructed illustrations, not practitioner-validated cases.

\textit{Use mode 1: Explanation requirements specification.} During AI system design, a developer or UX designer walks through the post-reliability codes systematically and determines applicability based on system architecture and deployment context. If inputs are machine-generated, PoRC~2.7 What-if forward and PoRC~2.8 What-if backward are often not applicable in practice; if inputs are user-modifiable and the hypothetical space can be bounded meaningfully, both What-if codes may apply. If the system produces ranked recommendations, the epistemic-group codes for actual outcomes apply. If the system encodes explicit decision logic, the rule-based group codes apply.

\textit{Use mode 2: Explanation content audit.} For an existing AI system with an implemented explanation interface, an evaluator applies the post-reliability codebook to the interface's explanation elements and records which codes are present, absent, or only partially implemented. The reliability evidence reported in the main paper demonstrates that independent evaluators can reach consistent coding decisions, suggesting that the blueprint could function as an audit instrument across teams and organizations, pending empirical validation with practitioners.

\textit{Use mode 3: Cross-system gap analysis.} The coded corpus from Studies~1 to~6 constitutes a comparative reference dataset. A practitioner designing a system in a new domain can examine the distribution for a comparable prior study, identify which explanation-content types were present and absent, and assess whether absences reflect structural constraints or potential design gaps.

A compact illustration of Use Mode~1 is helpful because it shows how the blueprint functions as a completeness checklist rather than as a taxonomy alone. Consider a building-management AI system that flags an anomalous energy-consumption pattern and displays an explanation to a facility manager. The explanation interface shows the current consumption value and sensor readings (PoRC~2.2 Input), the historical baseline (PoRC~2.1 Context), the top contributing factors ranked by influence (PoRC~3.1 Set of ranked elements with ranking mechanism, actual), each factor with its current value and normal range (PoRC~3.2 Individual element with ranking properties, evidence, and/or attributes, actual), and the predicted trajectory if no action is taken (PoRC~2.5 Future state). The What-if codes (PoRC~2.7 and PoRC~2.8) are not applicable because the inputs are machine-generated and cannot readily be varied by the user. The rule-based group codes (PoRC~1.1 and PoRC~1.2) are not applicable because the system does not expose explicit decision logic. The similar-outcome epistemic codes (PoRC~4.1 and PoRC~4.2) are not applicable because the system does not maintain a historical case base for analogical comparison. In this illustration, five of fourteen post-reliability codes are instantiated, and the inapplicability of the remaining nine codes is determined by documented system properties rather than arbitrary omission.

A separate line of work is still needed to validate all three use modes empirically, including controlled studies that measure completeness, decision accuracy, and time compared to baseline methods, as well as studies that test whether practitioners can use the blueprint without researcher mediation.

\clearpage


\section{Illustrative Example: Applying the Explanation Content Model to the SHIELD Cybersecurity Project}
\label{sec:Example}

\begin{figure}[ht!]
	\centering
	\includegraphics[width=1.0\linewidth]{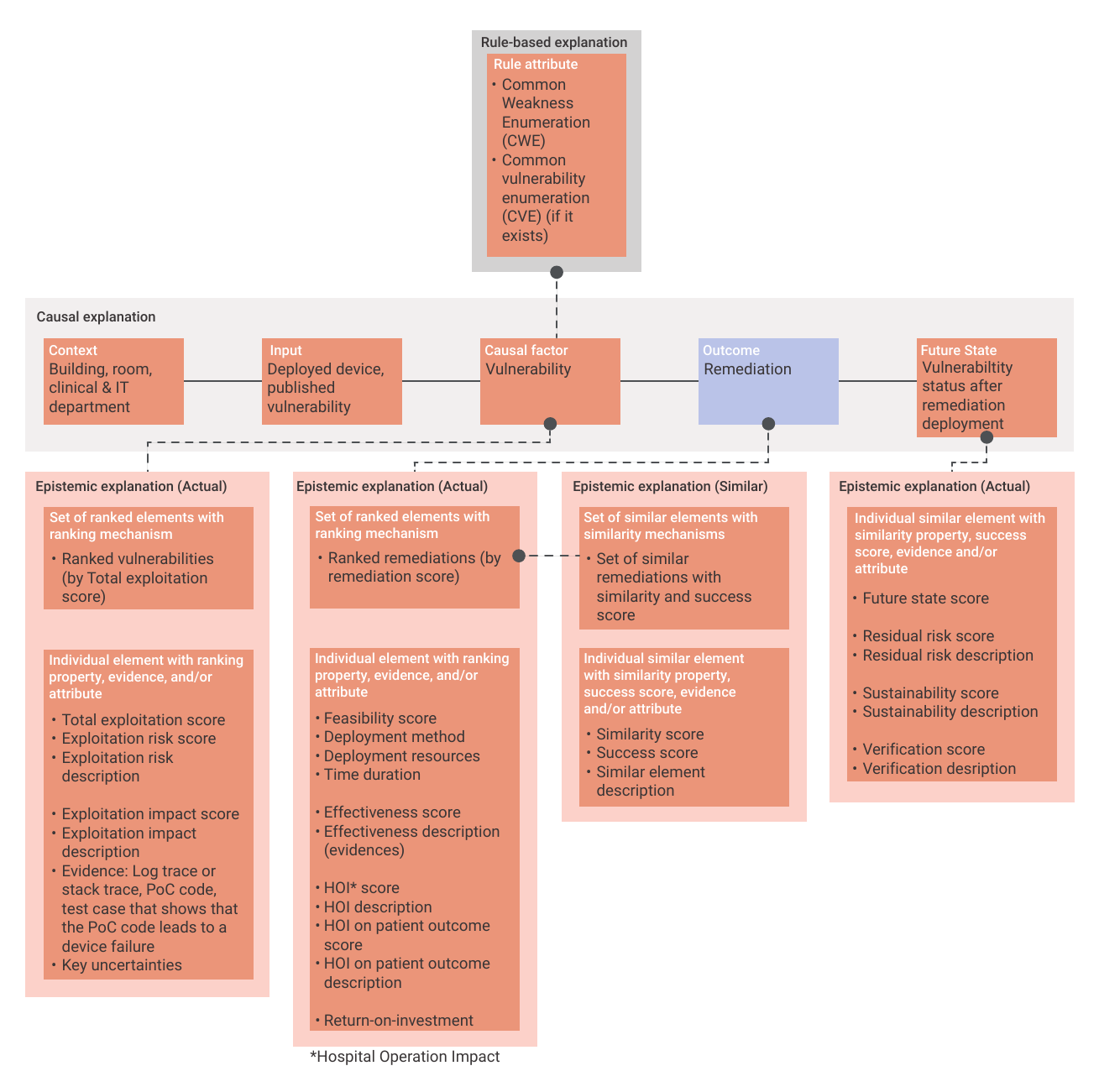}
	\caption{Explanation content for SHIELD organized by the four explanation-content groups.}
	\label{fig:XAIModel6-SHIELD_Revised}
\end{figure}

The following example illustrates Use Mode 1 (explanation requirements specification) and Use Mode 3 (cross-system gap analysis) as described in Appendix~\ref{app:UseModesExtended}, demonstrating \emph{how} the model \emph{could} structure elicited explanation content and surface missing content. This example is drawn from Study~6. The explanation content shown in Figure~\ref{fig:XAIModel6-SHIELD_Revised} was elicited from hospital cybersecurity practitioners (Head of Cybersecurity and IT Specialty Lead) and was subsequently validated and revised based on their feedback. However, the mapping of this content to the model's categories was performed by the researcher for illustrative purposes; practitioners did not use the model as a requirements specification or gap analysis tool. Empirical validation of the three use modes with practitioners is therefore not provided here; see Appendix~\ref{app:UseModesExtended} for a discussion of the validation required as future work.

The project name is SHIELD (Secure Healthcare Infrastructure Enhancement and Defense). SHIELD is an ARPA-H funded project under the UPGRADE program \cite{ARPA-H_UPGRADE_2024}, with the aim of enabling hospitals to recognize and address cybersecurity vulnerabilities within days rather than months or even years, while minimizing the impact on hospital operations.

As part of a human-centered design approach, explanation content was elicited for two identified user roles: "Head of Cybersecurity" and "IT Specialty Lead." The explanation content is intended to assist these hospital stakeholders in prioritizing discovered vulnerabilities and selecting a remediation from a set of SHIELD-identified remediations per vulnerability, with confidence, accuracy, and efficiency. As shown in Figure~\ref{fig:XAIModel6-SHIELD_Revised}, the explanation content model provides a framework for organizing this elicited explanation content.

The four explanation groups align with distinct forms of reasoning: rule-based with deductive reasoning, causal with causal reasoning, epistemic (actual) with inductive and contrastive reasoning, and epistemic (similar) with analogical reasoning \cite{Russell2020, Pearl2018, Mill1843, Lipton1990, Gentner1983, Aamodt1994}.

Beyond structuring existing explanation content, the model can also surface potentially missing explanation content by making explanation content gaps visible. Importantly, because the model is domain-agnostic in its structure, the same four explanation groups can organize elicited explanation content regardless of the underlying AI system, the application domain, or the user role. The SHIELD example thus serves as one instance of this generalizable pattern. However, this generalizability claim is theoretical and based on the domain-agnostic nature of the category definitions (e.g., causal factors, outcomes, epistemic uncertainty exist across domains). Empirical validation of cross-domain transferability requires separate studies and is not provided here (see Limitation 7).

It is noteworthy that two elements of the explanation content model are absent from the SHIELD example: what-if explanations (both forward and backward) and the causal mechanism explanation content type. The explanation content model makes such absences visible and interpretable, but does not by itself resolve whether they reflect genuine irrelevance or elicitation gaps. In this illustration, the identification of these absences was performed by the researcher applying the model to the validated explanation content. Whether practitioners themselves would identify these as gaps or as intentional omissions was not tested. To distinguish between these two interpretations in practice, a practitioner applying the model in Use Mode 1 can consider the practical tractability of each content type in the deployment context. For what-if explanations, the absence is likely explained by the complexity of the SHIELD domain: cybersecurity vulnerability management involves a large number of interdependent variables, asset configurations, and remediation constraints, making hypothetical forward or backward reasoning over inputs difficult to bound in a way that is actionable and interpretable for hospital stakeholders. In industrial domains of this kind, what-if content is often not practical even when it is technically feasible, because the space of hypothetical inputs is too large and too interdependent for users to navigate meaningfully within the time and cognitive constraints of an operational decision. The absence of what-if content in the SHIELD example therefore most plausibly reflects this practical constraint rather than an elicitation gap. The absence of causal mechanism content, by contrast, is not readily explained by domain complexity: information about the underlying mechanism by which a vulnerability arises is in principle bounded and expressible, and this content type is applicable given the system architecture. Its absence therefore warrants revisiting during elicitation, either to confirm that hospital stakeholders do not require this content for confident, accurate, and efficient remediation decisions, or to identify whether it was inadvertently omitted. This distinction between practically constrained absence and potentially elicitation-driven absence illustrates how the model functions as a reflective instrument: it not only structures what is present but provides principled grounds for interpreting what is missing.


\end{document}